\documentclass[12pt,centertags,dvips,twoside]{report}
\usepackage{a4}
\usepackage{fancyheadings}
\newcommand{\clearemptydoublepage}{%
    \newpage{\pagestyle{empty}\cleardoublepage}}
\usepackage{amssymb}
\usepackage{amsfonts}
\usepackage{varioref}
\usepackage{citesort}
\usepackage{mcite}
\usepackage{amsmath}
\newlength{\diracchlen}

\newlength{\vecchlen}
\newlength{\vecchhgt}

\usepackage{subfigure}
\usepackage{array}
\usepackage{epic}
\usepackage{eepic}
\usepackage{epsfig}
\usepackage{afterpage}
\usepackage{axodraw}
\makeatletter
\renewcommand{\@makecaption}[2]{%
  \vskip\abovecaptionskip
  \sbox\@tempboxa{#1: \emph{#2}}%
  \ifdim \wd\@tempboxa >\hsize
    #1: \emph{#2}\par
  \else
    \global \@minipagefalse
    \hb@xt@\hsize{\hfil\box\@tempboxa\hfil}%
  \fi
  \vskip\belowcaptionskip}
\makeatother
\allowdisplaybreaks

\newcommand{\be}{\begin{eqnarray}}   
\newcommand{\ee}{\end{eqnarray}}

\newcommand{\si}{\sigma}

\newcommand{\gev    }{\ensuremath{\mathrm{GeV}}}
\newcommand{\gevsq  }{\ensuremath{\mathrm{GeV^2}}}
\newcommand{\Lumi}{\ensuremath{{\cal L}}}

\newcommand{\gaminelp}{\ensuremath{\gamma^p_{\rm inel}}}
\newcommand{\der}{\ensuremath{{\operatorname{d}}}}
\newcommand{\av}[1]{\ensuremath{\langle{#1}\rangle}}

\newcommand{\shat}{\ensuremath{\hat{s}}}
\newcommand{\that}{\ensuremath{\hat{t}}}
\newcommand{\uhat}{\ensuremath{\hat{u}}}
\newcommand{\vmin}[1]{\ensuremath{#1_{\rm min}}}
\newcommand{\vmax}[1]{\ensuremath{#1_{\rm max}}}

\begin{document}

\setlength{\baselineskip}{0.65cm}
\setlength{\parskip}{1ex}
\renewcommand{\arraystretch}{1.3}  

\begin{titlepage}
\begin{flushright}
\vspace*{-2.cm}
DO-TH 05/12  \\
hep-ph/0512306\\
\end{flushright}
\begin{center}
\vspace*{3cm}
{\LARGE 
{\bf The Polarized and Unpolarized  \\
Photon Content of the\\ 
\vspace*{0.2 cm}
Nucleon}}

\vspace*{3.0cm} 
{\Large 
{\bf Dissertation} \\
zur Erlangung des Grades eines\\
Doktors der Naturwissenschaften\\
der Abteilung Physik\\
der Universit\"{a}t Dortmund\\
}

\vspace*{3.cm}
vorgelegt von
\\
\vspace*{0.2cm}
{\Large{\bf Cristian Pisano}}

\vspace{1.cm}
{\large Mai 2005}
\end{center}
\end{titlepage}

\clearemptydoublepage
\clearemptydoublepage
\pagenumbering{roman} 
\tableofcontents
\clearemptydoublepage



\pagenumbering{arabic} 

\chapter{{\bf Introduction}}

In this thesis the  polarized  and unpolarized photon distributions  of 
the nucleon (proton, neutron),  evaluated in  the equivalent photon 
approximation, are computed theoretically and the possibility of their
experimental determination is demonstrated. 
The thesis is based on the following publications 
\cite{gpr1,gpr2,pap1,pp2,pol,cris}:

\begin{itemize}

\item   M. Gl\"uck, C. Pisano, E. Reya, 
        {\em The polarized and unpolarized photon content of the nucleon}, 
        Phys. Lett. {\bf B 540}, 75 (2002) [Chapter 3].
  
\item   M. Gl\"uck, C. Pisano, E. Reya, I. Schienbein,
        {\em Delineating the polarized and unpola\-rized photon distributions
        of the nucleon in $e N$ collisions},
        Eur. Phys. J. {\bf C 27}, 427 (2003) [Chapter 4].

\item   A. Mukherjee, C. Pisano, 
        {\em Manifestly covariant analysis of the QED Compton process in 
        $e p \rightarrow e\gamma p$ and $ep\rightarrow e\gamma X$},
        Eur. Phys. J. {\bf C 30}, 477 (2003) [Chapter 5].  

\item   A. Mukherjee, C. Pisano, 
        {\em Suppressing the background process to QED Compton scattering
        for delineating the photon content of the proton},
        Eur. Phys. J. {\bf C 35}, 509 (2004) [Chapter 6].

\item   A. Mukherjee, C. Pisano, 
        {\em Accessing the longitudinally polarized photon content of the
        proton}, 
        Phys. Rev. {\bf D 70 }, 034029  (2004) [Chapter 7].

\item   C. Pisano, 
        {\em Testing the equivalent photon approximation of the proton
        in the process $ep\rightarrow \nu W X$},        
        Eur. Phys. J. {\bf C 38}, 79 (2004) [Chapter 8]. 
\end{itemize}
The results of \cite{pol}  have  been summarized in \cite{pol2} as well.
Furthermore in Section \ref{sec:qcd} we 
shortly recall the main findings of the 
recent paper \cite{curv}, also concerning the structure of the nucleon: 
\begin{itemize}
\item  M. Gl\"uck, C. Pisano, E. Reya, {\em Probing the perturbative NLO
       parton evolution in the small-$x$ region},
       Eur. Phys. J. {\bf C 40}, 515 (2005).
\end{itemize}


The equivalent photon approximation (EPA) of a charged fermion is a 
technical device which allows for a rather simple and 
efficient calculation of any photon-induced subprocess.
The first explicit formulation and quantitative application 
of the EPA were given in 1924 by
Fermi \cite{fermi}, who utilized it  to
estimate the electro-excitation and electro-ionization of atoms, and also
the energy loss, due to ionization, of $\alpha$-particles travelling through
matter. In several cases, he obtained a satisfactory numerical agreement with
experimental data.  Ten years later,  in order to simplify calculations of
processes involving relativistic collisions of charged particles, 
Williams \cite{williams} and  Weizs\"acker \cite{weiz} further
developed  Fermi's semi-classical treatment and extended it to high-energy 
electrodynamics. They observed that the electromagnetic field generated 
at a given point by a fast charged particle passing close to it contains 
predominantly transverse components. By making a Fourier analysis of 
the field, they concluded that the incident particle would produce the same 
effects as a beam of photons and computed their distribution in energy.  
This model assumes that the particle motion is not appreciably affected 
during the interaction, in particular that its scattering angle is small. 

The first field-theoretical derivations were given in the 
fifties by Dalitz and Yennie \cite{dalitz}, Curtis \cite{curtis}, 
Kessler and Kessler \cite{dkessler}, and later by Chen and Zerwas
\cite{chen}. The equi\-valent photon method for pointlike fermions, like
the electron, has been investigated and utilized widely, for example it has
been applied to  pion production in 
electro-nucleon collisions \cite{dalitz,curtis}
and to two-photon processes for particle production at high energies 
\cite{terazawa,brotera,budnev}. A detailed history  of the method, 
its various formulations and its first applications are contained in Kessler's
review article \cite{pkessler}. In more recent times, the validity of this 
ap\-proximation has been examined by Bawa and Stirling \cite{bawa} in the 
context of the production of large transverse momentum photons
at the HERA collider, by comparing the exact and approximate cross sections. 
Frixione, Mangano, Nason and Ridolfi in a well-known paper \cite{ref8} further 
modified the method in order to improve its accuracy 
and tested it in the case of heavy-quark electroproduction at HERA. 
Their results were extended by De Florian and Frixione \cite{flo}  to
the case of a longitudinally polarized electron.     

\begin{figure}[t] 
\begin{center} 
\hspace*{0.6cm}
\epsfig{figure=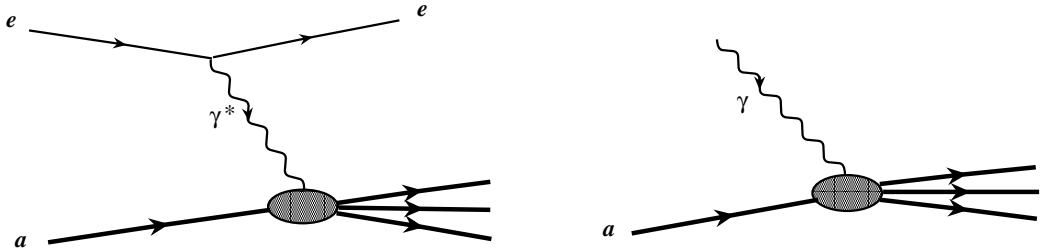, width= 15cm} 
\vspace{-2.5cm}
\caption{Connection between electroproduction and photoproduction
on a target $a$.}
\label{intro_epa}
\end{center}
\end{figure} 

As an example to illustrate the EPA,  we shall consider the process in which 
an electron $e$ of very high energy scatters from a target hadron
 $a$, e.g. a proton. 
At leading order  in $\alpha$, the fine structure constant of 
quantum electrodynamics (QED), the electron 
is connected to the target by one photon propagator, as depicted in 
the first diagram in Figure \ref{intro_epa}.
If we denote with $E$ and $E'$ respectively the initial and 
final energies of the electron, the photon will carry a 
momentum $q$ such that 
\be
q^2 \simeq - 2 E E'(1-\cos\theta ),
\ee  
where $\theta$ is the electron scattering angle.
In the limit of forward scattering, whatever the energy loss, the photon
momentum approaches $q^2 = 0$; therefore the reaction is highly peaked in 
the forward direction and the underlying dynamics is that
of a photoproduction process. It can be shown that the    
cross section $\sigma_{e a}$, integrated over $q^2$ with integration bounds
\be
q^2_{\rm{max}} = -\frac{m_e^2x^2}{1-x}~,~~~~~~~~~~~~~q^2_{\rm {min}}
\equiv -\mu^2,
\label{eq:bounds} 
\ee
is given by a convolution of the 
probability that the electron radiates  off a photon, the equivalent 
photon distribution $\gamma^e(x, \mu^2)$, with the 
corresponding real photoproduction cross section $\sigma_{\gamma a}$, 
which is in general
easier to calculate than the original one:
\begin{equation}
\der\sigma_{ea} = \gamma^e(x, \mu^2) \,\sigma_{\gamma a}\der x.
\label{eq:elsigma}
\end{equation}
The connection between  electroproduction and photoproduction on
a target hadron $a$ is shown, in terms of Feynman diagrams, in Figure 
\ref{intro_epa}.
An explicit expression of $\gamma^e(x, \mu^2)$ is given in \cite{ref8},
\be
\gamma^e(x, \mu^2) = \frac{\alpha}{2\pi} \bigg [\frac{1 + (1-x)^2}{x}\, 
\ln\frac{\mu^2(1-x)}{m_e^2x^2} + 2m_e^2x \bigg ( \frac{1}{\mu^2} - 
\frac{1-x}{m_e^2 x^2} \bigg ) \bigg ],
\label{eq:gamma_el}
\ee
where $x$ is the fraction of the electron energy carried
by the photon and $\mu$ has to be identified with a   momentum scale 
of the photon-induced subprocess. While $q^2_{\rm max}$ in
\eqref{eq:bounds} is the largest value of $q^2$ kinematically 
allowed, $q^2_{\rm min}$ is not a well defined quantity, since 
 when the  momentum transfer squared becomes too large (in absolute value), 
the EPA  breaks down and use of \eqref{eq:elsigma} would lead to huge 
errors. Critical examination of the EPA in electron-hadron collisions,
in connection with different choices for the scale $\mu$, can be found in
\cite{bawa,ref8}.

The longitudinally polarized electron-target $a$ cross 
section $\Delta\sigma_{e a}$ can be obtained from
\eqref{eq:elsigma} with the formal substitutions
\be
\sigma \rightarrow \Delta\sigma, ~~~\gamma^e \rightarrow \Delta\gamma^e .
\label{eq:polsub}
\ee 
The quantities $\Delta \sigma$ are defined in terms of cross sections 
$\sigma_{\lambda_1  \lambda_2}$ for incoming particles of definite helicities,
\begin{eqnarray}
\der\Delta\sigma & = &\frac{1}{4} \,\,(\der\sigma_{+ +} + \der\sigma_{- -} 
- \der\sigma_{+ -}-\der\sigma_{- +}) \nonumber \\
& = & \frac{1}{2}\,(\der\sigma_{+ +} - \der\sigma_{+ -}),
\label{eq:helicity1} 
\end{eqnarray}
where the second equality follows from parity invariance of the 
electromagnetic interaction. The corresponding unpolarized cross section 
is obtained by taking the sum instead, namely
\be
\sigma =\frac{1}{2}\,(\der\sigma_{+ +} + \der\sigma_{+ -}).
\label{eq:helicity2} 
\ee
The polarized equivalent photon distribution $\Delta\gamma^e$ are 
defined in terms of densities for photons of definite helicity in
electrons of definite helicity  
\be
\Delta\gamma^e = \gamma^{e +}_{+} -\gamma^{e +}_{-} = 
\gamma^{e -}_{-} -\gamma^{e -}_{+}, 
\ee
while the sum  will give the unpolarized photon distribution,
\be
\gamma^e = \gamma^{e +}_{+} +\gamma^{e +}_{-} = 
\gamma^{e -}_{-} +\gamma^{e -}_{+},
\label{eq:eunp} 
\ee
where the superscripts refer to the parent electron and the subscripts to
the photon. One can show that \cite{flo}
\be
\Delta\gamma^e(x, \mu^2) = \frac{\alpha}{2\pi} \bigg [\frac{1 - (1-x)^2}{x}\, 
 \ln\frac{\mu^2(1-x)}{m_e^2x^2} + 2m_e^2x^2 \bigg ( \frac{1}{\mu^2} - 
\frac{1-x}{m_e^2 x^2} \bigg ) \bigg ].
\label{eq:dgamma_el}
\ee

In the treatment of protons and neutrons, \eqref{eq:elsigma} and 
\eqref{eq:polsub}-\eqref{eq:eunp} still hold, with the replacement 
$e\rightarrow p$ or $n$, but a special situation arises
in the calculation of their photon distributions, due to the fact that
they are not pointlike particles.
Here it is necessary to distinguish between elastic and inelastic scattering. 
In the former case
the nucleon does not break up but is temporaly in an excited state, which
can be described in terms of certain form factors. 
In the latter case, appealing
to the parton model, only quarks, antiquarks and gluons
are usually considered as the (essentially free) constituents of the 
initial nucleon, which ceases to exist and its constituents finally hadronize.
The photons radiated off the quarks and antiquarks may be characterized by
\eqref{eq:gamma_el} and \eqref{eq:dgamma_el}, and due to their logarithmic
enhancement factor they are expected to become increasingly relevant  
at very high energies, when the momentum scale $\mu$ is also large.
At this level the ``inelastic''  photons can be included among the parton 
distributions of the nucleon.

Therefore  the total (polarized) unpolarized photon distribution 
of a  nucleon $N$ will be given by  
\begin{equation} 
(\Delta)\gamma(x,\mu^2) =(\Delta) \gamma_{\mathrm{el}}(x) + 
   (\Delta)\gamma_{\mathrm{inel}}(x,\mu^2), 
\end{equation} 
where the  elastic component $(\Delta)\gamma_{\rm el}$ is  due to 
$N\to \gamma N$ and the inelastic component
$(\Delta)\gamma_{\mathrm{inel}}$   is due to $N\to\gamma X$ with  
$X\neq N$. As pointed out by 
Drees and Zeppenfeld \cite{ref9} and by Kniehl \cite{kniehl}, 
$\gamma^p_{\rm el}$ {\em cannot} be 
obtained from \eqref{eq:gamma_el} by just replacing the electron mass 
with the proton mass: this would strongly overestimate it. One has to take
into account  the effects of the form factors, which determine the 
scale independence of $(\Delta)\gamma_{\rm el}$. 
The derivations of $\gamma^p_{\mathrm {el}}$ and $\gamma^p_{\mathrm{inel}}$,
performed in \cite{kniehl} and by Gl\"uck, Stratmann,
Vogelsang in \cite{gsv}, 
will be generalized to the neutron and to the polarized sector in Chapter 3.

The photon content of the nucleon  
$\gamma(x,\mu^2)$ can be utilized, instead of the
more common form factors and parton distributions, to calculate   
photon-induced subprocesses in elastic and 
deep inelastic $ep$ reactions, leading to  great simplifications.
For example, as shown by Bl\"umlein \cite{blu} and by
De R\'ujula and Vogelsang \cite{ruju},
 the analysis of the deep inelastic QED Compton 
scattering process $ep\to e\gamma X$ reduces to the calculation 
of the $2\to 2$ subprocess $e\gamma^p\to e\gamma$ instead of having 
to calculate the full $2\to 3$ subprocess $eq\to e\gamma q$.
Analogously one  can consider just the simple $2\to 2$ 
subprocess \mbox{$\gamma^e\gamma^p\to \mu^+\mu^-$}  (instead
of $\gamma^e q\rightarrow \mu^+\mu^-q$) 
for the analysis of  deep inelastic  $ep\to \mu^+\mu^- X$, 
or $e\gamma^p\to\nu W$ (instead of $e q\rightarrow \nu W q$) for associated 
$\nu W$ production in $ep\to\nu WX$.

Similarly,    $\gamma\gamma$ fusion processes like  
$\gamma^p\gamma^p\to\ell^+\ell^-,\,  
c\bar{c},\, H^+H^-,\, \tilde{\ell}^+\tilde{\ell}^-$  for (heavy) lepton ($\ell$), heavy quark  
($c$), charged Higgs ($H^{\pm}$) and slepton 
($\tilde{\ell}$) production can be easily analyzed in purely hadronic 
$pp$ reactions, providing also an interesting possibility 
of producing charged particles which do not have  
strong interactions.  Carlson and Lassila \cite{ref12},
Drees, Godbole, Nowakowski, Rindani \cite{drees},
and Ohnemus, Walsh, 
Zerwas \cite{ohn} initiated these studies;
in particular, in \cite{drees} it is shown that the cross section for the 
pair production
of heavy charged scalars or fermions via $\gamma\gamma$ fusion amounts 
to a few percent of the corresponding Drell-Yan 
$q\bar q$ annihilation cross sections, at energies reached at the 
CERN Large Hadron Collider (LHC). However, the disadvantage of the low
production rates is compensated by the simple and clean experimental situation 
encountered when the  photons are emitted from protons which 
do not break up (purely elastic processes) \cite{ohn}. 

It still remains
to extend the above-mentioned analysis to $pd$ and $dd$ reactions.
Moreover, analogous remarks hold for the longitudinally  
polarized $\vec{e}\vec{N}$ and $\vec{p}\vec{p},\, 
\vec{p}\vec{d}$, $\vec{d}\vec{d}$ reactions 
where the polarized photon content of the nucleon 
$\Delta\gamma(x,\mu^2)$ enters.  
   
In this thesis we shall concentrate on lepton-nucleon scattering, with 
special attention to the aforementioned QED Compton process, which is 
one of the most  important reactions for directly measuring 
the photon content of the nucleon and testing the 
reliability of the EPA. Furthermore, we shall discuss how such
measurements can provide  
additional, independent informations concerning the usual polarized and 
unpolarized structure functions, along the lines of 
\cite{blu,thesis,f2h1,lend}.

 In detail, the outline of the thesis will be 
as follows:
\begin{itemize}

\item Chapter 2 serves as an introduction into the concepts of electromagnetic
      form factors, structure functions and parton distributions of 
      the nucleon, 
      in terms of which elastic and inelastic $eN$ cross sections
      are commonly expressed. They are presented here because they enter in 
      the definition of the  photon content of the nucleon.  
      Notations and conventions used in this chapter as well as in the rest
      of the thesis are summarized in Appendix \ref{app:notations}.

\item In Chapter 3 a  new expression for the polarized equivalent photon
      distribution is explicitly derived. The calculation of the 
      unpolarized one, already performed in \cite{kniehl,gsv,ruju}, is also 
      shown in  detail for completeness and the resulting photon asymmetries 
      are presented for some typical relevant momentum scales.

\item The production rates of lepton-photon  
      and dimuon pairs  at the HERA collider and HERMES experiment
      are evaluated  in Chapter 4, utilizing the photon distributions
      previously 
      derived, convoluted with the cross sections relative to the 
      subprocesses $e\gamma\to e\gamma$ and $\gamma\gamma\to\mu^+\mu^-$,
      given in Appendix B.
      It is shown that the production 
      rates are sufficient to measure the polarized and unpolarized photon
      content of the nucleon.  

\item Chapter 5 is devoted to the unpolarized QED Compton scattering   in $e p 
      \rightarrow e\gamma p $ and $e p \rightarrow e \gamma X$,
      with the photon emitted from the lepton. 
      The   full $2 \rightarrow 3$  process  is 
      calculated in a manifestly covariant way  by employing appropriate 
      parametrizations of the proton's structure functions.
      These  results are compared with the  ones based on the 
      EPA, as well as with the experimental data  and theoretical estimates 
      for the 
      HERA collider given in \cite{thesis}.  It is  shown that the  
      cross section is reasonably well described by the EPA of the proton, 
      also in the  inelastic channel. 
      In addition it turns out that the results obtained in \cite{thesis}, 
      based on  an iterative approximation procedure 
      proposed  by Courau and Kessler \cite{kessler},
      deviate appreciably from our analysis in certain kinematical regions. 
      Details about the kinematics of the process are given in Appendix C.

\item In Chapter 6 the virtual Compton scattering process  
      in $e p \rightarrow e\gamma p $ 
      and $e p \rightarrow e \gamma X$, where the photon is emitted from the 
      hadronic vertex, is  investigated. It represents the major background 
      process to QED Compton scattering. New kinematical cuts are suggested 
      in order to suppress the virtual Compton scattering  
      background and facilitate the extraction of the equivalent photon 
      distribution of the proton at the HERA collider.  
      The analytic expressions of the matrix elements of 
      unpolarized QED and virtual
      Compton scattering can be found in Appendix D.    

\item Chapter 7 is devoted  to the QED Compton process in longitudinally 
      polarized lepton-proton scattering.
      The kinematical cuts  necessary to measure the po\-la\-ri\-zed 
      photon content of the proton  and to
      suppress the major background process coming from Virtual
      Compton scattering are provided
      for  HERMES, COMPASS and future eRHIC experiments. 
      We point out that these measurements  
      will also give access to the spin dependent structure function  
      $g_1$ in a kinematical region not well covered by inclusive
      measurements. 
      The analytic expressions of the matrix elements of 
      polarized QED and virtual
      Compton scattering are relegated to Appendix E.   

\item The accuracy of the EPA  of 
      the proton in describing the inelastic process 
      $ep\rightarrow \nu W X$ is
      investigated in Chapter 8. In particular, the scale dependence 
      of the correspon\-ding 
      inelastic photon distribution is discussed. Furthermore, an estimate 
      of the total number 
      of events, including the ones coming from the elastic and 
      quasi-elastic channels of the reaction, is given for the 
      HERA collider.       

\item Finally, summary and conclusions can be found in Chapter 9.

\end{itemize}

\clearemptydoublepage

\chapter{{\bf The Structure of the Nucleon}}

In physics, one of the most common ways of getting information on the 
structure of extended objects like hadrons  is to use  structureless
particles as projectiles which scatter off the hadron in question. 
To probe the inside of a nucleon one naturally uses charged lepton
(electron or muon) beams: such reactions dominantly take place by the exchange
of a single virtual photon, therefore they can be studied with
relative ease and clarity. Effects of higher order virtual photon
exchange are believed to be less than a few percent.\\  
According to Heisenberg's uncertainty 
principle, one can resolve the target structure 
 down to the scale $\lambda \sim \hbar/Q$, where $Q$ is the
momentum transfer from the lepton to the nucleon. 
Hence, the higher the energy loss of the lepton, the finer the structure 
that can be resolved. The study of such reactions represent the main subject
of the present chapter.

In Section \ref{sec:formfact} we examine electron-nucleon 
elastic scattering and discuss the significance of the electromagnetic 
form factors. In Section \ref{sec:inel} inelastic electron-nucleon 
scattering is described in terms of the structure functions 
$F_1$ and $F_2$. The spin dependent structure functions $g_1$ and $g_2$
are introduced  in Section \ref{sec:poldis}. The parton model and 
quantum chromodynamics description of inelastic electron-nucleon
collisions can  be found in  Section \ref{sec:parton} and 
in Section \ref{sec:qcd} respectively.

\begin{figure}[t] 
\begin{center} 
\epsfig{figure=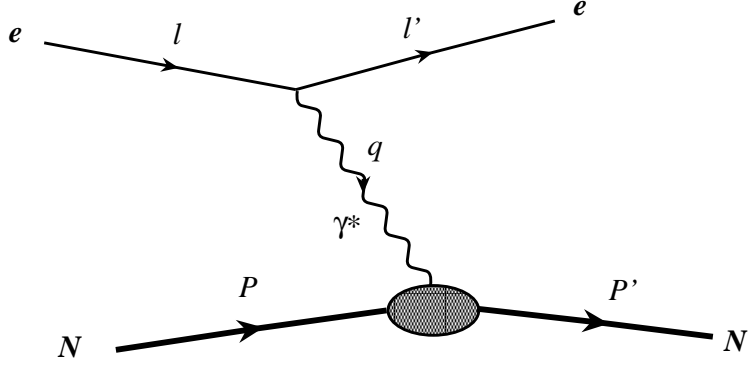, width= 13cm}
\vspace{-2.5cm} 
\caption{The one-photon exchange diagram for elastic electron-nucleon
scattering.}
\label{elastic}
\end{center}
\end{figure} 

\section{Elastic Form Factors}
\label{sec:formfact}

We consider the elastic electron-nucleon scattering process:
\begin{equation}
e(l) + N(P) \rightarrow e(l') + N(P'),
\label{eq:elastic}
\end{equation}
where the four-momenta of the particles are given in the brackets. At lowest 
order in perturbation theory of QED the reaction is described by
a one-photon exchange diagram, depicted in Figure \ref{elastic}. 
In terms of the Dirac 
$\gamma$ matrices and  spinors, the electron transition 
current is
\begin{equation}
\langle\, e(l')\,|\,J^{\alpha}_{\mathrm{em}}(0)\,|\, e(l)\,\rangle 
 = - e \, [\,\bar{u}(l')\, \gamma^{\alpha}\,u(l)\,],
\label{elmatrix}
\end{equation}
with $e$ denoting the electric charge of the proton. The vertex linking the 
photon with the nucleon is not  point-like, therefore the nucleon
transition current is different from (\ref{elmatrix}); it is
the most general Lorentz four-vector that can be constructed from $P$, $P'$
and the $\gamma$ matrices,
 sandwiched between $\bar u(P)$ and $u(P)$:
\begin{equation}
\langle\, N(P')\,|\,J^{\alpha}(0)\,|\, N(P)\,\rangle   =   e \, [\,\bar{u}(P') \,\Gamma^{\alpha}\,u(P)\,].
\label{eq:matrix}
\end{equation}
In $\Gamma^{\alpha}$ terms involving $\gamma^5$ are ruled 
out by the conservation of parity.  Furhermore, using the Dirac equations 
$ P\!\!\!\!/ \, u(P) = m u(P)$ and 
$\bar{u}(P') P' \!\!\!\!\!/\, =
 m \bar{u}(P')$,  where $m$ is the nucleon mass, one can show that there 
are only three independent terms, $\gamma^\alpha$, $i\sigma^{\alpha\beta}q_{\beta}$ and $q^{\alpha}$, with
\begin{equation}
\sigma^{\alpha\beta} = \frac{i}{2}\,[\gamma^\alpha,\,\gamma^\beta]
\end{equation} 
and $q$ being the momentum transfer,
\be
q = l-l'.
\label{eq:qiu}
\ee
Therefore, quite generally, 
\begin{equation}
\Gamma^{\alpha} = F_1(q^2)\gamma^{\alpha}+\frac{1}{2m} 
  F_2(q^2)i\sigma^{\alpha\beta}q_{\beta} + F_3(q^2)q^{\alpha}. 
\label{eq:Gammagen}
\end{equation} 
The coefficients $F_k(q^2)$ ($k = 1, 2, 3$) are the electromagnetic 
{\em elastic form factors} of the 
nucleon, which are functions of the momentum transfer squared $q^2$, the 
only independent scalar variable at the nucleon vertex 
(being $P\cdot q = -q^2/2$).
For the electromagnetic case we are interested in, $J^\alpha_{\mathrm{em}}$
is a conserved current and therefore the matrix element (\ref{eq:matrix}) 
has to satisfy the condition 
\begin{equation}
q_\alpha\, \langle\,N(P')\,|\,J^{\alpha}_{\mathrm{em}}\,|\, N(P)\,\,\rangle 
= 0,
\label{eq:gauge}
\end{equation}
which implies $F_3(q^2) = 0$ in (\ref{eq:Gammagen}). 
The Dirac equation can be used again in order to replace
$\bar{u}(P') i\,\sigma^{\alpha\beta}q_{\beta} u(P)$ by $\bar{u}(P')
[2 \,m\gamma^\alpha -(P+P')^\alpha] u(P)$  in (\ref{eq:Gammagen}) so that
one can write
\begin{equation} 
\Gamma_{\mathrm{em}}^{\alpha}=(\,F_1(q^2)+F_2(q^2)\,)\,\gamma^{\alpha}-\frac{1}{2m} \,
  F_2(q^2)\,(P+{P'})^{\alpha}\,. 
\end{equation} 
At $q^2 = 0$, which physically 
corresponds to the nucleon interacting with a static electro-magnetic field,
the form factors are related to the electric charge $Q$ and the magnetic 
dipole moment $\mu$ of the nucleon:
\begin{equation}
eF_1(0) = Q, ~~~~~~~\frac{e}{2m}\,[F_1(0) + F_2(0)] = \mu .
\end{equation}
For an electrically neutral particle, like the neutron, one has $F_1(0)$ = 0. 
If a particle  has no anomalous magnetic moment $\kappa$,
defined such that $\mu =  (1 + \kappa) \,e/(2m)$, then $F_2(0) = 0$.
Experimentally, 
\be
F_1^p(0) = 1, ~~~~~~~F_2^p(0)=\kappa_p \simeq 1.79
\ee 
for the proton, and 
\be
F_1^n(0) = 0,~~~~~~~ F_2^n(0) =\kappa_n \simeq -1.91
\ee
for the neutron. In the following, we will always express $\mu$ in units
of the nucleon magneton $e \hbar/(2 m c)$, that is $\mu_p = 2.79$, 
$\mu_n = -1.91$.

The  amplitude $M$ relative to the process (\ref{eq:elastic}) has the form
\begin{equation}
M = -e^2\,[\bar{u}(l') \gamma_{\alpha}u(l)]\,\frac{1}{q^2}\,
[\bar{u}(P') \Gamma_{\mathrm{em}}^{\alpha}u(P)].
\label{eq:amplitude}
\end{equation}
Taking the modulus squared of the amplitude and multiplying by the appropriate
phase space and flux factors,  one finds that the differential cross section
can be written as
\begin{eqnarray}
\der\sigma = \frac{1}{4\sqrt{ ( l\cdot P)^2 -m^2m_e^2}} \,\overline{|M|^2}\,(2\pi)^4\delta^4(l+P-l'-P')\, 
\frac{\der^3{\mbox{\boldmath $l$}}'}{(2\pi)^3\,2l'_0 }\,\frac{\der^3 {\mbox{\boldmath $P$}}'}{(2\pi)^3\,2P'_0 }~,
\label{eq:initial}
\end{eqnarray}
which holds with the normalization of the spinors given in 
\eqref{eq:norm_spinor} and where  $m_e$ denotes
 the electron mass. After integrating over the phase space of the scattered
nucleon, one can use the condition ${\mbox{\boldmath $P$}}' = {\mbox{\boldmath $P$}} +  {\mbox{\boldmath $q$}}$  to rewrite the energy conserving 
$\delta$-function as
\be
\frac{1}{2P'_0 }\,\delta(l_0+P_0-l'_0-P'_0)& =& \delta ({P'_0}^2-(P_0+q_0)^2)\,
=\, \delta(P'^2-(P+q)^2)\nonumber \\
& = & \delta (q^2+ 2 P\cdot q).
\ee
Therefore \eqref{eq:initial} reduces to
\begin{eqnarray}
\der\sigma =  \frac{1}{4\sqrt{ ( l\cdot P)^2 -m^2m_e^2}}   \,\overline{|M|^2}\, 
\frac{\der^3{\mbox{\boldmath $l$}}'}{(2\pi)^2\,2l'_0 }\,
\delta (q^2+ 2 P\cdot q),
\label{eq:sigmael}
\end{eqnarray}
with
\begin{equation}
\overline{|M|^2} = \frac{1}{q^4}\,L_{\alpha\beta}(l;l') \,H_{\mathrm{el}}^{\alpha\beta}(P;P'),
\label{eq:sigmael2}
\end{equation}
where the tensors $L_{\alpha\beta}(l;l')$ and $H_{\mathrm{el}}^{\alpha\beta}(P;P')$
come from averaging over initial spins and summing over final spins in the 
products of the electron and  nucleon matrix elements (\ref{elmatrix}) and 
(\ref{eq:matrix}), when (\ref{eq:amplitude}) is squared.
The completeness relation \eqref{eq:compl_u} is  commonly used to
perform the spin sums.  More explicitly:
\begin{eqnarray}
L^{\alpha\beta}(l;l') &= &\frac{1}{2} \sum_{\mathrm{spins}} \,{\langle 
e(l') \mid J_{\mathrm{em}}^\alpha (0)\mid e(l) \rangle}^* \langle e(l') 
\mid J_{\mathrm{em}}^\beta(0)  \mid e(l) \rangle\nonumber \\
 &= & \frac{1}{2}\,e^2\,\mathrm{Tr} [(l'\!\!\!\!/ +m_e)\gamma^{\alpha}(l\!\!\!/+m_e)\gamma^{\beta}]\nonumber \\
& = & 2 \,e^2\{ l^{\alpha}l'^{\beta} + l'^{\alpha}l^{\beta} -g^{\alpha\beta}
(l\cdot l'-m_e^2)\}
\label{eq:leptonic}
\end{eqnarray}
and
\begin{eqnarray}
H^{\alpha\beta}_{\mathrm{el}}(P;P') &=&  \frac{1}{2}\sum_{\mathrm{{spins}}}\, 
{\langle 
N(P') \mid J_{\mathrm{em}}^\alpha (0)\mid N(P) \rangle}^* \langle N(P') 
\mid J_{\mathrm{em}}^\beta(0)  \mid N(P) \rangle \nonumber \\
& = & \frac{1}{2}\,e^2\,\mathrm{Tr} [(P'\!\!\!\!\!/ +m)\Gamma^{\alpha}(P\!\!\!
\!/+m)\Gamma^{\beta}]\nonumber \\
& = & e^2\,\bigg[ \bigg (F_1^2-\frac{q^2}{4m^2}F_2^2\bigg ) (P+P')^\alpha(P+P')^\beta \nonumber \\
&& ~~~~~~~~~~~~~~~~~~+ (F_1+F_2)^2 (q^2 g^{\alpha\beta} -q^\alpha q^\beta)\bigg ].
\label{eq:hadronic}
\end{eqnarray}
In the kinematical region of high energies,
 the electron mass can be neglected
and then the contraction of (\ref{eq:leptonic}) and (\ref{eq:hadronic}) gives
\begin{eqnarray}
L_{\alpha\beta}(l;l')H_{\mathrm{el}}^{\alpha\beta}(P;P') &= &2e^4\bigg \{\,4 \bigg (F_1^2 -\frac{q^2}{4m^2}F_2^2\bigg )[\, 2 (P\cdot l') (P\cdot l) - (l\cdot l') m^2]\nonumber \\
 &&~~~~~~~~~~~~~~~~~~~-2  (F_1 + F_2)^2 \,q^2 \,(l\cdot l')\,\bigg \}.
\label{eq:ampl}
\end{eqnarray} 
In a frame in which the nucleon is at rest and the electron moves along the
$z$ axis with energy $E$ and is scattered into a solid angle 
$\Omega = (\theta,\varphi)$ with final energy $E'$, i.e.
\begin{equation}
P = (m,0,0,0,),~~~l=E(1,0,0,1), ~~~l' = E'(1,\sin\theta\cos\varphi,\sin\theta\sin\varphi,\cos\theta), \label{eq:kinema}
\end{equation}
then
\begin{equation}
q^2 = (l-l')^2 = -4EE'\sin^2\frac{\theta}{2}
\label{eq:q2}
\end{equation}
and one can define the energy transfer from the electron to the target
\begin{equation}
\nu = E-E'= \frac{q\cdot P}{m}~.
\label{eq:n}
\end{equation}
The differential cross section (\ref{eq:sigmael}) can be rewritten as
\begin{equation}
\frac{\der\sigma}{\der E'\der\Omega} = \frac{\alpha^2}{4E'^2\sin^4\frac{\theta}{2}}\,
\bigg [ \bigg (F_1^2-\frac{q^2}{4m^2}F_2^2\bigg ) \cos^2\frac{\theta}{2} -\frac{q^2}{2 m^2}(F_1+F_2)^2\sin^2
\frac{\theta}{2}\, \bigg ]\delta \bigg (\nu + \frac{q^2}{2m}\bigg ),
\label{eq:rosenbluth1}
\end{equation}
with $\alpha = e^2/(4\pi)$.
The form factors $F_1$ and $F_2$, usually referred to as the Dirac 
and Pauli form factors respectively, parametrize our ignorance of the 
complicated structure of the nucleon.  
In practice, however, it is better to use linear combinations 
of them, the Sachs electric and magnetic form factors 
\begin{eqnarray}
        G_E(q^2) & = & F_1(q^2) +\frac{q^2}{4m^2}F_2(q^2),\nonumber\\ 
        G_M(q^2) & = & F_1(q^2) + F_2(q^2),
\label{eq:sachs}
\end{eqnarray}
defined so that no interference terms, $G_EG_M$, occur in 
(\ref{eq:rosenbluth1}). At $q^2 = 0$, one has
\be
G_E^p(0) &= & 1,~~~~~~~G_M^p(0)\, = \,1 + \kappa_p \simeq 2.79,\nonumber \\
G_E^n(0) & = & 0,~~~~~~~G_M^n(0) \,= \,\kappa_n \simeq -1.91.
\ee
Integration  of \eqref{eq:rosenbluth1}     
over $E'$, taking into account that $q^2$ depends on $E'$
when $\theta$ is held fixed, see (\ref{eq:q2}), gives the Rosenbluth formula
\cite{rosen}
\begin{equation}
\frac{\der\sigma}{\der\Omega} = \frac{\alpha^2}{4E^2\sin^4\frac{\theta}{2} }\,
\frac{E'}{E}\,\bigg ( \frac{G_E^2 + \tau G_M^2}{1+\tau} 
\cos^2\frac{\theta}{2}+2\tau G_M^2\sin^2\frac{\theta}{2}\bigg ),
\label{eq:rosenbluth}
\end{equation}
with $\tau \equiv -q^2/4m^2$ and the factor
\begin{equation}
\frac{E'}{E} = \bigg (1 + \frac{2E}{m}\sin^2\frac{\theta}{2}\bigg )^{-1}
\label{eq:recoil}
\end{equation}
arises from the recoil of the target.
The Rosenbluth formula is the basis of all experimental studies of the
electromagnetic structure of the nucleon and allows to determine
the  form factors by measuring $\der \sigma /\der \Omega$ as a function
of $\theta$ and $q^2$. Experimentally it turns out that the form factors
drop rapidly as $-q^2$ increases:
\be
G^p_E(q^2)  =\frac{G_M^p(q^2)}{\mu_p} = 
 \frac{G_M^n(q^2)}{\mu_n} = 
\bigg ( 1 - \frac{q^2}{0.71 \, \mathrm{GeV}^2}\bigg )^{-2},\label{eq:dipole} \\
0 \le G_E^n(q^2)\le 0.1~.~~~~~~~~~~~~~~~~~~~~~~~~~~~~~~~~~~~~~~~~~~~~~
\ee
The function given in \eqref{eq:dipole} is merely empirical and is often called
a {\em dipole} fit; a monopole fit would be $(1 - cq^2)^{-1}$.  Also there
is no fundamental theoretical reason for $G_E^p$, $G_M^p$, $G_E^n$ to have 
the same $q^2$ behaviour: this likeness is expressed by saying that  the
three form factors {\em scale} together in $q^2$.

In the non-relativistic limit, $E\ll m$,   one can see from   
\eqref{eq:recoil} that $E=E'$; therefore $-q^2 =|${\mbox{\boldmath $q$}}$|^2$ $\ll m^2$. In this limit  the form factors $G_E$ and $G_M$
are the Fourier transforms of the nucleon's charge and magnetic moment 
density distributions, respectively \cite{perl}. 
Assuming, for example, that the charge density distribution of the proton is 
spherically 
symmetric, i.e. a function of $r \equiv |{\mbox{\boldmath$r$}}|$ alone, and   
that it is normalized such that
\be
\int\der^3 {\mbox{\boldmath$r$}} \,\rho({\mbox{\boldmath$r$}}) = 1,
\ee
then the exponential in its Fourier transform can be expanded for
small $|{\mbox{\boldmath$q$}}|$ as follows 
\be
G^p_E(q^2) = \int\der^3 \,{\mbox{\boldmath$r$}} \rho(r) e^{i\,
{{\mbox{\boldmath$q$}}\cdot {\mbox{\boldmath$r$}}}
} & = & \int\der^3 \,{\mbox{\boldmath$r$}} \rho(r) \bigg ( 1 + i {
{\mbox{\boldmath$q$}}\cdot {\mbox{\boldmath$r$}}}
 - \frac{{( {\mbox{\boldmath$q$}}
\cdot {\mbox{\boldmath$r$}})^2}}{2} + ...\bigg ) \nonumber\\
& = &1 -\frac{1}{6}\,|{\mbox{\boldmath$q$}}|^2 \, \langle r^2 \rangle 
+ ...~,
\label{eq:taylor}
\ee
where the mean square charge radius of the proton is defined by
\be
\langle r^2 \rangle  = \int \der^3{\mbox{\boldmath$r$}}\,\rho (r)r^2.
\ee
Identifying \eqref{eq:taylor} with the expansion of \eqref{eq:dipole},
\be
G_E^p(q^2) = G_E^p(0) + q^2\,\bigg ( \frac{\der G_E^p(q^2)}
{\der q^2}\bigg )_{q^2=0} + ... = 1 -|{\mbox{\boldmath$q$}}|^2\,
\bigg ( \frac{\der G_E^p(q^2)}{\der q^2}\bigg )_{q^2=0} + ...~,
\ee
one gets
 \be
\langle r^2 \rangle & = &\int \der^3{\mbox{\boldmath$r$}}\,\rho (r)r^2
\, =\, 6\,\bigg ( \frac{\der G_E^p(q^2)}{\der q^2}\bigg )_{q^2=0} \nonumber \\
& =& (0.81 \times 10^{-13} \mathrm{cm})^2.
\ee 
The same radius of about 0.8 fm is obtained for the magnetic moment 
distribution. Finally, the charge distribution of the nucleon 
has an exponential shape in configuration space: the Fourier 
transform of $\rho (r) = e^{-m r}$, with $m =0.84 $ ${\rm GeV}$, gives 
the result \eqref{eq:dipole} for $G_E^p$.

\begin{figure}[t] 
\begin{center} 
\epsfig{figure=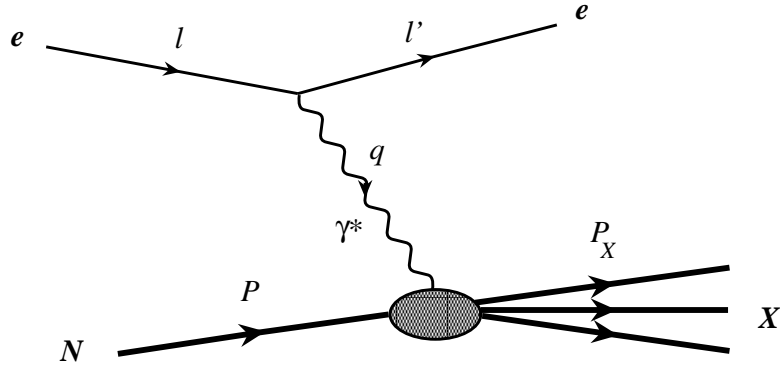, width= 13cm}
\vspace{-2.5cm} 
\caption{The one-photon exchange diagram for  inelastic electron-nucleon
scattering.}
\label{inelastic}
\end{center}
\end{figure} 


\section{Unpolarized Structure Functions}
\label{sec:inel}

Probing the nucleon with a large wavelength photon (small  momentum transfer
squared $-q^2$)  can only provide information about its dimension.
It is  possible to have a better spatial resolution by increasing $-q^2$,
but already for $-q^2 \gtrsim m^2$ the elastic process \eqref{eq:elastic} 
is not dominant any more and the nucleon often breaks up into 
hadronic debris. 
The inelastic reaction
\begin{equation}
e(l) + N(P) \rightarrow e(l') + X(P_X),
\label{eq:dis}
\end{equation}
where $X$ is the undetected hadronic system, represents the most
direct way to explore the internal structure of the nucleon;
it  is {\em fully inclusive} with respect to
the hadronic state and, when the momentum transfer is not ultra-high,
is dominated  by one-photon exchange, as  described by the diagram 
in Figure \ref{inelastic}.
The  reaction \eqref{eq:dis} is described by three kinematic
variables. One of them, the incoming lepton energy $E$, or alternatively
the center-of-mass energy squared $s = (l + P)^2$, is fixed by the 
experimental conditions. The other two independent variables can be 
chosen among the following invariants: $Q^2\equiv -q^2$ and $\nu$,  
already defined in \eqref{eq:q2} and \eqref{eq:n},
 the center-of-mass 
energy squared of the 
$\gamma^*N$ system (that is the invariant mass squared of the hadronic
system $X$)
\be
W^2 & = & (P+q)^2,
\ee
the Bjorken variable
\be
x_B & = & \frac{Q^2}{2P\cdot q} 
= \frac{Q^2}{Q^2 + W^2 -m^2}~,
\label{bjorkenv}
\ee
and the ``inelasticity''
\be
y  & = & \frac{P\cdot q}{P\cdot l} = \frac{W^2 + Q^2 -m^2}{s-m^2}~.
\ee
In the target
rest frame, where  $\nu$ is the transferred energy from the electron to
the target, 
$y$ is the fraction of the incoming electron energy carried
by the exchanged photon, $ y = \nu/E $. The relation connecting 
$x_B$, $y$ and $Q^2$ is given by 
\be
x_B y = \frac{Q^2}{s-m^2} \simeq \frac{Q^2}{s}~.   
\ee
Since $W^2 \ge m^2$  the Bjorken variable $x_B$ takes values between 0 and 1, 
and so does $y$. 
The measurement of the cross section corresponding to the 
process \eqref{eq:dis} shows a peak when the nucleon does not break up 
($W\simeq m$) and broader peaks when the target is excited to resonant
barion states, most of them concentrated in the range 
$m \lesssim  W \lesssim 1.8$  ${\rm GeV}$.  As $W$ increases
one reaches a region where a smooth behaviour is set in. This is the 
{\em deep inelastic region}, where both $Q$ and $W$ are large compared
to the typical hadron masses. In this kinematical domain, the reaction
\eqref{eq:dis} is known as deep inelastic scattering.

If we neglect the electron mass, the cross section for deep inelastic 
scattering (DIS) can be written as
\be
\der \sigma = \frac{1}{4P\cdot l}  \,\frac{1}{Q^4}\,
L_{\alpha\beta}(l;l')W^{\alpha\beta}\,\frac{\der^3\mbox{\boldmath $l$}'}
{(2\pi)^32l_0'}~,
\label{eq:DIScross}
\ee 
where, as before,  we have averaged
over the initial electron  and  nucleon spins, and summed over the 
final electron spin. 
The leptonic tensor  $L_{\alpha\beta}(l;l')$ was calculated
in \eqref{eq:leptonic} and the hadronic tensor, corresponding 
to the electromagnetic transitions of the target nucleon to all possible
final states, is defined as
\be
W^{\alpha \beta}= \sum_X \int \frac{\der^3{\mbox{\boldmath $P$}}_X}{(2\pi)^32P^0_X}\,
 H^{\alpha \beta}_{\mathrm{inel}}(P;P_X)(2\pi)^4 \delta^4 (l+P-l'-P_X),
\label{eq:hadr}
\ee
with
\be
H^{\alpha \beta}_{\mathrm{inel}}(P;P_X) = {1\over 2}\sum_{\mathrm{{spins}}} 
{\langle X(P_X) \mid J_{\mathrm{em}}^\alpha 
(0)\mid N(P) \rangle}^* \langle X(P_X) \mid J_{\mathrm{em}}^\beta(0) \mid N(P) \rangle
\label{eq:hinelas}
\ee
and  $P_X$ denoting the total four-momentum of the state $|X\rangle$.
The definition \eqref{eq:hadr} holds with states normalized as in 
\eqref{eq:norma}. The cross section \eqref{eq:DIScross} reduces to 
\eqref{eq:initial}, with $m_e=0$, when $X$ is restricted to be also a 
nucleon.
  
In general, if we do not average over the initial nucleon spin $S$,
the electromagnetic hadronic tensor  
$W_{\alpha\beta}$ consists of  a symmetric and an antisymmetric (spin 
dependent) part under $\alpha \leftrightarrow \beta $, 
\be
W_{\alpha\beta}  = W^{\mathrm{S}}_{\alpha\beta}(q, P) 
+ W^{\mathrm{A}}_{\alpha\beta}(q; P, S).
\label{eq:hadronictensor} 
\ee
Since $J^{\alpha}_{\rm {em}}$ is hermitian, $W^*_{\alpha\beta} 
= W_{\beta\alpha}$, and \eqref{eq:hadronictensor} corresponds also to break 
$W_{\alpha\beta}$ into its real and immaginary parts.
As $L_{\alpha\beta}$ is symmetric under 
$\alpha\leftrightarrow \beta$,
when contracted with $W_{\alpha\beta}$ only the symmetric piece of 
$W_{\alpha\beta}$ will contribute. Furthermore, the condition 
\be
q^{\alpha}W_{\alpha\beta} = W_{\alpha\beta}q^{\beta}= 0
\label{eq:conser}
\ee
must hold for both real and imaginary parts of $W_{\alpha\beta}$
due to current conservation, see \eqref{eq:gauge}. 
The most general form of 
$W_{\alpha\beta}^{\rm S}$ compatible
with parity conservation and with  \eqref{eq:conser} is
\be
\frac{1}{4\pi e^2m} \,W^{\rm{S}}_{\alpha\beta} & =& \bigg (-g_{\alpha\beta} 
+ \frac{q_\alpha q_\beta}{q^2} \bigg ) 
W_1(P\cdot q, q^2) \nonumber \\
&&~~~~+ \frac{1}{m^2} \, \bigg ( P_{\alpha} - \frac{P\cdot q}{q^2}\,
 q_{\alpha} \bigg )
\bigg ( P_{\beta} - \frac{P\cdot q}{q^2} \,q_{\beta} \bigg ) 
W_2(P\cdot q, q^2),
\label{eq:hadron_symm}
\ee
where $W_{1, 2}(P\cdot q, q^2)$ are known as the {\em structure functions} of the 
nucleon.
They are the generalization to the inelastic case of the elastic form
factors. If we substitute \eqref{eq:hadron_symm} and \eqref{eq:leptonic}
in \eqref{eq:DIScross}, then
\be
\frac{1}{4\pi e^2m}\, L^{\alpha\beta}W_{\alpha\beta} = 4\, (l\cdot l')W_1 
+ 2 [\,2(P\cdot l) (P\cdot l') -(l\cdot l')m^2\,]\,\frac{W_2}{m^2}
\label{eq:cont_symm}
\ee
and, in the nucleon rest frame, where $P\cdot l=mE$, $P\cdot l' = m E'$,
one gets for the cross section \eqref{eq:DIScross}
\be
\frac{\der\sigma}{\der\Omega\der E'} = \frac{4\alpha^2}{Q^4}\,E'^2\,\bigg 
(2 W_1\sin^2\frac{\theta}{2} +
 W_2\cos^2\frac{\theta}{2} \bigg ).
\label{eq:DIS_unpol}
\ee
The cross section has again the characteristic angular
dependence that was found for $ep\rightarrow ep$. 
One usually introduces the longitudinal and transverse structure functions
\be
F_T(x_B, Q^2) & = & 2 x_BF_1(x_B, Q^2),\label{eq:ft} \\
F_L(x_B, Q^2) & = & F_2(x_B, Q^2) - 2 x_B F_1(x_B, Q^2),\label{eq:fl}
\ee
which correspond to the absorption of transversely and longitudinally
polarized virtual photons respectively, and in \eqref{eq:ft}, \eqref{eq:fl}
are expressed in terms of 
the following dimensionless structure functions
\be
F_1(x_B, Q^2) & = & m W_1(P\cdot q, q^2)\label{eq:f1}\\
F_2(x_B, Q^2) & = & \nu W_2(P\cdot q, q^2).\label{eq:f2}
\ee
Bjorken \cite{paschos} argued that in the limit 
\be
\nu, \,Q^2 \rightarrow \infty, ~~~x_B = \frac{Q^2}{2 m \nu} ~~{\rm fixed},
\ee
now referred to as {\em Bjorken limit}, $F_1$ and $F_2$ approximately
scale, namely depend on $x_B$ only. This behaviour is  
already present for $Q^2 \ge 1$ ${\rm GeV}^2$ and is 
surprisingly in contrast to the strong $Q^2$ dependence, roughly as $Q^{-4}$, 
of the elastic form factors of the nucleon. On the other hand
the elastic form factors of a {\em pointlike} particle like the muon are 
constants independent of $Q^2$: the $e\mu \rightarrow e\mu$ scattering cross 
section is given by \eqref{eq:rosenbluth1} with $F_1 = 1$ and $F_2 = 0$. 
Hence scaling seems to be an indication
of scattering from charged  pointlike constituents of the nucleon, 
the {\em partons}, and
historically its observation  \cite{blooms,briedenbach,friedman}
inspired the so-called {\em parton model} \cite{paschos,feynman}.

Moreover, structure function measurements show that $F_L \ll F_2$, 
suggesting the spin-1/2 property of  partons, since a (massless) 
spin-1/2 particle
cannot absorb a longitudinally polarized photon \cite{reyaqcd}.
In contrast, spin-0 (scalar) partons could not absorb transversely polarized 
photons and so we would have $F_1 = 0$, i.e. $F_L = F_2$, 
in the Bjorken limit. Partons are nowadays identified with the quarks of 
quantum chromodynamics (QCD).
    

To conclude, the hadronic tensor $W_{\alpha\beta}$ describes the unknown 
coupling of the virtual photon to the nucleon in terms of the structure 
functions, which can be extracted from experiments. 
However, the hadronic tensor can also be computed from models; 
in this case it is 
useful to develope a technique which allows one to extract the
structure functions from a knowledge of $W_{\alpha\beta}$. Defining
the projection operators \cite{anselmino}  
\be
{\cal{P}}_1^{\alpha\beta}& =& \frac{1}{8\pi e^2} \,\bigg [\frac{1}{a}
P^{\alpha}P^{\beta} - g^{\alpha\beta} \bigg ], \nonumber \\
{\cal{P}}_2^{\alpha\beta}& = &\frac{3P\cdot q}{8\pi e^2 a}\, 
\bigg [\frac{P^{\alpha}P^{\beta}}{a} -\frac{1}{3}g^{\alpha\beta} \bigg  ],
\label{eq:proj_unp}
\ee
with
\be
a = \frac{P\cdot q}{2x_B}+ m^2,
\label{eq:a}
\ee
and using \eqref{eq:hadron_symm}, \eqref{eq:f1}, \eqref{eq:f2}, one can see 
that
\be
{\cal{P}}_1^{\alpha\beta}W_{\alpha\beta} = F_1
\label{eq:projf1}
\ee
and
\be
{\cal{P}}_2^{\alpha\beta} W_{\alpha\beta}= F_2~.
\label{eq:projf2} 
\ee

\section{Polarized Structure Functions}
\label{sec:poldis}

Polarized DIS, involving the collision of a longitudinally polarized
electron on a polarized (either longitudinally or transversely) nucleon, 
provides a different, but  equally important insight into the structure of 
the nucleon. As mentioned in the previous section, if we do not
average over the nucleon spin, the hadronic tensor will consist also of 
an antisymmetric part,  $W_{\alpha\beta}^{\rm A}$. 
Imposing \eqref{eq:conser}, $W_{\alpha\beta}^{\rm A}$ can be expressed
in terms of the polarized structure functions $G_1$ and $G_2$ as follows
\cite{bjorken66,bjorken70}
\be
\frac{1}{4\pi e^2 m}\,W^A_{\alpha\beta} = i\varepsilon_{\alpha\beta\rho\sigma}
  q^{\rho} \bigg \{mS^{\sigma} G_1(P\cdot q, q^2) + 
    [(P\cdot q) S^{\sigma} -(S\cdot q)P^{\sigma}] \,
     \frac{G_2(P\cdot q, q^2)}{m} \bigg \},\nonumber \\
\label{eq:polhadrtens}
\ee
where $S$ is the covariant spin vector of the nucleon, whose essential
properties are
\be
S\cdot P = 0,~~~~~~~S^2 = -1.
\ee
 
Clearly $W^{\rm A}_{\alpha\beta}$ changes sign under reversal of the
nucleon's polarization. From the cross section formula \eqref{eq:DIScross},
one notices that $G_1$ and $G_2$ cannot be obtained from an experiment
with just  a polarized target. Both the electron and the nucleon must be
polarized, otherwise the term $L_{\alpha\beta}W^{\rm{A}}_{\alpha\beta}$ 
drops out.
Analogously to \eqref{eq:hadronictensor}, the
leptonic tensor has to be generalized to 
\be
L_{\alpha\beta}= L^{\rm {S}}_{\alpha\beta}(l;l') + 
L^{\rm {A}}_{\alpha\beta}(l, s;l'),
\ee
where $s$ is the spin four-vector of the electron, defined such that 
$s\cdot l = 0$, $s^2 = -1$, and $L_{\alpha\beta}^{\rm S}$  is
given in \eqref{eq:leptonic}.
The additional, antisymmetric, term can be calculated using the spin 
projector operator \eqref{eq:spin_prj}.
If  in \eqref{elmatrix} we make the replacement
\be
u(l) \longrightarrow \frac{1}{2}\, (1 + \gamma^5 s\!\!\!/  )u(l, s),
\label{eq:spinlepton}
\ee
then we can again utilize  \eqref{eq:leptonic}, without the 
factor $1/2$ alredy included in \eqref{eq:spinlepton}, to compute the leptonic
tensor and perform the sum over the initial electron spins
with help of the completeness relation.
Having inserted the projection operator, only one of the two possible 
polarizations will contribute. We get 
\be
L_{\alpha\beta} = \frac{1}{2} \,e^2 {\rm{Tr}}[(1 + \gamma^5 s\!\!\!/)
\,(l\!\!/+m_e)
\gamma_{\alpha}(l'\!\!\!\!/+m_e)\gamma_{\beta}\,]
\label{eq:pollepton}
\ee
and the term proportional to $\gamma^5$ will give
\be
L^A_{\alpha\beta} = 2ie^2m_e\varepsilon_{\alpha\beta\rho\sigma}s^{\rho}
(l-l')^{\sigma}~.
\label{eq:leptonictensor}
\ee
For a high energy ($E\gg m_e$), longitudinally polarized electron, the 
spin vector is 
\be  
s^{\alpha} = \frac{2\lambda_e}{m_e}\, l^{\alpha},~~~~~~\lambda_e= \pm 
\frac{1}{2} ~,
\ee
and \eqref{eq:leptonictensor} becomes
\be
L^{\rm A}_{\alpha\beta} = 2 ie^2 \varepsilon_{\alpha\beta\rho\sigma}
l^{\rho}q^{\sigma},
\label{eq:antilept}
\ee
where the helicity of the electron $\lambda_e$ has been fixed to be
$+1/2$. The amplitude squared in the cross  section \eqref{eq:DIScross} will
have the form 
\be
L_{\alpha\beta}W^{\alpha\beta} = 
L_{\alpha\beta}^{\rm S}W^{\alpha\beta \rm S} +
L_{\alpha\beta}^{\rm A}W^{\alpha\beta \rm A},
\label{eq:sum_as}
\ee
with
\be
\frac{1}{4\pi e^4 m}\,L_{\alpha\beta}^{\rm A}W^{\alpha\beta \rm A}& =&
  4 \,\bigg \{ [\,(l\cdot q) (S\cdot q) - q^2 (S\cdot l)\,] 
mG_1 \nonumber \\
 &&~~~+\frac{q^2}{m}[\, (S\cdot q)(P\cdot l) -(P\cdot q)(S\cdot l)\,] G_2
 \bigg \}
\ee
and  $L_{\alpha\beta}^{\rm S}W^{\alpha\beta \rm S}$ given in 
\eqref{eq:cont_symm}. The difference of cross sections with nucleons
of opposite polarizations will single out only the antisymmetric part
of the leptonic and hadronic tensors, namely the  second term in
\eqref{eq:sum_as}. For a longitudinally polarized nucleon (that is polarized
along the incoming electron direction), with the kinematics 
specified in \eqref{eq:kinema},
the spin vector reads
\be
S = (0,0,0,1)
\ee 
and the polarized cross section is given by
\be
\frac{\der \Delta \sigma}{\der E'\der \Omega} & = &
\frac{1}{2}\,\bigg[\,\frac{\der \sigma_{+}}{\der E'\der \Omega}-
\frac{\der \sigma_{-}}{\der E'\der \Omega}\,\bigg ]\nonumber \\
 &=&- \frac{2\alpha^2E'}{E Q^2}\,
[\,(E+E'\cos\theta)m G_1 -Q^2G_2\,],
\label{eq:polDIS}
\ee
with the subscripts $\pm$ meaning $\pm S$. Taking the sum instead of 
the difference in the first line of \eqref{eq:polDIS}, one recovers the result
\eqref{eq:DIS_unpol}  for the unpolarized cross section.

Similarly to the unpolarized case, one introduces the structure functions
\be
g_1(x_B, Q^2) & = & m^2\nu G_1(P\cdot q, q^2), \label{eq:g1}  \\
g_2(x_B, Q^2) & = & m\nu^2 G_2(P\cdot q, q^2), \label{eq:g2}
\ee 
which are observed to approximately scale in the deep inelastic region,
and  the projectors \cite{anselmino}
\be
{\cal{P}}^{\alpha\beta}_3 & = &\frac{1}{2\pi e^2}\,
\frac{(P\cdot q)^2}{b m^2(q\cdot S)}\,[(q\cdot S)S_{\lambda} + q_{\lambda} ] 
P_{\eta}
\varepsilon^{\alpha\beta\lambda\eta}\nonumber \\
{\cal{P}}_4^{\alpha\beta}& =& \frac{1}{2\pi e^2b}\bigg \{ \bigg [ 
\frac{(P\cdot q)^2}{m^2} + 2 (P\cdot q)x_B\bigg ] S_{\lambda} + 
(q\cdot S)q_{\lambda} \bigg \} P_{\eta}\varepsilon^{\alpha\beta\lambda\eta}, 
\label{eq:proj_pol} 
\ee
with
\be
b = -4m\bigg [\frac{(P\cdot q)^2}{m^2} + 2(P\cdot q)x_B - (q\cdot S)^2 \bigg ],
\label{eq:b}
\ee
such that
\be
{\cal{P}}^{\alpha\beta}_3W_{\alpha\beta} &= &g_2
\label{eq:proj_g1}
\ee
and
\be
{\cal{P}}_4^{\alpha\beta}W_{\alpha\beta}& =& g_1 + g_2.
\label{eq:proj_g2}
\ee

\section{Parton Model}
\label{sec:parton}
 
In the parton model the nucleon is considered to be made of collinear,
free constituents, each carrying a fraction $\xi$ of the nucleon
four-momentum: the quarks and antiquarks. Here we limit the discussion
only to quarks, the extension to antiquarks being straightforward.
The cross section of deep inelastic 
scattering is then described as the incoherent sum of all the 
electron-quark cross sections $\der\hat\sigma$:
\be
\der\sigma = \sum_{q,s}\int_0^1 \der \xi \,q(\xi, s; S)\der\hat\sigma~,
\label{eq:DISparton}
\ee
where $q(\xi,s;S)$ is the number density of quarks $q$,  with charge $e_q$
in units of $e$,
four-momentum fraction $\xi$ and spin $s$ inside a nucleon with spin $S$
and four-momentum $P$.
The cross section $\der\hat \sigma$ refers to the electron-quark
scattering  subprocess
\be
e(l)+q(\xi P)\rightarrow e(l')+q(k')
\ee
and, similarly to \eqref{eq:sigmael}, \eqref{eq:sigmael2}, after integration
over the struck quark phase space, reads
\be
\der\hat\sigma = \frac{1}{4\xi l\cdot P}\,\frac{1}{Q^4}
\,L^{\alpha\beta}w_{\alpha\beta}
\,\delta((\xi P   + q)^2)\,\frac{\der^3\mbox{\boldmath $l$}'}{(2\pi)^2\,2l'_0}~,
\ee
where $q= l-l'$, $L_{\alpha\beta}$ is given in \eqref{eq:leptonic} and the
quark tensor $w_{\alpha\beta} = w_{\alpha\beta}(\xi, q, s)$ is the same as 
the leptonic tensor 
$L_{\alpha\beta}$, with the replacements
$l^{\alpha} \rightarrow \xi P^{\alpha}$, 
$l'^{\alpha}\rightarrow \xi P^{\alpha} + q^{\alpha}$. That is
\be
w_{\alpha\beta}(\xi, q, s) = w^{\rm{S}}_{\alpha\beta}(\xi, q) 
+ w^{\rm{A}}_{\alpha\beta}(\xi, q, s),
\label{eq:had2}  
\ee
with
\be
w^{\rm{S}}_{\alpha\beta}(\xi, q) & = &  
2e^2[2 \xi^2P_{\alpha}P_{\beta} + \xi P_{\alpha}q_{\beta} 
+ \xi q_{\alpha}P_{\beta} -\xi(P\cdot q)g_{\alpha\beta} ]\nonumber \\
w^{\rm A}_{\alpha\beta}(\xi, q, s) & = & - 2ie^2m_q
\varepsilon_{\alpha\beta\rho\sigma}s^{\rho}q^{\sigma}
\label{eq:had3}
\ee
and the quark mass  for consistency is taken to be $m_q = \xi m$,
before and after the interaction with the virtual photon.
By comparison of \eqref{eq:DISparton} with \eqref{eq:DIScross}, and 
using the relation 
\be
\delta((\xi P + q)^2) \simeq \delta(-Q^2 + 2\xi P\cdot q) 
= \frac{1}{2P\cdot q}\,\delta(\xi -x_B),
\ee
one can express the hadronic tensor
$W_{\alpha\beta}$  in terms of the quark tensor $w_{\alpha\beta}$ as follows
\be
\frac{1}{2\pi}\,W_{\alpha\beta}(q;P,S) = \sum_{q, s}e^2_q\,\frac{1}{2P\cdot q} \int_0^1
\frac{\der \xi} {\xi}\,\delta (\xi-x_B)q(\xi,s;S)w_{\alpha\beta}(\xi,q,s).
\label{eq:had1}
\ee
From \eqref{eq:proj_unp}-\eqref{eq:projf2} and 
\eqref{eq:had3}-\eqref{eq:had1}, one obtains the parton model
predictions for the unpolarized  structure functions:
\be
F_1(x_B) = \frac{1}{2}\sum_q e^2_q\, q(x_B),
\label{eq:f1par}
\ee
and
\be 
F_2(x_B) =
x_B \sum_qe^2_q \,q(x_B) = 2 x_B \,F_1(x_B).
\label{eq:f2par}
\ee
where the unpolarized quark number densities $q(x_B)$ are defined as 
\be
q(x_B) = \sum_s q(x_B, s; S).
\label{eq:unpoldistr}
\ee
From \eqref{eq:f1par} and \eqref{eq:f2par}, the Callan-Gross relation 
\cite{callan} follows
\be
F_L(x_B) = F_2(x_B) -2x_BF_1(x_B) = 0, 
\label{eq:callan}
\ee
and it turns out that a mesurement of the structure function 
$F_{2}(x_B)$ allows  us to 
determine the momentum distributions of partons in the nucleon. 

From \eqref{eq:proj_pol}-\eqref{eq:proj_g2} and 
\eqref{eq:had3}-\eqref{eq:had1}, 
the polarized nucleon structure functions are obtained:
\be
g_1(x_B) &=&\frac{1}{2} \sum_q e^2_q\, \Delta q(x_B),  \\
g_2(x_B) & =& 0,
\ee
where 
\be
\Delta q (x_B) = q(x_B, S; S) - q(x_B, -S; S)
\label{eq:pol_distr}
\ee
is the difference between the number densities of quarks with spin
parallel to the nucleon ($s = S$) and those with spin anti-parallel 
($s =-S$).  Fixing the nucleon to be  longitudinally
polarized with positive helicity, \eqref{eq:unpoldistr} and
\eqref{eq:pol_distr} can be rewritten in terms of
parton densities with definite helicity, with notation analogous to
\eqref{eq:helicity1} and \eqref{eq:helicity2}:  
\be
q(x_B) = q^+_{+}(x_B) +  q^+_{-}(x_B), ~~~~~~~~~\Delta q(x_B) = q^+_{+}(x_B) -  q^+_{-}(x_B),
\label{eq:quark_dis}
\ee
where $\Delta q (x_B)$ measures how much the parton $q$ ``remembers'' of  its
parent nucleon polarization.
    
The structure function  $g_1(x_B)$ yields information 
  on how the helicity of the  nucleon is 
distributed among its parton constituents, while $g_2(x_B)$ has not
a simple interpretation in the parton model. It can be shown 
\cite{anselmino} that, if we allow the partons to have some 
transverse momentum ${\mbox{\boldmath $k$}_{\perp}}$
inside the nucleon, then  $g_2(x_B)$ is non-zero. However, it cannot
be calculated without making some model of the ${\mbox{\boldmath $k$}_{\perp}}$
distribution.

\section{Structure Functions in QCD}
\label{sec:qcd}

The parton model is only the zero-th order approximation to  the real 
world: quarks and antiquarks  are not free particles,  
they interact by emitting and absorbing gluons. 
A detailed discussion of QCD,  the  theory 
which describes the strong intractions of quarks and gluons, can be found in
\cite{reyaqcd,Leader:1996hk,Ellis:1991qj}.

From an empirical point of view, one observes that the scaling predicted
by the parton model is violated. Structure functions appear to depend on $Q^2$,
although in a relatively mild way, logarithmically. This behaviour
arises from perturbative QCD and represents the original, and still one of 
the most powerful, quantitative test of the theory.
The radiation of gluons  produces the $Q^2$-evolution
of the quark (and antiquark) distributions in \eqref{eq:quark_dis},  
furthermore it
determines the appearence of the unpolarized and polarized gluon 
distributions, defined in a way similar  to \eqref{eq:quark_dis},
\be
g(x_B, Q^2)& =& g^+_+(x_B, Q^2) +g^+_-(x_B, Q^2), \nonumber \\
\Delta g(x_B, Q^2) &= &g^+_+(x_B, Q^2) -g^+_-(x_B, Q^2);
\label{eq:gluon_dis}
\ee
$g_{+}(x_B, Q^2)$ and $g_{-}(x_B, Q^2)$ 
being the  densities associated to the positive and negative circular
polarization states of the massless, spin-1 gluon. 
Moreover, Similarly to QED,  it is possible to define
an effective ``fine-structure constant'' for QCD, 
\be
\alpha_s = \frac{g_s^2}{4\pi}~,
\ee  
with $g_s$ being the strong coupling. 
Furthermore it  is  convenient to introduce
the dimensional parameter $\Lambda$, because it provides a description
of the dependence of $\alpha_s$ on the renormalization scale 
(in DIS usually identified with the scale of the probe $Q$). The
de\-fi\-ni\-tion of $\Lambda$ is arbitrary; one possibility is to write
$\alpha_s$ as an expansion in inverse powers of $\ln Q^2/\Lambda^2$, 
\be
\frac{\alpha_s(Q^2)}{4\pi} = \frac{1}{\beta_0\ln{Q^2}
/{\Lambda^2}} -\frac{\beta_1}{\beta_0^3} \,\frac{\ln\ln Q^2/\Lambda^2}
{(\ln Q^2/\Lambda^2)^2} + {\cal{O}}\bigg ( \frac{1}{\ln^3(Q^2/\Lambda^2) } 
\bigg ),
\label{eq:alphas}
\ee  
where $\beta_0= 11 -2\,n_f/3$, $\beta_1  = 102 -38\,n_f/3$ and $n_f$ is
the number of quarks with mass less than the momentum scale $Q$.
Equation \eqref{eq:alphas}  illustrates the {\em asymptotic freedom} property:
$\alpha_s\rightarrow 0$ as $Q^2\rightarrow \infty$ and shows that QCD
becomes strongly coupled at $Q\sim \Lambda$. Therefore perturbative 
calculations (and the parton model) are reliable only for large momentum 
transfer. 
The value of $\Lambda$ depends on the renormalization scheme adopted and
must be determined from experiment.

Once the parton distributions are fixed at a specific input scale 
$Q^2 = Q_0^2$, mainly by experiment, their evolution to any $Q^2> Q_0^2$
is predicted by perturbative QCD. If we define, in the unpolarized sector, 
the flavor nonsinglet distributions
\be
  q_{\rm NS\, -} &= &  u -   \bar{u},~~~d  -   \bar{d},\nonumber \\
  q_{\rm NS\, +} &= &  (u +  \bar{u})-(d + \bar d),~~~(u + \bar u) + 
 (d + \bar d) - 2(s+\bar s),
\label{eq:ns}
\ee
and the  singlet combination 
\be
\Sigma = \sum_{q=u, d, s} ( \,q  +   \bar q\,), 
\ee
at NLO QCD the evolution equations take the form
\be
\frac{\der}{\der t}\,   q_{{\rm NS}\,\pm}(x_B, Q^2) = 
  P_{\rm NS\,\pm}
\otimes   q_{\rm NS \,\pm} ~,
\label{eq:nonsingl}
\ee
and
\begin{figure}[ht] 
\begin{center} 
\epsfig{figure=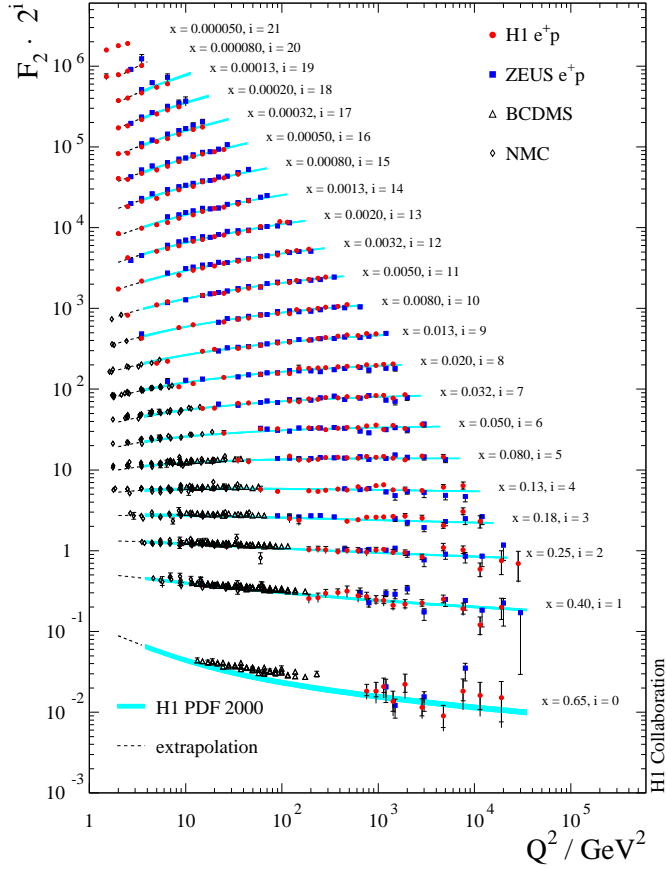, width= 9.cm} 
\caption{Measurements of the proton structure function $F_2(x_B, Q^2)$
from HERA col\-la\-bo\-rations and fixed target experiment. The data are shown
as a function of $Q^2$ for various fixed values of $x_B$ (denoted as $x$ in
the plot). Curves represent a NLO QCD fit to the data.}
\label{f2data}
\end{center}
\end{figure} 
\begin{figure}[ht] 
\begin{center} 
\epsfig{figure=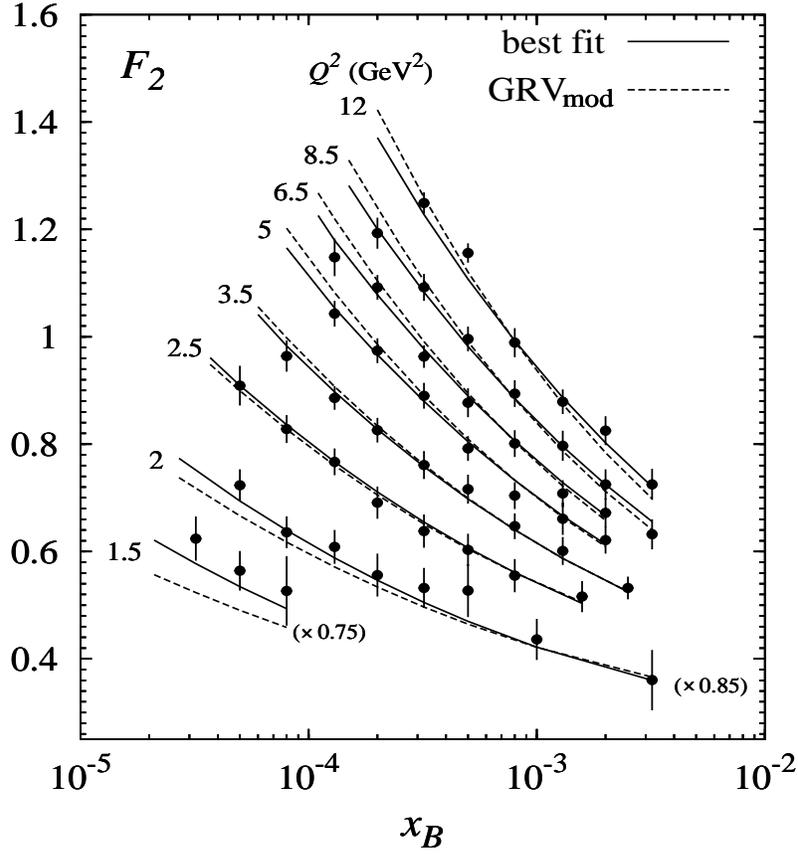, width= 12cm, height = 11cm} 
\caption{Comparison of two NLO QCD fits, corresponding to different 
input distributions [8], with the measurements of the proton structure 
function $F_2$ at small $x_B$ [62]. 
The results and data for the bins in $Q^2 = 1.5$ GeV$^2$ and 2 GeV$^2$
have been multiplied by 0.75 and 0.85, respectively, as indicated.}
\label{fig:curv}
\end{center}
\end{figure} 
\begin{figure}[ht] 
\begin{center} 
\epsfig{figure=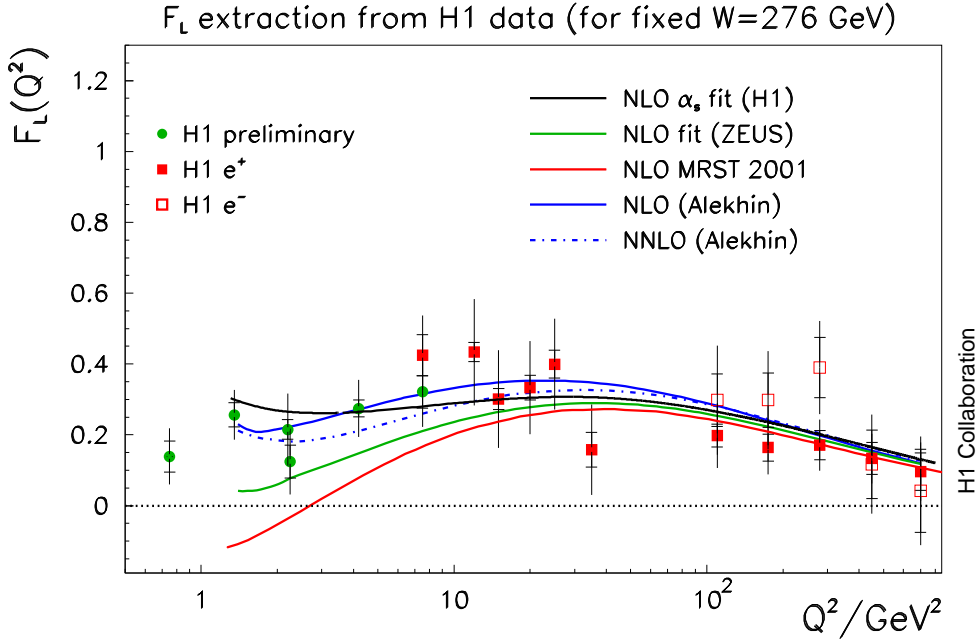, width= 13cm} 
\caption{Measurements of the proton structure function $F_L$ by the H1
collaboration at HERA at a fixed value of $W = 276$ GeV [65].}
\label{fl}
\end{center}
\end{figure} 
\begin{figure}[ht] 
\begin{center} 
\epsfig{figure=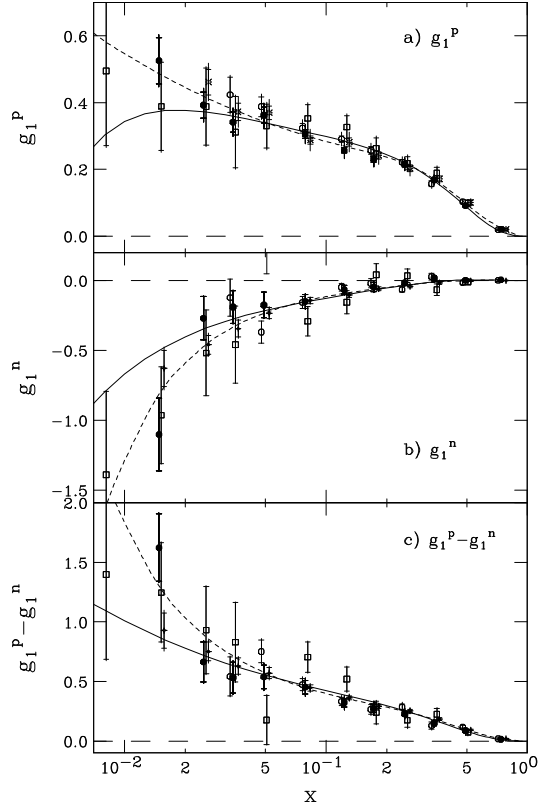, width= 7cm} 
\caption{Measurements of the nucleon structure function $g_1$ as 
a function of $x_B$ (denoted as $x$ in the plot) 
at $Q^2= 5$ GeV$^2$ [67]. The data are from the following experiments:
E155 (solid circles), E143 (open circles), SMC (squares), HERMES (stars)
and E154 (crosses). The solid curves correspond to a NLO QCD fit, while
the dashed curves are from a purely phenomenological fit.}
\label{g1data}
\end{center}
\end{figure} 
\be
\frac{\der}{\der t}
 \left( \begin{array}{c}   \Sigma(x_B, Q^2) \\    g(x_B, Q^2)
\end{array} \right  ) =   \hat{P} \otimes 
\left (\begin{array}{c}   \Sigma \\    g \end{array} \right),
\label{eq:singl}
\ee
where $t = \ln Q^2/Q^2_0$  and the convolution ($\otimes$) is defined by
\be
(  P\otimes   q)(x_B, Q^2) = \int_{x_B}^1\frac{\der y}{y}\,
  P\bigg 
(\frac{x_B}{y}\bigg )\,   q(y, Q^2).
\ee
The splitting functions are given by 
\be
   P_{\rm NS\,\pm} = \frac{\alpha_s(Q^2)}{2\pi} \, 
   P_{qq}^{(0)}(x_B)+ \bigg ( \frac{\alpha_s(Q^2)}{2\pi}\bigg )^2
  P^{(1)}_{{\rm NS \,\pm}}(x_B)
\label{eq:pns}
\ee 
and 
\be
  \hat P = \frac{\alpha_s(Q^2)}{2\pi} \,  \hat P^{(0)}(x_B) +
\bigg (\frac{\alpha_s(Q^2)}{2\pi} \bigg )^2    \hat P^{(1)}(x_B),
\label{eq:p}
\ee
where
\be
  \hat P^{(j)}(x_B) = \left( \begin{array}{cc} 
  P^{(j)}_{qq}  & 2 n_f P^{(j)}_{qg} \\  
  P^{(j)}_{gq}  &    P^{(j)}_{gg}   \end{array}\right ),~~~~
~~{\rm with }~~j = 0, 1;  
\ee 
 $\alpha_s(Q^2)$ being given in \eqref{eq:alphas}. The LO expressions are 
entailed in \eqref{eq:alphas} and in \eqref{eq:pns}-\eqref{eq:p}; they can be 
obtained  by simply dropping all higher order terms 
($ \beta_1$, $P^{(1)}$, $\hat{P}^{(1)}$). At  LO \eqref{eq:nonsingl},
\eqref{eq:singl}    reduce to the well-known DGLAP evolution 
equations \cite{dokshitzer,lipatov,lipatov2,altarelli}. The 
complete set of NLO splitting functions in the commonly used
 $\overline{\rm MS}$
factorization scheme has been calculated  \cite{curci,furpetr,floratos}
and can be found, for example, in \cite{Ellis:1991qj}.

Equations \eqref{eq:nonsingl}-\eqref{eq:p} hold also in the polarized 
sector, with the formal replacements $q\rightarrow \Delta q$ and 
$P\rightarrow \Delta P$. The polarized splitting functions are known
up to NLO in the  $\overline{\rm MS}$ scheme \cite{vanneerven,vogelsang1,vogelsang2}  and are listed in \cite{reya}.

The resulting NLO parton distributions are directly 
related to  physical quantities, such as structure functions, by a 
convolution with calculable, process dependent, coefficient functions.
For consistency, the choice of the factorization convention must be
the same for both the coefficient functions and the splitting functions
underlying the parton distributions.
Within the $\overline{ \rm MS}$ scheme, $F_2(x_B, Q^2)$ is given by
\be
\frac{1}{x_B}\,F_2(x_B, Q^2) & =& \,\sum_q\,e^2_q 
\bigg \{ q(x_B, Q^2) + \bar{q}(x_B, Q^2) \nonumber \\
&&~~~~~~~+ \frac{\alpha_s(Q^2)}{2\pi}\, \bigg [C_q\otimes (q+\bar{q}) + 
2 C_{g} \otimes g\,\bigg ]\bigg \},
\label{eq:f2nlo}  
\ee
where the unpolarized  coefficients functions $C_{q, g}$  are given, 
for example, in 
\cite{petronzio} and  in \cite{grv95}. Measurements of the proton structure 
function $F_2$ together with a NLO QCD fit, presented in \cite{f2data}, 
are shown in Figure \ref{f2data}. As one can see, QCD appears to 
predict correctly the $Q^2$ dependence of structure function over
four orders of magnitude.

Recently, a dedicated test of the validity
of \eqref{eq:f2nlo} at very low $x_B$ 
in the perturbative regime, $Q^2  \gtrsim 1$ ${\rm GeV}^2$,   has been 
performed \cite{curv} and the results are shown in Figure \ref{fig:curv}.
A good agreement with recent precision  data
for  $F_2$ \cite{h1curv}, restricted to
\be
3\times 10^{-5}\lesssim x_B \lesssim 3\times 10^{-3},~~~~~~~~~~
1.5 ~ {\rm GeV}^2 \le Q^2 \le 12 ~ {\rm GeV}^2,
\ee
has been found, as well as with the present experimental determination 
of the curvature of $F_2$ \cite{haidt}.

The structure function $F_L(x_B, Q^2)$
 is only non-zero at  order $\alpha_s$ in perturbation theory, i.e.
$F_L (x_B, Q^2)= {\cal{O}}(\alpha_s)$; deviations from the Callan-Gross
relation \eqref{eq:callan} are evident in Figure \ref{fl}, taken from 
\cite{fldata}. The data are much less precise than the ones 
for $F_2(x_B, Q^2)$, but QCD seems to work well also in this case.

Similarly to $F_2(x_B, Q^2)$, $g_1(x_B, Q^2)$ can be written as
\be
g_1(x_B, Q^2) &= &\frac{1}{2}\,\sum_q\,e^2_q \,\bigg \{\Delta  q(x_B, Q^2) + 
\Delta  \bar{q}(x_B, Q^2) \nonumber \\
&&~~~~~~+ \frac{\alpha_s(Q^2)}{2\pi} \bigg [\Delta  C_q\otimes
(\Delta  q + \Delta  \bar{q}) + 2\Delta  C_g\otimes \Delta  g \bigg ]
\bigg \},
\label{eq:g1nlo}
\ee
where  $\Delta C_{q, g}$ can be found in  \cite{gluck}.
In Figure \ref{g1data} results on $g_1$  presented in \cite{anthony}
are shown as a function of 
$x_B$ at $Q^2 = 5$ ${\rm GeV}^2$.

At  LO  the coefficients $(\Delta)C_{q, g}$
vanish; from \eqref{eq:f2nlo} and \eqref{eq:g1nlo} it turns out
that  the gluon  does {\em not directly} 
contribute to the structure functions, but only indirectly via the 
$Q^2$-evolution equations. 
The sums in \eqref{eq:f2nlo} and in \eqref{eq:g1nlo}  usually run over the 
light quark-flavors $q = u$, $d$, $s$,
since the heavy quark contributions ($c$, $b$, ...)  have preferrably
to be calculated perturbatively from the intrisic light quarks 
($u$, $d$, $s$) and gluon ($g$) partonic constituents of the nucleon.
This treatment of the heavy quarks underlies the 
unpolarized GRV98 
\cite{grv98} and polarized GRSV01 \cite{grsv}
parton distributions, which will be used in our numerical estimates
in the perturbative regime.
For our studies in the low-$Q^2$ region, where perturbative QCD is not
applicable, we resort to the purely phenomenological 
parametrization ALLM97 of $F_2$ \cite{allm97} and to the parametrization BKZ
of $g_1$ \cite{bad}, shortly described in Chapters \ref{ch:qedcs} and 7
respectively.  

To conclude, it has been shown that 
the inclusion of ${\cal{O}(\alpha)}$ QED corrections to the parton evolution
modifies only very slightly  equations \eqref{eq:nonsingl}-\eqref{eq:singl}
\cite{Spiesberger:1994dm,Roth:2004ti,Martin:2004dh}, so we will
not  consider such effect. We will concentrate on the other 
consequence \cite{Roth:2004ti,Martin:2004dh} of the emission of photons
from quarks, the appearence of the (inelastic)  photon  
distributions of  the proton and the neutron.

\clearemptydoublepage

\chapter{{\bf The Equivalent Photon Distributions of the Nucleon}} 


In this chapter the polarized and unpolarized photon content of protons and 
neutrons,  evaluated in the equivalent photon approximation,  
are presented. 
In particular, the universal and process independent
 elastic photon components turn out to be uniquely determined by the well-known
 electromagnetic form factors of the nucleon and their 
derivation  is shown in detail. 
The inelastic photon components  
are obtained from the corresponding momentum  
evolution equations subject to the boundary  
conditions of their vanishing at some low momentum  
scale.  The resulting photon asymmetries, important  
for estimating cross section asymmetries in photon-induced 
subprocesses are also presented for some  
typical relevant momentum scales.

\section{Unpolarized Photon Distributions}

As already mentioned in the Introduction, the concept of the photon 
content of (charged)  
fermions is based on the EPA, the equivalent photon  approximation.
Applied to the nucleon $N=p,\, n$ it consists of two  
parts, an elastic one due to $N\to \gamma N$ and  
an inelastic part due to $N\to\gamma X$ with  
$X\neq N$.  Accordingly the total photon distribution 
of the nucleon is given by  
\begin{equation} 
\gamma(x,\mu^2) = \gamma_{\mathrm{el}}(x) + \gamma_{\mathrm{inel}}(x,\mu^2),
\label{eqone} 
\end{equation} 
where $x$ is the fraction of the nucleon energy carried by 
the photon and     
$\mu$ is a momentum scale of the photon-induced subprocess. 
The two components are discussed separately 
in the following.

\subsection{Elastic Component}
The elastic photon distribution of the proton,  
$\gamma_{\mathrm{el}}^p$, has been presented in \cite{kniehl} 
and can be generally written as 
\begin{equation} 
\gamma_{\mathrm{el}}(x)=-\frac{\alpha}{2\pi}  
\int_{-\infty}^{-\frac{m^2x^2}{1-x}} \frac{\der t}{t} 
 \left\{ \left[2\left(\frac{1}{x}-1\right) 
  +\frac{2m^2x}{t}\right] H_1(t)+ xG_M^2(t)\right\}, 
\label{eqtwo}
\end{equation} 
where
\begin{equation} 
H_1(t)\equiv F_1^2(t)+\tau F_2^2(t) = 
  \frac{G_E^2(t) +\tau G_M^2(t)}{1+\tau}
\label{eq:h1} 
\end{equation} 
with $\tau\equiv -t/4m^2$, $m$ being the nucleon mass, 
and where $G_E$ and $G_M$ are  
the Sachs elastic  form factors  discussed in Section
\ref{sec:formfact}. 


The result (\ref{eqtwo}) can be obtained  extending to 
an unpolarized nucleon $N$ the analysis \cite{ref8} for a photon emitting 
unpolarized electron.
We consider the  process
\begin{equation}
N(P) + a(l) \rightarrow N(P') + X,
\end{equation}
where the target $a$ is a massless parton ($l^2  = 0$),  $P^2 = P'^2 = m^2$,  
and  $X$ is a generic hadronic system. 
The corresponding cross section can be written as
\begin{equation}
\der\sigma_{Na}(P,l) = \frac{1}{4 P\cdot l}\,\frac{1}{q^4}\,
H_{\mathrm{el}}^{\alpha\beta}(P;P')W_{\alpha\beta}(k,l)\, 
\frac{\der^3{\mbox{\boldmath$P$}}'}{(2\pi)^32E'}~,
\label{eq:crsec}
\end{equation}
where 
\begin{equation}
q \equiv -k= P'-P,
\label{eq:q}
\end{equation}
$k$ being the four-momentum of photon emitted by the nucleon.
The tensor $H_{\mathrm{el}}^{\alpha\beta}(P;P')$  is
defined in (\ref{eq:hadronic}) and, using \eqref{eq:sachs}, (\ref{eq:h1}) and
(\ref{eq:q}), can be rewritten in a more compact form as
\begin{eqnarray}
H^{\alpha\beta}_{\mathrm{el}}(P;P')= e^2\,[ H_1 (q^2) (2 P-k)^\alpha(2 P-k)^\beta + G_M^2(q^2) (q^2 g^{\alpha\beta}-k^\alpha k^\beta)].
\label{eq:hel}
\end{eqnarray}
The partonic tensor $W_{\alpha\beta}(k,l)$, 
as shown in  \eqref{eq:hadron_symm}, 
can be decomposed as 
\begin{eqnarray}
W_{\alpha\beta}(k,l) &= &4\pi e^2 \bigg [\bigg (-g_{\alpha\beta} + 
\frac{1}{q^2} \,k_\alpha k_\beta \bigg ) F_1(q^2, k\cdot l) \nonumber \\
& &-\frac{1}{(k\cdot l)} \,\bigg (l_\alpha -
\frac{k\cdot l}{q^2}\,k_{\alpha}\bigg )\bigg (l_\beta -
\frac{k\cdot l}{q^2}\,k_{\beta}\bigg ) F_2(q^2, k\cdot l) \bigg ],
\label{eq:hinel}
\end{eqnarray}
with the structure functions $F_1$ and $F_2$ defined in \eqref{eq:f1} 
and \eqref{eq:f2}. 
As pointed out in \cite{ref8}, in the limit $q^2\rightarrow 0$, 
$W_{\alpha\beta}(k,l)$ must be an
analytic function of $q^2$, therefore  $q^2 W_{\alpha\beta}$ has to
vanish for $q^2=0$. This implies  
\begin{equation}
\frac{(k\cdot l)}{q^2}\,F_2(q^2, k\cdot l) = F_1(0, k\cdot l) + {\cal{O}}(q^2),
\label{eq:w2}
\end{equation} 
and the terms ${\cal{O}}(q^2)$ will not be considered in the following.
Furthermore, one can introduce the variable 
\begin{equation}
x = \frac{k\cdot l  }{P\cdot l}~,
\label{eq:xg}
\end{equation}
which represents the fraction  of  
longitudinal momentum  carried by the emitted  photon, 
assumed to be real and collinear with the parent nucleon, that is
\be
k = x P. 
\label{eq:kappa}
\ee   
From (\ref{eq:hel})-(\ref{eq:xg}) one finds
\begin{eqnarray}
H_{\mathrm{el}}^{\alpha\beta}(P;P')W_{\alpha\beta}(k,l) = 4\pi e^4 \bigg \{ 4 H_1(q^2) 
\bigg [ -m^2+ \frac{q^2}{x} -\frac{q^2}{x^2}  \bigg ]
-2 q^2G_M^2(q^2)\bigg \} F_1(0, l\cdot k). \nonumber\\
\end{eqnarray}
If we write the four-momenta of the incoming and outgoing nucleons as 
\begin{equation}
P = (E, 0, 0, E\beta),~~~~~~~P' =  (E', 0, E'\beta'\sin\theta, 
E'\beta'\cos\theta),
\end{equation}
with
\begin{equation}
\beta= \sqrt{1-\frac{m^2}{E^2}}~,~~~~~~~~~~~\beta' = 
\sqrt{1-\frac{m^2}{E'^2}}~,
\end{equation}
then the phase space for the scattered nucleon will be given by
\begin{equation}
\frac{\der^3{\mbox{\boldmath$P$}}'}{E'} =  2\pi\beta'^2E'\der P'\der\cos\theta,
\label{eq:phase}
\end{equation}
where the azimuth integration has already been carried out. The
integration  variables ($P'$, $\cos\theta$) can be replaced by 
($q^2$, $x$);  using the following relations 
\begin{eqnarray}
q^2 & = & 2 m^2 - 2 E E'(1-\beta\beta'\cos\theta), \label{eq:qsq} \\
x & = & 1 - \frac{E'(1+\beta'\cos\theta)}{E(1+\beta)}~,\label{eq:ics}
\end{eqnarray}
it can be proved that the Jacobian of this change of variables is 
$2E'\beta'^2$ and (\ref{eq:phase}) becomes
\begin{equation}
\frac{\der^3{\mbox{\boldmath$P$}}'}{E'} = \pi \der q^2\der x.
\end{equation} 
The cross section  (\ref{eq:crsec})  can now be written as 
\begin{equation}
\der\sigma_{N a} (P,l)= -\frac{\alpha}{2\pi} \,\frac{1}{t} 
\bigg \{ \bigg [\frac{2m^2x}{t} + 2\bigg  ( \frac{1}{x} -1\bigg )\bigg ]\,
H_1(t) + x G_M^2(t)\bigg \} \,\sigma_{\gamma a}(k,l)\, \der t\, \der x,
\end{equation}
where $t\equiv q^2$ and
\be
\sigma_{\gamma a} (k,l)= -\frac{1}{8\,k\cdot l} \, g_{\alpha\beta}  
W^{\alpha\beta}
(k,l) = \frac{\pi e^2}{\,k\cdot l} \,F_1(0, k\cdot l) 
\ee
is the cross section for the process $\gamma (k) + N(P) \rightarrow X$
for a real photon. 
Integrating over $t$ one gets
\begin{equation}
\der\sigma_{N a} (P,l)=  \gamma_{\mathrm{el}}(x)\,\sigma_{\gamma a} (k,l)\,
\der x,
\label{eq:unpol}
\end{equation}
where $\gamma_{\mathrm{el}}(x)$ is given by
\begin{equation}
\gamma_{\mathrm{el}}(x)=-\frac{\alpha}{2\pi}  
\int_{t_{\mathrm{min}}}^{t_{\mathrm{max}}} \frac{\der t}{t} 
 \left\{ \left[2\left(\frac{1}{x}-1\right) 
  +\frac{2m^2x}{t}\right] H_1(t)+ xG_M^2(t)\right\}.
\label{eq:el2}
\end{equation}
 The extrema of integration can be determined from (\ref{eq:qsq}) and 
(\ref{eq:ics}); from the latter we have
\begin{equation}
E'  = \frac{A}{1-\cos^2\theta}\,\bigg ( 1 - 
        \sqrt{ 1 -\frac{(A^2 + m^2\cos^2\theta)(1-\cos^2\theta)}{A^2}} ~\bigg ),\label{eq:ener}
\end{equation}
where
\begin{equation}
A   = E(1+\beta)(1-x).  
\end{equation}
Expanding (\ref{eq:ener})  in  powers of $\theta$, for $\theta \ll 1$, one gets
\cite{ref8} 
\begin{equation}
E' = \frac{A^2 +m^2}{2 A} + \frac{(A^2 -m^2)^2}{8A^3}\, 
      \theta^2 + {\cal {O}}(\theta^4),
\end{equation}
and (\ref{eq:qsq}) becomes
\begin{equation}
q^2 = -\frac{m^2x^2}{1-x}-\frac{E(1+\beta)(A^2-m^2)^2}{4 A^3}\,\theta^2 + 
{\cal{O}}(\theta^4).
\end{equation}
The value $t_{\mathrm{max}}= q^2_{\mathrm{max}}$ is obtained by taking 
$\theta = 0$, namely 
\begin{equation}
t_{\mathrm{max}} = -\frac{m^2x^2}{1-x}~.
\label{eq:tmaximum}
\end{equation}
The minimum value of the photon virtuality $t_{\mathrm{min}}$ can be 
computed in a
simple way by imposing that the invariant mass $W^2$ of  the produced hadronic
system  $X$ be bounded from below \cite{ref8}:
\begin{equation}
W^2 \equiv (k+l)^2 > W^2_{\mathrm{min}},
\end{equation}
from which it comes, for $W^2_{\mathrm{min}} = m^2$,
\begin{equation}
t_{\mathrm{min}} = -2 k\cdot l + m^2 = -2 x P\cdot l+m^2 = - x (s-m^2)
+ m^2,
\end{equation}
where $s$ is the nucleon-parton centre-of-mass energy squared. 
 
The Sachs form factors which appear in  (\ref{eq:el2}) are   
conveniently parametrized by the dipole  
form proportional to $(1-t/0.71$ GeV$^2)^{-2}$ as 
extracted from experiment, see \eqref{eq:dipole}. 
As pointed out in \cite{kniehl}, this implies 
that the support from values $t\ll -0.71$ $\mathrm{GeV}^2$ to the integral
in  (\ref{eq:el2}) is suppressed, hence, in the kinematical region
$x s \gg m^2$, one may integrate from $t_{\mathrm{min}} = -\infty$ to
$t_{\max} = -m^2x^2/(1-x)$ so as to obtain the universal process independent
$\gamma_{\mathrm{el}}$ in (\ref{eqtwo}) .        

Equation (\ref{eqtwo}) can now be  
analytically integrated, remembering that, from \eqref{eq:dipole} and 
\eqref{eq:h1}, for the proton we have 
\begin{equation} 
G_E^p(t)=(1+a\tau)^{-2}\,,\quad\quad 
 G_M^p(t)\simeq \mu_p G_E^p(t)\,,\quad\quad 
  H_1^p(t)=\frac{1+\mu_p^2\tau}{1+\tau}(1+a\tau)^{-4} 
\label{eqfour}
\end{equation} 
with $\mu_p=1+\kappa_p\simeq 2.79$ and  
$a\equiv 4m^2/0.71$ GeV$^2\simeq 4.96$, while for  the  
neutron 
\begin{equation} 
G_E^n(t) = \kappa_n\tau (1+a\tau)^{-2}\,, \quad\quad 
 G_M^n(t)=\kappa_n(1+a\tau)^{-2}\,, \quad\quad 
  H_1^n(t)=\kappa_n^2\tau(1+a\tau)^{-4}\,, 
\label{eqfive} 
\end{equation} 
with $\kappa_n \simeq -1.79$.
After integration, the elastic component of the equivalent photon 
distribution reads \cite{gpr1}, for the proton 
\begin{eqnarray} 
\gamma_{\mathrm{el}}^p(x)& =& \frac{\alpha}{2\pi}\, 
 \frac{2}{x} \left\{  
  \left[ 1-x +\frac{x^2}{4}(1+4a+\mu_p^2)\right]  
   I + (\mu_p^2-1)  
    \left[1-x+\frac{x^2}{4}\right] \tilde{I} - 
      \frac{1-x}{z^3} \right\},\nonumber \\ 
\label{eqsix}
\end{eqnarray} 
and for the neutron 
\begin{equation} 
\gamma_{\mathrm{el}}^n(x) = \frac{\alpha}{2\pi}\,\kappa_n^2 
 \frac{x}{2}  
  \left\{ I +\frac{1}{3}\,\frac{1}{(z-1)z^3}\right\},
\label{eq7}
\end{equation} 
where $z\equiv 1+\frac{a}{4}\,\frac{x^2}{1-x}$ and 
\begin{eqnarray} 
I & = & \int_{\frac{x^2}{4(1-x)}}^{\infty} \der\tau 
 \frac{1}{\tau(1+a\tau)^4} = -\ln \left(1-\frac{1}{z}\right) 
  - \frac{1}{z}-\frac{1}{2z^2} -\frac{1}{3z^3}\label{eq8}~,\\ 
\tilde{I} & = & \int_{\frac{x^2}{4(1-x)}}^{\infty}\der\tau 
 \frac{1}{(1+\tau)(1+a\tau)^4} = -\frac{1}{a_-^4}\ln 
  \left(1+\frac{a_-}{z}\right) +\frac{1}{a_-^3z} - 
   \frac{1}{2a_-^2z^2} +\frac{1}{3a_-z^3}~, \label{eq9}\nonumber\\
\end{eqnarray} 
with $a_-=a-1$.  For arriving at  (\ref{eqsix}) we have  
utilized the relation 
\begin{displaymath} 
\int_{\frac{x^2}{4(1-x)}}^{\infty} \der\tau 
 \frac{1}{\tau^2(1+a\tau)^4} = -4aI +4\frac{1-x}{x^2z^3}~, 
\end{displaymath} 
which will be also relevant for the polarized photon 
contents to be presented below.  The result in  (\ref{eqsix}) agrees 
with the one presented in a somewhat different form in 
\cite{kniehl}. 

If we integrate \eqref{eq:unpol} taking $G_E^2 = G_M^2 = 1$ 
(pointlike particle), with 
the integration bounds \eqref{eq:tmaximum} and 
$t_{\rm min} \equiv -\mu^2$, 
we recover the result 
\eqref{eq:gamma_el} for the unpolarized photon content of the electron.

\subsection{Inelastic Component}

As pointed out in \cite{ruju}, the complete function 
$\gamma (x, \mu^2)$   in \eqref{eqone} could be built-up by adding to the 
elastic contribution all resonant  \cite{kessler} and non-resonant final 
hadronic states, and their interferences. Alternatively one could guess
an inclusive or 'continuous' $\gamma_{\mathrm{inel}}(x, \mu^2)$, based 
on the parton model, where the photon  is emitted by one of the quarks 
in the nucleon \cite{gsv}. In this latter picture, which we
adopt, the addition of the resonant and continuous contributions 
may be double-counting. The inelastic part in (\ref{eqone}) 
is then given by the leading order (LO) QED evolution equation \cite{gsv} 
\begin{equation} 
\frac{\der\gamma_{\mathrm{inel}}(x,\mu^2)}{\der\ln \mu^2} = 
 \frac{\alpha}{2\pi} \sum_{q=u,d,s} e_q^2 
  \int_x^1 \frac{dy}{y}\, P_{\gamma q} 
   \left(\frac{x}{y}\right)  
    \left[ q(y,\mu^2)+\bar{q}(y,\mu^2)\right], 
\label{eqten}
\end{equation} 
where 
\begin{equation}
P_{\gamma q}(y) = \frac{1+(1-y)^2}{y}
\end{equation}
is the quark-to-photon splitting function
and $q(y, \mu^2)$, $\bar{q}(y, \mu^2)$ are respectively the quark 
and antiquark distribution functions of the nucleon at LO QCD \cite{grv98},
with $\stackrel{\!\!\!(-)}{u^p} = \stackrel{\,(-)}{d^n}$, $\stackrel{\!\!\!(-)}{d^p} = \stackrel{\!\!\!(-)}{u^n}$,  $\stackrel{\!\!\!(-)}{s^p} =    
\stackrel{(-)}{s^n}$.  
Equation \eqref{eqten}, 
which states that the 
probability  to find a photon in the nucleon is given by the convolution of 
the probabilities to find first a quark inside the nucleon and then
a photon inside the quark, 
is integrated subject to the  `minimal' boundary condition
\begin{equation} 
\gamma_{\mathrm{inel}}(x,\mu_0^2)=0
\label{bound}
\end{equation}
at \cite{grv98} 
$\mu_0^2=0.26$ GeV$^2$.  The boundary condition \eqref{bound}   
is obviously not  
compelling and affords further theoretical and  
experimental studies.  Since for the time being 
there are no experimental measurements available, 
the  `minimal' boundary condition provides at  
present a rough estimate for the inelastic 
component at $\mu^2\gg \mu_0^2$.

\section{Polarized Photon Distributions}
 
The previous analysis can be extended  to the polarized sector, i.e., to 
\begin{equation} 
 \Delta\gamma(x,\mu^2)=\Delta\gamma_{\mathrm{el}}(x)+ 
  \Delta\gamma_{\mathrm{inel}}(x,\mu^2)\, .
\label{eqeleven} 
\end{equation} 
As before, the elastic and inelastic parts will be studied separately.

\subsection{Elastic Component}

The elastic part $\Delta\gamma_{\mathrm{el}}(x)$ in 
(\ref{eqeleven}) is determined via the antisymmetric part of the 
tensor describing the photon emitting nucleon $N$ 
\begin{equation} 
H_{\mathrm{el}}^{\alpha\beta}(P;P')= \frac{1}{2}\,e^2\,\mathrm{Tr} 
 \left[ 
(1+\gamma^5S\!\!\!/) 
  (P\!\!\!\!/+m)\Gamma^{\alpha}(P\!\!\!\!/\,'+m)\Gamma^{\beta} 
   \right]
\label{tensor} 
\end{equation} 
for the  process 
\begin{equation} 
  N(P;\, S)+a(l;\, s)\to  N({P'}) + X
\label{eq14} 
\end{equation} 
where $a$ being a parton with four-momentum $l$ initially kept 
off-shell and $S$, 
$s$ are the  polarization vectors \cite{flo}  
satisfying the transversality condition $S\cdot P=0$ and $s\cdot l=0$.
Equation tensor has been obtained in a way similar to \eqref{eq:pollepton},
with $\gamma^{\alpha} \to \Gamma^{\alpha}$.  
In terms 
of the Dirac and Pauli form factors $F_{1,2}(t)$  the elastic vertices 
$\Gamma^{\alpha}$ are 
given by 
\begin{equation} 
\Gamma^{\alpha}=(F_1+F_2)\gamma^{\alpha}-\frac{1}{2m} 
  F_2(P+{P'})^{\alpha}\,. 
\end{equation} 
The analysis has been carried out originally in \cite{gpr1} and is  a 
straightforward extension of 
the calculation \cite{flo} of the polarized 
equivalent photon distribution resulting from 
a photon emitting electron.  
The antisymmetric  part of \eqref{tensor}, as calculated 
in \cite{gpr1}, reads 
\begin{eqnarray} 
H_{\mathrm{el}}^{\alpha\beta {\rm A}}(P;P')&=& 2ie^2m\,G_M^2\,\varepsilon^
{\alpha\beta\rho\sigma} 
  S_{\rho}k_{\sigma}+2i\, G_M(F_2/2m)\nonumber \\ 
  &&~~~\times\,\left[(P+{P'})^{\alpha}\, 
   \varepsilon^{\beta\rho\sigma\sigma'} -(P+{P'})^{\beta} 
    \varepsilon^{\alpha\rho\sigma\sigma'}\right] 
     S_{\rho}P_{\sigma}{P'}_{\sigma'}
\label{eq:ant} 
\end{eqnarray} 
with, as before, $k=P-{P'}$.
One can show that
\begin{eqnarray}
\left [(P+P')^{\alpha}\varepsilon^{\beta\rho\sigma\sigma'} - 
  (P+P')^{\beta}\varepsilon^{\alpha\rho\sigma\sigma'} \right ] 
    S_{\rho}P_{\sigma}P'_{\sigma '} &= &(k\cdot S) P_\sigma k_{\sigma '}
          \varepsilon^{\beta\alpha\sigma\sigma'}~~~~~~~~~~~~~~~~~~~~\nonumber\\
           && + \,\,(2m^2 + P\cdot k)\varepsilon^{\beta\rho\alpha\sigma'}
              S_{\rho}k_{\sigma '},
\end{eqnarray}
where we made use of the $\varepsilon$-identity
\begin{equation}
g^{\alpha\mu} \varepsilon^{\beta\rho\sigma\sigma'} = 
   g^{\alpha\beta}\varepsilon^{\mu\rho\sigma\sigma'} + 
      g^{\alpha\rho}\varepsilon^{\beta\mu\sigma\sigma'} + 
        g^{\alpha\sigma}\varepsilon^{\beta\rho\mu\sigma'} +
          g^{\alpha\sigma'}\varepsilon^{\beta\rho\sigma\mu};
\end{equation}
hence (\ref{eq:ant}) can be rewritten in a more compact form \cite{ji} as
\be
H^{\alpha \beta {\rm A}}_{\mathrm{el}}(P;P')=-i e^2 m 
\varepsilon^{\alpha \beta \rho \si} 
 k_\rho \bigg [ 2 G_E G_M S_\si
-{G_M (G_M-G_E)\over  1+\tau} {k\cdot S\over m^2} P_\si \bigg ].
\label{poleltens}
\ee
The polarized cross section relative to the process (\ref{eq14}) is
given by
\begin{equation}
\der\Delta\sigma_{Na} (P, l) = \frac{1}{4\, P\cdot l}\,\frac{1}{q^4} 
\,H_{\mathrm{el}}^{\alpha\beta {\rm A}}(P;P')\,
W_{\alpha\beta}^{\rm A}(k,l)\,\frac{\der^3{\mbox{\boldmath$P$}}'}{(2 \pi)^3 2 E'}~,
\label{eq:pol}
\end{equation}
where $W_A^{\alpha\beta}$ is the antisymmetric part of the partonic tensor,
describing the polarized target $a(l;\,s)$, 
which is expressed in terms of the usual polarized structure functions 
$g_1$ and $g_2$, see \eqref{eq:polhadrtens} together with  
\eqref{eq:g1} and \eqref{eq:g2},
\be
W^{\rm A}_{\alpha \beta}(k,l)= 2i\pi e^2 \frac{ \sqrt{|l^2|}}{ k \cdot l} \,
\varepsilon_{\alpha \beta
\rho \sigma} k^\rho \bigg [ g_1(q^2, k \cdot l) s^\sigma +g_2 (q^2, k \cdot
l) \bigg (s^\sigma-{k \cdot s\over k \cdot l} \,l^{\sigma}  \bigg ) \bigg ]. 
\ee
The spin vectors for the incoming nucleon and parton can be written as 
\begin{equation}
S_{\alpha} = N_S \bigg (P_{\alpha} - \frac{m^2}{l\cdot P}\,l_{\alpha} \bigg ),
~~~~~s_{\alpha} = N_s \bigg (l_{\alpha} -\frac{l^2}{l\cdot P}
\,P_{\alpha}\bigg ),
\label{eq:spin}
\end{equation}
where the normalization factors  $N_S$ and $N_s$ are related to $S^2$
and $s^2$ by \cite{gabrid}
\begin{equation}
N^2_S = -\frac{S^2}{m^2\tilde{\beta}^2} \, ,~~~
N^2_s = -\frac{s^2}{l^2\tilde{\beta}^2} \, ,
\label{eq:normals}
\end{equation}
with 
\begin{equation}
\tilde{\beta} = \sqrt {1 - \frac {m^2 l^2}{(P\cdot l)^2}}\, ,
\label{eq:normals2}
\end{equation}
and   are fixed in order to satisfy the condition
$|S^2| = |s^2| = 1$. \\
Putting the parton on-shell ($l^2 = 0$) and using the definition 
(\ref{eq:xg}), together with (\ref{eq:spin}), we have  
\begin{eqnarray}
H_{\mathrm{el}}^{\alpha\beta {\rm A}}(P;P') W^{\rm A}_{\alpha\beta}(k,l) &=&  -4\pi e^4q^2 
  \bigg \{ \bigg [  \frac{2m^2x}{q^2} -1 +\frac{2}{x} \bigg ]\, G_M^2
   ~~~~~~~~~~~~~~~~~~~~~~~~~~~~~~~~~~\nonumber \\
&&~~~   -2\bigg [ \frac{m^2x}{q^2}-1 +\frac{1}{x}\bigg ] 
\frac{G_M(G_M-G_E)}{1 +  \tau} \bigg \}\,g_1(q^2, k\cdot l),
\end{eqnarray}
where all the terms proportional to $g_2$ drop from this equation.
Then (\ref{eq:pol}) becomes
\begin{eqnarray}
\der\Delta\sigma_{Na}(P,l) &= &-\frac{\alpha}{2\pi}\, \frac{1}{q^2} 
  \bigg \{ \bigg [ 1 + \frac{2m^2x}{q^2} -\frac{2}{x} \bigg ]\, G_M^2\nonumber \\
&&~~~   -2\bigg [ 1 + \frac{m^2x}{q^2} -\frac{1}{x}\bigg ] 
\frac{G_M(G_M-G_E)}{1 +  \tau} \bigg \}\,2\pi^2 \alpha\,\frac{g_1(0, k\cdot l)}
{ P\cdot l}~,
\label{eq:sigmapol1}
\end{eqnarray}
which holds in the limit $q^2\rightarrow 0$, after changing the variables of 
integration as for the unpolarized case.  
This expression can be related to the polarized cross section for real
photon-parton scattering, which can be computed by convoluting the partonic
tensor with the antisymmetric part of the photon polarization density matrix
\cite{gabrid}
\begin{equation}
P^{\rm A}_{\alpha\beta} = 
\frac{1}{2}\, (\,\epsilon_{\alpha}\epsilon_{\beta}^ * 
   -\epsilon_{\beta}\epsilon_{\alpha}^ *) = \frac{i}{2\sqrt{|q^2|}}\,\varepsilon_{\alpha\beta\rho\sigma}k^{\rho}t^{\sigma},
\label{eq:density}
\end{equation}          
where $\epsilon^{\mu}$ is the photon polarization vector and $t^{\alpha}$ is
its spin vector
\begin{equation}
t_{\alpha} = N_t \,\bigg (k_{\alpha} - \frac{q^2}{k\cdot  l}\, 
l_{\alpha}\bigg ),
\label{eq:dens2}
\end{equation}
with $N_t$ chosen so that $|t^2|=1$.
We get ($q^2\rightarrow 0 $)
\begin{equation}
\Delta\sigma_{\gamma a}(k, l) = \frac{1}{4 \,k\cdot l} \,
W^{\alpha\beta {\rm A}}P^{\rm A}_{\alpha\beta}  = 
   2\pi^2\alpha \,\frac{g_1(0, k\cdot l)}{k\cdot l}~.
\end{equation}
Combining the last equation with (\ref{eq:sigmapol1}) and integrating
over $q^2$, one gets the analogous of (\ref{eq:unpol}) 
for a  polarized process
\begin{equation}
\der\Delta\sigma_{Na} (P,l) = \Delta\gamma_{\mathrm{el}}(x)
\Delta\sigma_{\gamma a}(k,l)\der x,
\label{factor:pol}
\end{equation}
with \cite{gpr1}
\begin{eqnarray} 
\Delta\gamma_{\mathrm{el}}(x) & = & -\frac{\alpha}{2\pi} 
   \int_{t_{\rm min}}^{t_{\rm max}} \frac{\der t}{t} 
    \left\{ \left[2-x + \frac{2m^2x^2}{t}\right] 
     G_M^2(t)-2\left[1-x + \frac{m^2x^2}{t}\right] 
      G_M(t)F_2(t)\right\}\nonumber\\ 
& = & -\frac{\alpha}{2\pi}  
     \int_{t_{\rm min}}^{t_{\rm max}} \frac{\der t}{t}\, 
      G_M(t)\left\{ \left[2-x +\frac{2m^2x^2}{t}\right] 
       F_1(t) + xF_2(t)\right\} 
\label{eq17}
\end{eqnarray} 
where the first term  
proportional to $G_M^2$ in the first line corresponds 
to the pointlike result of \cite{flo}. Following 
\cite{kniehl}, we again approximate the integration 
bounds by $t_{\rm min}=-\infty$ and $t_{\rm max}= 
-m^2x^2/(1-x)$ as in (\ref{eqtwo}) in order to obtain an  
universal process independent polarized elastic  
distribution.  Using, in addition to (\ref{eqfour}) and (\ref{eqfive}), 
\begin{eqnarray} 
F_1^p(t) & = & \frac{1+\mu_p\tau}{1+\tau}\, 
               (1+a\tau)^{-2},\quad 
      F_2^p(t) = \frac{\kappa_p}{1+\tau} 
                  (1+a\tau)^{-2}\\ 
F_1^n(t) & = & 2\kappa_n \frac{\tau}{1+\tau} 
               (1+a\tau)^{-2},\quad 
      F_2^n(t) = \kappa_n \frac{1-\tau}{1+\tau} 
                  (1+a\tau)^{-2}\, , 
\end{eqnarray} 
equation (\ref{eq17}) yields for the proton 
\begin{equation} 
\Delta\gamma_{\mathrm{el}}^p(x) = \frac{\alpha}{2\pi}\,\mu_p 
 \left\{ \left[ (2-x) 
  \left( 1+\kappa_p\frac{x}{2} \right)  
   +2ax^2\right] I +2\kappa_p \left (1-x+\frac{x^2}{4} \right ) 
    \tilde{I} -2\, \frac{1-x}{z^3}\right\}\, ,
\label{eq20} 
\end{equation} 
and for the neutron  
\begin{equation} 
\Delta\gamma_{\mathrm{el}}^n(x) = \frac{\alpha}{2\pi}  
 \kappa_n^2 \left\{ x(1-x) I +4 \left( 1-x + 
  \frac{x^2}{4} \right  )\tilde{I}\right\} 
\label{eq21}
\end{equation} 
with $I$ and $\tilde{I}$ being given in  (¸\ref{eq8}) and (\ref{eq9}).

Integrating \eqref{eq17} between 
$t_{\rm min} = -\mu^2$ and $t_{\rm max}$ given in  \eqref{eq:tmaximum},  
with the form factors $G_M = F_1 = -1$ and $F_2=0$,
we obtain the polarized photon distribution of the electron
\eqref{eq:dgamma_el}. 

\begin{figure}[t] 
\begin{center} 
\epsfig{figure=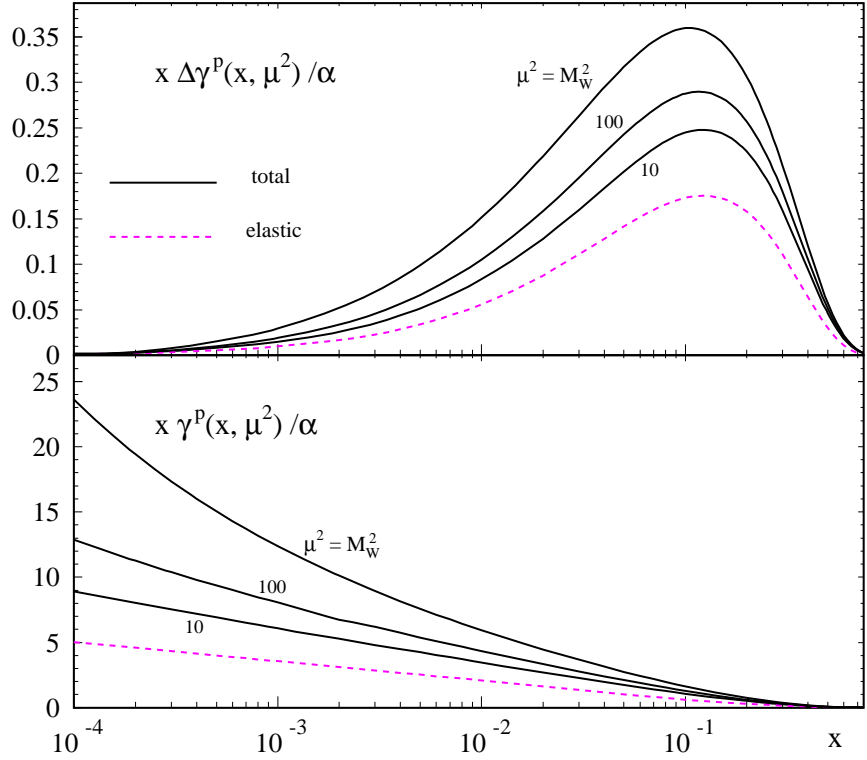, width= 13.5cm} 
\caption{ {The polarized and unpolarized 
     total photon contents of the proton,  
      $\Delta\gamma^p$ and $\gamma^p$, according to 
       (\ref{eqone}) and (\ref{eqeleven}) at some typical fixed values 
      of $\mu^2$ (in GeV$^2$).  The $\mu^2$-independent 
      elastic contributions are given by  (\ref{eq20}) 
      and (\ref{eqsix}).}}
\label{fig1}
\end{center}
\end{figure} 

\begin{figure}[ht] 
\begin{center}  
\epsfig{figure=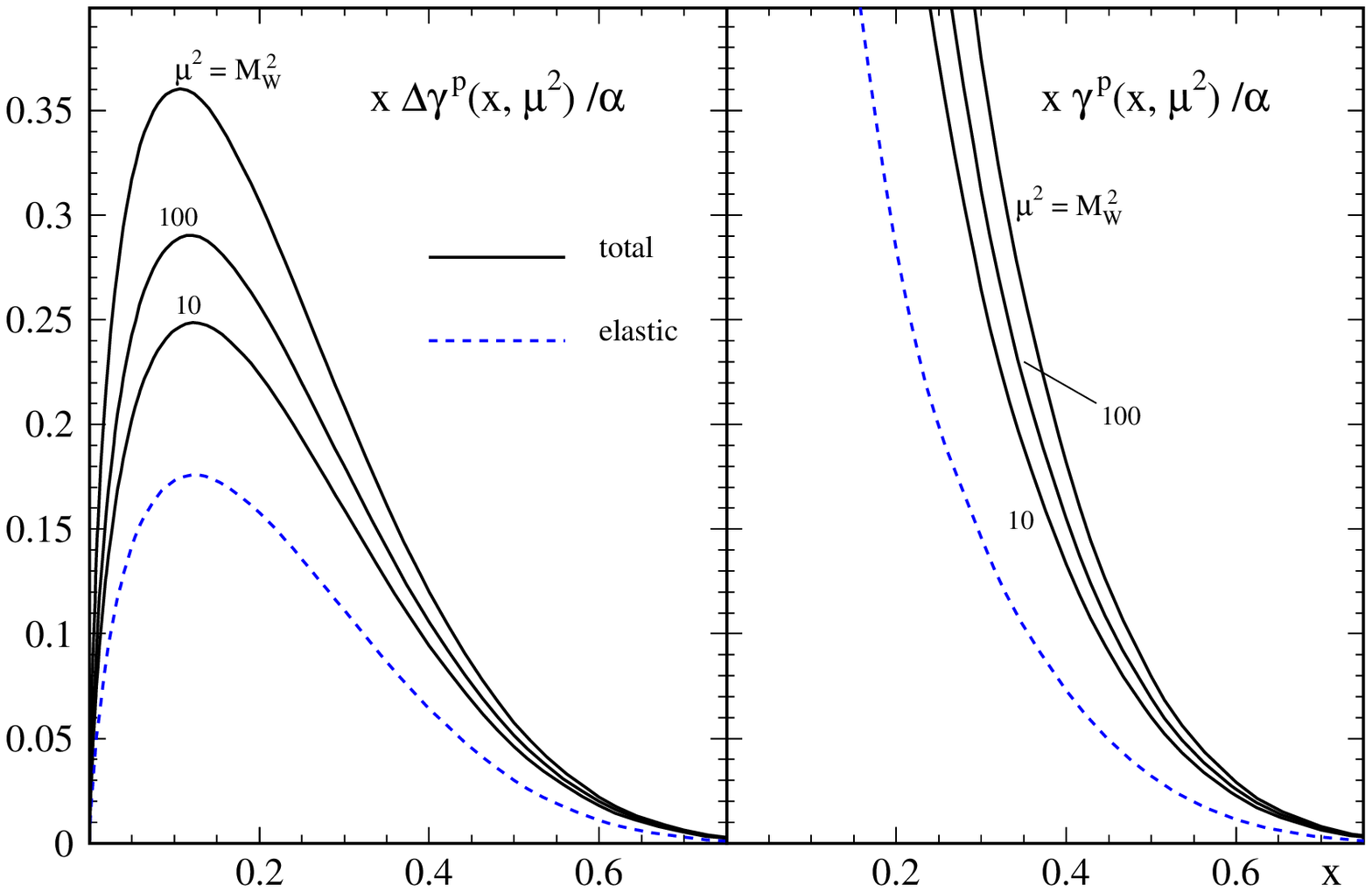,width=14.5cm}  
\caption{{As in Figure  \ref{fig1} but for a linear $x$ scale.}}
\label{fig2}
\end{center} 
\end{figure}  
\begin{figure}[ht] 
\begin{center}
\epsfig{figure=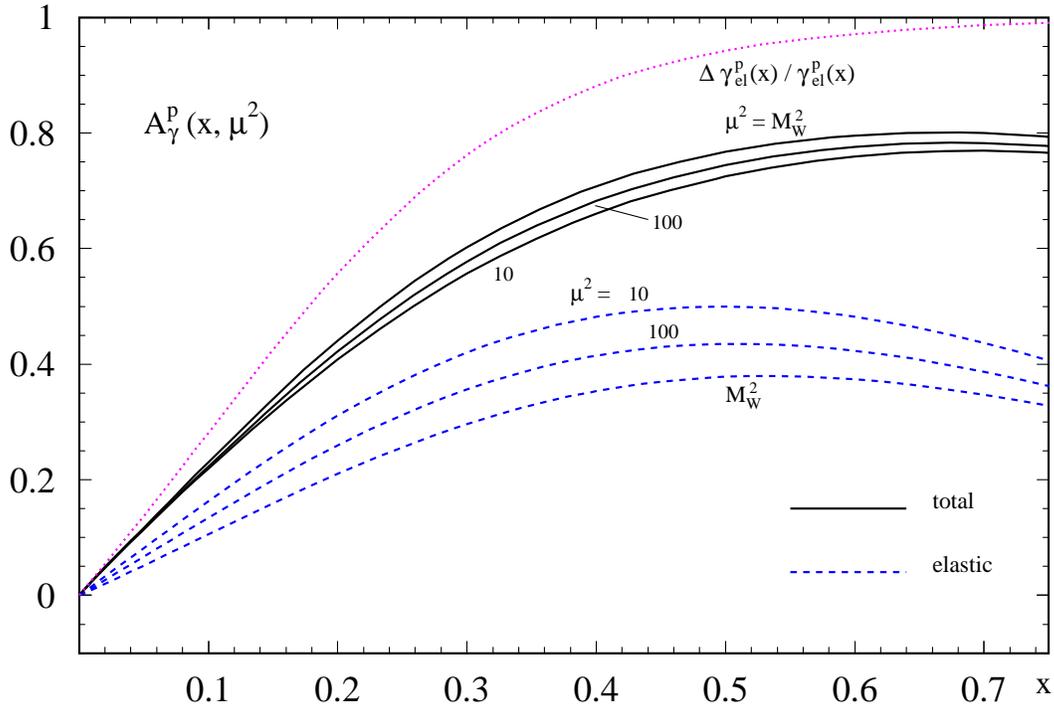,width=14.5cm} 
\caption{ {The asymmetry of the polarized 
      to the unpolarized photon content of the proton 
      as defined in  (\ref{eq22}) at various fixed values of  
      $\mu^2$ (in GeV$^2$) according to the results in 
      Figure  \ref{fig1}.  The $\mu^2$-dependence of the elastic 
      contribution to $A_{\gamma}^p$ is caused by 
      the $\mu^2$-dependent total unpolarized photon 
      content in the denominator of  (\ref{eq22}).  For 
      illustration the $\mu^2$-independent elastic ratio 
      $\Delta\gamma_{\mathrm{el}}^p/\gamma_{\mathrm{el}}^p$ is 
      shown as well.}}
\label{fig3} 
\end{center}
\end{figure}
\begin{figure}[ht] 
\begin{center}
\epsfig{figure=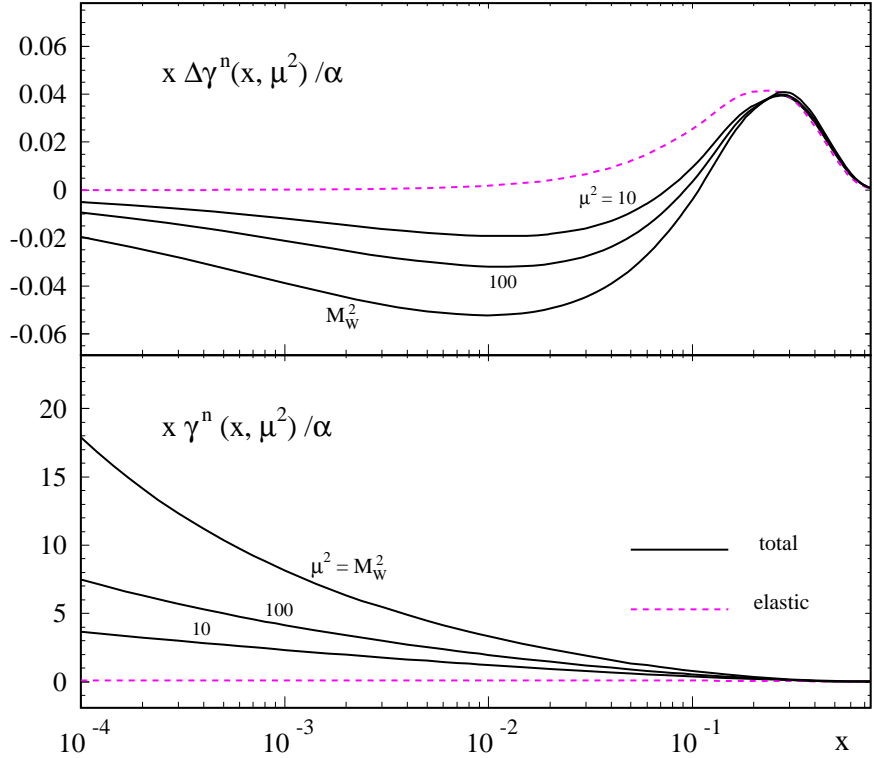, width = 13.5cm} 
\caption{ {As Figure  \ref{fig1} but for the neutron, 
      with elastic polarized and unpolarized contributions 
      being given by  (\ref{eq21}) and (\ref{eq7}).}} 
\label{fig4}
\end{center}
\end{figure} 
\begin{figure}[ht] 
\begin{center}
\epsfig{figure = 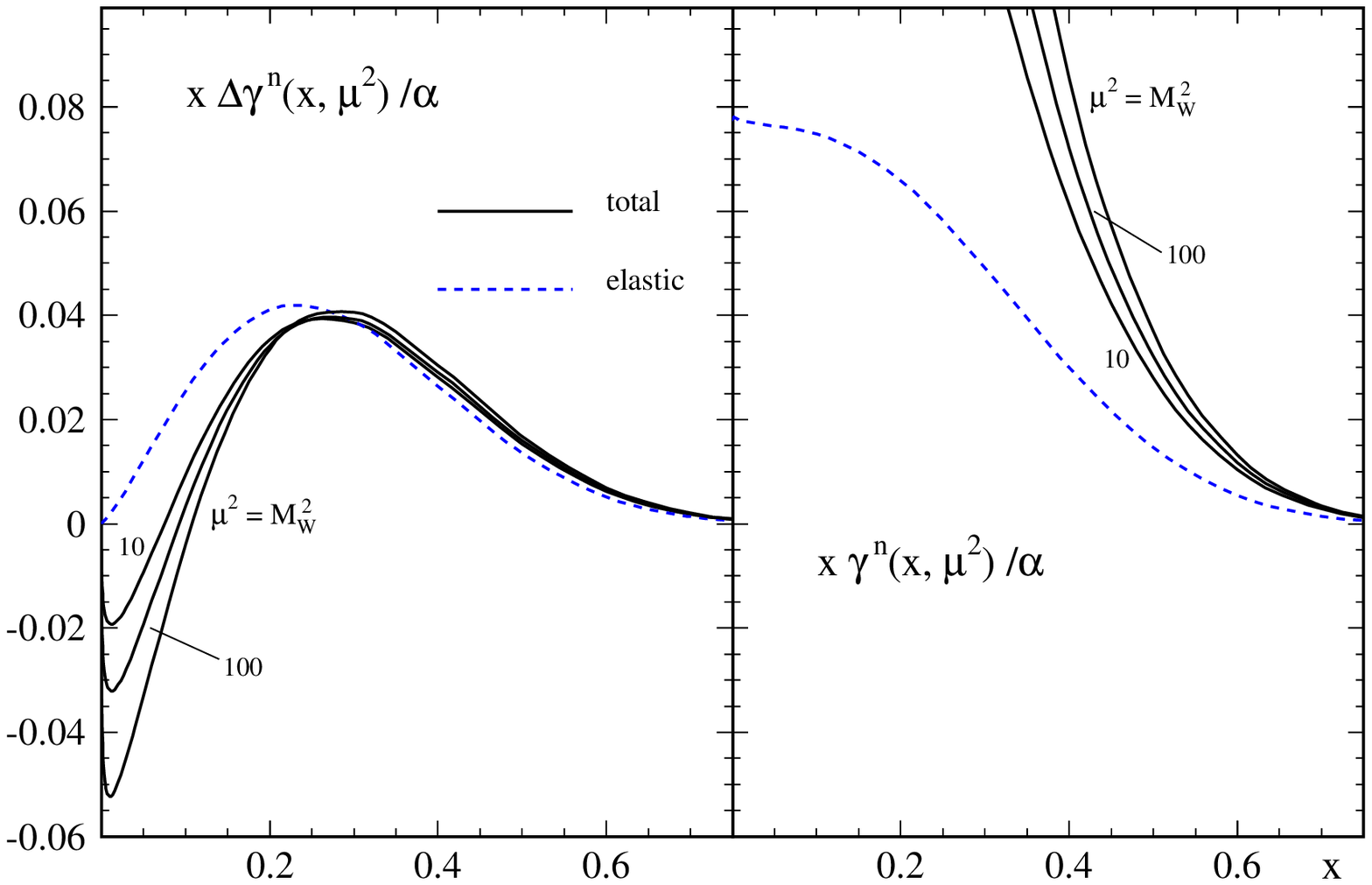, width = 14.5cm} 
\caption{{As in Figure  \ref{fig4} but for a linear $x$ scale.}}
\label{fig5}
\end{center} 
\end{figure}
\begin{figure} [ht] 
\begin{center}
\epsfig{figure= 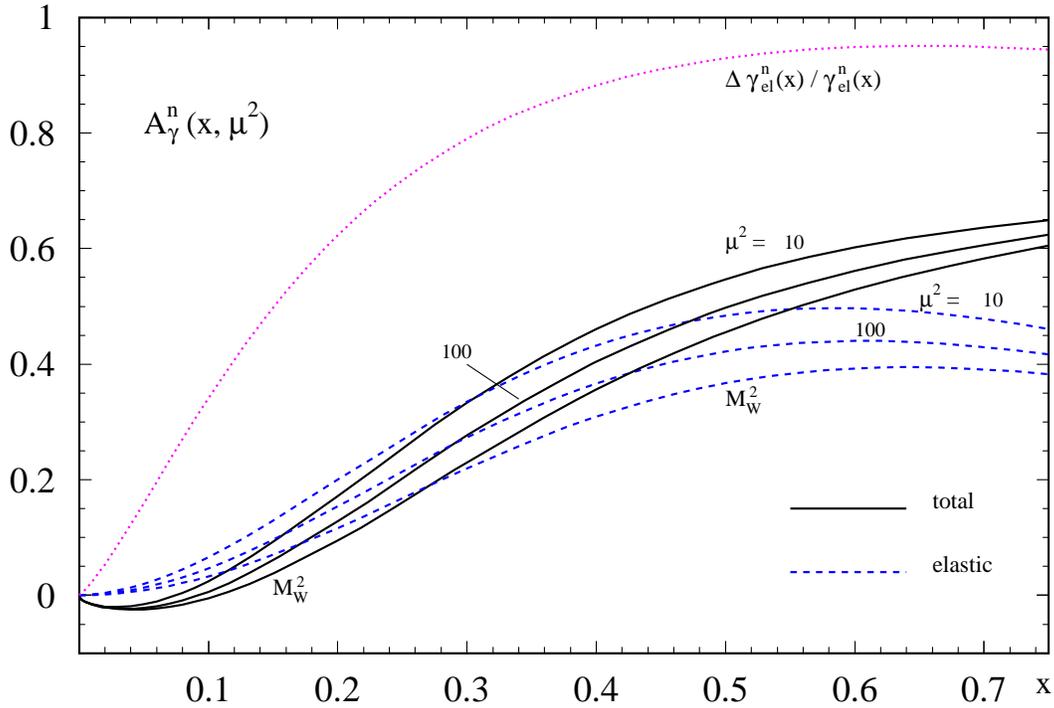, width = 14.5cm} 
 \caption{ {As Figure  \ref{fig3} but for the neutron 
      asymmetry according to the results in Figure  \ref{fig4}.}} 
\label{fig6}
\end{center}
\end{figure}

\subsection{Inelastic Component}
The inelastic contribution \cite{gpr1} derives from a  
straightforward extension of (\ref{eqten}), 
\begin{equation} 
\frac{\der\Delta\gamma_{\mathrm{inel}}(x,\mu^2)}{\der\ln \mu^2} 
 = \frac{\alpha}{2\pi} \sum_{q=u,d,s} e_q^2 
  \int_x^1\, \frac{\der y}{y}\, \Delta P_{\gamma q} 
   \left( \frac{x}{y}\right)  
    \left[\Delta q(y,\mu^2)+\Delta\bar{q}(y,\mu^2) 
      \right] 
\label{eq12}
\end{equation} 
where 
\begin{equation}
\Delta P_{\gamma q}(y)=\frac{1-(1-y)^2}{y}=2-y
\end{equation} 
is the polarized quark-to-photon splitting function and 
$\Delta q(y,\mu^2)$, $\Delta \bar{q}(y, \mu^2)$ are the 
po\-la\-ri\-zed quark and antiquark distribution functions of the nucleon.
We integrate this evolution equation assuming again 
the not necessarily compelling  `minimal' boundary 
condition 
\begin{equation}
\Delta\gamma_{\mathrm{inel}}(x,\mu_0^2)=0,
\end{equation} 
according to $|\Delta\gamma_{\mathrm{inel}}(x,\mu_0^2)|\leq 
\gamma_{\mathrm{inel}}(x,\mu_0^2)=0$, at $\mu_0^2=0.26$ GeV$^2$ 
using the  LO polarized parton densities of 
\cite{grsv}.  
These latter two equations together with   (\ref{eq12}) yield now the 
total photon content  
$\Delta\gamma(x,\mu^2)$ of a polarized nucleon in  (\ref{eqeleven}).

\section{Numerical Results} 
Our results for $\Delta\gamma^p(x,\mu^2)$ in  (\ref{eqeleven}) are 
shown in Figure  \ref{fig1} for some typical values of $\mu^2$ up 
to $\mu^2=M_W^2=6467$ GeV$^2$.
For comparison the 
expectations for the unpolarized $\gamma^p(x,\mu^2)$ 
in  (\ref{eqone}) are depicted as well.  The $\mu^2$-independent 
polarized and unpolarized elastic contributions in 
 (\ref{eq20}) and (\ref{eqsix}), respectively, are also shown 
separately. Due to the singular small-$y$ behavior 
of the unpolarized parton distributions  
$y\!\!\stackrel{(-)}{q}\!\!(y,\mu^2)$ in  (\ref{eqten}) as well as of 
the singular $x\gamma_{\mathrm{el}}^p(x)$ in  (\ref{eqsix}) as  
$x\to 0$, the total $x\gamma^p(x,\mu^2)$ in Figure \ref{fig1}  
increases as $x\to 0$, whereas the polarized  
$x\Delta\gamma^p(x,\mu^2)\to 0$ as $x\to 0$ because of  
the vanishing of the polarized parton distributions 
$y\Delta\!\!\stackrel{(-)}{q}\!\!(y,\mu^2)$ in  (\ref{eq12}) 
at small $y$ and of the vanishing  
$x\Delta\gamma_{\mathrm{el}}^p(x)$ in  (\ref{eq20}) at small $x$. 
In fact, $x\Delta\gamma^p(x,\mu^2)$ is negligibly 
small for $x$  
\raisebox{-0.1cm}{$\stackrel{<}{\sim}$} $10^{-3}$ 
as compared to $x\gamma^p(x,\mu^2)$.
  
For larger values of $x$, $x>10^{-2}$, $x\Delta\gamma^p(x,\mu^2)$  
becomes sizeable and in particular is dominated by  
the $\mu^2$-independent elastic contribution  
$x\Delta\gamma_{\mathrm{el}}^p(x)$ at moderate values of 
$\mu^2$, $\mu^2$ 
\raisebox{-0.1cm}{$\stackrel{<}{\sim}$} 100 GeV$^2$  
(with a similar behavior in the unpolarized sector).
This is evident from Figure  \ref{fig2} where the  
results of Figure  \ref{fig1} are plotted versus a linear 
$x$ scale.  

The asymmetry $A_{\gamma}^p(x,\mu^2)$ 
is shown in Figure  \ref{fig3} where 
\begin{equation} 
A_{\gamma}(x,\mu^2)\equiv  
  \left[ \Delta\gamma_{\mathrm{el}}(x) 
   +\Delta\gamma_{\mathrm{inel}} (x,\mu^2)\right] 
     /\gamma(x,\mu^2) 
\label{eq22}
\end{equation} 
with the total unpolarized photon content of the  
nucleon being given by  (\ref{eqone}).  To illustrate the size 
of $\Delta\gamma_{\mathrm{el}}^p$ relative to the  
unpolarized $\gamma_{\mathrm{el}}^p$, we also show the 
$\mu^2$-independent ratio  
$\Delta\gamma_{\mathrm{el}}^p(x)/\gamma_{\mathrm{el}}^p(x)$ 
in Figure  \ref{fig3} which approaches 1 as $x\to 1$.

The polarized photon distributions $\Delta\gamma^p 
(x,\mu^2)$ shown thus far always refer to the  
so called  `valence' scenario \cite{grsv} where the 
polarized parton distributions in  (\ref{eq12}) have  
flavor-broken light sea components $\Delta\bar{u} 
\neq \Delta\bar{d}\neq\Delta\bar{s}$, as is the  
case (as well as experimentally required) for the 
unpolarized ones in  (\ref{eqten}) where $\bar{u}\neq\bar{d} 
\neq\bar{s}$.  Using instead the somehow unrealistic 
`standard' scenario \cite{grsv} for the polarized 
parton distributions with a flavor-unbroken sea 
component $\Delta\bar{u}=\Delta\bar{d}=\Delta\bar{s}$, 
all results shown in Figures \ref{fig1}-\ref{fig3} remain practically 
almost undistinguishable.  The same holds true for 
the photon content of a polarized neutron to which 
we now turn. 
 
The results for $\Delta\gamma^n(x,\mu^2)$ are  
shown in Figure  \ref{fig4} which are sizeably smaller than 
the ones for the photon in Figure  \ref{fig1} and, furthermore, 
the elastic contribution is dominant while 
the  
inelastic ones become marginal at  
$x$ \raisebox{-0.1cm}{$\stackrel{>}{\sim}$} 0.2. 
For comparison the unpolarized $\gamma^n(x,\mu^2)$ 
in  (\ref{eqone}) is shown in Figure  \ref{fig4} as well.  Here, 
$\gamma_{\mathrm{el}}^n$ in  (\ref{eq7}) is marginal and  
$x\gamma_{\mathrm{el}}^n(x)$ is non-singular as $x\to 0$ 
with a limiting value $x\gamma_{\mathrm{el}}^n(x)/\alpha 
=\kappa_n^2/(3\pi a)\simeq 0.078$.  Thus the  
increase of $x\gamma^n(x,\mu^2)$ at small $x$ is  
entirely caused by inelastic component 
$x\gamma_{\mathrm{inel}}^n(x,\mu^2)$ in  (\ref{eqten}), due to the 
singular small-$y$ behavior of  
$y\!\!\stackrel{(-)}{q}\!\!(y,\mu^2)$, which is in  contrast to 
$x\gamma^p(x,\mu^2)$ in Figure  \ref{fig1}.  

These facts are more clearly displayed in Figure  \ref{fig5} where 
the results of Figure  \ref{fig4} are presented for a linear 
$x$ scale.  Notice that again the polarized 
$x\Delta\gamma^n(x,\mu^2)\to 0$ as $x\to 0$ because 
of the vanishing of the polarized parton  
distributions  
$y\Delta\!\!\stackrel{(-)}{q}\!\!(y,\mu^2)$ in  (\ref{eq12})  
at small $y$ and of the vanishing of 
$x\Delta\gamma_{\mathrm{el}}^n(x)$ in  (\ref{eq21}) at small $x$.
 
Finally, the asymmetry $A_{\gamma}^n(x,\mu^2)$ 
defined in  (\ref{eq22}) is shown in Figure  \ref{fig6} which is  
entirely dominated by the elastic contribution for 
$x$ \raisebox{-0.1cm}{$\stackrel{>}{\sim}$} 0.2. 
As in Figure  \ref{fig3} we illustrate the size of the elastic 
$\Delta\gamma_{\mathrm{el}}^n(x)$ relative to the 
unpolarized $\gamma_{\mathrm{el}}^n(x)$ by showing the 
ratio $\Delta\gamma_{\mathrm{el}}^n/\gamma_{\mathrm{el}}^n$ 
in Figure   \ref{fig6} as well.  Notice that  
$\Delta\gamma_{\mathrm{el}}^n/\gamma_{\mathrm{el}}^n\to\frac 
{6}{7}$ as $x\to 1$ in contrast to the case of  
the proton.  
 
Clearly, the nucleon's photon content $\gamma 
(x,\mu^2)$ is not such a fundamental quantity as are  
its underlying parton distributions $f(x,\mu^2)=q,\, 
\bar{q},\, g$ or the parton distributions $f^{\gamma} 
(x,\mu^2)$ of the photon, since $\gamma^p(x,\mu^2)$  
is being derived from these more fundamental 
quantities.  Moreover, its reliability remains to be
studied.  
We shall try to carry out this task starting on
Chapter 5.

A FORTRAN package (grids) containing our results for 
$\Delta \gamma (x, \mu^2)$ as well as those for 
$\gamma (x, \mu^2)$ can be obtained by electronic 
mail.

\clearemptydoublepage

\chapter{ \bf{Measurement of the Equivalent Photon Distributions}}

In the previous chapter we estimated the polarized and unpolarized 
equivalent photon distributions of the nucleon $(\Delta)\gamma (x, \mu^2)$,
 consisting of two components,
\begin{equation}
(\Delta)\gamma(x, \mu^2) = (\Delta)\gamma_{\mathrm{el}}(x) + 
(\Delta)\gamma_{\mathrm{inel}}(x, \mu^2),
\end{equation} 
where the elastic parts $(\Delta)\gamma_{\mathrm{el}}$ are uniquely 
determined by the well-known 
electromagnetic form factors $F_{1, 2}(q^2)$ of the nucleon. 
The inelastic components
are fixed via the boundary conditions 
\begin{equation}
(\Delta)\gamma_{\mathrm{inel}}(x, \mu_0^2) = 0 
\end{equation}
at $\mu_0^2= 0.26$ $\mathrm{GeV}^2$, evolved for $\mu^2 > \mu_0^2$, according
to the LO equations
\begin{equation} 
\frac{\der(\Delta)\gamma_{\mathrm{inel}}(x,\mu^2)}{\der\ln \mu^2} = 
 \frac{\alpha}{2\pi} \sum_{q=u,d,s} e_q^2 
  \int_x^1 \frac{\der y}{y}\, (\Delta)P_{\gamma q} 
   \left(\frac{x}{y}\right)  
    \left[ (\Delta)q(y,\mu^2)+ (\Delta)\bar{q}(y,\mu^2)\right], 
\label{eq:evolution}
\end{equation} 
with the unpolarized and polarized parton distributions in LO taken
from \cite{grv98,grsv}. 
As already stated, the boundary conditions
are not compelling but should be tested expe\-rimentally.
However at large scales $\mu^2$ the results
become rather insensitive to details at the input scale $\mu_0^2$ and
thus the vanishing boundary conditions  yield reasonable
results for $(\Delta )\gamma_{\mathrm{inel}}$ which are essentially 
determined by the quark
and antiquark (sea) distributions of the nucleon in \eqref{eq:evolution}.
At low scales $\mu^2$, however, $(\Delta)\gamma_{\mathrm{inel}}(x,\mu^2)$ 
depend obviously
on the assumed details at the input scale $\mu_0^2$.
Such a situation is encountered at a fixed target experiment, typically
HERMES at DESY. 
At present it would be too speculative and arbitrary
to study the effects due to a non-vanishing boundary 
$\gamma_{\mathrm{inel}}(x,\mu_0^2) \ne 0$.
Rather this should be examined experimentally if our expectations 
based on the  vanishing boundary  turn out to be
in disagreement with observations.

In the present chapter  we consider two processes which offer a clear
opportunity to gain information on the photonic structure of the nucleon:
 muon pair production in electron-nucleon 
collisions $e N \to e \mu^+ \mu^- X$  via the subprocess 
$\gamma^e \gamma^N \to \mu^+ \mu^-$ and the QED Compton process
$e N \to e \gamma X$ via the subprocess 
$e \gamma^N \to e \gamma$ for both the HERA collider experiments
and the polarized and unpolarized fixed target HERMES experiment at
DESY. The  production rates of lepton-photon and dimuon pairs
are evaluated in the leading order equivalent photon approximation and
it is shown that they are sufficient to 
facilitate the extraction of the polarized and unpolarized
photon distributions of the nucleon in the available kinematical regions 
\cite{gpr2}.
On the other hand, it should be noted that a study of 
$N N \to \mu^+ \mu^- X$ via 
$\gamma^N \gamma^N \to \mu^+ \mu^-$
in hadron-hadron collisions is impossible 
\cite{drees,ohn}
due to the dominance of the Drell-Yan subprocess 
$q^N \bar{q}^N \to \mu^+ \mu^-$. The logarithmic enhancement of the
photon densities is not enough to overcome
completely the extra factor $\alpha^2$ in the
$\gamma\gamma$ fusion process.

Measurements of $(\Delta)\gamma(x,\mu^2)$ are not only 
interesting on their own,
but may provide additional and independent informations concerning 
$(\Delta) \overset{(-)}{q}$ 
in \eqref{eq:evolution}, in particular
about the polarized parton distributions which are not well determined
at present. 

\section{Theoretical Framework}
In this section we present  the kinematics and cross section
formulae of the reactions under study, following the lines of
Appendix D in \protect\cite{Brock:1995sz}. The unpolarized and 
polarized  subprocess cross sections for  $\gamma\gamma \to \mu^+\mu^-$ 
and $e\gamma\to e\gamma$ can be found, for example, in \cite{Gastmans:1990xh};
their derivation is shown in detail in Appendix B.

\subsection{Dimuon Production}
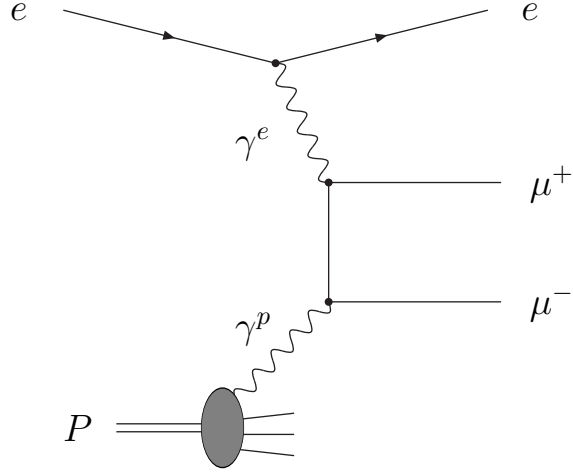
\begin{figure}[t]
\begin{center} 
\begin{picture}(210,200)(0,0)
\ArrowLine(20,180)(100,160)
\ArrowLine(100,160)(180,180)
\Vertex(100,160){1.5}
\Photon(100,160)(120,115){3}{5}
\Photon(85,33)(120,70){3}{5}
\Line(40,21)(80,21)
\Line(40,24)(80,24)
\Line(80,25)(107,28)
\Line(80,20)(107,20)
\Line(80,15)(107,12)
\GOval(80,22.5)(15,8)(0){0.5}
\Line(185,70)(120,70)
\Line(120,70)(120,115)
\Line(120,115)(185,115)
\Vertex(120,70){1.5}
\Vertex(120,115){1.5}
%
\Text(0,180)[cl]{\large $e$}
\Text(200,180)[cr]{\large $e$}
\Text(205,115)[c]{\large $\mu^+$}
\Text(205,70)[c]{\large $\mu^-$}
\Text(20,22.5)[cl]{\large $P$}
\Text(85,60)[l]{\large $\gamma^p$}
\Text(85,130)[l]{\large $\gamma^e$}
\end{picture}

\end{center}
\caption{ Lowest-order Feynman diagram for dimuon production
in $ep$ collisions. The crossed $\uhat$-channel diagram
is not shown.}
\label{fig:dimuon}
\end{figure}

We consider first deep inelastic dimuon production
$e p \to e \mu^+ \mu^- X$ 
via the subprocess
\begin{equation}
\gamma^e(k_1)+ \gamma^p(k_2) \to \mu^+(l_1)+ \mu^-(l_2)
\label{eq:submuon}
\end{equation}
as depicted in 
Figure \ref{fig:dimuon}.  With the four-momenta of the particles given in 
the brackets in
\eqref{eq:submuon}, the Mandelstam variables are defined as
\begin{equation}
\hat{s} = (k_1+k_2)^2,~~~~~~\hat t = (k_1-l_1)^2,~~~~~~\hat u = (k_2-l_1)^2.
\label{eq:3mandel}
\end{equation}
Suppose that the photon $\gamma^e$ carries a fraction $\xi$ of the
electron's momentum and that a similar definition for $x$
exists for the photon $\gamma^p$. Then in the $e-p$ center-of-mass system
the four-momenta $k_1$ and $k_2$ of the colliding photons, 
assumed to be collinear with the parent particles, can be written as
\begin{equation}
k_1 = \frac{\xi \,\sqrt{s}}{2}\, (1, 0,0,1),~~~~
k_2 = \frac{x \,\sqrt{ s}}{2} \,(1, 0,0,-1),~~~~
\label{eq:uno}
\end{equation}
where the positive $z$ axis is taken to be along the direction of the 
incident electron and $s$ is the squared center-of-mass energy, which 
satisfies the condition
\begin{equation}
\hat s = \xi x s.
\end{equation}
The four-momenta of the outgoing muons can be written in terms of 
their rapidities $y_{1, 2}$ and momentum components $l_{T_{1, 2}}$, 
transverse with respect to the $z$ axis,
\begin{equation}
l_1 = l_{T_1}(\cosh y_1, 1,0, \sinh y_1),~~~~
   l_2 = l_{T_2}(\cosh y_2, -1,0, \sinh y_2).~~~~
\label{eta00}
\end{equation}
In general, an outgoing  particle with energy $E$ has 
component of the velocity 
along the  $z$ axis given by
\begin{equation}
\beta = \frac{l_{z}}{E}~,
\label{eq:beta}
\end{equation}
and its rapidity $y$ can be defined so that
\begin{equation}
E = l_T \cosh y,~~~~~~~l_z = l_T \sinh y,
\end{equation}
where $l_z$ and $l_T$ are respectively the longitudinal and 
transverse components of its momentum $l$. Therefore, the relation between
$y$ and $\beta$ is given by  
\begin{equation}
y = \arctan \beta = 
\frac{1}{2} \ln \frac{1+\beta}{1-\beta}~
\label{eq:eta}
\end{equation}
and, substituting  \eqref{eq:beta} in \eqref{eq:eta}, one has also 
\begin{equation}
y = \frac{1}{2}\,\ln\frac{E+l_{z}}{E-l_{z}}~.
\label{eq:eta2}
\end{equation}
For a massless particle, $l_{z} = E \cos\theta$, $\theta$ being
the center-of-mass scattering angle, and \eqref{eq:eta2} assumes the
much simpler form
\begin{equation}
y = -\ln\tan\frac{\theta}{2}~,
\end{equation}
called pseudorapidity and very convenient experimentally, since one
needs to measure only $\theta$ in order to determine it. 

The cross section relative to the inclusive production 
of the two muons, in the most differential form, is given by 
\begin{eqnarray}
\frac{\der\sigma}{\der\xi\der x} & = & \gamma^e(\xi, \hat s)\,
 \gamma^p(x, \hat s) \,\der \hat\sigma \nonumber \\
   & = & \frac{1}{2 \hat s}\,\gamma^e(\xi, \hat s)\,
 \gamma^p(x, \hat s) \,\overline{|\hat M|^2} \,
   (2\pi)^4 \delta^4 (k_1 + k_2 -l_1-l_2)\,\frac{\der^3
{\mbox{\boldmath $l$}}_1}
    {(2\pi)^32E_1}\frac{\der^3{\mbox{\boldmath $l$}}_2}
     {(2\pi)^32E_2}~,\nonumber \\ 
\label{eq:dimuon-cm} 
\end{eqnarray}
where $\hat \sigma$ is the cross section of the subprocess
 $ \gamma^e\gamma^p\rightarrow \mu^+\mu^-$. 
In the spirit of the leading order equivalent photon
approximation underlying \eqref{eq:dimuon-cm}, we shall adopt
the LO photon distribution of the proton $\gamma^p(x,\shat)$ given by
\eqref{eqone}, together with \eqref{eqtwo} and \eqref{eqten},  
  as well as the LO equivalent photon distribution
of the electron $\gamma^e(\xi,\shat)$,
\be
\gamma^e(\xi,\shat) = \frac{\alpha}{2 \pi} \frac{1+(1-\xi)^2}{\xi}
\ln \frac{\shat}{m_e^2}~ ,
\label{eq:gamma_e}
\ee
where $m_e$ is the electron mass. 
Equation \eqref{eq:gamma_e} is obtained from 
\eqref{eq:gamma_el}, retaining only the 
leading logarithmic term and identifying the scale $\mu^2$ with $\hat s$.
The phase space elements of the two muons can be written as
\begin{eqnarray}
\frac{\der^3 {\mbox{\boldmath $l$}}_i}{2 E_i} &=&  \frac{1}{2}\,
\der^2 {\mbox{\boldmath $l$}}_{T_i} \,\der y_i = \pi\, 
\der l^2_{T{_i}}\, \der y_i,
\label{eq:phase1} 
\end{eqnarray}
where $l_{T_i} = |{\mbox{\boldmath $l$}}_{T_i}|$, $i = 1, 2$, 
  and the azimuth integrations have been carried out
in the last equality.
Furthermore, the original four-dimensional 
$\delta$-function in \eqref{eq:dimuon-cm} can be split into its
energy, transverse momentum and  longitudinal momentum parts:
\begin{eqnarray}
\delta^4 (k_1 + k_2 -l_1-l_2) & = & \delta \bigg 
( \frac{\xi \,\sqrt{ s}}{2} +   
  \frac{x \,\sqrt{ s}}{2} -l_{T_1}\cosh y_1 - l_{T_2} \cosh y_2 \bigg ) \,
   \delta^2 ({\mbox{\boldmath $l$}}_{T_{1}} - {\mbox{\boldmath $l$}}_{T_{2}})
\nonumber \\
 & &~~~~~~\times\, \delta \bigg (  \frac{\xi \,\sqrt{ s}}{2} -   
  \frac{x \,\sqrt{ s}}{2} -l_{T_1} \sinh y_1 -l_{T_2}\sinh y_2  \bigg ),
\label{eq:deltas}
\end{eqnarray}
therefore, at lowest order, the transverse momentum components of the 
$\delta$-function ensure that the muons are produced with equal and opposite
transverse momenta. We define 
\begin{equation}
p_{T} \equiv  l_{T_1} = l_{T_2}
\label{eq:pt}
\end{equation}
and the integrations over $\xi$ and $x$ in \eqref{eq:dimuon-cm} can be 
carried out using the two remaining $\delta$-functions in \eqref{eq:deltas}.
Finally, making use of  \eqref{eq:phase1}, \eqref{eq:pt}  
and  the definition
\begin{equation}
\frac{\der\hat\sigma}{\der\hat t} = \frac{1}{16\pi\hat s^2}
\,\overline{ |\hat{M}|^2},
\label{eq:dsdt2}
\end{equation}
  we get \cite{Brock:1995sz} 
\begin{equation}
\frac{\der\sigma}{\der y_1 \der y_2 \der p^2_T} = \xi \gamma^e(\xi, \hat s)
 \, x \gamma^p(x, \hat s) \, \frac{\der\hat \sigma}{\der \hat t}~,
\label{eq:diff}
\end{equation}
where the dependence of momentum fractions $\xi$ and $x$ on the 
variables $y_1$, $y_2$, $p_T$ is given by 
\begin{eqnarray}
\xi = \frac{p_T}{\sqrt{s}}\, (e^{y_1} + e^{y_2}), 
\label{eq:fraction1} 
\end{eqnarray}
and
\begin{eqnarray}
x = \frac{p_T}{\sqrt{s}}\, (e^{-y_1} + e^{-y_2}).
\label{eq:fraction2}
\end{eqnarray}
The dimuon invariant mass squared $\hat{s}$ in \eqref{eq:3mandel} can 
also be expressed in terms of the rapidities of the two muons and the 
transverse momentum of one of them as 
\begin{equation}
\hat s = 2\, p_T^2[ 1+ \cosh (y_1 -y_2)];
\label{eq:hats}
\end{equation}
using this relation one can derive, from  \eqref{eq:diff}, the cross 
section differential in $y_1$, $y_2$ and $\hat s$: 
\begin{equation}
\frac{\der\sigma}{\der y_1 \der y_2 \der \hat s} = \frac{1}{2 [1 + 
\cosh(y_1-y_2)]}\,\xi \gamma^e(\xi, \hat s)
 \, x \gamma^p(x, \hat s) \, \frac{\der\hat \sigma}{\der \hat t}~,
\label{eq:diff2}
\end{equation}
with
\begin{equation}
\hat t = -\xi p_T \sqrt{s} e^{-y_1},~~~~~~~~ \hat u = -\xi p_T \sqrt{s} e^{-y_2}.
\label{eq:cm}
\end{equation}

At HERA ($s=4 E_e E_p$) rapidities are commonly measured along the proton
beam direction, hence one should replace $y_i$ with $-y_i$ 
(or, equivalently, exchange  $\xi$ with $x$) in 
\eqref{eq:fraction1}, \eqref{eq:fraction2} and \eqref{eq:cm}, 
since the $e-p$ center-of-mass
rapidities $y_i$ were defined to be positive in the electron forward
direction.  
Being  rapidities additive quantities under successive boosts, 
the laboratory-frame rapidities of $\mu^+$ and $\mu^-$, $\eta_1$ and $\eta_2$, 
are related to  $y_1$ and $y_2$ by
\begin{equation}
\eta_i = y_i + \ln \sqrt{\frac{E_p}{E_e}}~, 
\label{eq:boost1}
\end{equation}
where the last term in \eqref{eq:boost1} is  the rapidity
relative to the boost along the $z$ axis from the laboratory to the
center-of-mass frame, calculated according to \eqref{eq:eta}  with velocity 
$\beta = (E_p-E_e)/(E_e + E_p)$, $E_p$
and $E_e$ being the colliding proton and electron energies. 
In terms of $\eta_1$ and $\eta_2$,  \eqref{eq:fraction1} and  
\eqref{eq:fraction2} are given by
\be
\xi &=& \frac{\sqrt{\shat}}{2 E_e} 
\left(\frac{e^{-\eta_1}+e^{-\eta_2}}{e^{\eta_1}+e^{\eta_2}} \right)^{1/2},
\label{eq:xi}
\\
x &=& \frac{\sqrt{\shat}}{2 E_p} 
\left(\frac{e^{\eta_1}+e^{\eta_2}}{e^{-\eta_1}+e^{-\eta_2}} \right)^{1/2},
\label{eq:x}
\ee
and can be used to estimate the dimuon production process in the laboratory-frame, together with \eqref{eq:diff} and \eqref{eq:diff2}-\eqref{eq:boost1}.
Alternatively, we can choose as independent variables $\eta_1$, $\eta_2$ and
$\xi$; using the relation
\begin{equation}
\hat s = 4 \xi^2 E_e^2\, \bigg ( \frac{e^{\eta_1} + e^{\eta_2}}{e^{-\eta_1} +
e^{-\eta_2}} \bigg )
\label{eq:shat}
\end{equation} 
obtained from \eqref{eq:xi}, we are able to calculate the Jacobian of 
this change of variables and finally we get \cite{gpr2}
\be
\frac{\der \sigma}{\der \eta_1 \der \eta_2 \der \xi}
= \frac{4 \xi E_e^2}{1+\cosh(\eta_1-\eta_2)}
\frac{e^{\eta_1}+e^{\eta_2}}{e^{-\eta_1}+e^{-\eta_2}}\,
\xi \gamma^e(\xi,\shat) x \gamma^p(x,\shat)
\frac{\der \hat \sigma}{\der \that}~,
\label{eq:dimuon-hera}
\ee
where the cross section 
for the subprocess $\gamma^e \gamma^p \to \mu^+ \mu^-$ given in 
\eqref{eq:dsdt} reads,
in the laboratory-frame,
\be
\frac{\der \hat \sigma}{\der \that}
= \frac{2 \pi \alpha^2}{\shat^2} 
  \left ( \frac{\hat t}{\hat u} + \frac{\hat u }{\hat t}\right ) 
= \frac{4 \pi \alpha^2}{\shat^2} \cosh(\eta_1 - \eta_2).\ 
\label{eq:sub-dimuon}
\ee
The last equality in \eqref{eq:sub-dimuon} follows from \eqref{eq:cm}
with $y_i\to-y_{i}$ and  \eqref{eq:boost1}, that is 
\begin{equation}
\hat t = -2\xi p_T E_e e^{\eta_1},~~~~~~~ \hat u = -2 \xi p_T E_e e^{\eta_2}.
\label{eq:mandelstam}
\end{equation}
Furthermore, from \eqref{eq:hats} and  \eqref{eq:shat},
 remembering that $y_i-y_j = \eta_i-\eta_j$, one finds 
\begin{equation}
p_T = \frac{2\xi E_e}{e^{-\eta_1} + e^{-\eta_{2}}}~.
\label{eq:pthera}
\end{equation}

At the fixed-target experiment HERMES ($s=2 m E_e$),
where the $z$ axis is chosen to be along the electron beam,
\eqref{eq:uno}-\eqref{eq:cm} still hold  and  \eqref{eq:boost1} has to be 
replaced by
\begin{equation}
\eta_i = y_i + \ln\sqrt{\frac{2 E_e}{m}}~,
\end{equation}
as now $E_e/(E_e+m)$ is the velocity of the boost from the laboratory 
to the center-of-mass frame.  Therefore, in 
\eqref{eq:xi}-\eqref{eq:mandelstam} one has to make the following
replacements 
\begin{equation}
E_p \to \frac{m}{2}~, ~~~~~~~~ \eta_i \to - \eta_i,
\end{equation}
with $\eta_i$ now corresponding to the rapidities of the observed particles
with respect to the electron beam direction.

Furthermore, at HERMES one may study also $\gamma^n(x,\shat)$
as well as the polarized $\Delta \gamma^{p, n}(x,\shat)$, given in 
\eqref{eqeleven}, \eqref{eq17} and \eqref{eq12}, by utilizing,
from \eqref{eq:dgamma_el},
\be
\Delta \gamma^e(\xi,\shat) = 
\frac{\alpha}{2 \pi} \frac{1-(1-\xi)^2}{\xi}
\ln \frac{\shat}{m_e^2}
\label{eq:polgamma_e}
\ee
in the spin dependent counterpart of \eqref{eq:dimuon-hera},
while the relevant LO cross section for the polarized subprocess
is given by \eqref{eq:pol_final}, namely
\be
\frac{\der \Delta \hat \sigma}{\der \that}
&=&
-\frac{\der \hat \sigma}{\der \that}~.
\label{eq:sub-poldimuon}
\ee

\begin{figure}[t]
\begin{center}
\vspace*{1 cm}
\begin{picture}(240,100)(10,0)
\ArrowLine(60,90)(120,70)
\ArrowLine(120,70)(160,70)
\ArrowLine(160,70)(220,90)
\Vertex(120,70){1.5}
\Photon(160,70)(210,40){3}{5}
\Vertex(160,70){1.5}
\Photon(85,33)(120,70){3}{5}
\Line(40,21)(80,21)
\Line(40,24)(80,24)
\Line(80,25)(107,28)
\Line(80,20)(107,20)
\Line(80,15)(107,12)
\GOval(80,22.5)(15,8)(0){0.5}
%
\Text(40,90)[cl]{\large $e$}
\Text(240,90)[cr]{\large $e$}
\Text(230,40)[cr]{\large $\gamma$}
\Text(20,22.5)[cl]{\large $P$}
\Text(83,58)[l]{\large $\gamma^p$}
\end{picture}
\end{center}
\caption{ Lowest-order Feynman diagram for Compton scattering
in $ep$ collisions. The crossed $\uhat$-channel contribution
is not shown.}
\label{fig:compton}
\end{figure}
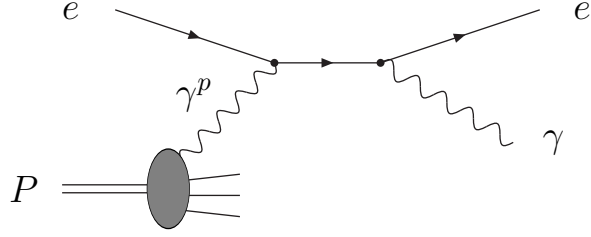

\subsection{Electron-Photon Production}
For the Compton process  $e p \to e \gamma X$ proceeding via the
subprocess 
\begin{equation}
e(l) + \gamma^p(k) \to e(l') + \gamma(k'),
\label{eq:subcom}
\end{equation}
as depicted in Figure \ref{fig:compton}, we define the variables
\begin{equation} 
\hat{s} = (l+k)^2,~~~\hat t = (l-l')^2,~~~\hat u = (l-k')^2.
\label{comptonmandel}
\end{equation}
The kinematics of the process is quite similar to the one of the reaction
 $ep\to \mu^+\mu^-X$, discussed above. In particular
if one  fixes $\xi = 1$ and drops the terms $\der\xi$ and 
$\gamma^e(x, \hat s)$,  \eqref{eq:uno}-\eqref{eta00}, 
\eqref{eq:dimuon-cm}, \eqref{eq:phase1}-\eqref{eq:pthera}  are still valid,
with  the obvious replacements 
\be
k_1\to l,~~~~~~ k_2\to k,~~~~~~ l_1\to l',~~~~~~l_2\to k', \nonumber \\
y_1\to y_e\,,~~~~~~ y_2\to y_{\gamma}\,, ~~~~~~
\eta_1\to \eta_e,~~~~~~\eta_2\to\eta_{\gamma}\,,
\ee
where $y_{e, \gamma}$ and  $\eta_{e,\gamma}$  are
respectively the center-of-mass and laboratory  rapidities of the
produced (outgoing) electron and photon.
Hence at HERA \eqref{eq:dimuon-hera} is substituted by \cite{gpr2}
\be
\frac{\der \sigma}{\der \eta_e \der \eta_\gamma}
= \frac{4 E_e^2}{1+\cosh(\eta_e-\eta_\gamma)}
\frac{e^{\eta_e}+e^{\eta_\gamma}}{e^{-\eta_e}+e^{-\eta_\gamma}}
x \gamma^p(x,\shat)
\frac{\der \hat \sigma}{\der \that}\ ,
\label{eq:dics-hera}
\ee
where  $x$ is fixed by \eqref{eq:x}, that is 
$\shat = 4 x E_e E_p $.  According to \eqref{eq:sub-compton},
\be
\frac{\der \hat \sigma}{\der \that} 
= - \frac{2 \pi \alpha^2}{\shat^2}
\left(\frac{\shat}{\uhat}+ \frac{\uhat}{\shat}\right),
\label{eq:subcrosss}
\ee
with
\be
-\frac{\shat}{\hat{u}}=1+e^{\eta_e-\eta_\gamma},
\label{eq.labmand}
\ee 
which can be derived from \eqref{eq:shat}, \eqref{eq:mandelstam} and 
\eqref{eq:pthera}. 
 
At the HERMES experiment $\hat s = 2 x m E_e$, and 
\eqref{eq:dics-hera}-\eqref{eq.labmand}
still hold, but with $\eta_{e, \gamma} \to -\eta_{e, \gamma}$.
The equivalent of  \eqref{eq:dics-hera} for longitudinally polarized
incoming particles is obtained by
replacing the  photon distribution $\gamma^p(x, \hat s)$  and 
subprocess cross section $\der\hat \sigma /\der\hat t$   
with their spin dependent counterparts $\Delta\gamma^p(x, \hat s)$  and 
$\der\Delta\hat \sigma /\der\hat t$. 
From  \eqref{eq:polcompt}, 
\be
\frac{\der \Delta \hat \sigma}{\der \that}
\hspace*{5mm}
&=& - \frac{2 \pi \alpha^2}{\shat^2}
\left(\frac{\shat}{\uhat}- \frac{\uhat}{\shat}\right).
\label{eq:sub-polcompton}
\ee
 These expressions, as well as the ones relative to dimuon production at
the  HERMES experiment presented in the previous section,
 apply obviously also to the COMPASS $\mu p$ experiment
at CERN whose higher incoming lepton energies
($E_\mu = 50 - 200\ \gev$) enable the determination of
$\Delta \gamma^p(x,\mu^2)$ at lower values of $x$ as compared to the
corresponding measurements at HERMES.
(Notice that for a muon beam one has obviously to replace
$m_e$ by $m_\mu$ in \eqref{eq:gamma_e} and \eqref{eq:polgamma_e}).

\section{Numerical Results}
We shall present here the expected number of events for the
accessible $x$-bins at HERA collider experiments and at 
the fixed-target HERMES experiment subject to some 
representative kinematical cuts which, of course, may be slightly
modified in the actual experiments.
These cuts entail $\shat \ge \vmin{\shat}$, 
$\vmin{\eta} \le \eta_i \le \vmax{\eta}$ and 
$E_i \ge \vmin{E}$ where $E_i$ are the energies of the observed
outgoing particles.

\begin{figure}[t]
\centering
\epsfig{figure=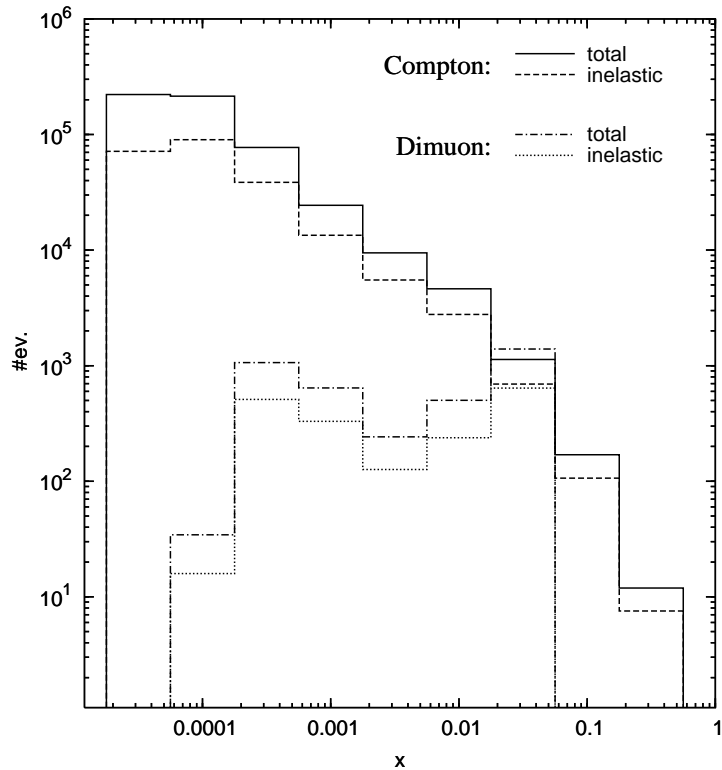, width = 10cm}
\caption{%
Event rates for QED Compton ($e\gamma \to e \gamma$) and dimuon 
production ($\gamma \gamma \to \mu^+ \mu^-$) processes at the 
HERA collider. The cuts applied are as described in the text.}
\label{fig:fig3}
\end{figure}
\begin{figure}[ht]
\centering
\epsfig{figure=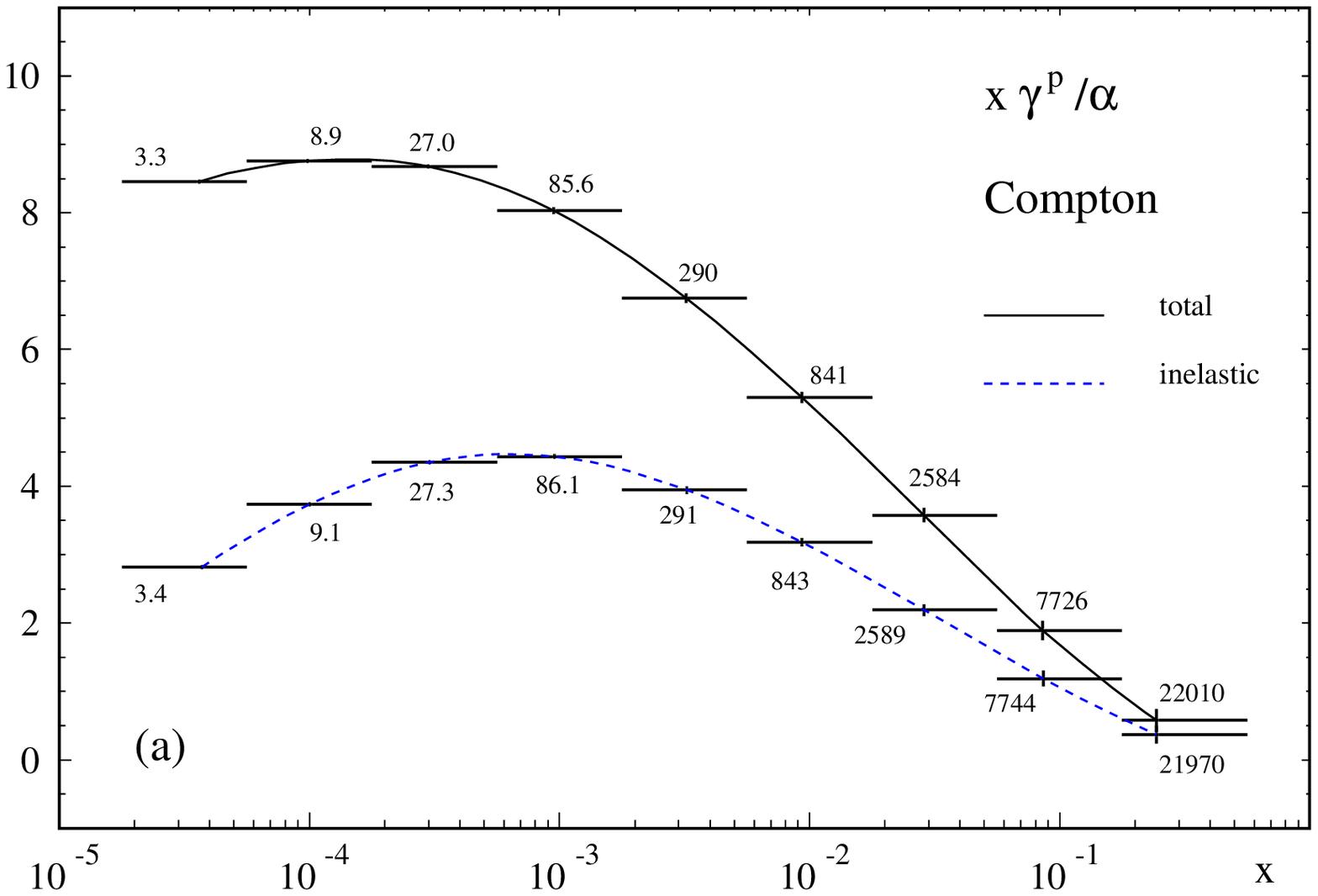, width = 11cm, height = 6.68cm }
\epsfig{figure=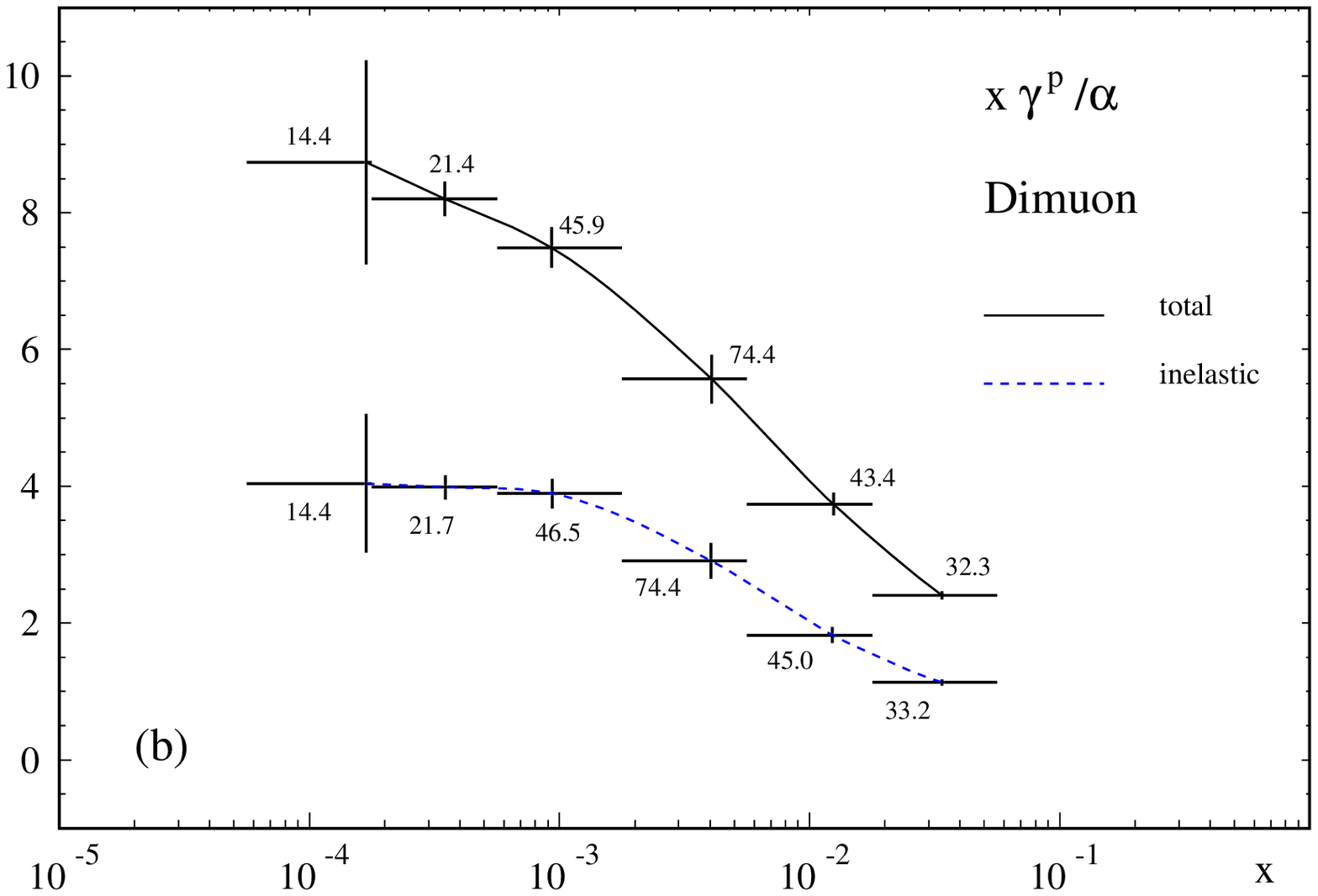, width = 11cm, height = 6.68cm}
\caption{%
Expected statistical accuracy of the determination
of $\gamma^p(\av{x},\av{\shat})$ via the (a) QED Compton process
and (b) the dimuon production process at the HERA collider.
The numbers indicate the average scale $\av{\shat}$ (in  GeV$^2$ units)
for each $x$-bin.}
\label{fig:fig4}
\end{figure}
\begin{figure}[ht]
\centering
\epsfig{figure=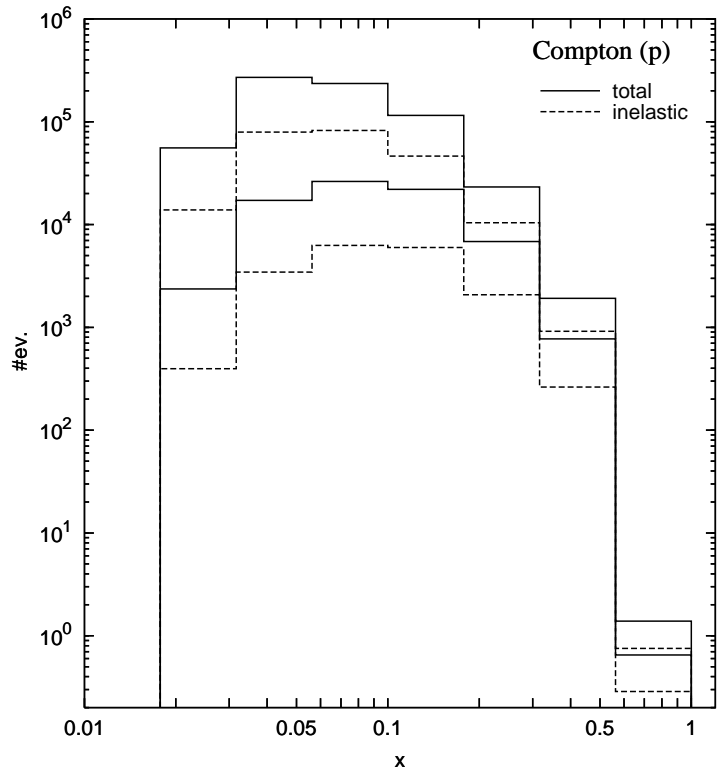, width = 10cm}

\caption{%
Event rates for the QED Compton process at HERMES using an
(un)polarized proton target. The upper (solid and dashed)
curves refer to an unpolarized proton, whereas the lower ones
refer to a polarized proton target. The cuts applied are as described
in the text.}
\label{fig:fig5}
\end{figure}
\begin{figure}[ht]
\centering
\epsfig{figure=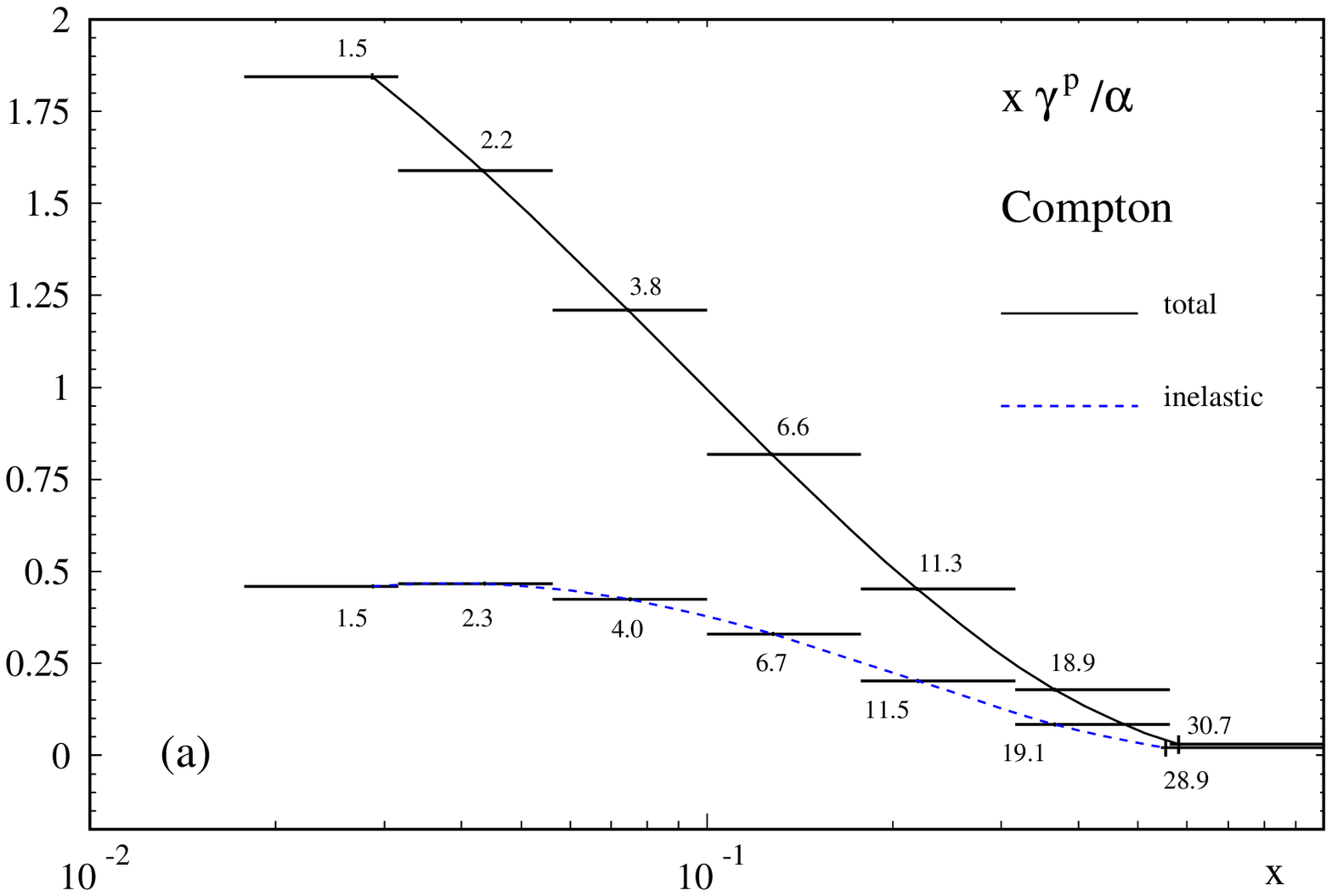, width = 11cm, height = 6.68cm}
\epsfig{figure=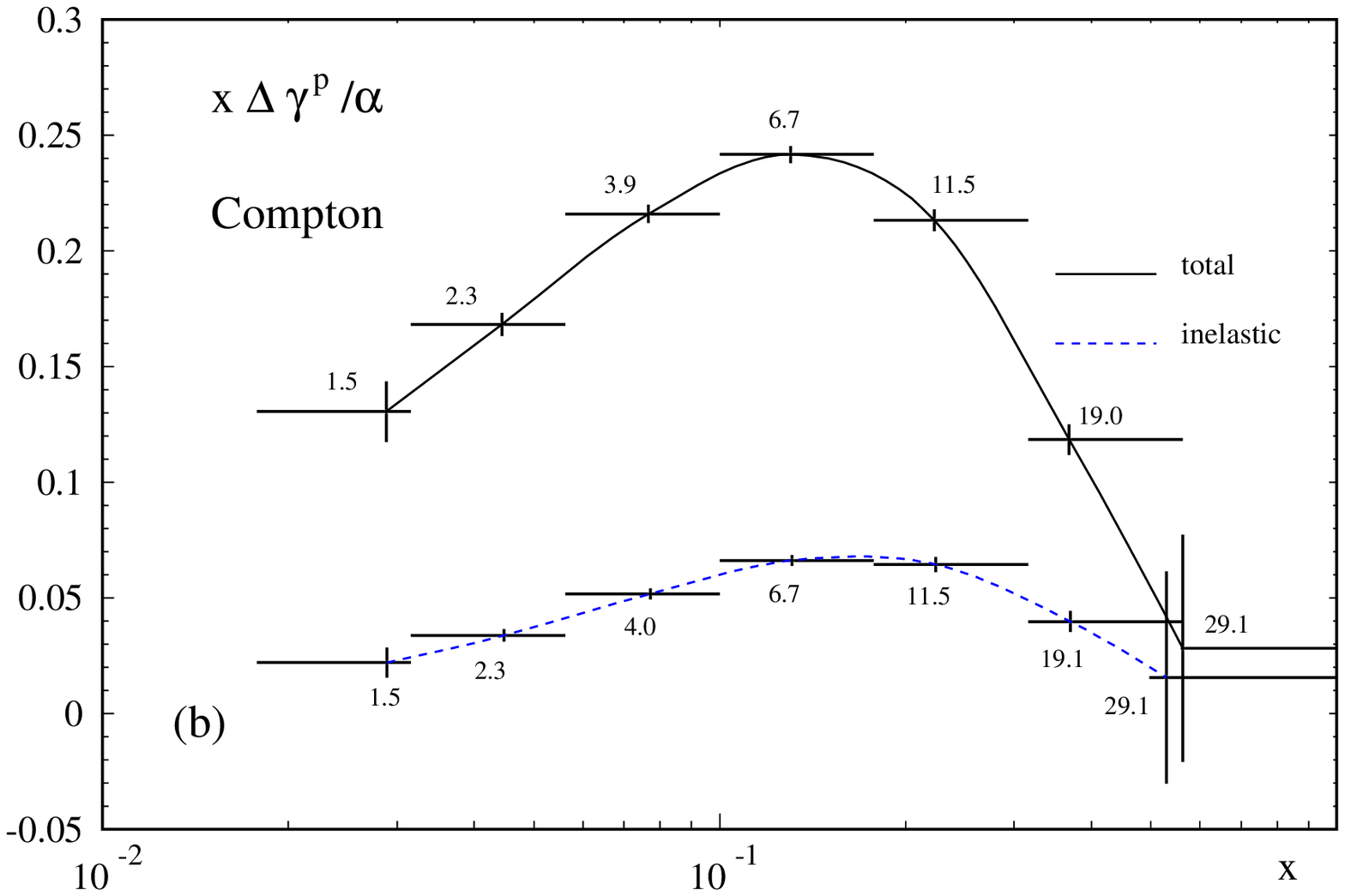, width = 11cm, height = 6.68cm}
\caption{%
Expected statistical accuracy of the determination
of (a) $\gamma^p(\av{x},\av{\shat})$ 
and (b) $\Delta \gamma^p(\av{x},\av{\shat})$ 
via the QED Compton process at HERMES using an (un)polarized
proton target.
The numbers indicate the average scale $\av{\shat}$ 
(in GeV$^2$ units) for each bin.}
\label{fig:fig6}
\end{figure}

\begin{figure}[ht]
\centering
\epsfig{figure=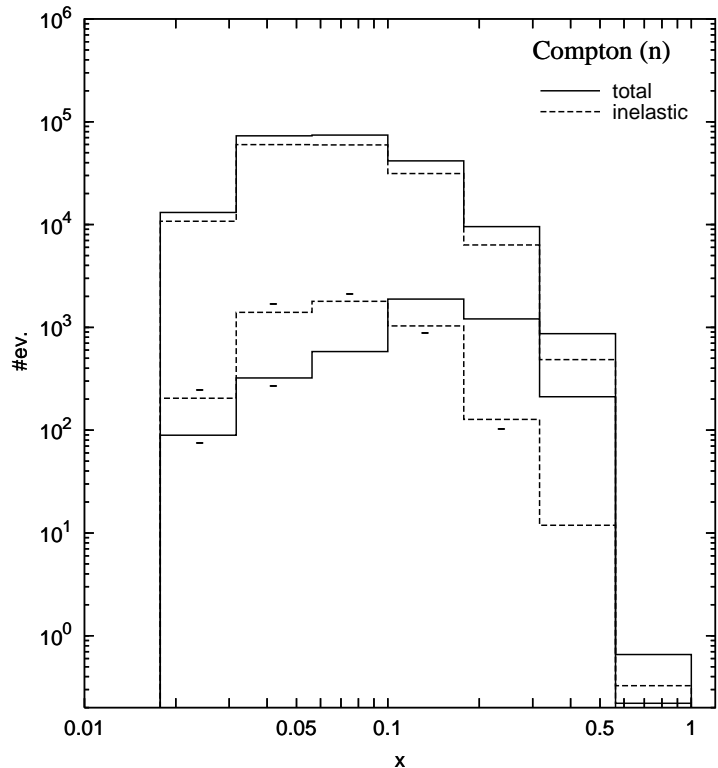, width = 10cm}

\caption{%
As in Figure \protect\ref{fig:fig5} but for a neutron target.
The negative signs at some lower-$x$ bins indicate that the 
polarized total cross section and/or inelastic contribution
is negative.}
\label{fig:fig7}
\end{figure}
\begin{figure}[ht]
\centering
\epsfig{figure=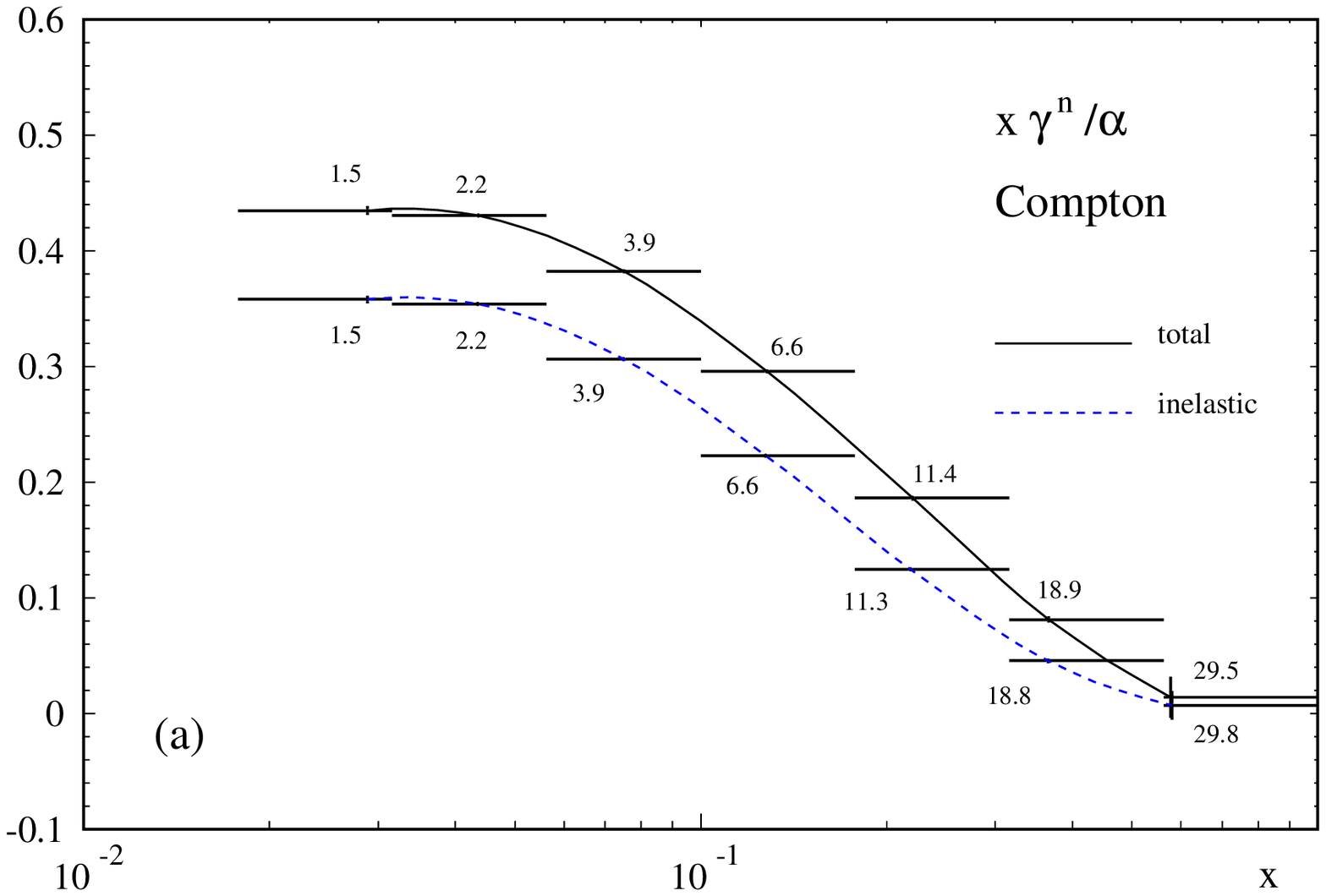, width = 11cm, height = 6.68cm}
\epsfig{figure=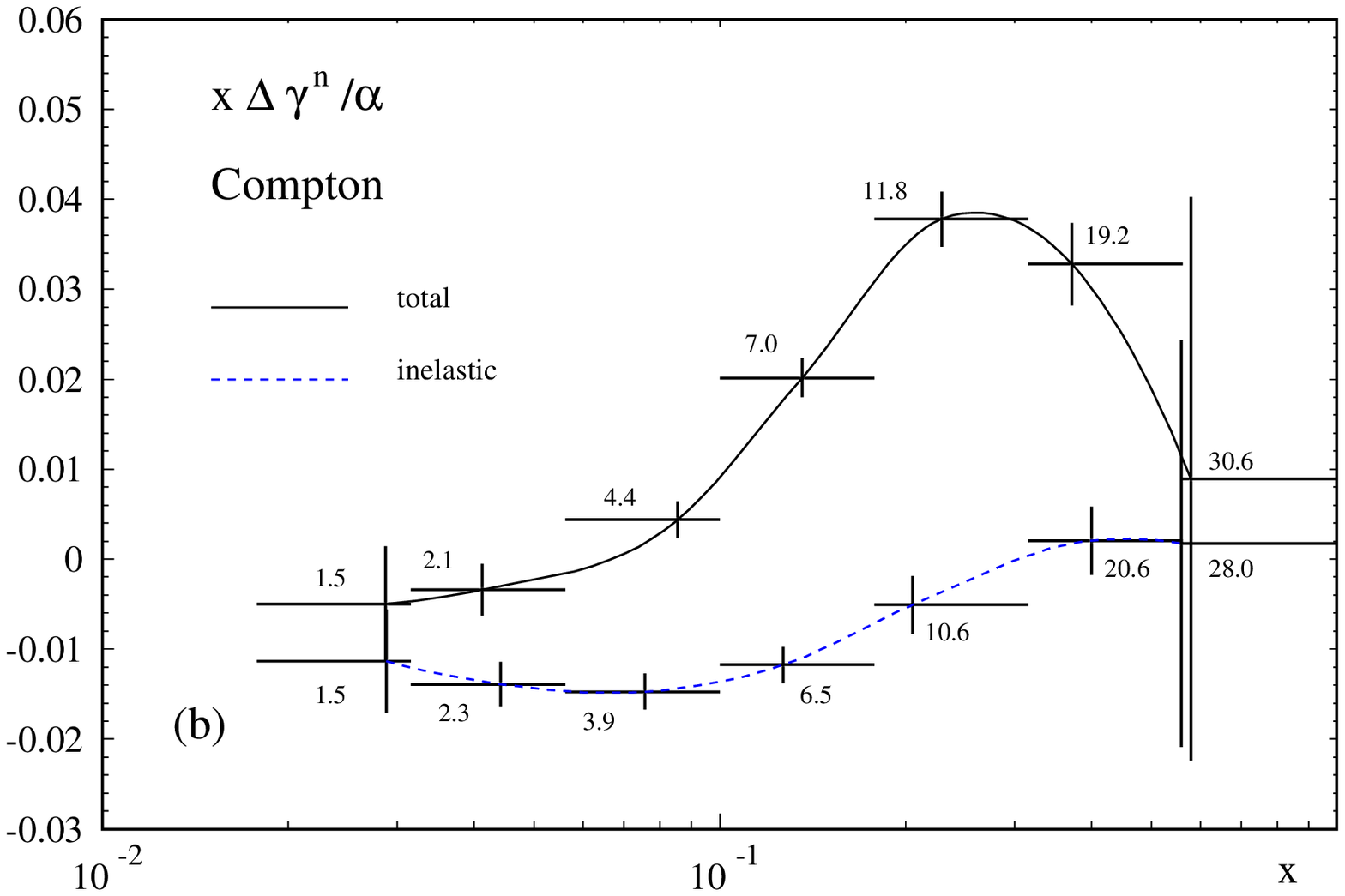, width = 11cm, height = 6.68cm}
\caption{%
As in Figure \protect\ref{fig:fig6} but for a neutron target.}
\label{fig:fig8}
\end{figure}
\begin{figure}[ht]
\centering
\epsfig{figure=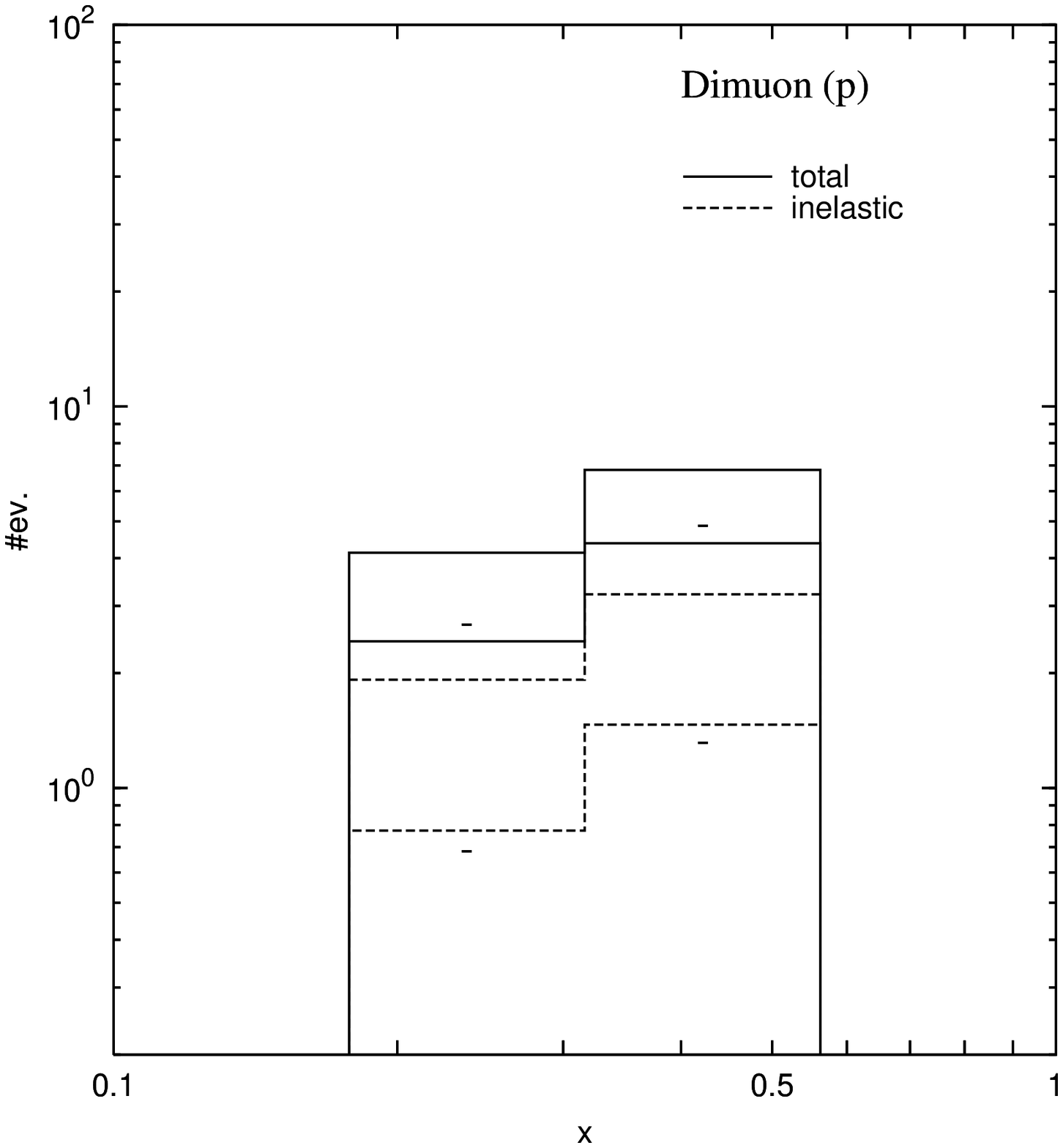, width = 7.9cm}
\epsfig{figure=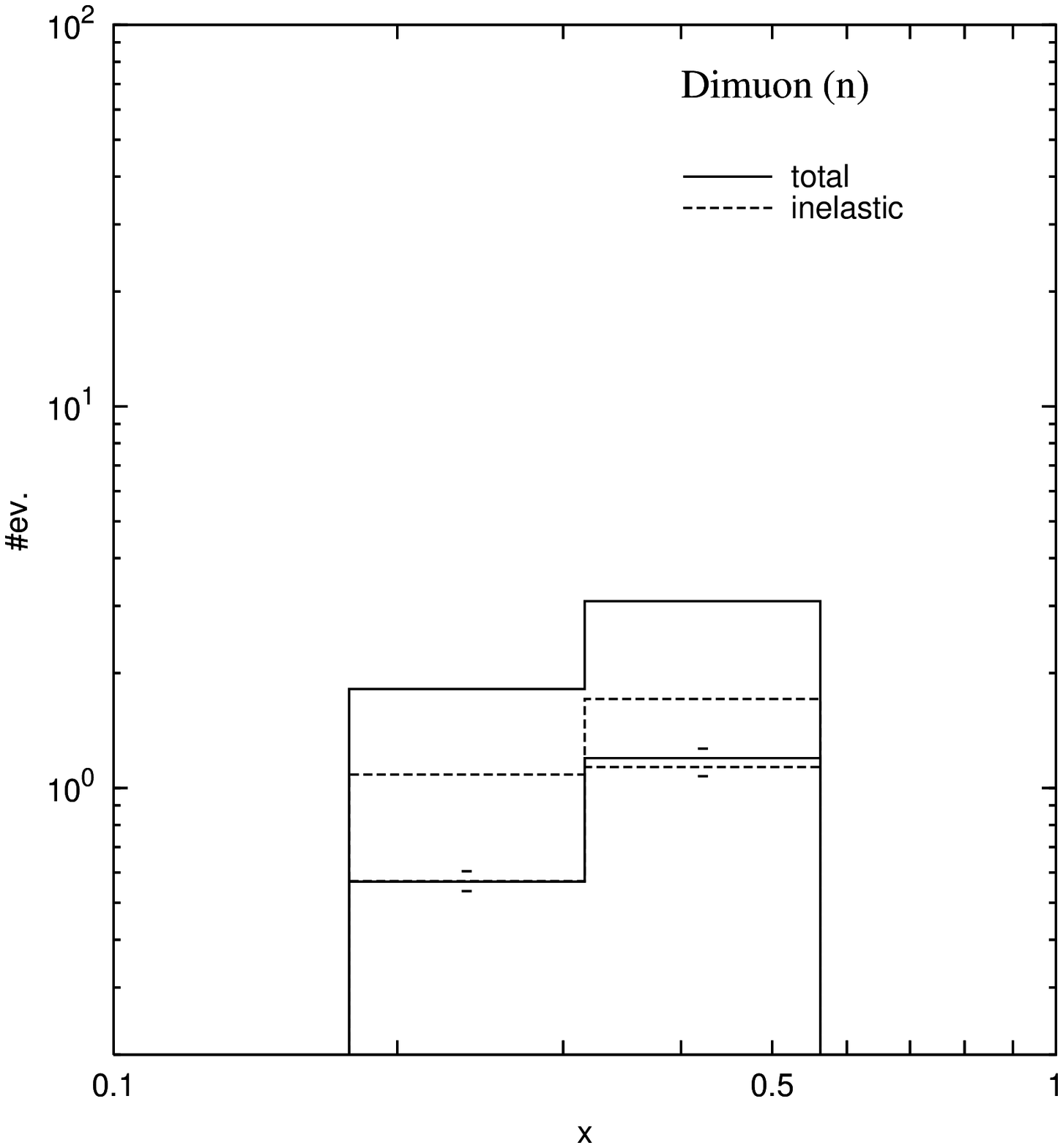, width = 7.9cm}
\caption{%
As in Figure \protect\ref{fig:fig5} but for 
dimuon production at HERMES using (un)polarized proton and
neutron targets. The lower solid and dashed curves refer to
a polarized nucleon target and the negative signs indicate that
the polarized cross sections are negative.}
\label{fig:fig9}
\end{figure}

\begin{figure}[ht]
\centering
\epsfig{figure=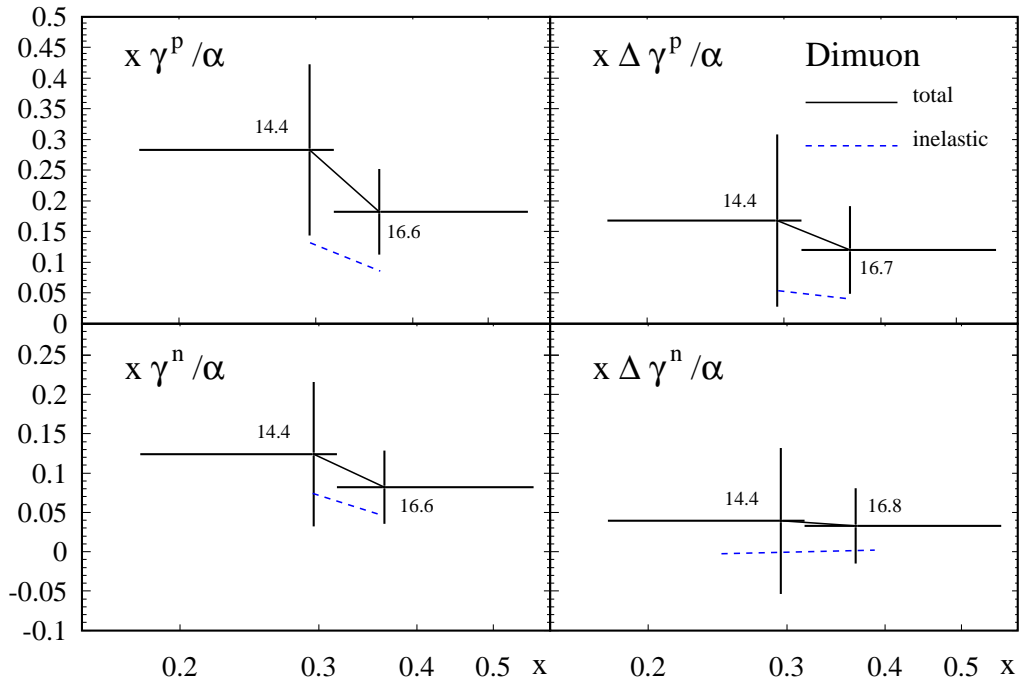, width = 14cm}
\caption{
As in Figure \protect\ref{fig:fig6} but for 
dimuon production at HERMES for (un)polarized proton
and neutron targets. The statistical accuracy for the inelastic
contributions is similar to those shown for the total result, except
for the almost vanishing $\Delta \gamma^n_{\rm inel}$.}
\label{fig:fig10}
\end{figure}

The relevant integration ranges at HERA are fixed via
\be
0 \le \xi \le 1,~~~~~~~~~~~~ \frac{\vmin{\shat}}{4\xi E_e E_p} \le x \le 1,
\label{eq:boundshera}
\ee
with $\shat$ given by $\shat = 4 x \xi E_e E_p$ while
$\eta_i$ are constrained by 
\be
\eta_1 + \eta_2 = \ln \frac{x E_p}{\xi E_e}~,
\ee
which follows from   \eqref{eq:xi} and \eqref{eq:x}.
The relation 
\be
\eta_i - \eta_j = \ln\left [\frac{\xi E_e}{E_i}(1+ e^{2 \eta_i})-1\right ],
\label{eq:etahera}
\ee
as obtained from the outgoing particle energy 
$E_i= p_T\cosh\eta_i$ and its transverse
momentum \eqref{eq:pthera}, further restricts the integration range of
$\eta_{i,j}$ as dictated by $E_i \ge \vmin{(E_i)}$.
For
the QED Compton scattering process \eqref{eq:dics-hera},
$\xi = 1$, $\eta_1 = \eta_e$, $\eta_2 = \eta_{\gamma}$ in 
\eqref{eq:boundshera}-\eqref{eq:etahera}.
At HERMES $E_p \to m/2$ and $\eta_i \to - \eta_i$ in the above 
expressions with $\eta_i$ the outgoing particle rapidity
with respect to the ingoing lepton direction.

In the following we shall consider $\vmin{E} =4\ \gev$.
For the QED Compton scattering process we further employ
$\vmin{\shat} = 1\ \gevsq$ so as to guarantee the applicability of 
perturbative QCD, i.e., the relevance of the utilized
$\gamma^{p}(x,\shat)$, see \eqref{eq:evolution} with $\mu^2 = \hat s$.
For the dimuon production process we shall impose 
$\vmin{\shat} = m^2[\Psi(2 S)] = (3.7\ \gev)^2$ 
so as to evade the dimuon background
induced by charmonium decays at HERMES (higher charmonium states have 
negligible branching ratios into dimuons);
for HERA we impose in addition 
$\vmax{\shat} = m^2[\Upsilon(1S)]= (9.4\ \gev)^2$ in order to avoid
the dimuon events induced by bottomium decays.
Finally, at HERA we consider $\vmin{\eta}=-3.8$, $\vmax{\eta}=3.8$
and at HERMES $\vmin{\eta}= 2.3$, $\vmax{\eta}=3.9$.
The integrated luminosities considered are
$\Lumi_{\rm HERA} = 100\ \mathrm{pb}^{-1}$ and
$\Lumi_{\rm HERMES} = 1\ \mathrm{fb}^{-1}$.

In Figure \ref{fig:fig3} the histograms depict the expected number
of dimuon and QED Compton events at HERA found by integrating 
\eqref{eq:dimuon-hera} and \eqref{eq:dics-hera} 
applying the aforementioned cuts 
and constraints.
The important inelastic contribution due to $\gaminelp$,
 being calculated according to \eqref{eq:evolution}
using the minimal boundary condition, is shown
separately by the dashed curves.

To illustrate the experimental extraction of $\gamma^p(x,\shat)$ we
translate the information in Figure \ref{fig:fig3} into a statement
on the accuracy of a possible measurement by evaluating 
$\gamma^p(\av{x},\av{\shat})$ at the averages
$\av{x}$, $\av{\shat}$ determined from the event sample in
Figure \ref{fig:fig3}.
Assuming that in each bin the error is only statistical, i.e.
$\delta \gamma^p = \pm \gamma^p/ \sqrt{N_{\rm bin}}$,
the results for $x \gamma^p/\alpha$ are shown in
Figure \ref{fig:fig4}. It should be noticed that the statistical
accuracy shown will increase if $\gaminelp(x,\mu_0^2) \ne 0$ in
contrast to our vanishing boundary condition  used
in all our present calculations. Our results for the QED Compton process
in Figures \ref{fig:fig3} and \ref{fig:fig4} are, apart from our somewhat
different cut requirements, similar to the ones presented in 
{\cite{ruju}}.

Apart from testing $\gamma^p (x,\shat)$ at larger values of $x$, the
fixed-target HERMES experiment can measure the polarized
$\Delta \gamma^p (x,\shat)$ as well.
In Figure \ref{fig:fig5} we show the expected number of QED Compton events
for an (un)polarized proton target. 

The accuracy of a possible 
measurement of $\gamma^p(\av{x},\av{\shat})$ and
$\Delta \gamma^p(\av{x},\av{\shat})$ is illustrated in 
Figure \ref{fig:fig6} where the averages
$\av{x}$, $\av{\shat}$ are determined from the event sample in
Figure \ref{fig:fig5} by assuming that the error is only statistical
also for the polarized photon distribution, i.e.
$\delta(\Delta \gamma^p) = \pm (\sqrt{N_{\rm bin}}/|\Delta N_{\rm bin}|)
\Delta \gamma^p$.

The analogous expectations for an (un)polarized neutron target are
shown in Figures \ref{fig:fig7} and \ref{fig:fig8}.
It should be pointed out that, according to Figures \ref{fig:fig6}(b)
and \ref{fig:fig8}(b), HERMES measurements will be sufficiently
accurate to delineate even the polarized $\Delta \gamma^{p,n}$ 
distributions in the medium- to small-$x$ region, in particular the
theoretically more spe\-cu\-lative inelastic contributions.

For completeness, in Figures \ref{fig:fig9} and \ref{fig:fig10}
we also show the results for dimuon production at HERMES for
(un)polarized proton and neutron targets despite the fact that the
statistics will be far inferior to the Compton process.

The dimuon production can obviously proceed also via the genuine Drell-
Yan subprocess $q \bar{q}\to \mu^+ \mu^-$ where one of the (anti)quarks
resides in the resolved component of the photon emitted by the electron.
However, as already noted in \cite{Arteaga-Romero:1991wn}, this 
contribution is negligible as compared to the one due to the Bethe-
Heitler subprocess $\gamma\gamma \to \mu^+ \mu^-$. The unpolarized dimuon 
production rates at HERA where also studied in \cite{{Arteaga-Romero:1991wn},
{Bussey:1996vq}} utilizing, however, different prescriptions for
the photon content of the nucleon.\\
Exact expressions for the Bethe-Heitler contribution to the longitudinally
polarized $\gamma N \to \mu^+ \mu^- X$ process are presented in 
 \cite{Gehrmann:1997qh} but no estimates for the expected production
rates at, say, HERMES or COMPASS are given.

\section{Summary}
The analysis of the production rates of lepton-photon and muon pairs
at the colliding beam experiments at HERA and the fixed-target 
HERMES facility,
as evaluated in the leading order equivalent photon approximation,
demonstrates the feasibility of de\-ter\-mi\-ning the polarized and
unpolarized equivalent photon distributions of the nucleon in the
available kinematical regions. The above mentioned production rates
can obviously be determined in a more accurate calculation along the
lines of \cite{kessler}, involving the polarized 
and unpolarized
structure functions $g_{1,2}$ and $F_{1,2}$, respectively, of
the nucleon. The expected production rates are similar to those
obtained in our equivalent photon approximation, as discussed 
in detail in the next chapters. 
It thus turns out that lepton-photon and muon pair production
at HERA and HERMES may provide an additional and independent
source of information concerning these structure functions.

\clearemptydoublepage


\chapter{{\bf The Unpolarized QED Compton Scattering Process}}
\label{ch:qedcs}


The QED Compton scattering in high 
energy electron-proton collisions $ ep\rightarrow e\gamma X$ is
one of the most important processes for an understanding of the photon
content of the proton. 
In addition, it can also shed some light on the proton structure functions 
$F_{1,2}(x_B,Q^2)$ \cite{thesis,kessler,f2h1}  in the low-$Q^2$
region, where they are presently poorly known \cite{blu}.

The QED Compton scattering has
been recently analyzed in \cite{thesis}, where the above mentioned
alternative
descriptions were confronted with the experimental data and it was found that 
the description
in terms of  $F_{1,2}$, i.e. for $X \not= p$, is superior to the one
in terms of the inelastic photon distribution 
$\gamma^p_{\mathrm{inel}} (x_B,Q^2)$.
Henceforth we shall refer to the description in terms of $F_{1,2}$ as
exact to distinguish it  from the approximations involved in the EPA. 

It should be noted, however,  that 
the analysis in \cite{thesis,kessler} utilized the Callan-Gross
relation \cite{callan}
$ F_L(x_B,Q^2)=F_2(x_B,Q^2)-2 x_B F_1 (x_B,Q^2)=0$. This relation is
contaminated by higher order (NLO) QCD corrections as well as by higher
twist contributions relevant in the low-$Q^2$ region which may invalidate     
the assumptions underlying the exact analysis.
Furthermore, the analysis in \cite{thesis,kessler} was carried out
within the framework of the helicity amplitude  formalism \cite{kessler}.
The implementation of experimental cuts within this formalism is nontrivial
and affords therefore an iterative numerical approximation procedure
\cite{kessler,h1} whose first step  corresponds to $-k^2=Q^2=0$, where $k$
is the momentum of the virtual photon.

It is this second issue that we intend to study here. We shall replace the
noncovariant helicity amplitude analysis of \cite{kessler} by a standard
covariant tensor analysis whose main advantage, besides compactness and
transparency, is the possibility to implement the experimental cuts directly
and thus avoid the necessity of employing an iterative approximation of
limited accuracy. The first issue concerning the $F_L$ contributions affords
some estimates of this poorly known structure function and we refrain  from
its study here.   

In  Section \ref{sec:radiative} it is specified what  QED Compton scattering
is and how it can be selected from the reaction $ep\rightarrow e\gamma X$. 
In Sections \ref{sec:elastic} and \ref{sec:inelastic}, we calculate its exact 
cross section  for the elastic and inelastic  channels.
The numerical results are discussed in
Section \ref{sec:results}. The summary is given in Section 
\ref{sec:summary} and 
the kinematics  in Appendix \ref{app:kinel1}. 

\section{Radiative Corrections to Electron-Proton  Collisions}
\label{sec:radiative}

The lowest order Feynman diagrams describing  the process 
$ep \rightarrow e \gamma X$,
with a real photon emission from the electron side, are shown in Figure 
\ref{fig:one}. The  corresponding amplitudes  contain the
denominators $k^2 (k'+l')^2$ and  $k^2(l-k')^2$, therefore the main
contributions to the cross section  come from  configurations where
one or both denominators tend to zero. These configurations have different
experimental signatures and they are  described in the following 
\cite{kessler,thesis}.

\begin{itemize}
\item The dominant contribution stems from the so-called {\em bremsstrahlung 
      process}, which corresponds to a configuration where
      $k^2$, $(l'+k')^2$, $(l-k')^2$ stay  all close to zero. 
      It involves 
      quite large counting rates but in order to be measured, requires a 
      specific small-angle detector, because the outgoing electron and 
      photon have small polar angles and escape through the beam pipe. 
\begin{figure}[t]
\begin{center}
\epsfig{figure= 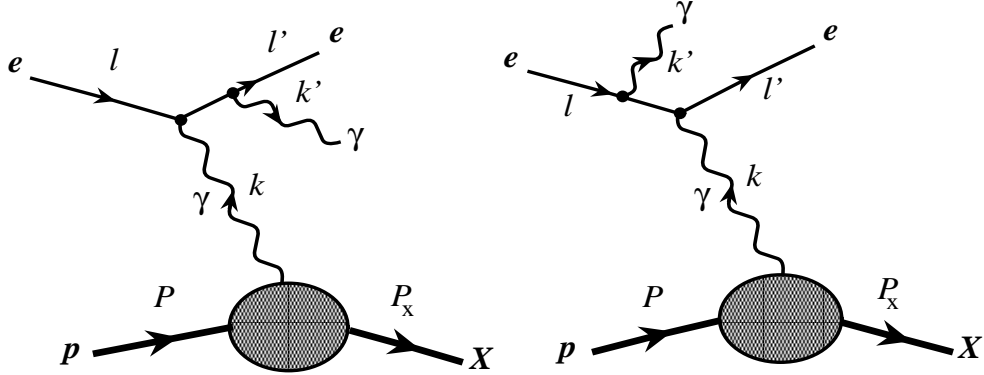, width=16cm, height= 8cm}\\
\end{center}
\caption{Feynman diagrams considered for $ep \rightarrow e\gamma X$, with  a 
real final state photon $(k'^2=0$).}
\label{fig:one}
\end{figure}

\item Either $(l'+k')^2 \simeq 0$ or $(l-k')^2 \simeq 0$, but 
      the momentum squared of the exchanged virtual photon $k^2$ is 
      finite: the outgoing photon is emitted either along the final
      or the incoming electron line and this configuration corresponds 
      to the so called {\em radiative corrections} to
      electron-proton scattering.
  
      In the first case, $(l'+k')^2 \simeq 0$, the cross section
      is dominated by the contribution given by the first diagram, and
      this kind of events is called {\em Final State Radiation}  (FSR).
      It is usually very difficult to distinguish this process experimentally 
      from non-radiative deep  inelastic scattering $ep\rightarrow eX$,
      since the outgoing electron and photon are almost collinear.	
      
      In the second case, $(l-k')^2\simeq 0$, the main contribution to 
      the cross  section 
      is given by the second Feynman diagram in Figure \ref{fig:one} 
      and such events are classified as {\em Initial State Radiation} (ISR).
      In the detector one observes only the outgoing electron, 
      the final photon being emitted along the incident electron line.

\item The virtuality of the exchanged photon is small, $k^2 \simeq 0$, 
      but both 
      $(l'+k')^2$ and $(l-k')^2$ are finite: the produced hadronic system
      goes straightforwardly along the incident proton line,
      the outgoing electron and photon are detected under large polar angles
      and almost back-to-back in azimuth, so that their total transverse 
      momentum is close to zero. This configuration is referred to as 
      {\em QED Compton scattering}, since it involves the scattering
      of a quasi-real photon on an electron. This process will thus be
      selected by performing a cut on the total transverse momentum of 
      the outgoing electron and photon or on the acoplanarity 
      \eqref{eq:acopla} of the electron-photon system. 

      As pointed out in
      \cite{blu}, the corresponding cross section is large, despite the 
      fact that it contains an additional factor $\alpha = 1/137$ compared
      to the tree-level cross section for $ep\rightarrow eX$. This 
      follows since the emission of a large transverse momentum photon 
      can lead to a reduction of the true momentum squared transferred to the 
      proton: $(l-l')^2$ in $ep\rightarrow eX$  is shifted 
      to $(l-l'-k')^2$ in $ep\rightarrow e\gamma X$, which can become of the 
      order of 
      the proton mass squared or even smaller, as will be discussed below, see
      \eqref{eq:q2min}. 
      The reduction in $Q^2$ and the corresponding
      increase of the cross section in \eqref{siin}, compensates for
      the smallness  of the additional factor $\alpha$. Hence QED Compton
      scattering can provide a tool to investigate the small-$Q^2$ behaviour
      of the proton structure functions. The first measurements of $F_2$
      using QED Compton scattering have been  published  by the H1 
      collaboration at HERA \cite{f2h1}.

\end{itemize}

\section{Elastic QED Compton Scattering}
\label{sec:elastic}
We consider elastic QED Compton scattering:
\be
e(l)+p(P) \rightarrow e(l')+\gamma(k')+ p(P'),
\ee
where the four-momenta of the particles are given in the brackets. 
We introduce the invariants 
\be
s=(P+l)^2, ~~~~~~~~~ t=k^2,
\label{pap1_invar}
\ee
where  $k=P-P'$ is the four-momentum of the 
exchanged virtual photon. Moreover,  we will make use of the Mandelstam
variables   \eqref{comptonmandel} relative to the subprocess 
$e(l)\gamma(k)\to e(l')\gamma(k')$.
The photon in the final state is real, $k'^2=0$. We neglect the
electron mass everywhere except when it is necessary to avoid divergences in
the formulae and take the proton to be massive, $ P^2=P'^2=m^2 $. The
relevant Feynman diagrams for this process are shown in Figure \ref{fig:one}, 
with $X$
being a proton and $P_X=P'$. The
squared matrix element can be written as
\be
\overline{{\mid M_{\mathrm{el}} \mid }^2}= \frac{1} {t^2} H^{\alpha \beta}_
{\mathrm{el}}(P,P') 
T_{\alpha \beta}(l;l',k'),
\label{eq:amplsqel}
\ee
where $T_{\alpha \beta}(l;l',k')$ is the leptonic tensor 
\eqref{eq:leptonict}, given also in  \cite{ji,anf} 
and $H^{\alpha \beta}_{\mathrm{el}}(P,P') $ is the 
hadronic tensor defined 
in the first line of \eqref{eq:hadronic}, with $N = p$, in terms of
the electromagnetic current $J^{\alpha}_{\mathrm{em}}$.
If we use the notation 
\be
\der PS_N(p;p_1,...,p_N)=(2 \pi)^4 \delta \bigg (p-\sum_{i=1}^N p_i \bigg ) 
\prod_{i=1}^N
\frac{\der^3p_i}{ (2 \pi)^3 2 p_i^0}
\label{eq:lorentz}
\ee
for the Lorentz invariant $N$-particle phase-space element, the total 
cross section will be
\be
\sigma_{\mathrm{el}}(s)=\frac{1}{2 (s-m^2)} \int \der PS_{2+1}(l+P;l',k',P')
\overline{{\mid M_{\mathrm{el}} \mid }^2}~.
\label{pap1_sigmael}
\ee
Equation (\ref{pap1_sigmael}) can be rewritten following the technique of
 \cite{kniehl},
which we slightly modify to implement the experimental cuts and 
constraints; in particular all the integrations will be  performed
numerically. Rearranging the ($2+1$)-particle phase space into a sequence
of two $2$-particle ones,  (\ref{pap1_sigmael}) becomes:
\be
\sigma_{\mathrm{el}}(s)=\frac{1}{2 (s-m^2)} \int \frac{\der\hat s} {2 \pi} 
\,\der PS_2
(l+P;l'+k',P') \frac{1}{t^2} H^{\alpha \beta}_{\mathrm{el}} (P,P') 
X_{\alpha \beta}(l,k)~.
\label{eq:sigmaintel}
\ee
The tensor $X_{\alpha \beta}$ contains all the informations about the 
leptonic part of the process and is defined as
\be
X_{\alpha \beta}(l,k)=\int \der PS_2(l+k;l',k') T_{\alpha \beta}(l;l',k')
\label{eq:xmunu}
\ee
and $T_{\alpha \beta}$ can be written as \cite{pap1}
\be
T_{\alpha \beta}(l;l',k') =\,\frac{4 e^4} {\hat{s}\hat{u}}\,\bigg
\{ \,\frac{1}{2}\,
g_{\alpha\beta}\,(\hat s^2 + \hat u^2 + 2 \hat t t) + 2\hat s\, l_{\alpha}l_{\beta}
+ 2 \hat u\,l_{\alpha}'l_{\beta}'~~~~~~~~~~~~~~~~~~\nonumber \\
~~~+\,(\hat t+t)(l_{\alpha}l_{\beta}'+l_{\beta} l_{\alpha}') 
-({\hat s}-t)\,(l_{\alpha}k_{\beta}'+l_{\beta}k_{\alpha}')~~~~~~~~\nonumber \\ 
+\,(\hat u -t) \, (l_{\alpha}'k_{\beta}' +l_{\beta}'k_{\alpha}')\bigg\}.
\label{lept}    
\ee
It can be shown that
\be
\der PS_2(l+k;l',k')=\frac{\der\hat t\, \der\varphi^*}{16 \,\pi^2 (\hat s-t)}~,
\label{eq:phasel1}
\ee
with  $\varphi^*$ denoting the azimuthal angle of the outgoing $e-\gamma$ 
system in the $e-\gamma$ center-of-mass frame.
For unpolarized scattering, $X_{\alpha \beta}$ is symmetric in the indices
$\alpha$, $\beta$  and can be expressed in terms of  two 
Lorentz scalars, $\tilde{X}_1$ and $\tilde{X}_2$:
\be
X_{\alpha \beta}(l,k)&=&\frac{1}{2 t} \bigg \{ [ 3\tilde{X}_1(\hat s,t
)+\tilde{X}_2 (\hat s,t) ] \bigg
(\frac{2 t} {\hat s-t}\, l-k \bigg )_\alpha \bigg (\frac{2 t}{ \hat s-t} \,
l-k\bigg )_\beta
\nonumber \\&&~~~~~+[ \tilde{X}_1(\hat s,t)+\tilde{X}_2(\hat s,t)] (t g_{\alpha \beta}-k_\alpha k_\beta)
\bigg \},
\label{pap1_xmunu}
\ee
with
\be
\tilde{X}_1(\hat s,t)= \frac{4 t} {(\hat s-t)^2} l^\alpha l^\beta X_{\alpha \beta}(l,k),
\ee
\be
\tilde{X}_2(\hat s,t)=g^{\alpha \beta} X_{\alpha \beta}(l,k).
\ee
Using the leptonic tensor (\ref{lept}) and also the relations
\be
l \cdot k= \frac{1}{2}\, (\hat s-t), ~~~~l \cdot P=\frac{1} {2} \,(s-m^2),~~~k
\cdot P=\frac{1} {2} \,t,
\ee
we obtain
\be
\frac{t\, l^\alpha l^\beta T_{\alpha \beta}}{4 \pi^2 (\hat s-t)^3}=\,
{e^4}\, \frac{-t \hat t}{2 \pi^2 (\hat s-t)^3} \equiv X_1(\hat s,t,\hat t),
\label{x1}
\ee
\be
\frac{g^{\alpha \beta} T_{\alpha \beta}} {16 \pi^2 (\hat s-t)}= \,{e^4}\, \frac{(t^2-2 t
\hat s+2 \hat s^2+2 \hat s \hat t+\hat t^2)} {4 \pi^2 \hat s (\hat s-t)
(t-\hat s-\hat t)} \equiv X_2(\hat s,t, \hat t).
\label{x2}
\ee
The invariants $X_i(\hat s,t,\hat t)$, with $i=1,2$, are related to $\tilde{X}
_i(\hat s,t)$ by
\be
\tilde{X}_i(\hat s,t)=2 \pi \int_{\hat t_{\mathrm{min}}}^
{\hat t_{\mathrm{max}}} \der \hat t\,  {X}_i(\hat s,t,\hat t).
\ee
The integration limits of $\hat t$ are:
\be
\hat t_{\mathrm{max}}=0, ~~~~\hat t_{\mathrm{min}} = -\hat s +t+\frac{\hat s }
{\hat s-t} \, m_e^2,
\label{thatlim}
\ee
where $m_e$ is the mass of the electron. We point out that 
the kinematical cuts 
employed by us prevent the electron propagators to become too small and 
thus the divergences are avoided, so we can safely
neglect the electron mass in the numerial calculation. 
The hadronic tensor in the case of elastic scattering can be expressed in 
terms of the common proton form factors as in \eqref{eq:hel}, namely
\be
H^{\alpha \beta}_{\mathrm{el}}(P,P')=  e^2\,[ H_1(t) (2 P-k)^\alpha (2 P-k)^\beta +
G_M^2(t) (t g^{\alpha
\beta}-k^\alpha k^\beta)],
\label{eq:hadronictens}
\ee
where $H_1(t)$, already introduced in \eqref{eq:h1}, is given by
\be
H_1(t)=\frac{G_E^2(t)- (t/4 \,m^2)\, G_M^2(t)}{1-{t/4\, m^2}}~.
\label{eq:h12}
\ee
Using 
\be
\der PS_2(l+P;l'+k',P')=\frac{\der t} {8 \pi (s-m^2)}~,
\label{eq:phasel2} 
\ee
finally we get \cite{pap1}
\be
\sigma_{\mathrm{el}}(s)&=&\frac{\alpha}{ 8 \pi (s-m^2)^2} \int_{\hat s_{\mathrm
{min}}}^{(\sqrt
s-m)^2} \der\hat s \int_{t_{\mathrm{min}}}^{t_{\mathrm{max}}}{\der t \over t} 
\int_
{\hat t_{\mathrm{min}}}^{\hat t_{{\mathrm{max}}}} \der \hat t \int_0^{2 \pi} 
\der \varphi^*\bigg\{ \bigg [ 2\, {s-m^2\over \hat s-t} \nonumber \\ 
&& \times
\bigg ( {s-m^2\over \hat s-t}-1 \bigg )  [3 X_1(\hat s,t,\hat
t)+X_2 (\hat s,t,\hat t)] +{2 m^2\over t} [X_1(\hat s,t,\hat t)+X_2(\hat
s,t,\hat t)]\nonumber \\&& ~~~~~~~~~~~~~~~~~~~~~~~~~+X_1(\hat s,t,\hat t)
\bigg ]
H_1(t) + X_2(\hat s,t,\hat t) G_M^2(t) \bigg \},
\label{sigel}
\ee
where $\hat s_{\mathrm{min}}$ denotes the minimum of $\hat s$ and $t_{\mathrm
{min},\, {\mathrm{max}}}$ are given by
\be
t_{\mathrm{min},\,{\mathrm{max}}}=2 m^2-{1\over 2  s} \Big [ (s+m^2)  (s-\hat 
s+m^2) \pm (s-m^2) \sqrt
{(s-\hat s+m^2)^2-4 s m^2}\,\Big ].
\label{eq:tlimits}
\nonumber \\
\ee 
It is to be noted that in  (\ref{sigel}) we have shown the integration
over $\varphi^*$ explicitly, because of the cuts that we shall impose on the 
integration variables for the numerical calculation of the cross section. 
The cuts are discussed in Section \ref{sec:results}.
  
The EPA consists of considering the exchanged
photon as real, so it is particularly good for the elastic process in which
the virtuality of the photon $|t|$ is constrained to be small ($\lesssim 1\, 
\mathrm{GeV}^2$) by the form factors. It is possible to get the approximated
cross section $\sigma_{\mathrm{el}}^{\mathrm{EPA}}$ from the exact one in a 
straightforward way, following again {\cite{kniehl}}. If the invariant mass
of the system $e-\gamma$ is large compared to the proton mass, $\hat s_{\mathrm
{min}} \gg m^2$, one can neglect $ |t|  $ versus $\hat s $, $m^2$ versus $s$, 
then
\be
{X}_1(\hat s, t,\hat t) \approx {X}_1(\hat s,0, \hat t)=0,
\label{xone}
\ee
and \be
{X}_2(\hat s, t, \hat t)\approx {X}_2(\hat s,0, \hat t) = -{{2 \hat s}\over{\pi}}\, {{\der\hat{\sigma}(\hat s, \hat t)}\over{\der\hat t}}~,
\label{xtwo}
\ee
where the differential cross section for the real 
photoproduction process $e \gamma \rightarrow e \gamma$  is given in
\eqref{eq:sub-compton}.                        
We get:
\be
\sigma_{\mathrm{el}}(s) \approx \sigma_{\mathrm{el}}^{\mathrm{EPA}} =
\int_{x_{\mathrm{min}}}^{(1-{m/ 
\sqrt s})^2}\, \der x \,\int_{m_e^2 -\hat s}^0
\der\hat t  \,\gamma^p_{\mathrm{el}} (x) \,{\der \hat \sigma (x s, \hat t)\over \der\hat t}~ ,
\label{epael}
\ee
where $x={\hat s/ s}$ and $\gamma^p_{\mathrm{el}}(x)$ is the elastic
contribution to the equivalent 
photon distribution of the proton \eqref{eqtwo}:
\be
\gamma^p_{\mathrm{el}}(x) =-{\alpha\over 2 \pi} x \int_{t_{\mathrm{min}}}^{t_{\mathrm
{max}}} {\der t\over t} \bigg \{ 2 \bigg [ {1 \over x}\bigg  ( {1\over x}-1 \bigg ) 
+{m^2\over t}\bigg ] H_1(t) + G_M^2(t) \bigg \},
\ee
with
\be
{t_{\mathrm{min}} \approx -\infty} ~~~~~~~~~~~~t_{\mathrm{max}} \approx 
-{{m^2 x^2}\over{1-x}}~.
\ee
To clarify the physical meaning of $x$, let us introduce the variable $x_{\gamma}$:
\be
x_{\gamma} = {{l \cdot k} \over {P \cdot l}}~.
\label{eq:icsgamma}
\ee
It is possible to show that $x_{\gamma}$ represents the fraction  of the 
longitudinal momentum of the proton carried by the virtual photon, 
so that one can write
\be
k = x_{\gamma} P + \hat k,
\ee   
with $\hat k \cdot l = 0$. Using (\ref{pap1_invar}) one gets
\be
x_{\gamma} = {{\hat s - t}\over{s-m^2}}~,
\label{xgamma}
\ee
which reduces to $x$ in the EPA limit, see also \eqref{eq:xg} 
and \eqref{eq:kappa}. One can also 
define the leptonic variable $x_l$:
\be
x_l = {{Q^2_l}\over{2 P \cdot (l - l')}}~,
\label{xl}
\ee
where $Q^2_l = - \hat t$. When $t \simeq 0$, it turns out that also
$x_l \simeq x$.


\section{Inelastic QED Compton Scattering}
\label{sec:inelastic}
To calculate the inelastic QED Compton scattering cross section, we extend the
approach discussed in the previous section. In this case, an electron and a
photon are produced in the final state with a general hadronic system $X$.  
In other words, we consider the process
\be
e(l)+p(P) \rightarrow e(l')+\gamma(k')+X(P_{X}),
\ee
where $P_{X}=\sum_{X_i} P_{X_i}$ is the sum over all momenta of the produced 
hadrons.
Let the invariant mass of the produced hadronic state $X$ to be $W$; (\ref
{pap1_invar}) still holds with $Q^2 = -t$.
The cross section for inelastic scattering will be
\be
\sigma_{\mathrm{inel}}(s)={1\over {2 (s-m^2)}} \int \der PS_{2+N}(l+P;l',k',P_{X_1}, 
..., P_{X_N})
\overline{{\mid M_{\mathrm{inel}} \mid }^2},
\label{sigmainel}
\ee
where 
\be
\overline{{\mid M_{\mathrm{inel}} \mid }^2}={1\over Q^4} H^{\alpha \beta}_
{\mathrm{inel}}(P,P_X) 
T_{\alpha \beta}(l,k;l',k')
\ee
is the squared matrix element and the tensor $H^{\alpha \beta}_{\mathrm{inel}}
(P,P_X) $ has already been introduced in \eqref{eq:hinelas}.
If we rearrange the  ($2+N$)-particle space phase into a sequence of a $2$-particle and a $N$-particle one, we get  
\be
\sigma_{\mathrm{inel}}(s)={1\over 2 (s-m^2)} \int {\der W^2\over 2 \pi} 
\int {\der 
\hat s\over 2 \pi } \int \der PS_2(l+P;l'+k',P_{X}) {1\over Q^4} W^{\alpha \beta}
(P,k) 
X_{\alpha \beta} (l,k),\nonumber \\
\ee
where $X_{\alpha \beta}$ is given by  (\ref{pap1_xmunu}) and 
$W^{\alpha \beta}$ is the
hadronic tensor for inelastic scattering defined in \eqref{eq:hadr}, that is,
using the notation \eqref{eq:lorentz},
\be
W^{\alpha \beta}= \int \der PS_{N}(P-k;P_{X_1},....,P_{X_N})\,H^{\alpha \beta}_{\mathrm{inel}}~.
\ee 
The hadronic tensor is parametrized in terms of  $F_1$ and  $F_2$,  the usual 
structure functions of the proton, as in \eqref{eq:hinel}, i.e.
\be
W^{\alpha \beta}&=&{4 \pi e^2 \over Q^2} \bigg [- (Q^2 g^{\alpha \beta}+ k^{\alpha} k^{\beta}) F_1(x_B,Q^2) 
\nonumber \\&&~~~~~~~~~~~~~~~~~~~~~~~~~~
+(2x_B P^{\alpha}-k^{\alpha}) (2 x_B P^{\beta}-k^{\beta}){F_2(x_B,Q^2)\over 
2 x_B} \bigg ], 
\ee
where $x_B$ is the Bjorken variable \eqref{bjorkenv},
\be
x_B= {Q^2\over 2 P\cdot (-k)} = {Q^2\over Q^2+W^2-m^2}~.
\label{eq:bjorkenx}
\ee
Using 
\be
\der PS_2(l+P;l'+k',P_X)= {\der Q^2\over 8 \pi (s-m^2)} 
\ee
as before, we get \cite{pap1}
\be
\sigma_{\mathrm{inel}}(s)&=&{\alpha\over 4 \pi (s-m^2)^2} \int_{W^2_{{\mathrm
{min}}}}^{W^2_{\mathrm{max}}}
 \der W^2 \int_{\hat s_{\mathrm{min}}}^{(\sqrt
s-W)^2} \der\hat s \int_{Q^2_{\mathrm{min}}}^{Q^2_{\mathrm{max}}} {dQ^2 \over 
Q^4} \int_{\hat t_{\mathrm{min}}}^{\hat
t_{\rm max}} \der \hat t \int_0^{2 \pi} \der \varphi^*\nonumber \\ 
&&~~\times \bigg \{ \bigg [\bigg ( 2 \,{{s-m^2}\over {\hat s+Q^2}}
\bigg (1-{{s-m^2}\over {\hat s+Q^2}}\bigg )  + (W^2-m^2) \bigg ( {2\,
(s-m^2)\over {Q^2 (\hat s + Q^2)}}-{1\over Q^2} 
\nonumber \\ &&~~~~~+{m^2-W^2\over 2 \,Q^4}\bigg ) 
\bigg ) [3
X_1(\hat s,Q^2,\hat t)+X_2(\hat s,Q^2,\hat t)]+\bigg ({1\over
Q^2}(W^2-m^2)
\nonumber \\&&~~~~~~~~+{(W^2-m^2)^2\over 2\, Q^4}+{2 m^2\over Q^2} \bigg  )
 [X_1(\hat s,Q^2,\hat t)+X_2(\hat s,Q^2,\hat t)]-X_1(\hat
s,Q^2,\hat t)\bigg ] \nonumber \\ & &~~~~~~~~~~~ \times F_2(x_B,Q^2) {x_B\over
2}-X_2(\hat s,Q^2,\hat t)F_1(x_B,Q^2)\bigg\}.
\label{siin}
\ee 
Here $X_i(\hat s,Q^2,\hat t)$, with $\,i=1,2 $, are given by  (\ref{x1})-(\ref{x2}) with $t$ replaced by $-Q^2$. The limits of the integration over
$Q^2$ are:
\be
Q^2_{{{\mathrm{min}},{\mathrm{max}}}}&=&-m^2-W^2+{1\over 2 s} \Big [(s+m^2) (s-\hat s+W^2) \nonumber \\ && ~~~~~~~~~~~~~~~~~~~~~~~~\mp (s-m^2) {\sqrt
{(s-\hat s+W^2)^2-4 s W^2}} \Big ],
\label{eq:q2limit}
\ee  
while the extrema of $\hat t$ are the same as  (\ref{thatlim}).
The limits $W^2_{\mathrm{min},\mathrm{max}}$ are given by:
\be
W^2_{\mathrm{min}}=(m+m_{\pi})^2,~~~~~~~~W^2_{\mathrm{max}}=
(\sqrt s-\sqrt {\hat s_{\mathrm{min}}}\, )^2,
\label{eq:w2limit}
\ee
where $m_{\pi}$ is the mass of the pion. 

In the EPA, we neglect $m^2$ compared to $s$
and $Q^2$ compared to $\hat s$ as before. Using  (\ref{xone}) and 
({\ref{xtwo}), we can write
\be
\sigma_{\mathrm{inel}}(s) \approx \sigma_{\mathrm{inel}}^{\mathrm{EPA}} =
\int_{x_{\mathrm{min}}}^{(1-m/\sqrt s)^2} \der x \, \int_{m_e^2 -\hat s}^0
\der\hat t~
\gamma^p_{\mathrm{inel}}(x, x s) \,{{\der\hat\sigma(x s, \hat t)}
\over{\der\hat t}}~,
\label{pap1_epain}
\ee
where again $x={\hat s/s}$  and $\gamma^p_{\mathrm{inel}}(x, x s)$ is the 
inelastic contribution to the equivalent photon distribution of the proton
\cite{anlf}:
\be
\gamma^p_{\mathrm{inel}} (x, x s)&=&{\alpha\over 2 \pi} \int_x^1\, \der y 
\int_{Q^2_{\mathrm{min}}}^{Q^2_{\mathrm{max}}} {\der Q^2\over Q^2}\,
{y\over x}  
 \bigg [F_2\bigg ({x\over y},Q^2\bigg )\bigg ({{1+(1-y)^2}\over y^2} -
{2 m^2 x^2\over y^2 Q^2} 
\bigg )\nonumber\\&&~~~~~~~~~~~~~~-F_L \bigg({x\over y},Q^2 \bigg) \bigg ],
\label{gammain}
\ee
The limits of  the $Q^2$ integration can be approximated as
\be
Q^2_{\mathrm{min}}={x^2 m^2\over 1-x}~, 
\label{eq:q2min}
\ee
and we choose the scale $Q^2_{\rm{max}}$ to be $\hat s$.
Our expression of $\gamma^p_{\mathrm{inel}}(x, xs)$ differs from 
 \cite{blu} by a (negligible) term proportional to $m^2$.
Following \cite{thesis,kessler} we shall use the  LO Callan-Gross relation
\be
F_L(x_B,Q^2)~ =~ F_2(x_B,Q^2)-2 x_B F_1(x_B,Q^2)~=~0
\label{eq:callangross}
\ee
in our numerical calculations.

\section{Numerical Results}
\label{sec:results}

In this section, we present an estimate of the cross section, calculated
both exactly and using the equivalent photon approximation of the proton.
We have used the same kinematical cuts as used in \cite{thesis}
for the HERA collider, which are 
slightly different from the ones in \cite{kessler}. 
They are imposed on the following laboratory frame variables: energy of the 
final electron $E_e'$, energy of the final  photon $E_{\gamma}'$, 
polar angles  of the outgoing electron and photon, $ \theta_e$ and 
$\theta_{\gamma}$ respectively,  and  acoplanarity angle $\phi$, 
which is defined as 
\begin{equation}
\phi=|\,\pi-|\phi_{\gamma} -\phi_{e}|\,|,
\label{eq:acopla}
\end{equation} 
where $\phi_{\gamma}$ and $\phi_{e}$ are the azimuthal angles of 
the outgoing photon and electron respectively ($0 \le \phi_{\gamma},\, 
\phi_e \le 2 \,\pi$). 
The cuts are given by:  
\be
E_e',E_{\gamma}' > 4\, {\mathrm{GeV}}, ~~~~~~~~E_e' + E_{\gamma}'> 20\, 
{\mathrm{GeV}},
\label{cut1}
\ee
\be
0.06 < \theta_e, \theta_{\gamma} <  \pi-0.06,
\label{cut2}
\ee
\be
0 < \phi < {\pi\over 4}~.
\label{cut4_4}
\ee
The energies of the incoming particles are: $E_e = 27.5\,\, 
\mathrm{GeV}$ (electron) and $E_p = 820 \,\,\mathrm{GeV}$ (proton). 
In our conventions, we fix the laboratory frame such that $\phi_e = 0$, so the 
acoplanarity will be $\phi=|\,\pi-\phi_{\gamma}\,|$.
These cuts reflect experimental acceptance constraints as well as the
reduction of the background events due  to emitted photons with $ (l'+k')^2
\approx 0$ and/or $(l-k')^2 \approx 0 $, which are
unrelated to the QED Compton scattering process (for which $ -k^2=Q^2 
\approx 0$ but
with both $(l'+k')^2$ and $(l-k')^2$ finite), 
i.e. photons emitted parallel to the ingoing (outgoing) electron  
 \cite{blu,kessler}, as  discussed in Section 
\ref{sec:radiative}.

In contrast to \cite{kessler}, we find that the cuts 
\eqref{cut1}-\eqref{cut4_4} are {\em not} sufficient to suppress
the background due to photon emission at the hadron vertex, discussed in 
the next chapter.  Such a background has been subtracted from 
the measurements  in 
\cite{thesis} and from the numerical estimates in \cite{thesis,lend}, which
will be shown for comparison with our results in Figure \ref{fig:two}
 and  in the Tables \ref{tab:one}  and \ref{tab:two}.

We numerically integrate the elastic and inelastic cross sections  
given by   (\ref{sigel}) and (\ref{siin}). To implement the cuts in  
(\ref{cut1})-(\ref{cut4_4}), we express $E_e'$, $E_{\gamma}'$, $\cos{
\theta_e}$, $\cos {\theta_{\gamma}}$ and $\cos {\phi}$ in 
terms of our integration variables $ \hat s$, $t$, $\hat t$, $\varphi^*$
(and  $W^2$ in the inelastic channel), as explained in Appendix 
\ref{app:kinel1}.  
More explicitly, we use  (\ref{thetae})-(\ref{phig}), (\ref{cmT}), 
(\ref{cmU}), together with  (\ref{cmenergy})-(\ref{cmangle2}) for the
elastic channel and  (\ref{cmenergyin})-(\ref{cmangle2in}) for the 
inelastic one. The cuts imposed on the laboratory frame variables  
restrict the range of our integrations numerically. In this way, 
we are able to remove  the contributions from  outside the 
considered kinematical region.

In the calculation of the elastic cross section, the electric and magnetic 
form factors are empirically parametrized as dipoles
\be
G_E(t)= \frac{1}{ [1 - t/(0.71\,\mathrm{GeV}^2)]^{2}}~,~~~~G_M(t)=2.79~G_E(t).
\label{eq:emp_dipole}
\ee
Following \cite{thesis}, in the evaluation of \eqref{siin}   
we have used  the ALLM97 parametrization of the proton structure function 
$F_2(x_B, Q^2)$ \cite{allm97}, which provides a purely phenomenological, 
Regge model inspired, description of $F_2(x_B, Q^2)$, including its vanishing 
in the $Q^2 = 0$ limit as well as its scaling behaviour at large $Q^2$. 
The ALLM97 parametrization is supposed to hold over the entire 
range of $x_B$ and $Q^2$ studied so far, namely 
$3 \times 10^{-6} < x_B < 0.85 $ 
and $ 0 \le Q^2 < 5000$ $\mathrm{GeV}^2$, above the quasi-elastic 
region ($W^2 > 3$ 
$\mathrm{GeV}^2$) dominated by resonances. 
We do not consider the resonance contribution 
separately but, using the so-called local duality \cite{bloom}, 
we extend the ALLM97 
parametrization from the continuous ($W^2 > 3$ $\mathrm{GeV}^2$) down to
the resonance  domain ($(m_{\pi}+ m)^2<W^2< 3$ $\mathrm{GeV}^2$): 
in this way it is possible to agree with the experimental data averaged 
over each resonance, as pointed out in \cite{thesis}. 

The elastic contribution 
to the EPA was calculated using (\ref{epael}) subject to the additional
kinematical restrictions given by
 (\ref{cut1})-(\ref{cut2}). For the inelastic channel we
have used  (\ref{pap1_epain}) together with  (\ref{gammain}), the cuts
being the same as in the elastic case. We have taken   
$F_L=0$ and used  the ALLM97 parametrization
of $F_2$, in order to compare consistently with the exact cross section.
We point out that in  \eqref{eqtwo}, as well as in \cite{ruju,gpr1,gpr2}, $F_2 (x_B, Q^2)$ in  
$\gamma^p_{\mathrm{inel}}(x,x s)$ was expressed in terms of parton
distributions for which the LO GRV98 parametrization \cite{grv98} was used, 
together with 
$Q^2_{\mathrm{min}}=0.26~ {\mathrm{GeV}}^2 $ so as to guarantee the
applicability of perturbative QCD \cite{gsv}.    
The new $\gamma^p_{\mathrm{inel}}(x,x s)$   gives slightly higher results
than the ones obtained with the photon distribution in \eqref{eqtwo}.    

\begin{figure}[t]
\begin{center}
\epsfig{figure= 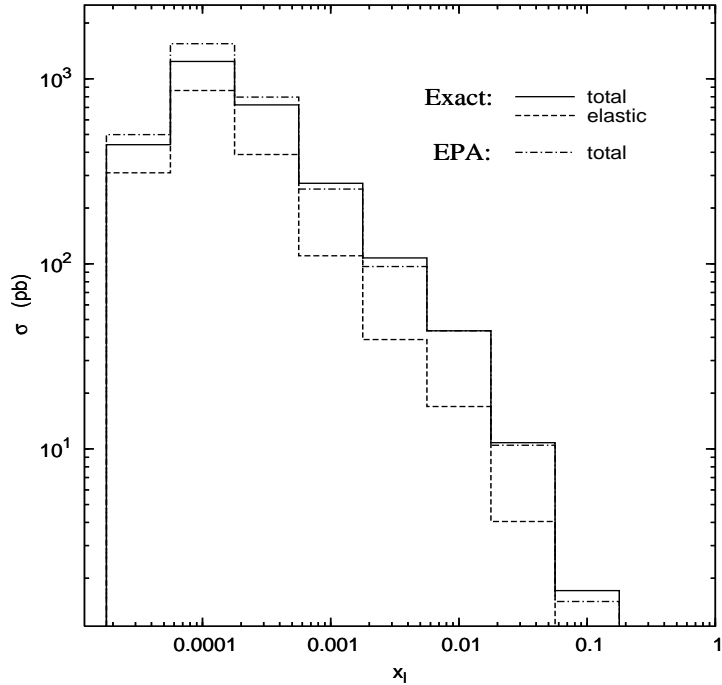,width=10.0cm,height=9.0cm}
\caption{Cross section for Compton process at HERA-H1. The cuts applied are 
given in \eqref{cut1}-\eqref{cut4_4}.}
\label{fig:two}
\end{center}
\end{figure}
\begin{figure}[ht]
\begin{center}
\epsfig{figure= 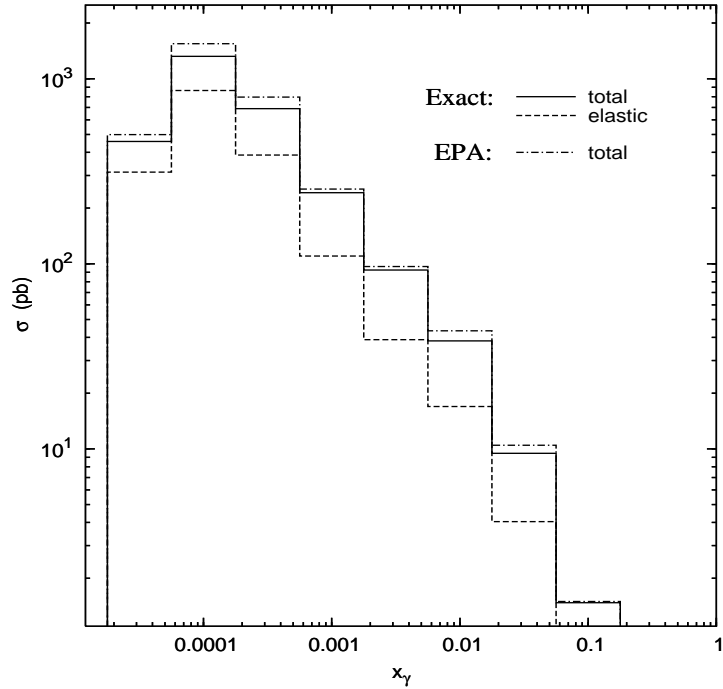,width=10.0cm,height=9cm}
\caption{Cross section for Compton process at HERA-H1. The bins are in
$x_{\gamma}$. The cuts applied are given in \eqref{cut1}- \eqref{cut4_4}.}
\label{fig:three}
\end{center}
\end{figure}

\begin{figure}
\begin{center}
\epsfig{figure=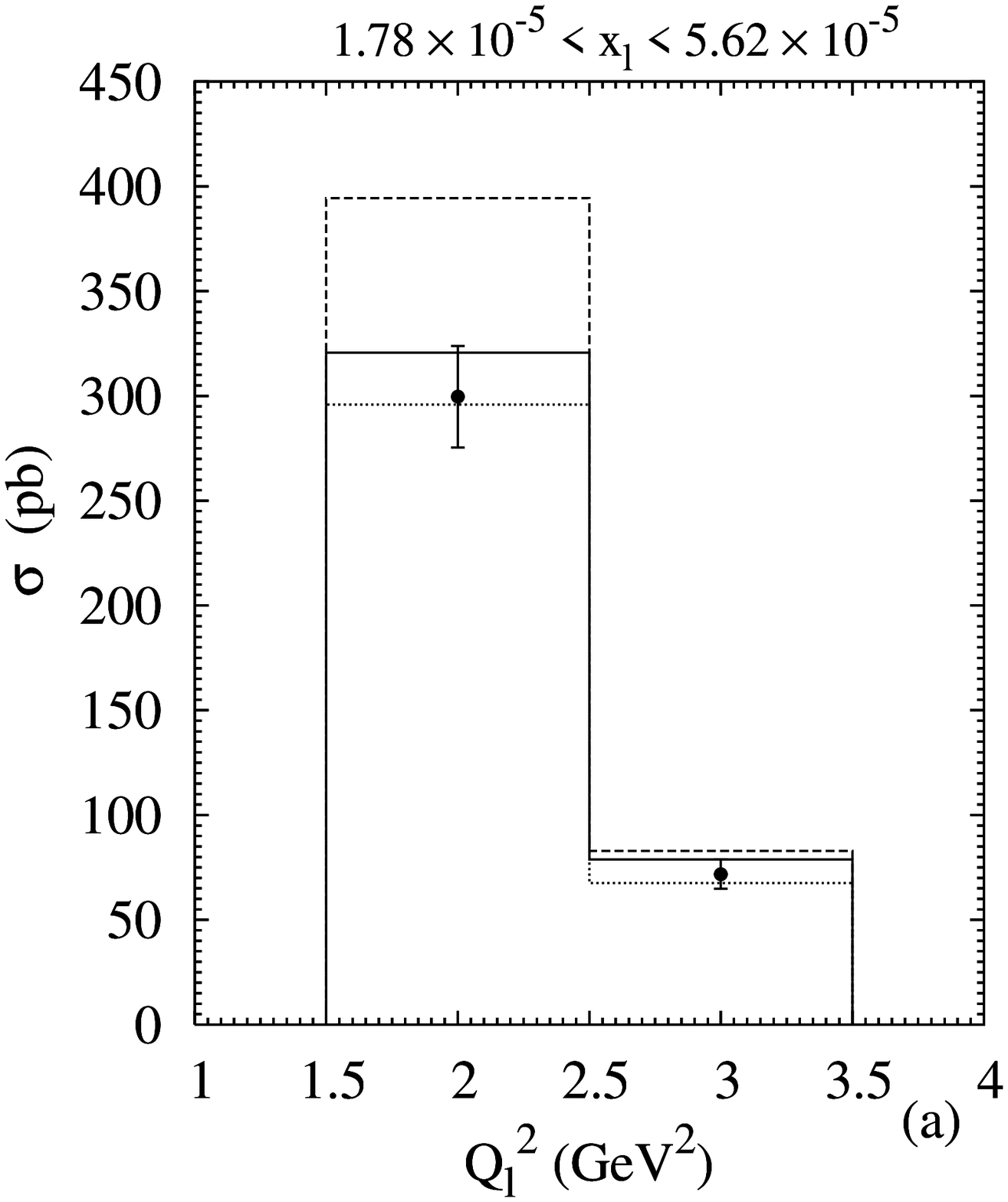,width=7.5 cm,height=7.5 cm}
\epsfig{figure=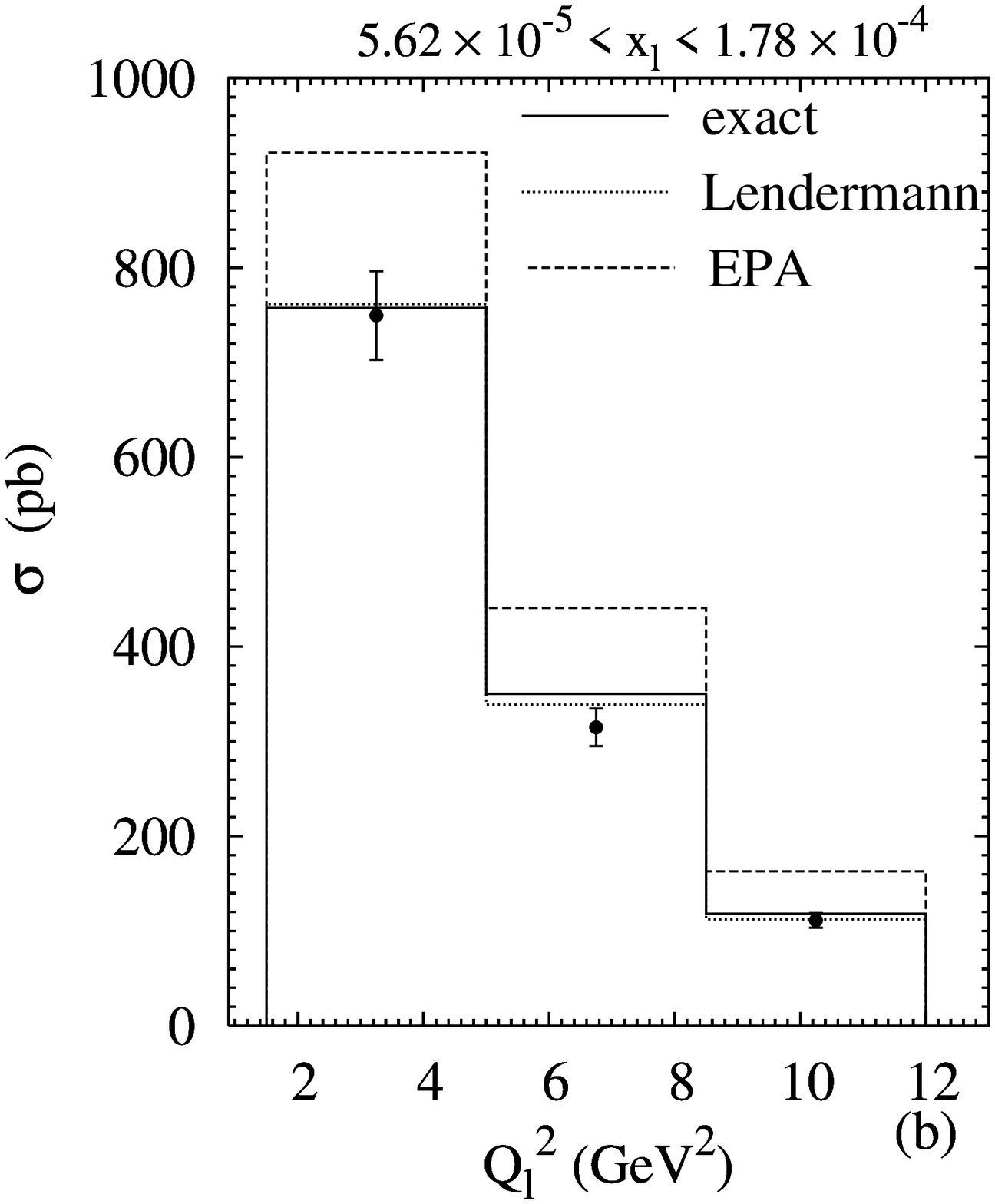,width=7.5 cm,height=7.5 cm}

\vspace{1.5cm}
\epsfig{figure=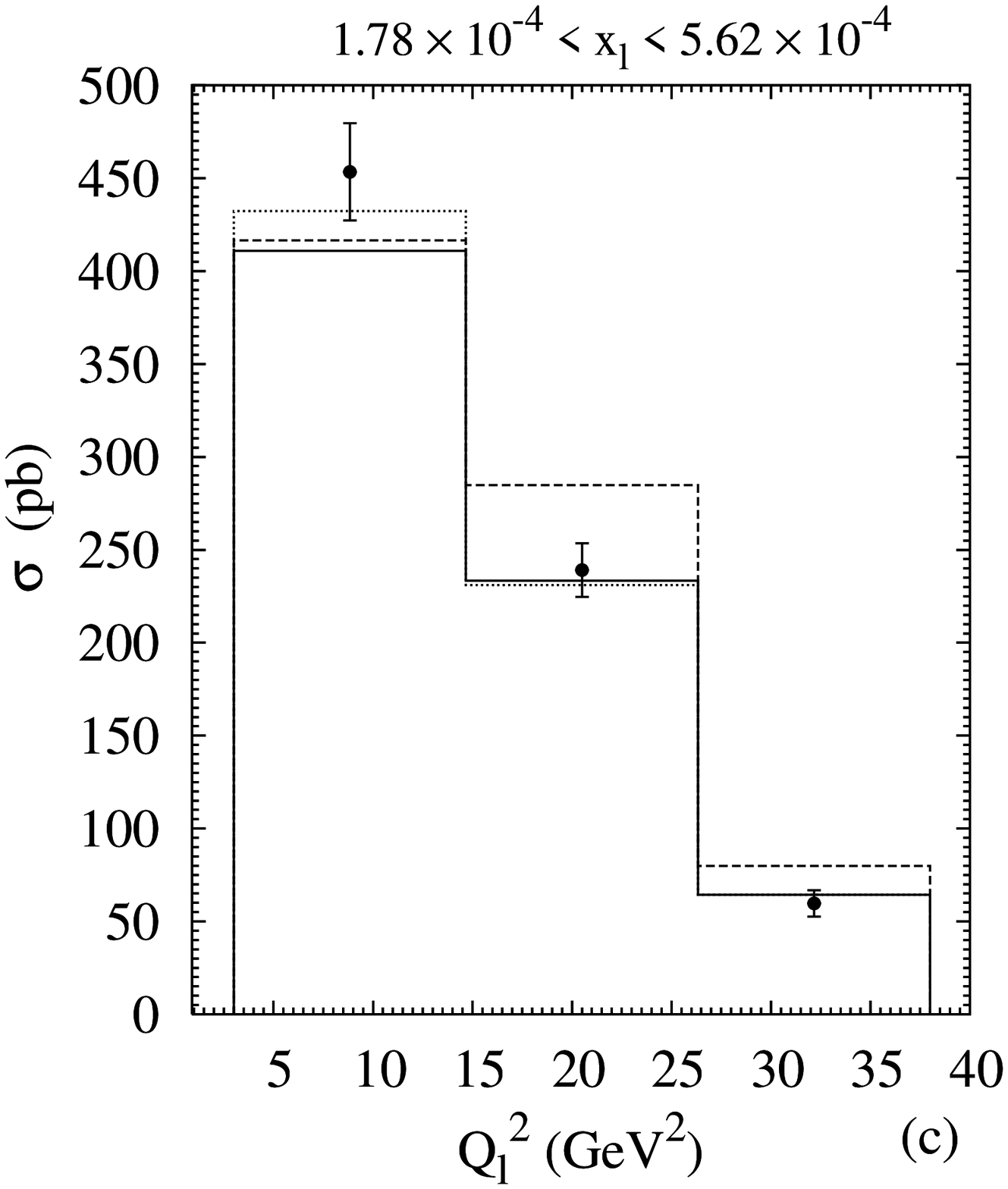,width=7.5 cm,height=7.5 cm}
\epsfig{figure=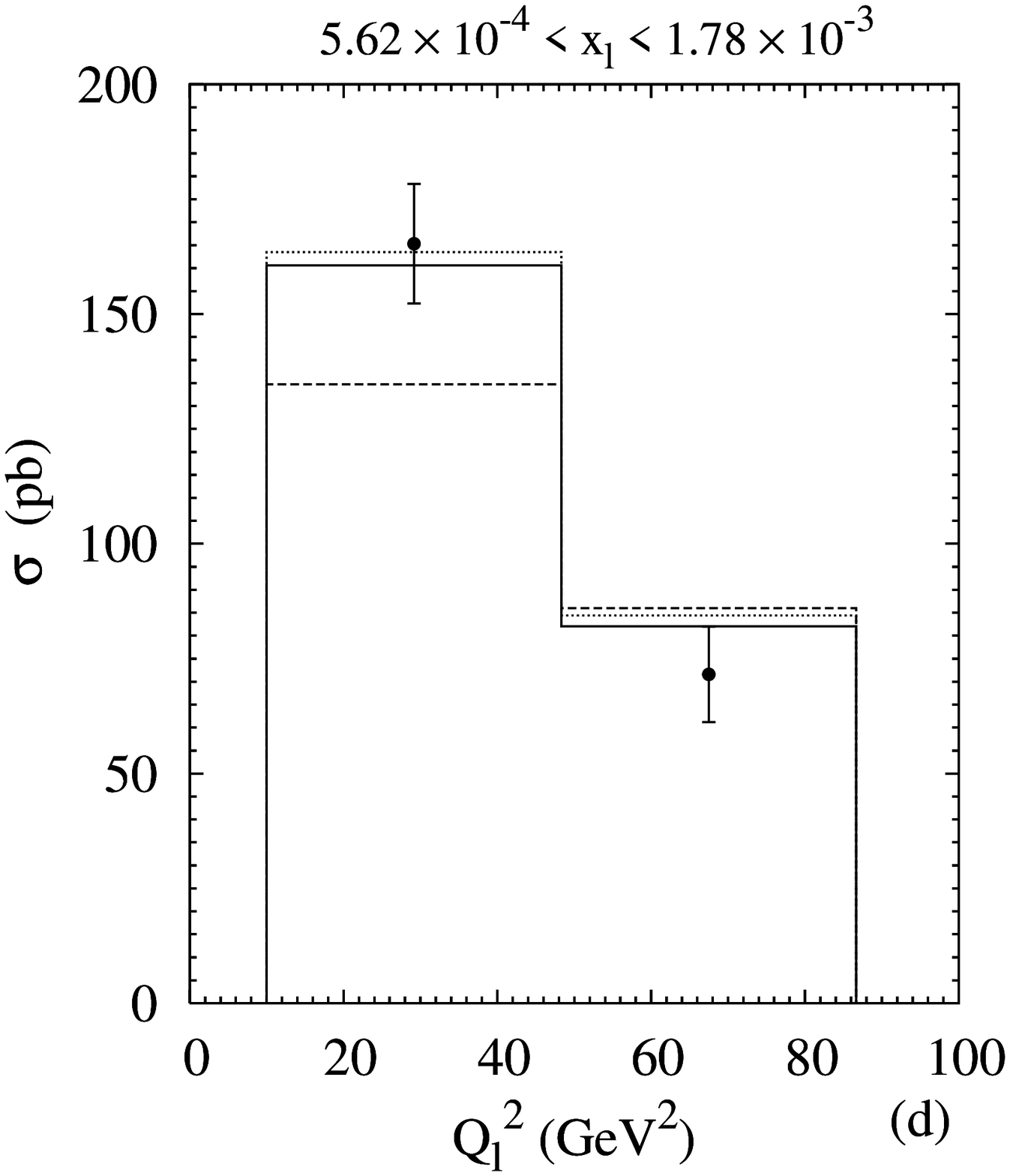,width=7.5 cm,height=7.5 cm}

\end{center}

\caption{Double differential cross section for QED Compton scattering at 
HERA-H1. The data are taken from [31]. The kinematical bins
correspond to Table \ref{tab:one}. The continuous line corresponds to our exact
calculation, the dotted line to the calculation in [31] and the dashed line to 
the EPA.}
\label{fig:four}
\end{figure}

The Compton process turns out to be dominated by the elastic channel, in fact
after Monte Carlo integration, we find that $\sigma_{\mathrm{el}} = 1.7346$ 
nb, while $\sigma_{\mathrm{inel}} = 1.1191$ nb. The approximated calculation
gives the results: $\sigma_{\mathrm{el}}^{\mathrm{EPA}}= 1.7296$ nb and     
$\sigma_{\mathrm{inel}}^{\mathrm{EPA}}= 1.5183$ nb. This means that in the 
kinematical region under consideration, the total (elastic + inelastic) 
cross section calculated using the EPA agrees with the exact one 
within 14\% and that the approximation  turns out to be particularly good in 
describing the elastic process, for which the agreement is within 0.3\%.  
This is not surprising since in the EPA one
assumes $Q^2 =0$, which is not true especially in the inelastic channel and
the inelastic cross section receives substantial contribution from the
non-zero $Q^2$ region. In terms of the kinematical cuts, the EPA corresponds 
to the situation when the outgoing electron and
the final photon are observed under large polar angles and almost opposite
to each other in azimuth, so that the acoplanarity is approximately zero.
For elastic scattering there is a sharp peak of the exact cross section
for $\phi=0$, contributions from non-zero $\phi$ are very small in this case.
But the inelastic cross section receives contribution even from non-zero
$\phi$, so that in this case the discrepancy from the approximated result is 
 higher. The discrepancy of the total cross section with the approximate
one is thus entirely due to the inelastic part.

In Figure \ref{fig:two}   we have compared the total cross sections (exact and EPA) in 
different $x_l $ bins, in the region $1.78 \times 10^{-5} < x_l < 1.78\times
10^{-1}$. Figure \ref{fig:three}  shows that the agreement improves  slightly for bins in 
the variable $x_{\gamma}$. Since $x_{\gamma} \simeq x_l$ for $Q^2\simeq 0$,
the elastic process is not sensitive to this change of variables. 
We point out again that in the EPA limit ($Q^2=0$) $x_l = x_{\gamma} = x$.    
  
In Figure \ref{fig:four} we show the exact and the EPA cross section in $x_l$ 
and $Q_l^2$ bins together with the experimental results and the 
estimates of the Compton event generator, 
already presented in \cite{thesis}.  Except for 
two bins, our exact result agrees with the experiment within the error bars.
The slight difference of our exact result and the one of \cite{thesis} 
may be due to the fact that in \cite{thesis} the cross section is calculated 
using a Monte Carlo generator in a step by step iteration \cite{kessler,h1} which starts by 
assuming $Q^2=0$, while we did not use any approximation. Our exact result 
is closer to  the EPA in most of the kinematical 
bins as compared to \cite{thesis}. The total cross section in the EPA lies above
the exact one in most of the bins.
  
For completeness, we have shown the numerical values of the exact 
and EPA double differential cross sections, 
both for the elastic (Table \ref{tab:one}) and inelastic 
(Table \ref{tab:two}) contributions. The kinematical bins are the same as 
in \cite{thesis}. 
The exact results when the bins are in $x_\gamma$ instead of $x_l$ 
are also shown. 

\begin{table}[h]
\small
\begin{center}
\begin{tabular}{|c|c|c|c|c|c|c|}
\hline 
$x$ bin & $Q^2_l$ bin & $\sigma_{\mathrm{el}}$  & 
$\sigma_{\mathrm{el}}^{\mathrm{Len}}$ &$\sigma_{\mathrm{el}}^*$ &
$\sigma_{\mathrm{el}}^{\mathrm{EPA}} $   \\ 
\hline \hline
          &           &             &     &      &  \\
$1.78\times 10^{-5}-5.62 \times 10^{-5}$ & $1.5 -2.5$ &$2.428\times 10^2$ & $2.342\times 10^2$   &$ 2.446\times 10^2  $ & $2.461 \times 10^2$    \\ 
$1.78\times 10^{-5}-5.62 \times 10^{-5}$ & $2.5 - 3.5$&$5.099\times 10^1$ & $4.71\times 10^1$ &$5.201\times 10^1$& $5.051\times 10^1$    \\ \hline
$5.62\times 10^{-5}-1.78\times 10^{-4} $ & $1.5 - 5.0$&$5.279\times 10^2$  &$5.319\times 10^2$ & $5.259\times 10^2$ & $5.247\times 10^2$ \\
$5.62\times 10^{-5}-1.78 \times 10^{-4}$&$ 5.0-8.5$&$2.396\times 10^2 $&$2.327\times 10^2$ & $2.404\times 10^2$ &  $2.395\times 10^2$ \\ 
$5.62 \times 10^{-5}-1.78\times 10^{-4}$ & $8.5-12.0$&$8.496\times 10^1 $ & $8.32\times 10^1$ & $8.559\times 10^1$ &$8.571\times 10^1 $ \\ \hline
$1.78\times 10^{-4}-5.62\times 10^{-4}$ & $3.0-14.67 $ & $2.080\times 10^2$ & $2.036\times 10^2 $ & $2.056\times 10^2$ &  $ 2.061\times 10^2$ \\ 
$1.78\times 10^{-4}-5.62\times 10^{-4}$ & $14.67-26.33 $ & $1.373\times 10^2$ & $1.388\times 10^2$ & $1.373 \times 10^2$ & $ 1.372\times 10^2$ \\ 
$1.78\times 10^{-4}-5.62\times 10^{-4}$ & $26.33-38.0 $ & $3.712\times 10^1$ & $3.86\times 10^1$ & $3.720 \times 10^1$   &$ 3.695\times 10^1$ \\ \hline
$5.62\times 10^{-4}-1.78\times 10^{-3}$ & $10.0-48.33 $ & $5.947\times 10^1$ & $5.71\times 10^1$ & $ 5.918\times 10^1$ & $ 5.921\times 10^1$ \\ 
$5.62\times 10^{-4}-1.78\times 10^{-3}$ & $48.33-86.67 $ & $3.714\times 10^1$ & $3.85\times 10^1$ &$3.715 \times 10^1$   & $ 3.704\times 10^1$ \\ 
$5.62\times 10^{-4}-1.78\times 10^{-3}$ & $86.67-125.0 $ & $1.056\times 10^1  $ & $1.028\times 10^1$ & $1.057\times 10^1$ & $1.054\times 10^1 $ \\ \hline
$1.78\times 10^{-3}-5.62\times 10^{-3}$ & $22-168 $ & $1.913\times 10^1$ & $1.877\times 10^1$ & $ 1.909\times 10^1 $ & $ 1.909\times 10^{1}$ \\ 
$1.78\times 10^{-3}-5.62\times 10^{-3}$ & $168-314 $ & $1.239\times 10^1$ & $1.229\times 10^1$ & $1.239 \times 10^1$& $ 1.238\times 10^1$ \\ 
$1.78\times 10^{-3}-5.62\times 10^{-3}$ & $314-460 $ & $5.917$ & $6.02$ & $5.915 $ & $ 5.914$ \\ \hline
$5.62\times 10^{-3}-1.78\times 10^{-2}$ & $0-500 $ & $4.811$ & $5.76$ & $4.890$ &  $4.890 $ \\ 
$5.62\times 10^{-3}-1.78\times 10^{-2}$ & $500-1000 $ & $9.271$ & $9.22$ & $9.264$ & $ 9.271$ \\ 
$5.62\times 10^{-3}-1.78\times 10^{-2}$ & $1000-1500  $ & $2.572 $ & $2.65$ & $2.571 $ & $ 2.573$ \\ \hline
$1.78\times 10^{-2}-5.62\times 10^{-2}$ & $0-1500 $ & $8.238\times 10^{-1}$ & $6.8\times 10^{-1}$ & $ 9.085\times 10^{-1}$ & $9.086\times 10^{-1}$ \\ 
$1.78\times 10^{-2}-5.62\times 10^{-2}$ & $1500-3000 $ & $2.431$ & $2.69$ & $2.430 $ &  $ 2.434$ \\ 
$1.78\times 10^{-2}-5.62\times 10^{-2}$ & $3000-4500  $ & $6.336\times 10^{-1}$ & $7.7\times 10^{-1}$ & $6.328 \times 10^{-1}$ & $6.345\times 10^{-1}$ \\ \hline
$5.62\times 10^{-2}-1.78\times 10^{-1}$ & $10-6005 $ & $3.120\times 10^{-1}$ & $4.27\times 10^{-1}$ & $3.120\times 10^{-1} $ & $ 3.117\times 10^{-1}$ \\ 
$5.62\times 10^{-2}-1.78\times 10^{-1}$ & $6005-12000 $ & $2.437\times 10^{-1}$ & $2.13\times 10^{-1}$ & $ 2.438\times 10^{-1} $ & $2.436\times 10^{-1} $ \\ 
$5.62\times 10^{-2}-1.78\times 10^{-1}$ & $12000-17995$ & $0.000$ &$0.000$ & $0.000$ & $2.461\times 10^{-2} $ \\ 
     &    &          &        &         &    \\
\hline
\end{tabular}
\end{center}

\caption {Double differential (elastic) QED Compton scattering cross section.
$\sigma_{\mathrm{el}}$ is the exact result in  (\ref{sigel}), 
$\sigma_{\mathrm{el}}^{\mathrm{Len}}$ corresponds to the results in 
[31].   
The $x$-bins refer to $x_l$ in
 (\ref{xl}) except for $\sigma_{\mathrm{el}}^*$ where they refer to 
$x_{\gamma}$ in  (\ref{xgamma}). $\sigma_{\mathrm{el}}^{\mathrm{EPA}}$ 
is given in  (\ref{epael}) where $x \equiv x_\gamma$. $Q^2_l$ is expressed in 
GeV$^2$ and the cross sections are in pb.}
\label{tab:one}
\end{table}
\begin{table}[h]
\small
\begin{center}
\begin{tabular}{|c|c|c|c|c|c|c|}
\hline 
$x$ bin & $Q^2_l$ bin & $\sigma_{\mathrm{inel}}$  & 
$\sigma_{\mathrm{inel}}^{\mathrm{Len}}$ &$\sigma_{\mathrm{inel}}^*$ &
$\sigma_{\mathrm{inel}}^{\mathrm{EPA}} $    \\ 
\hline \hline
          &           &             &     &      &  \\
$1.78\times 10^{-5}-5.62 \times 10^{-5}$ & $1.5 -2.5$ &$7.802\times 10^1$ & $6.170\times 10^1$ & $ 7.367\times 10^1  $ & $1.483 \times 10^2$    \\ 
$1.78\times 10^{-5}-5.62 \times 10^{-5}$ & $2.5 - 3.5$&$2.799\times 10^1$ & $2.050\times 10^1$ &$4.029\times 10^1$& $3.255\times 10^1$    \\ \hline
$5.62\times 10^{-5}-1.78\times 10^{-4} $ & $1.5 - 5.0$&$2.299\times 10^2$  &$2.296\times 10^2$ & $2.298\times 10^2$ & $3.967\times 10^2$ \\
$5.62\times 10^{-5}-1.78 \times 10^{-4}$&$ 5.0-8.5$&$1.108\times 10^2 $&$1.062\times 10^2$ & $1.450\times 10^2$ &  $2.016\times 10^2$ \\ 
$5.62 \times 10^{-5}-1.78\times 10^{-4}$ & $8.5-12.0$&$3.340\times 10^1 $ & $2.890\times 10^1$ & $6.048\times 10^1$ &$7.751\times 10^1 $ \\ \hline
$1.78\times 10^{-4}-5.62\times 10^{-4}$ & $3.0-14.67 $ & $2.029\times 10^2$ & $2.287\times 10^2 $ & $1.228\times 10^2$ &  $ 2.104\times 10^2$ \\ 
$1.78\times 10^{-4}-5.62\times 10^{-4}$ & $14.67-26.33 $ & $9.644\times 10^1$ & $9.230\times 10^1$ & $1.164 \times 10^2$ & $ 1.476\times 10^2$ \\ 
$1.78\times 10^{-4}-5.62\times 10^{-4}$ & $26.33-38.0 $ & $2.742\times 10^1$ & $2.570\times 10^1$ & $4.431 \times 10^1$   &$ 4.298\times 10^1$ \\ \hline
$5.62\times 10^{-4}-1.78\times 10^{-3}$ & $10.0-48.33 $ & $1.011\times 10^2$ & $1.064\times 10^2$ & $ 5.077\times 10^1$ & $ 7.555\times 10^1$ \\ 
$5.62\times 10^{-4}-1.78\times 10^{-3}$ & $48.33-86.67 $ & $4.485\times 10^1$ & $4.590\times 10^1$ &$5.304 \times 10^1$   & $ 4.897\times 10^1$ \\ 
$5.62\times 10^{-4}-1.78\times 10^{-3}$ & $86.67-125.0 $ & $1.228\times 10^1  $ & $1.132\times 10^1$ & $1.887\times 10^1$ & $1.462\times 10^1 $ \\ \hline
$1.78\times 10^{-3}-5.62\times 10^{-3}$ & $22-168 $ & $4.320\times 10^1$ & $4.917\times 10^1$ & $ 2.225\times 10^1 $ & $ 2.791\times 10^1$ \\ 
$1.78\times 10^{-3}-5.62\times 10^{-3}$ & $168-314 $ & $1.831\times 10^1$ & $1.735\times 10^1$ & $2.117\times 10^1 $& $ 1.849\times 10^1$ \\ 
$1.78\times 10^{-3}-5.62\times 10^{-3}$ & $314-460 $ & $6.314$ & $5.760$ & $8.303 $ & $ 9.046$ \\ \hline
$5.62\times 10^{-3}-1.78\times 10^{-2}$ & $0-500 $ & $1.277\times 10^1$ & $1.432\times 10^1$ & $6.438$ &  $7.627$ \\ 
$5.62\times 10^{-3}-1.78\times 10^{-2}$ & $500-1000 $ & $1.086\times 10^1$ & $9.890 $ & $1.152\times 10^1$ & $ 1.450\times 10^1$ \\ 
$5.62\times 10^{-3}-1.78\times 10^{-2}$ & $1000-1500  $ & $2.734 $ & $2.600$ & $3.201 $ & $ 4.067$ \\ \hline
$1.78\times 10^{-2}-5.62\times 10^{-2}$ & $0-1500 $ & $2.787$ & $2.500$ & $ 1.321$ & $1.439$ \\ 
$1.78\times 10^{-2}-5.62\times 10^{-2}$ & $1500-3000 $ & $3.118$ & $2.150$ & $3.149$ &  $ 3.855$ \\ 
$1.78\times 10^{-2}-5.62\times 10^{-2}$ & $3000-4500  $ & $7.718\times 10^{-1}$ & $6.600\times 10^{-1}$ & $8.421\times 10^{-1}$ & $1.004$ \\ \hline
$5.62\times 10^{-2}-1.78\times 10^{-1}$ & $10-6005 $ & $7.203\times 10^{-1}$ & $1.460\times 10^{-1}$ & $4.677\times 10^{-1} $ & $ 4.924\times 10^{-1}$ \\ 
$5.62\times 10^{-2}-1.78\times 10^{-1}$ & $6005-12000 $ & $3.739\times 10^{-1}$ & $2.110\times 10^{-1}$ & $ 3.830\times 10^{-1} $ & $3.845\times 10^{-1} $ \\ 
$5.62\times 10^{-2}-1.78\times 10^{-1}$ & $12000-17995$ & $3.738\times 10^{-2}$ &$4.300\times 10^{-2}$ & $4.182\times 10^{-2}$ & $3.849\times 10^{-2} $ \\ 
     &    &          &        &         &    \\
\hline

\end{tabular}
\end{center}
\caption{Double differential (inelastic) QED Compton scattering cross section.
$\sigma_{\mathrm{inel}}$ is the exact result in  (\ref{siin}),
$\sigma_{\mathrm{inel}}^{\mathrm{Len}}$ 
corresponds to the results in [31]. The $x$-bins are 
as in Table \ref{tab:one},
 i.e. refer to $x_l$ in
 (\ref{xl}) except for $\sigma_{\mathrm{inel}}^*$ where they refer to $x_\gamma$ in
 (\ref{xgamma}). $\sigma_{\mathrm{inel}}^{\mathrm{EPA}}$ is given in 
(\ref{pap1_epain}) where $x \equiv x_\gamma$. $Q^2_l$ is expressed in 
GeV$^2$ and the cross sections are in pb.} 

\label{pap1_tabletwo} 
\label{tab:two}
\end{table}

The EPA  elastic cross section agrees within  $ 1 \% $ with 
the exact one for  all the $x_l$ bins. The agreement becomes slightly 
better if we consider $x_{\gamma}$ bins. For the inelastic channel, the
discrepancies from the EPA results are considerably higher. Our 'exact'
results lie closer to the EPA compared to \cite{thesis} in almost all the
bins. The result in $x_{\gamma}$ bins shows better agreement with the EPA
compared to the $x_l$ bins, especially for higher $x_\gamma$. The
discrepancy with the EPA is about $20-30 \%$ in most of the bins, higher in
some cases. 

\section{Summary}
\label{sec:summary}

In this chapter, we have estimated both the elastic and the inelastic 
QED Compton scattering cross sections for unpolarized 
incoming electron and proton. Our approach for the
calculation of the total
cross section is manifestly covariant and we have used the same cuts as in
\cite{thesis}. The numerical estimates of the exact cross section for
different kinematical bins are presented and compared with the EPA
and the experimental results. The exact cross section in
the elastic channel agrees within $1 \%$ with the approximate one. The
discrepancy is thus due to the inelastic channel and, in the next chapter,
new cuts will be suggested in order to reduce it.
 A comparison with the
Monte Carlo results of \cite{thesis} is also shown.
For both elastic and inelastic cross sections, our exact results are  closer 
to the EPA as compared to \cite{thesis}. The agreement is even better if the
bins are in $x_{\gamma}$ instead of $x_l$.

\clearemptydoublepage

\chapter{{\bf Suppression  of the Background to  QED
Compton scattering}}

In this chapter we perform a detailed study  of the QED Compton scattering
process (QEDCS) in 
$e p \rightarrow e 
\gamma p$ and $e p \rightarrow e \gamma X$, depicted in Figure 
\ref{fig:qedcs}, together with the major 
background coming from the virtual Compton scattering (VCS), where 
the photon is emitted from the hadronic vertex, shown in Figure 
\ref{fig:vcs}. The two processes can be distinguished experimentally 
because they differ in the kinematic distributions of the outgoing
electron and photon. We suggest  new kinematical 
cuts to suppress the VCS  background,  which turns out to be 
important in the phase space domain of the HERA experiment.
We also study the impact 
of these constraints  on the QEDCS cross section. We show that in the 
phase space region suggested  and accessible at HERA, 
the photon content of the proton provides a reasonably good 
description of the QEDCS cross section, also in the inelastic channel. 
           
The VCS cross section in the inelastic channel is 
estimated utilizing  
effective parton distributions of the
proton. In the elastic channel, to make a relative
estimate of the VCS, we take the proton to be pointlike and replace
the vertex by an effective vertex \cite{pp2}.  

In Sections \ref{sec:vcsel}  and \ref{sec:vcsinel} we present 
the cross
section in the elastic and inelastic channels respectively. Numerical results 
are 
given in Section \ref{vcs:num}.
The summary  is presented in Section \ref{vcs:summ}. 
The 
matrix elements are explicitly shown in Appendix D.

\begin{figure}[ht]
\begin{center}
\epsfig{figure= 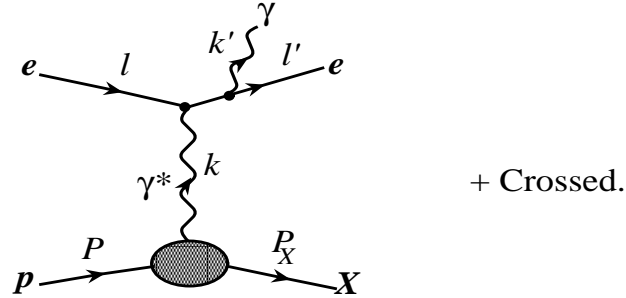, width=16cm, height= 5.cm}
\end{center}
\caption{ Feynman diagrams for the QED Compton process (QEDCS). $X \equiv p$
(and $P_X \equiv P'$) 
corresponds to elastic scattering.}
\label{fig:qedcs}
\end{figure}
\begin{figure}[ht]
\begin{center}
\epsfig{figure= 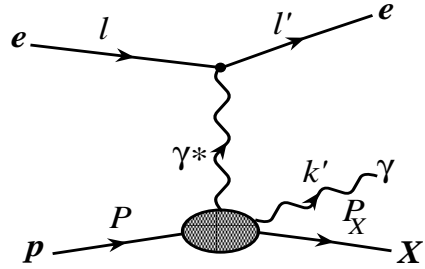, width=16cm, height= 5.cm}
\end{center}
\caption{As in Figure \ref{fig:qedcs}   but for the virtual Compton 
scattering (VCS)
background process. }
\label{fig:vcs}
\end{figure}

\section{Elastic Channel}
\label{sec:vcsel}

The elastic channel of the process under study,
\be
e(l)+p(P) \rightarrow e(l')+p(P')+\gamma(k'),
\label{eq1}
\ee
is described by the Feynman diagrams in Figures \ref{fig:qedcs} and
\ref{fig:vcs}, with $X$ being a proton and $P_X \equiv P'$.
As in the previous chapter, we neglect the
electron mass $m_e$ everywhere except when it is necessary to avoid 
divergences in
the formulae and take the proton to be massive, $ P^2=P'^2=m^2 $.  
The corresponding cross section, using the notation  \eqref{eq:lorentz} for 
the Lorentz invariant 
$N$-particle phase-space element and the variables defined in 
\eqref{comptonmandel} and in   \eqref{pap1_invar}, 
is given by (\ref{pap1_sigmael}): 
\be
\sigma_{\mathrm{el}}(s)=\frac{1}{2 (s-m^2)} \int \der PS_{2+1}(l+P;l',k',P')
\overline{{\mid M_{\mathrm{el}} \mid }^2}~.
\ee 
Rearranging the ($2+1$)-particle phase space into a sequence
of two $2$-particle ones as we did in Section \ref{sec:elastic}, see \eqref{eq:sigmaintel}, \eqref{eq:xmunu} together with \eqref{eq:amplsqel},
we get
\be
\sigma_{\mathrm{el}}(s) &= & \frac {1}{2 (s-m^2)} \int {\der\hat s\over 2 \pi} \,\der PS_2
(l+P;l'+k',P')\, \der PS_2(l+k;l',k') \,\overline {{\mid {M_{\mathrm{el}}}\mid }^2},  
\label{elcross}
\ee
where, as before, $k=P-P'$ is the momentum transfer between the 
initial and the final proton. 
Substituting  (\ref{eq:phasel1}) and (\ref{eq:phasel2}) in  (\ref{elcross}) 
we obtain the final formula
\be
\sigma_{\mathrm{el}}(s)=\frac{1}{2 (4\pi)^4 (s-m^2)^2}\int_{m_e^2}^{(\sqrt{s}-m)^2} \der\hat s \int_{{t}_{\mathrm{min}}}^{t_{\mathrm{max}}}
\der t \int_{\hat t_{\mathrm{min}}}^{\hat t_{\mathrm{max}}} \der\hat t \int_0^{2\pi} \der\varphi^* {1\over {(\hat s-t)}}\,
\overline {{\mid {M_{\mathrm{el}}}\mid }^2},
\label{elsig_4}
\ee
where $\varphi^*$ is the azimuthal angle of the outgoing $e-\gamma$ system in 
the $e-\gamma$ center-of-mass frame.
The limits of integrations in  (\ref{elsig_4}) follow from kinematics and
are given explicitly by   (\ref{thatlim}) 
and (\ref{eq:tlimits}). However, as
it will be discussed in Section \ref{vcs:summ}, we will impose additional 
 kinematical cuts
relevant to the experiment at HERA. 
 
As already shown in \eqref{eq:amplsqel}, the  amplitude squared of 
the QED Compton scattering can be written as
\be
\overline {{\mid {M_{\mathrm{el}}^{\rm{QEDCS}}}\mid }^2} = \frac{1}{t^2}\, T_{\alpha \beta}(l;l',k')\, H^{\alpha \beta}_{\mathrm{el}} (P,P') ~,
\label{eq:amplqedcs}
\ee
where $T_{\alpha \beta}$ is the leptonic tensor given by  (\ref{lept}) and
$H^{\alpha \beta}_{\mathrm{el}}$ is the hadronic tensor 
\eqref{eq:hadronictens},
expressed in terms of the  electromagnetic form factors of the 
proton. The full cross section for the process given by  (\ref{elsig_4}) also
receives a contribution from the VCS in Figure \ref{fig:vcs}. 
The cross section for this process can be expressed in terms of off-forward or
generalized parton distributions \cite{ji,gpd}.  In addition, there are
contributions due to the interference between the QEDCS and VCS.  In order to 
make a numerical estimate of these effects, one needs some 
realistic parametrization of the off-forward  distributions.
Our aim is to  estimate the VCS background so as to find the kinematical cuts
necessary to suppress it. We make a simplified
approximation to calculate the VCS cross section. We take the proton to be a
massive pointlike fermion, with the  equivalent $\gamma^* p$ vertex described
by a factor $-i\gamma^\alpha F_1(t)$. 
Incorporating the background effects, the cross section of the process in 
(\ref{eq1}) is given by (\ref{elsig_4}), where $\overline {{\mid {M_{\mathrm
{el}}}\mid }^2}$ now becomes
\be
\overline {{\mid {M_{\mathrm{el}}}\mid }^2}= \overline {{\mid {M^{\rm{QEDCS}}_{\mathrm{el}}}\mid
}^2}
+\overline {{\mid {M^{\rm{VCS}}_{\mathrm{el}}}\mid }^2}- 2\, {\Re{\bf{\it e}}}\,\overline {M^{\rm{QEDCS}}_{\mathrm{el}}
M^{\rm{VCS} *}_{\mathrm{el}}}.
\label{amplielasunp}
\ee
The interference term will have opposite sign if we consider a positron
instead of an electron.

The explicit expressions of 
$\overline {{\mid {M^{\rm{QEDCS}}_{\mathrm{el}}}\mid }^2}$, 
$\overline
{{\mid {M^{\rm{VCS}}_{\mathrm{el}}}\mid }^2}$ and 
$2\, \Re {\bf\it{e}} \,\overline {M^{\rm{QEDCS}}_{\mathrm{el}}M^{\rm{VCS} *}_{\mathrm{el}}}$ are given 
in Appendix \ref{sec:matrix_el}.
The effect of the proton mass is small in the kinematical range of HERA.

\section{Inelastic Channel}
\label{sec:vcsinel}

We next consider the corresponding inelastic process, where an electron and 
a photon are produced in the final state  together with
a general hadronic system $X$:
\be
e(l)+p(P) \rightarrow e(l')+\gamma(k')+X(P_X).
\label{eqin}
\ee
The exact calculation of the
QEDCS rates follows our treatment in Chapter \ref{ch:qedcs} based on the ALLM97
parametrization \cite{allm97} of the proton structure function $F_2(x_B, Q^2)$.
For the purpose of evaluating the relative importance of the VCS background 
we resort to a unified parton model estimate of the VCS and QEDCS rates. 
The cross section within the parton model is given by
\be
\frac{\der \sigma_{\rm{inel}}}{\der x_B \,\der Q^2 \,\der \hat s\, \der \hat t \,\der \varphi^*}=\sum_{q} \, q(x_B,Q^2)\, \frac{\der \hat
\sigma^q}{\der \hat s\, \der Q^2 \,\der \hat t\, \der \varphi^*}~,
\label{eq:parton}
\ee
where $q(x_B,Q^2)$  are the  quark and antiquark distributions of the initial 
proton, $q = u,\, d, \,s, \,\bar{u}, \,\bar{d}, \,\bar{s}$. Furthermore, $Q^2 
=- k^2 = -(l' + k' -l)^2$, $x_B= \frac{Q^2} {2 P \cdot (-k)}$  and  $d \hat \sigma^q$ is the 
differential cross section of the subprocess
\be
e(l)+q(p) \rightarrow e(l')+\gamma(k')+q(p'),
\label{eqsub_4}
\ee
which, similarly to \eqref{elsig_4}, can be written as
\be
\frac{\der \hat\sigma^q}{\der \hat s\, \der Q^2 \,\der \hat t\, \der \varphi^*}
 = \frac{1}{2 (4\pi)^4 (s-m^2)^2}\, \frac{1}{(\hat s-t)}\,
\overline {{\mid {\hat{M}^q_{\mathrm{ }}}\mid }^2}.
\label{eq:subq}
\ee
The relevant integrated cross section is obtained inserting \eqref{eq:subq}
into \eqref{eq:parton}, 
\be 
\sigma_{\mathrm{inel}}(s)&=& 
\frac{1}{2(4\pi)^4  (s-m^2)^2} \sum_q \int_{W^2_{\mathrm{min}}}^{W^2_{\mathrm{max}}}
\der W^2 \int_{m_e^2}^{(\sqrt{s}-W)^2} \der\hat s \int_{Q^2_{\mathrm{min}}}^{Q^2_{\mathrm{max}}}
 {\der Q^2 \over Q^2}
\int_{{\hat t}_{\mathrm{min}}}^ {{\hat t}_{\mathrm{max}}}
\der\hat t \int_0^{2\pi} \der\varphi^* 
\nonumber\\&&~~~~~~~~~~~~\times \frac{1}{(\hat s+Q^2)}\,\overline {{\mid 
{\hat{M}^q_{\mathrm{ }}}
\mid }^2} ~~q(x_B,Q^2),
\label{insig_4}
\ee
where we have traded the integration variable $x_B$ with
\be  
W^2 = (p-k)^2 = m^2 + Q^2 \,\frac{1-x_B}{x_B}~. 
\ee
The limits of integration
are given explicitly by   (\ref{thatlim}), (\ref{eq:q2limit}) and 
(\ref{eq:w2limit})  with
$\hat s_{\mathrm{min}}=m_e^2$. 
Further constraints, related to the HERA kinematics, will be discussed in
the numerical section.   
Similar to the elastic channel, we have
\be 
\overline {{\mid {\hat{M}^q\mid} }^2}= \overline {{\mid {\hat{M}^{q \,{\rm QEDCS}}_{\mathrm{ }}}\mid}^2}
+\overline {{\mid {\hat{M}^{q \,{\rm VCS}}_{\mathrm{ }}}\mid }^2}  -2\, {\Re{\it e}} \overline {\hat{M}^{q \,{\rm QEDCS}}_{\mathrm{ }}
\hat{M}^{q \,{\rm VCS} *}_{\mathrm{ }}}.
\label{eq:inelasticvcs}
\ee 
Again, the interference term will have opposite sign for a positron. 
Furthermore, we  introduce the auxiliary invariants 
$\hat S=(p'+k')^2$ 
and $\hat U= (p'-k)^2$, which can be written in terms of measurable 
quantities, 
\be
\hat S={\hat t (x_l-x_B)\over x_l}, ~~~~~~\hat U=\hat t- \hat S+Q^2,
\label{eq:auxiliary}
\ee
with $x_l$ already defined in \eqref{xl}.  
The explicit expression of \eqref{eq:inelasticvcs} is
deferred to Appendix \ref{sec:matrix_inel}. 
It is also given in
\cite{brodsky,metz} for a massless proton.

\section{Numerical Results}
\label{vcs:num}

In this section the numerical results are presented.  In
order to select the QEDCS events, certain kinematical constraints are imposed 
in the Monte Carlo studies in \cite{thesis,
lend}. As in Section \ref{sec:results} the following laboratory 
frame variables are used: energy of the final electron $E_e'$, 
energy of the final  
photon $E_{\gamma}'$,    
polar angles  of the outgoing electron and photon, $ \theta_e$ and   
$\theta_{\gamma}$ respectively,  and  acoplanarity angle $\phi$,     
which is defined as $\phi=|\,\pi-|\phi_{\gamma} -\phi_{e}|\,|$,     
where $\phi_{\gamma}$ and $\phi_{e}$ are the azimuthal angles of  
the outgoing photon and electron respectively ($0 \le \phi_{\gamma},\,
\phi_e \le 2 \,\pi$). The cuts are given in column A of Table 
\ref{tableone_4} (from
hereafter, they will be referred to as the set A). The energies of the 
incoming particles are: $E_e = 27.5\,\,
\mathrm{GeV}$ (electron) and $E_p = 820 \,\,\mathrm{GeV}$ (proton).
\begin{table}[t]
\begin{center}
\begin{tabular}{|c|c|}
\hline 
$ A $ & $B$ \\
\hline\hline

$~~E_e', \, E_{\gamma}' > 4 \, \mathrm{GeV}~~ $ & $~~E_e', \, E_{\gamma}' > 4 \,\mathrm{GeV}~~ $ \\
$E_e' + E_{\gamma}' > 20 \, \mathrm{GeV}$ & $ E_e' + E_{\gamma}' > 20\, \mathrm{GeV}$ \\
$~~~ 0.06 < \theta_e, \, \theta_{\gamma} < \pi - 0.06~~~$ & 
$~~~ 0.06 < \theta_e, \, \theta_{\gamma} < \pi - 0.06~~~$ \\
$ \phi < {\pi}/ {4}   $  & $ \hat s > Q^2 $ \\
$ \theta_h < {\pi}/{2}  $  & $ \hat S > \hat s $ \\

\hline

\end{tabular}
\end{center}
\caption{A: cuts to simulate HERA-H1 detector. B: cuts introduced in this 
chapter.}
\label{tableone_4}

\end{table}

So far the photon and  the electron in the final state have been identified 
only in the backward part of the H1 detector at HERA.  
To select signals where there 
are no hadronic activities near the two electromagnetic clusters,  the final 
hadronic state must not be found above the polar angle $\theta_h^{\mathrm{max}}
= \pi/2$ \cite{lend}.  
Motivated by this experimental arrangement, we have identified $\theta_h$ 
with the polar angle of the final quark 
$q'$ in the subprocess $e q \rightarrow e \gamma q'$.  It can be shown that
$\theta_h$ is  given by
\be
\cos \theta_h \equiv \cos\theta_{q'} = \frac{1}{E_{q'}} \,(x_B E_p-E_e-E_e' 
\cos \theta_e-E_\gamma'
\cos\theta_\gamma)
\ee
and  $E_{q'}  = x_B E_p+E_e-E_e'-E_\gamma'$  being the 
energy of the 
final parton. Here we have assumed that the final hadrons are emitted 
collinearly with the struck quark $q'$. For the elastic process 
$\theta_h \equiv \theta_{p'}$, the polar angle of the scattered proton, 
can be obtained
by substituting $x_B=1$ in the above expression. 
Thus we impose the additional condition \cite{lend}
\be
\theta_h < \pi /2
\label{cut4}
\ee
on the cross section. However, no constraint on the hadronic final state was
used in the cross section calculation presented in \cite{lend}. 
Inclusion of  (\ref{cut4}) reduces the QEDCS cross section by about $10\%$.

 In the
kinematical region defined by the constraints mentioned above, the contributions from the 
initial and final state 
radiations, unrelated to QED Compton scattering, are suppressed, see
Section \ref{sec:results}  and  \cite{kessler,blu,ruju,thesis}. 
Furthermore, we checked that the event rates related to the elastic VCS 
process and  its interference with elastic QEDCS are negligible compared 
to the ones
corresponding to pure elastic QEDCS. This is expected because the elastic QEDCS cross 
section is very much dominated by the small values of the variable $-t$, 
compared to $-\hat t$, see   (\ref{qel_unp}) and (\ref{vel}).  
Such an observation is similar to that of
\cite{thesis}, where the elastic DVCS background was calculated using a Regge
model in different kinematical bins.    
Our estimate was done  taking the 
proton to be pointlike with an effective vertex, as discussed in 
Section \ref{sec:vcsel}. 
We find that, in this approximation, the elastic QEDCS cross section differs 
from the actual one  calculated in Chapter \ref{ch:qedcs} by about $3 \% $ within the range defined by
the kinematical constraints.  

\begin{figure}
\begin{center}
\epsfig{figure=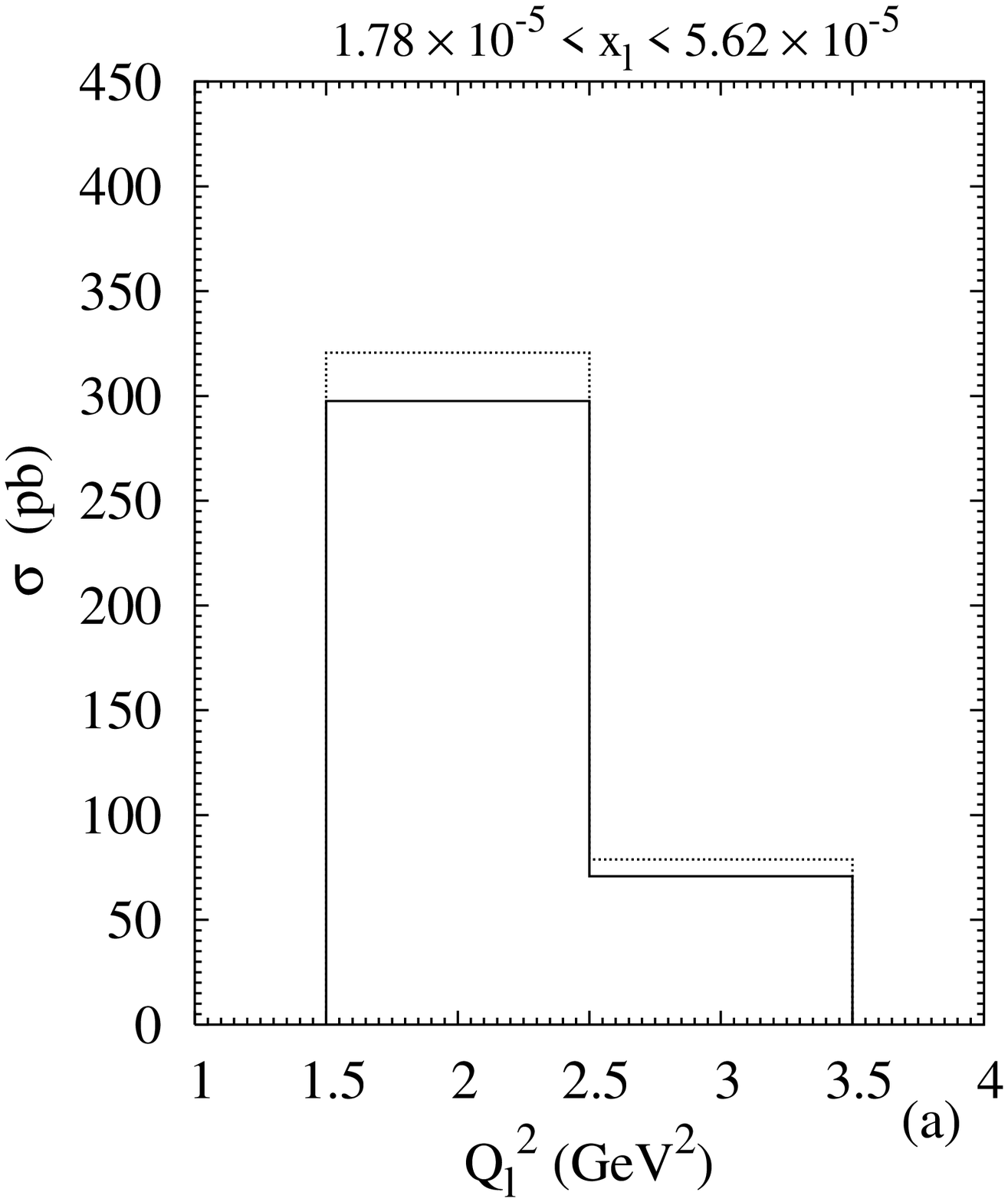,width=7.5 cm,height=7.5 cm}
\epsfig{figure=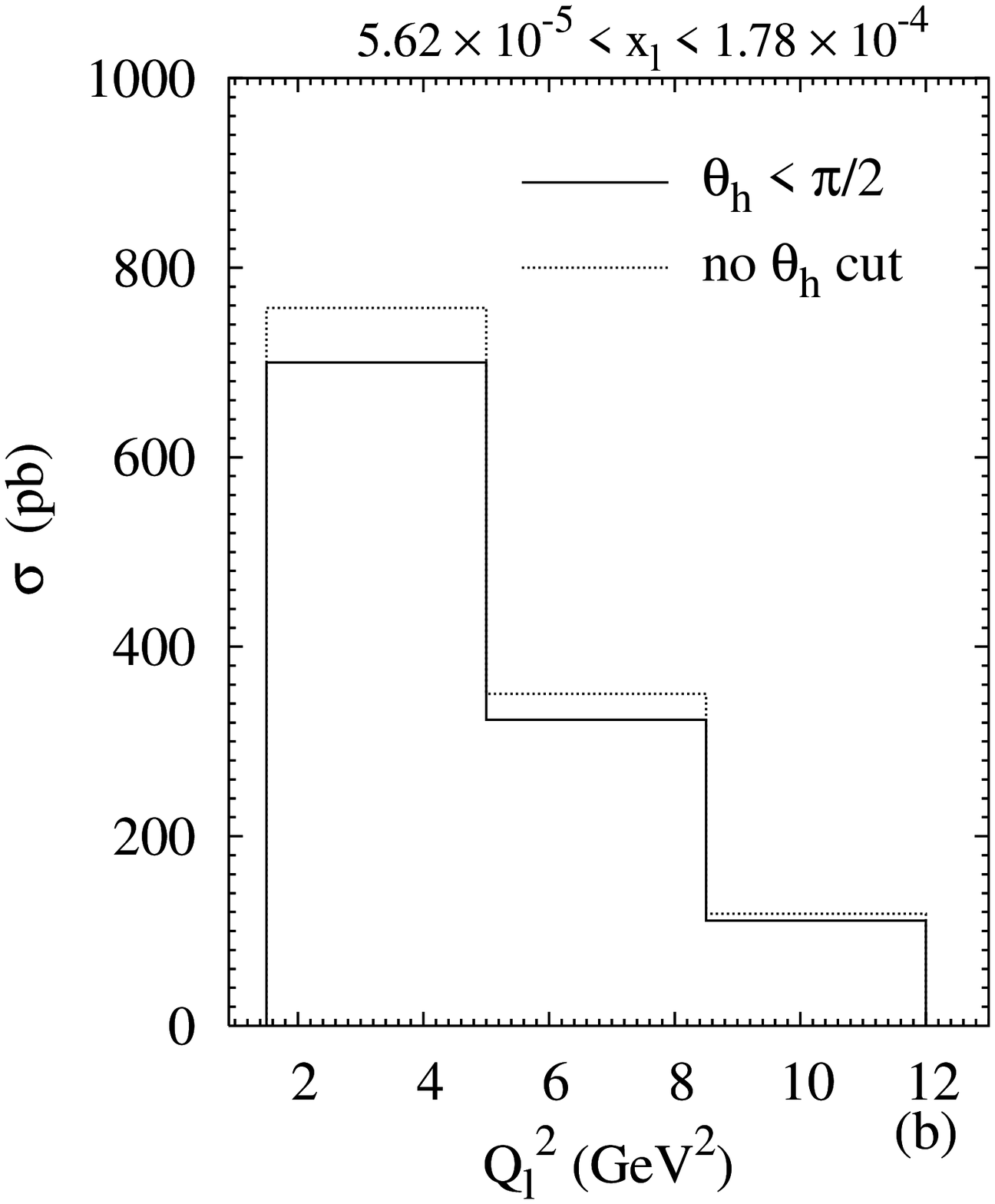,width=7.5 cm,height=7.5 cm}

\vspace{1.5cm}
\epsfig{figure=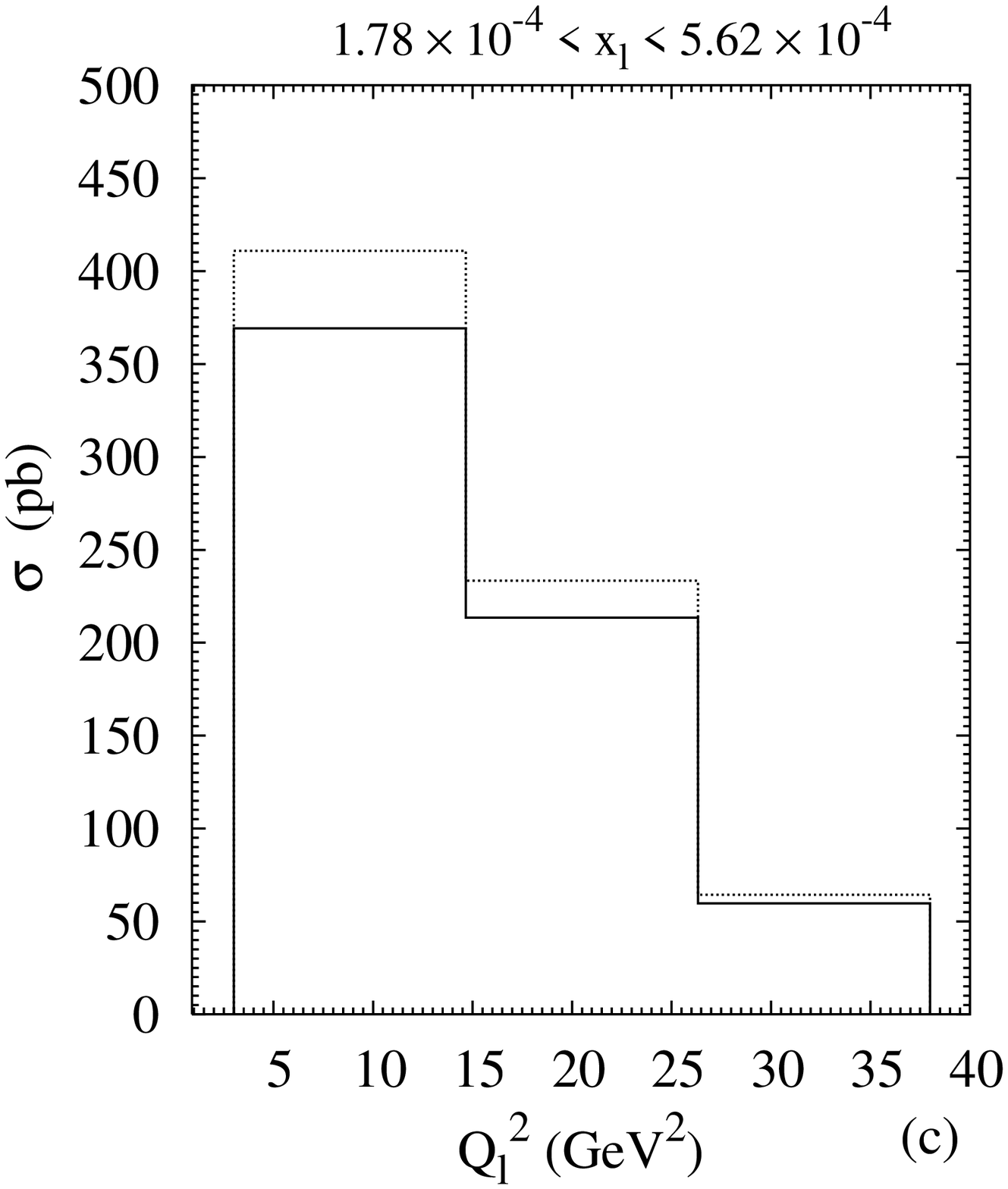,width=7.5 cm,height= 7.5 cm}
\epsfig{figure=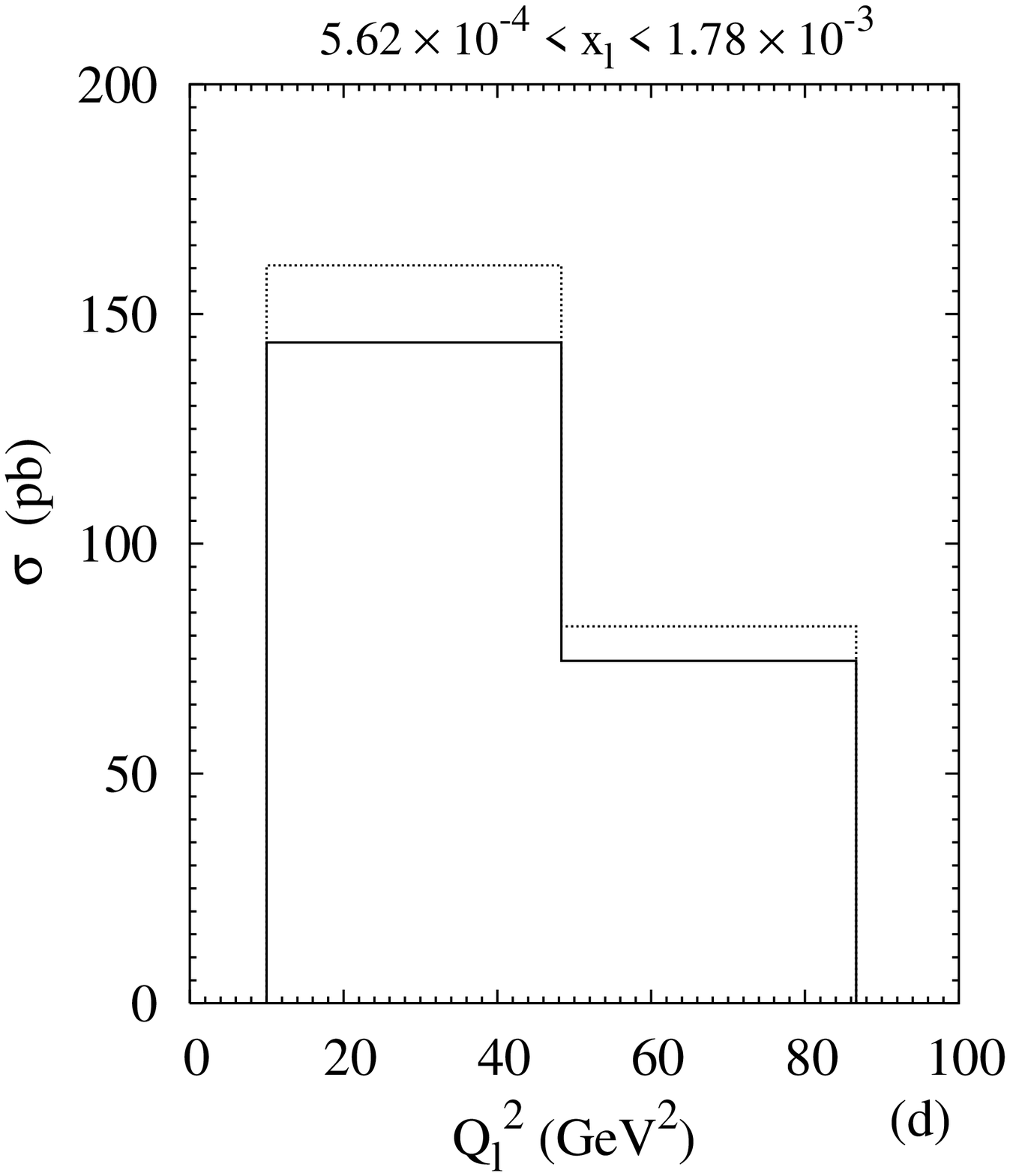,width=7.5 cm,height= 7.5 cm}

\end{center}
\caption{Double differential cross section for QED Compton scattering at
HERA-H1.  The kinematical bins correspond to Table \ref{tab:one}. 
The continuous line describes the total (elastic + inelastic) cross section 
subject to the set of cuts A in Table \ref{tableone_4}. The dotted line 
shows the same results when the constraint on $\theta_h$ is removed.}
\label{fig:q2lxl}
\end{figure}
\begin{figure}[ht]
\begin{center}
\epsfig{figure= 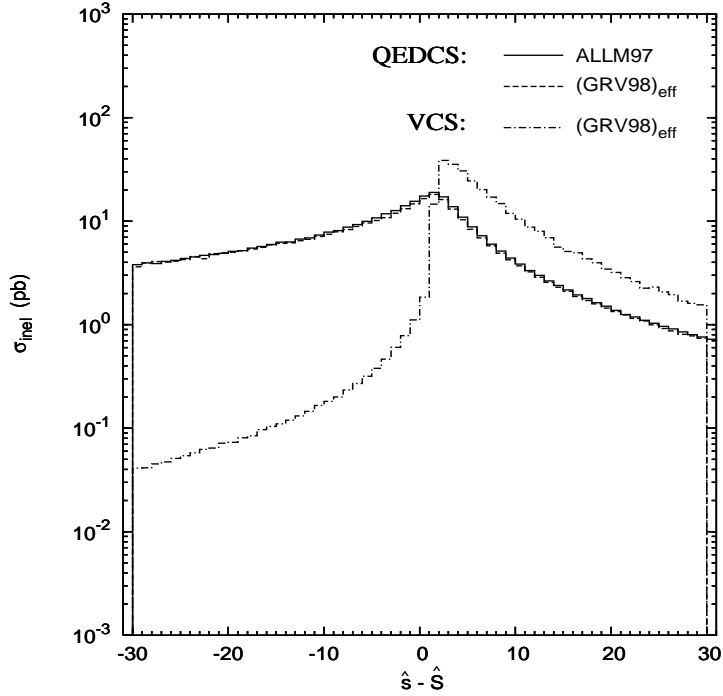,width=10 cm, height = 9cm}
\end{center}

\caption{Cross section for the QEDCS and VCS processes (inelastic) 
at HERA-H1. 
The bins are in $\hat s- \hat S$, expressed in GeV$^2$. The cuts applied are listed in Table \ref{tableone_4}, set B (except $\hat S \gtrsim \hat s$). The continuous line corresponds to the QEDCS cross section with
ALLM97 parametrization of $F_2(x_B, Q^2)$, the dashed line corresponds to
the  
QEDCS cross section using the effective GRV98 parton distributions in 
 (\ref{fq}) and  the dashed dotted line corresponds to the
VCS cross section using the same effective distributions.}
\label{fig:cutbins}
\end{figure}
\begin{figure}[ht]
\begin{center}
\epsfig{figure=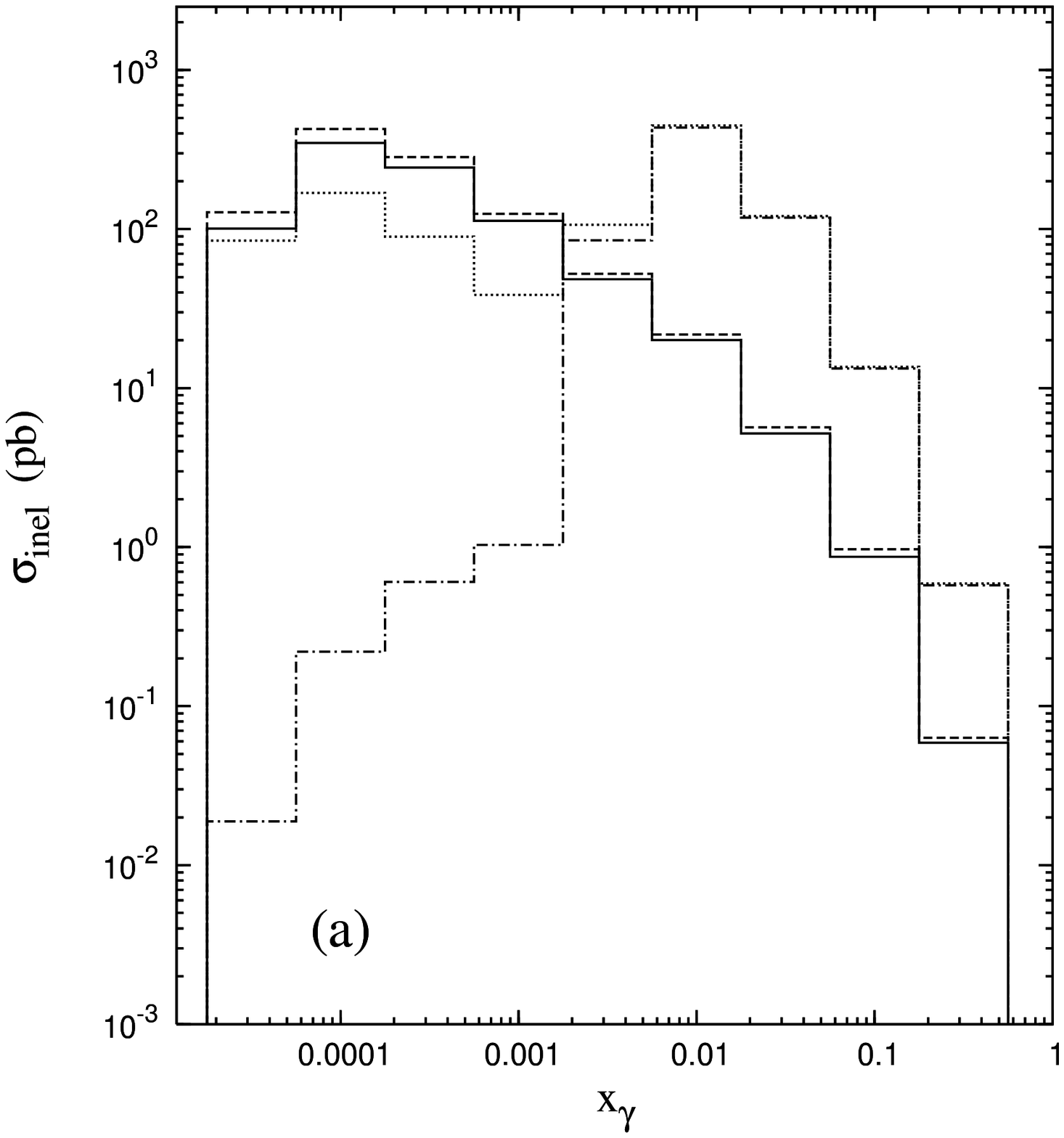,width= 7.9cm}
\epsfig{figure=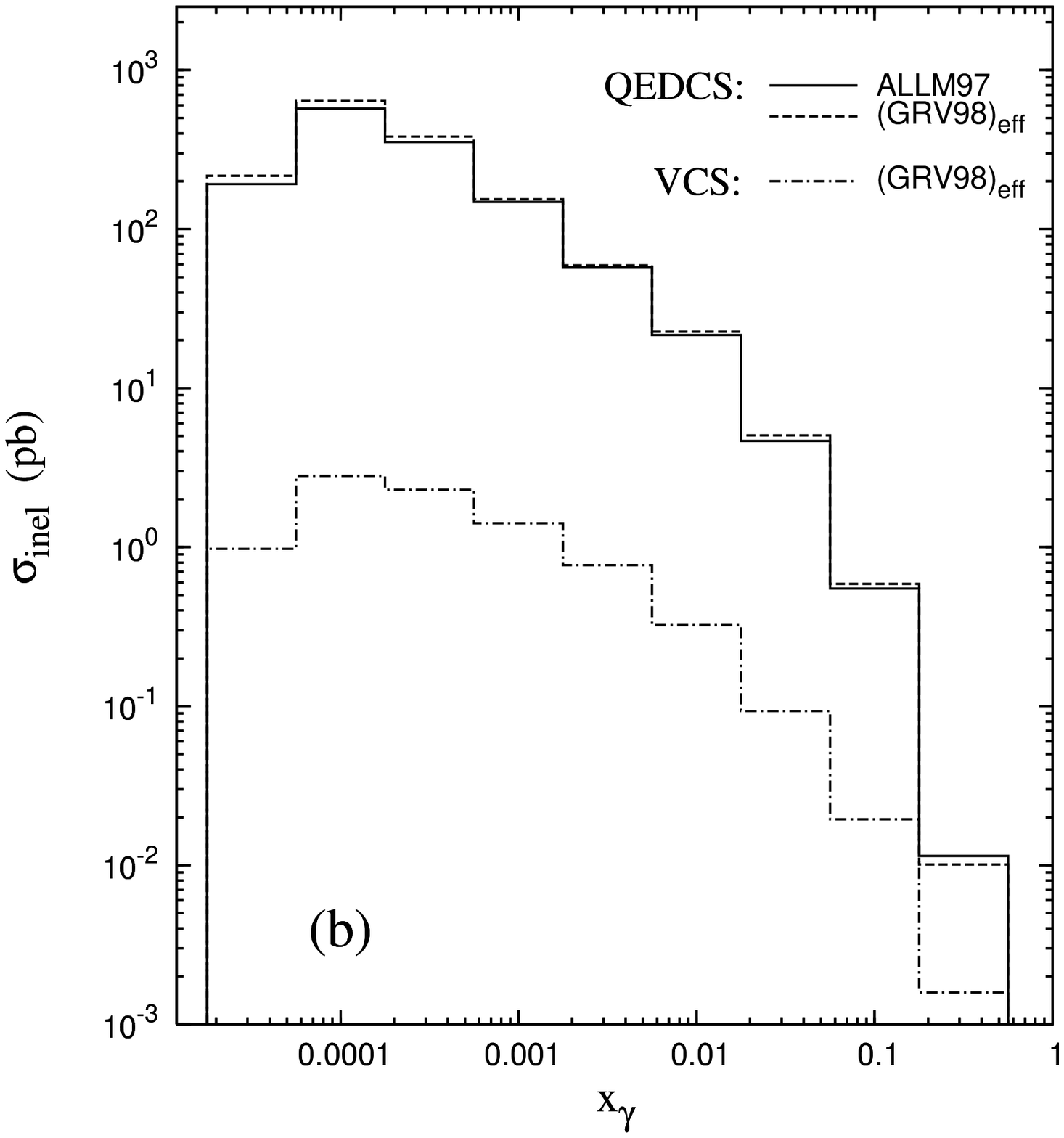,width= 7.9cm}
\end{center}

\caption{Cross section for QED Compton scattering in bins of $x_{\gamma}$
as calculated with the ALLM97 (full line) and the 
$\mathrm{(GRV98)}_{\mathrm{eff}}$ (dashed line) parametrization of 
$F_2(x_B, Q^2)$, respectively,  as compared to the VCS background cross section (dot-dashed line). The cuts employed are: a) as in set A, b) as in set B of 
Table \ref{tableone_4}. The dotted line in Figure \ref{fig:xgammabns} (a) 
shows the VCS cross section subject to
the set of cuts A without the constraint on $\theta_h$.}
\label{fig:xgammabns}
\end{figure}

\begin{figure}[ht]
\begin{center}
\epsfig{figure=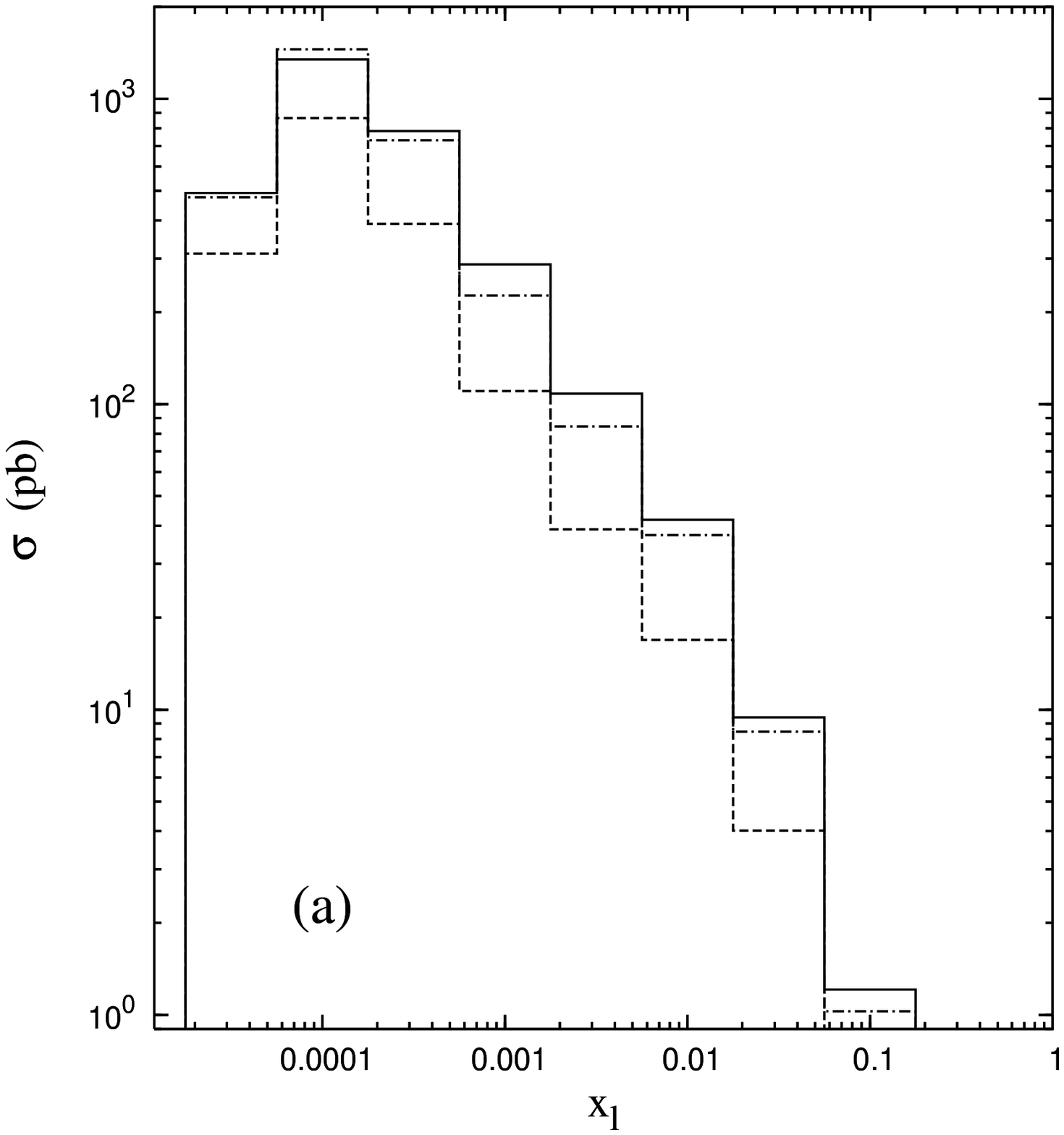,width=7.9 cm}
\epsfig{figure=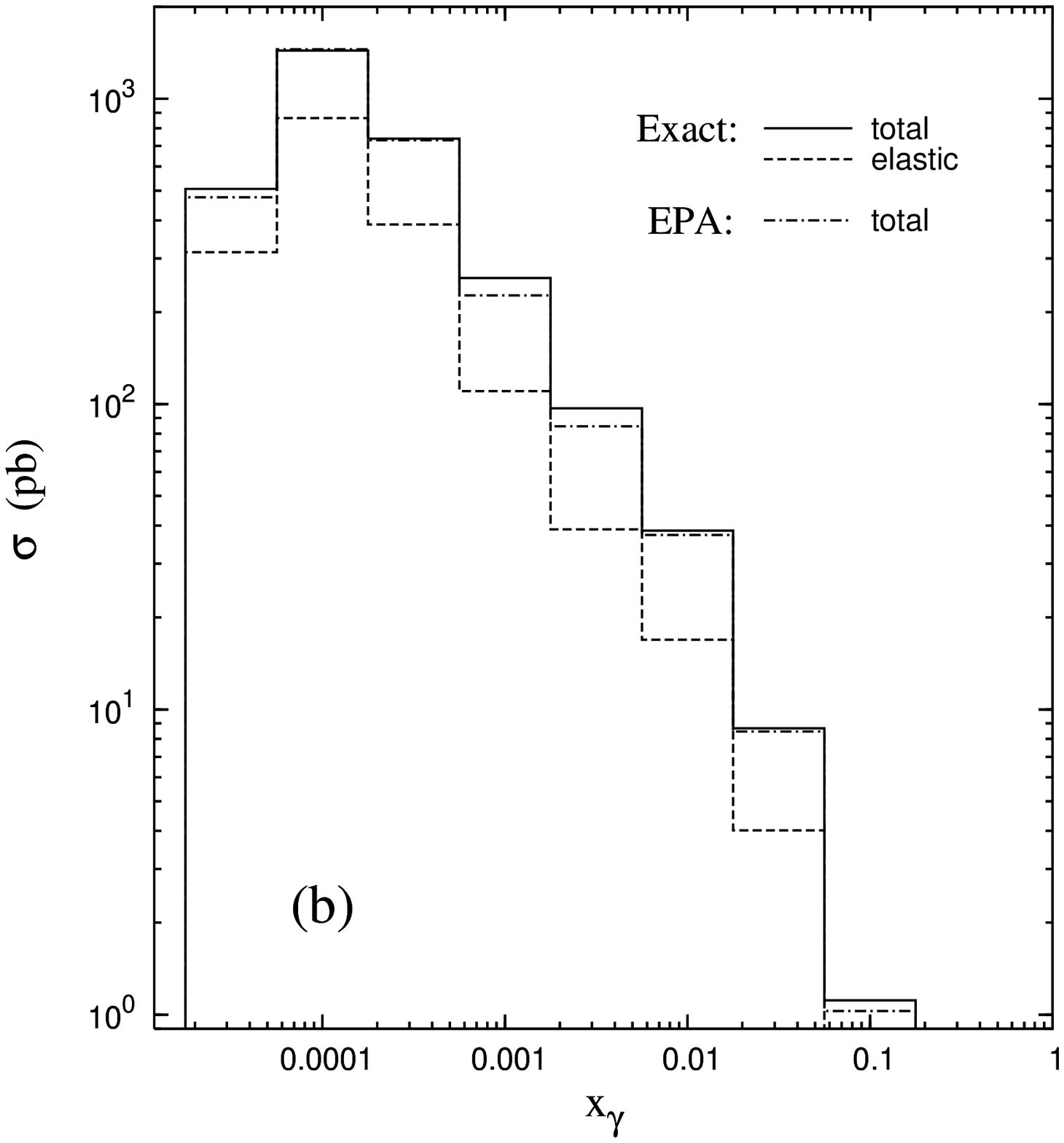,width=7.9cm}
\end{center}

\caption{Cross section for QED Compton scattering at HERA-H1 subject to the 
cuts of set B in Table \ref{tableone_4}, in (a) $x_l$ bins, 
(b) $x_{\gamma}$ bins. 
The continuous line corresponds to our exact calculation using ALLM97 
parametrization of $F_2(x_B, Q^2)$, the dot-dashed line corresponds to the 
same in the EPA, the dashed line shows the elastic contribution. }
\label{figxlbins}
\end{figure}

Figure \ref{fig:q2lxl} shows the total (elastic + inelastic) QEDCS cross section in 
$x_l-Q_l^2$ bins with $Q_l^2=-\hat t$, subject to the cuts of set A. 
For comparison we have also plotted the cross section without the cut 
on $\theta_h$, similar to the analysis in Chapter \ref{ch:qedcs}. 
This additional 
constraint affects the result only in the inelastic channel.

We  checked that the upper limit in  (\ref{cut4})  reduces the contribution 
from the inelastic VCS reaction. In order to calculate it, one needs a model 
for the parton distributions $q(x_B,Q^2)$.
However, in the relevant kinematical region, $Q^2$ can be very small and
may become close to zero, where the parton picture is not applicable.
Hence, in our estimate, we replace the parton distribution
$q(x_B,Q^2)$ by an effective parton distribution \cite{pp2} 
\be
\tilde{q}(x_B,Q^2)={Q^2\over Q^2+a\, Q_0^2}\, q(x_B,Q^2+Q_0^2),
\label{fq}
\ee
where $a=1/4$ and $Q_0^2=0.4 \, \mathrm{GeV}^2$  are two parameters and 
$q(x_B,Q^2)$ is the NLO GRV98 \cite{grv98} parton distribution. 
$Q_0^2$ prevents the scale in the distribution to become too 
low. Equation (\ref{fq}) is  motivated by a similar form used in 
\cite{bad,allm97} 
for the parametrization of the structure function $F_2(x_B,Q^2)$ in the low 
$Q^2$ region. It is
clear that at high $Q^2$, $\tilde{q}(x_B,Q^2) \rightarrow q(x_B,Q^2)$.

In this chapter, we introduce a new set of cuts, which are given in the
column B of Table \ref{tableone_4} (and will be referred to as the set B) 
for a 
better extraction of the equivalent photon distribution of the proton as 
well as to suppress the VCS background. These cuts
will be compared to the set A in the following. Instead of the constraint 
on the acoplanarity, namely $\phi < \pi/4$,
where the upper limit is actually ambiguous, we impose $\hat s > Q^2$. 
The relevance of the cut $\hat S \gtrsim \hat s $  can be seen from 
Figure \ref{fig:cutbins}. This shows the cross sections of  the QEDCS and VCS 
processes 
in the inelastic channel, calculated using  (\ref{insig_4}) and subject to the
kinematical limits of set B (except $ \hat S \gtrsim \hat s $), 
in bins of $\hat s-\hat S$.
Figure \ref{fig:cutbins}  shows that the VCS cross section is higher than QEDCS
for bins with $\hat s \gtrsim \hat S$  but falls sharply in bins
for which $\hat s $ is close to $\hat S$ and becomes much suppressed for
$\hat S \gtrsim \hat s$ . This is expected because $\hat S$ corresponds to the
quark propagator in the VCS cross section, see  (\ref{vin}), and a lower 
value enhances this
contribution. In fact the sharp drop of the VCS cross section in bins where 
$\hat S \gtrsim \hat s$ is due to the fact that both the propagators 
$\hat s, \,\hat u$ in the QEDCS cross section are constrained to be 
smaller than 
$\hat S, \,\hat U$ for  VCS in these bins, see   (\ref{qin}), (\ref{vin}). 
The QEDCS cross section is always
enhanced by the factor  $Q^2$ in the denominator of  (\ref{qin}) coming 
from the virtual photon,  which can be very small in the
kinematical region of interest here. This plot shows that {\em{imposing}} 
{\em{a cut on $\hat S$ can be very effective in reducing the background 
contribution from VCS}}. The interference between inelastic QEDCS
and VCS gives negligible contribution. We have also shown the QEDCS cross
section using the ALLM97 parametrization of $F_2(x_B,Q^2)$ as calculated
in Chapter \ref{ch:qedcs}. 
The discrepancy between this and the one calculated using the
parametrization in  (\ref{fq}) is less than $5 \%$ in almost all the bins, and 
maximally $7\%$ in two bins.  

In Figure \ref{fig:xgammabns} (a) we have shown the inelastic QEDCS and VCS cross 
sections in bins
of $x_\gamma$, defined in \eqref{eq:icsgamma}, subject to the cuts of set A. 
The VCS
cross section is much suppressed in the smaller $x_\gamma$ bins but becomes
enhanced as $x_\gamma$ increases, which indicates that such a set of cuts is 
{\em{not}}
suitable  to remove the background at higher $x_\gamma$. The situation will
be the same in $x_l$ bins, with $x_l$ defined in \eqref{xl}. \\ 
Figure \ref{fig:xgammabns} (b) shows the
cross sections but with the set B. The background in this case is suppressed 
for all $x_\gamma$ bins,  which means that such a cut is more effective in 
extracting QEDCS
events also for higher $x_\gamma$. In addition, we have plotted the QEDCS
cross section in terms of the structure function $F_2(x_B, Q^2)$, using 
the ALLM97 parametrization. Figure \ref{fig:xgammabns} shows that our 
parametrizaton gives 
a reasonably good description of the proton, at least for the QEDCS process, 
in all the bins. However, 
this parametrization has been used only to make a relative estimate of the 
 background events. In fact, a quantitative estimate of the inelastic VCS
events has not been presented in \cite{thesis,lend}. 

Figures  \ref{figxlbins} (a) and \ref{figxlbins} (b) show the  QEDCS cross section in bins of $x_l$ and 
$x_\gamma$, respectively, subject to 
the constraints of set B. The elastic cross section has been calculated
 using   (\ref{elsig_4})-(\ref{eq:amplqedcs}), as in Chapter \ref{ch:qedcs}. The inelastic cross 
section is given by 
\eqref{siin} in terms of the structure functions $F_1(x_B, Q^2)$ 
and $F_2(x_B, Q^2)$.
We have assumed the Callan-Gross relation and used the ALLM97 parametrization
\cite{allm97} for $F_2(x_B,Q^2)$. In this way the results presented in Figure 
\ref{figxlbins}, labelled as exact, are free from the parton model approximations 
in Figures \ref{fig:cutbins} and \ref{fig:xgammabns}. 
In the same plot, we have also 
shown the total cross section calculated in terms of the EPA,
according to   (\ref{epael}) and (\ref{pap1_epain}). 
Figure \ref{figxlbins} (b)  shows much better agreement between the 
approximate 
cross section based on the EPA and the exact one.
For Figure \ref{figxlbins} (a), the discrepancy 
is about $3-8 \%$ in
the first three bins, between $10-20 \%$ in three other bins and $15 \%$ 
in the last bin. 
In Figure \ref{figxlbins} (b) it is
$1-6 \%$ in five  bins, $12-13 \%$ in two bins and about $8 \%$ in the 
last bin. 
The discrepancy of the exact cross section, integrated over $x_{\gamma}$, 
with the approximate one, when subject to the 
constraints  of set B is $0.38\%$ in the elastic channel and $4.5\%$ 
in the inelastic one. The total (elastic + inelastic) discrepancy turns out to
be $2.26\%$, which should be compared to the  values $14 \%$,  already
observed in Chapter \ref{ch:qedcs} when subject to the set A, except the one on
$\theta_h$, and  $24 \%$ when this one is imposed too. 


\begin{table}
\small
\begin{center}
\begin{tabular}{|c|c|c|c||c|c|c|}
\hline
$ x_l$ bin & $Q^2_l$ bin & $\sigma_{\mathrm{inel}}$  & 
$\sigma_{\mathrm{inel}}^{\mathrm{EPA}}$ &$\sigma_{\mathrm{inel}}^*$ &
$\sigma_{\mathrm{inel}}^{\mathrm{EPA *}} $   \\ 
\hline \hline
          &           &             &     &      &  \\
$ 1.78\times 10^{-5}-5.62 \times 10^{-5} $ & $1.5 -2.5$ &$ 5.511\times 10^1 $  & $1.483 \times 10^2 $ & $ 1.062\times 10^2 $ & $  1.298\times 10^2 $    \\ 
$1.78\times 10^{-5}-5.62 \times 10^{-5}$ & $2.5 - 3.5$&$1.992\times 10^1$ &  $ 3.257\times 10^1 $ & $3.925\times 10^1$  & $ 2.888\times 10^1 $   \\ \hline
$5.62\times 10^{-5}-1.78\times 10^{-4} $ & $1.5 - 5.0$&$1.720\times 10^2$  & $3.967\times 10^2$  & $2.937\times 10^2$   & $3.369\times 10^2$ \\
$5.62\times 10^{-5}-1.78 \times 10^{-4}$&$ 5.0-8.5$&$8.355\times 10^1 $ &  $2.015\times 10^2$ & $1.407\times 10^2 $ &$1.764\times 10^2$  \\ 
$5.62 \times 10^{-5}-1.78\times 10^{-4}$ & $8.5-12.0$&$2.609\times 10^1 $&$7.752\times 10^1 $ & $4.334\times 10^1$  & $6.880\times 10^1$ \\ \hline
$1.78\times 10^{-4}-5.62\times 10^{-4}$ & $3.0-14.67 $ & $1.613\times 10^2$  &  $ 2.103\times 10^2$ & $2.330 \times 10^2$  &  $1.720\times 10^2$\\ 
$1.78\times 10^{-4}-5.62\times 10^{-4}$ & $14.67-26.33 $ & $7.639\times 10^1$ &$ 1.477\times 10^2$ & $1.194\times 10^2$ &  $1.283\times 10^2$\\ 
$1.78\times 10^{-4}-5.62\times 10^{-4}$ & $26.33-38.0 $ & $2.269\times 10^1$   &$ 4.229\times 10^1$ & $3.554\times 10^1$ & $3.759\times 10^1$ \\ \hline
$5.62\times 10^{-4}-1.78\times 10^{-3}$ & $10.0-48.33 $ & $8.425\times 10^1$ & $ 7.555\times 10^1$ & $9.953\times 10^1$  & $5.980\times 10^1$ \\ 
$5.62\times 10^{-4}-1.78\times 10^{-3}$ & $48.33-86.67 $ & $3.745\times 10^1$  & $ 4.897\times 10^1$  & $5.638\times 10^1$ &  $4.168\times 10^1$\\ 
$5.62\times 10^{-4}-1.78\times 10^{-3}$ & $86.67-125.0 $ & $1.066\times 10^1 $& $1.462\times 10^1 $ & $1.644\times 10^1$  & $1.253\times 10^1$ \\ \hline
$1.78\times 10^{-3}-5.62\times 10^{-3}$ & $22-168 $ & $3.846\times 10^1$ & $ 2.791\times 10^1$ & $3.773\times 10^1$ & $2.104\times 10^1$ \\ 
$1.78\times 10^{-3}-5.62\times 10^{-3}$ & $168-314 $ & $1.622\times 10^1$ & $ 1.849\times 10^1$  &$2.289\times 10^1$ & $1.543\times 10^1$ \\ 
$1.78\times 10^{-3}-5.62\times 10^{-3}$ & $314-460 $ & $5.836$ & $ 9.043$ &$7.827$ & $7.641$ \\ \hline
$5.62\times 10^{-3}-1.78\times 10^{-2}$ & $0-500 $ & $1.202\times 10^1$ &  $7.624 $ & $9.281 $  & $4.923$ \\ 
$5.62\times 10^{-3}-1.78\times 10^{-2}$ & $500-1000 $ & $1.010\times 10^1$ & $ 1.450\times 10^1$ & $1.242\times 10^1$ & $1.190 \times 10^1$  \\ 
$5.62\times 10^{-3}-1.78\times 10^{-2}$ & $1000-1500  $ & $2.584 $ & $ 4.067$ 
& $3.033$  &  $3.281$ \\ \hline
$1.78\times 10^{-2}-5.62\times 10^{-2}$ & $0-1500 $ & $2.712$ & $1.439$  & $1.536$ & $7.261\times 10^{-1}$ \\ 
 $1.78\times 10^{-2}-5.62\times 10^{-2}$ & $1500-3000 $ & $2.967$ & $ 3.855$
 & $3.091$ &  $2.922$ \\ 
$1.78\times 10^{-2}-5.62\times 10^{-2}$ & $3000-4500  $ & $7.423\times 10^{-1}$ &  $1.004$ & $7.210\times 10^{-1}$ & $7.204\times 10^{-1}$ \\ \hline
$5.62\times 10^{-2}-1.78\times 10^{-1}$ & $10-6005 $ & $7.083\times 10{-1}$ & $ 4.923\times 10^{-1}$ & $3.684\times 10^{-1}$ & $2.354 \times 10^{-1}$  \\ 
$5.62\times 10^{-2}-1.78\times 10^{-1}$ & $6005-12000 $ & $3.637\times 10^{-1}$  & $3.845\times 10^{-1}$  & $ 2.611\times 10^{-1} $ &  $2.124\times 10^{-1}$ \\ 
 $  5.62\times 10^{-2}-1.78\times 10^{-1}  $  & $ 12000-17995 $ & $ 3.638\times 10^{-2}  $& $ 3.847\times 10^{-2} $ & $ 1.884\times 10^{-2} $& $ 1.537\times 10^{-2} $   \\ 
     &    &          &        &         &    \\
\hline
\end{tabular}
\end{center}
\caption{Double differential QED Compton scattering cross section
(inelastic) in $x_l$ and $Q_l^2$ bins. $\sigma_{\mathrm{inel}}$ and 
$\sigma_{\mathrm{inel}}^*$
correspond to the exact   cross section  subject to the 
cuts A and B of Table \ref{tableone_4}
respectively. $\sigma_{\mathrm{inel}}^{\mathrm{EPA}}$ and  $\sigma_{\mathrm{inel}}^{\mathrm{EPA
*}}$  correspond to the cross sections in the EPA and  subject to 
the cuts A and B
respectively. $Q^2_l$ is expressed in 
GeV$^2$ and the cross sections are in pb.}
\label{tabletwo}
\end{table} 

\begin{table}
\small
\begin{center}
\begin{tabular}{|c|c|c|c||c|c|c|}
\hline 
$x_{\gamma}$ bin & $Q^2_l$ bin & $\sigma_{\mathrm{inel}}$  & 
$\sigma_{\mathrm{inel}}^{\mathrm{EPA}}$ &$\sigma_{\mathrm{inel}}^*$ &
$\sigma_{\mathrm{inel}}^{\mathrm{EPA *}} $    \\ 
\hline \hline
          &           &             &     &      &  \\
$1.78\times 10^{-5}-5.62 \times 10^{-5}$ & $1.5 -2.5$ &$5.191\times 10^1$  & $1.483 \times 10^2 $ & $9.932\times 10^1$ & $ 1.298\times 10^2$    \\ 
$1.78\times 10^{-5}-5.62 \times 10^{-5}$ & $2.5 - 3.5$&$2.839\times 10^1$ &  $3.257\times 10^1$ & $5.176\times 10^1$  & $2.888\times 10^1$   \\ \hline
$5.62\times 10^{-5}-1.78\times 10^{-4} $ & $1.5 - 5.0$&$1.761\times 10^2$  & $3.967\times 10^2$  & $2.992\times 10^2$   & $3.369\times 10^2$ \\
$5.62\times 10^{-5}-1.78 \times 10^{-4}$&$ 5.0-8.5$&$1.101\times 10^2 $ &  $2.015\times 10^2$ & $1.775\times 10^2 $ &$1.764\times 10^2$  \\ 
$5.62 \times 10^{-5}-1.78\times 10^{-4}$ & $8.5-12.0$&$4.573\times 10^1 $&$7.752\times 10^1 $ & $7.189\times 10^1$  & $6.880\times 10^1$ \\ \hline
$1.78\times 10^{-4}-5.62\times 10^{-4}$ & $3.0-14.67 $ & $1.006\times 10^2$  &  $ 2.103\times 10^2$ & $1.442 \times 10^2$  &  $1.720\times 10^2$\\ 
$1.78\times 10^{-4}-5.62\times 10^{-4}$ & $14.67-26.33 $ & $9.299\times 10^1$ &$ 1.477\times 10^2$ & $1.347\times 10^2$ &  $1.283\times 10^2$\\ 
$1.78\times 10^{-4}-5.62\times 10^{-4}$ & $26.33-38.0 $ & $3.564\times 10^1$   &$ 4.299\times 10^1$ & $5.166\times 10^1$ & $3.759\times 10^1$ \\ \hline
$5.62\times 10^{-4}-1.78\times 10^{-3}$ & $10.0-48.33 $ & $4.408\times 10^1$ & $ 7.555\times 10^1$ & $5.386\times 10^1$  & $5.980\times 10^1$ \\ 
$5.62\times 10^{-4}-1.78\times 10^{-3}$ & $48.33-86.67 $ & $4.476\times 10^1$  & $ 4.897\times 10^1$  & $6.082\times 10^1$ &  $4.168\times 10^1$\\ 
$5.62\times 10^{-4}-1.78\times 10^{-3}$ & $86.67-125.0 $ & $1.611\times 10^1 $& $1.462\times 10^1 $ & $2.222\times 10^1$  & $1.253\times 10^1$ \\ \hline
$1.78\times 10^{-3}-5.62\times 10^{-3}$ & $22-168 $ & $2.019\times 10^1$ & $ 2.791\times 10^1$ & $2.174\times 10^1$ & $2.104\times 10^1$ \\ 
$1.78\times 10^{-3}-5.62\times 10^{-3}$ & $168-314 $ & $1.896\times 10^1$ & $ 1.849\times 10^1$  &$2.418\times 10^1$ & $1.543\times 10^1$ \\ 
$1.78\times 10^{-3}-5.62\times 10^{-3}$ & $314-460 $ & $7.594$ & $ 9.043$ &$9.766$ & $7.641$ \\ \hline
$5.62\times 10^{-3}-1.78\times 10^{-2}$ & $0-500 $ & $6.058$ &  $7.624 $ & $5.219$  & $4.923$ \\ 
$5.62\times 10^{-3}-1.78\times 10^{-2}$ & $500-1000 $ & $1.077\times 10^1$ & $ 1.450\times 10^1$ & $1.263\times 10^1$ & $1.190\times 10^1$  \\ 
$5.62\times 10^{-3}-1.78\times 10^{-2}$ & $1000-1500  $ & $3.019 $ & $ 4.067$ 
& $3.449$  &  $3.281$ \\ \hline
$1.78\times 10^{-2}-5.62\times 10^{-2}$ & $0-1500 $ & $1.267$ & $1.439$  & $7.874\times 10^{-1}$ & $7.261 \times 10^{-1}$ \\ 
 $1.78\times 10^{-2}-5.62\times 10^{-2}$ & $1500-3000 $ & $3.005$ & $ 3.855$
 & $3.018$ &  $2.922$ \\ 
$1.78\times 10^{-2}-5.62\times 10^{-2}$ & $3000-4500  $ & $8.121\times 10^{-1}$ &  $1.004$ & $7.650\times 10^{-1}$ & $7.204\times 10^{-1}$ \\ \hline
$5.62\times 10^{-2}-1.78\times 10^{-1}$ & $10-6005 $ & $4.550\times 10^{-1}$ & $ 4.923\times 10^{-1}$ & $2.665\times 10^{-1}$ & $2.354\times 10^{-1}$  \\ 
$5.62\times 10^{-2}-1.78\times 10^{-1}$ & $6005-12000 $ & $3.729\times 10^{-1}$  & $3.845\times 10^{-1}$  & $2.619\times 10^{-1}$ &  $2.124\times 10^{-1}$ \\ 
 $ 5.62\times 10^{-2}-1.78\times 10^{-1} $  & $12000-17995$ & $4.083\times 10^{-2} $& $3.847\times 10^{-2}$ & $2.117\times 10^{-2}$& $1.537\times 10^{-2}$   \\ 
     &    &          &        &         &    \\

\hline

\end{tabular}
\end{center}
\caption{As in table \ref{tabletwo} but for $x_{\gamma}$ bins.}
\label{tablethree}
\end{table} 


As shown in Section \ref{sec:results}, the elastic QEDCS cross section is 
described very accurately 
by the EPA. It is thus more interesting to investigate
the inelastic channel in this context.  
The elastic QEDCS events can be separated from the inelastic ones by
applying a cut on $\theta_h$. We have found that, with the restriction
$\theta_h \ge 0.1^\circ$, the elastic events are rejected and 
all the inelastic events are retained in the cross section. A lower limit on
$\theta_h $ higher than $1^\circ$ removes a substantial part (more than 
$30\%$) of the inelastic events.   

Table \ref{tabletwo} shows the exact inelastic QEDCS cross section  
in $x_l$ and $Q_l^2$ bins, subject to the cuts A. 
We have also shown the cross section in
the EPA with the same constraints (the last two cuts of set A are not relevant in this case). The discrepancy with the EPA is quite substantial. We have also
shown the results with the cuts B, both the exact and the one in terms of 
the EPA, in the
same table (the constraint $\hat s > Q^2$ is not relevant for the EPA). The
discrepancy between the exact and the EPA here is much less and on the
average it is $24 \%$. 

Table \ref{tablethree} is almost similar, the only 
difference is that
the bins are now in $x_\gamma$. With the cuts of set A, the discrepancy now
is on the average $44 \% $, whereas, with the cuts B, the average
discrepancy is $17 \%$.    
   
Our results  show that the extraction of the equivalent photon
distribution $\gamma^p(x,\mu^2)$ is very much dependent on the kinematical 
constraints utilized to single out QEDCS events, in particular on the 
one on acoplanarity. The kinematical limits  presented here are much more appropriate than
those suggested in \cite{thesis} 
for a reliable extraction of $\gamma^p(x,\mu^2)$.  It is also clear that 
this  discrepancy is  entirely due to  the inelastic channel.

\section{Summary}
\label{vcs:summ}
 
To summarize, in this chapter we have analyzed the QED Compton process,
relevant for the experimental determination of the equivalent photon
distribution of the proton $\gamma^p(x,\mu^2)$. We have also calculated 
the major
background process, namely virtual Compton scattering, assuming an effective
parametrization of the parton distributions of the proton, both in 
the elastic and inelastic channels.
The elastic VCS is suppressed compared to the QEDCS, in the phase space 
region accessible at HERA. We have shown that a constraint on the
invariants $\hat S \gtrsim \hat s $ is very effective in removing the 
inelastic VCS background. Furthermore, the
selection of the QEDCS events in the process $e p \rightarrow e \gamma X$
is sensitive to the specific kinematical limits, in particular to the upper
limit  of the acoplanarity angle $\phi$, which was used in the recent 
analysis \cite{lend,thesis} of events as observed with the HERA-H1 detector. 
Instead of the acoplanarity, one can
also directly impose cuts on the invariants, like $\hat s > Q^2$ (both of
them are measurable quantities), which directly restricts one to the
range of validity of the EPA. With these constraints, the total (elastic +
inelastic) cross section agrees with the EPA within $3 \%$. Thus, we conclude 
that by choosing the kinematical domain relevant for this approximation 
carefully, it is possible
to have a more accurate extraction of $\gamma^p(x,\mu^2)$.

\clearemptydoublepage

\chapter{{\bf The Polarized QED Compton Scattering Process}}
\label{cap:polarized}

This chapter, based on \cite{pol,pol2}, is devoted to
the study the QED Compton scattering process in
 ${\vec \ell} {\vec p}\rightarrow e \gamma p $ 
and ${\vec \ell} {\vec p}\rightarrow e \gamma X $, where the initial lepton
and proton are longitudinally polarized. We show that, when the virtuality 
of the exchanged photon is
not too large, the cross section  can be expressed in terms of
the polarized equivalent photon distribution of the proton. 
We provide the necessary kinematical cuts to extract
the polarized photon content of the proton at HERMES, COMPASS and  eRHIC 
(the future polarized $ep$ collider planned at BNL). In addition, we
show that such an experiment can also access the polarized structure
function $g_1(x_B, Q^2)$ at HERMES in the low $Q^2$ region and at  eRHIC
over a wide range of the Bjorken scaling variable $x_B$ and $Q^2$.

The structure function
$g_1(x_B, Q^2)$ and its first $x_B$  moment in the low $Q^2$ region have been 
studied in fully inclusive measurements at SLAC \cite{slac}, HERMES
\cite{her1,her2} and JLab \cite{clas1,clas2}. 
The most recent measurements by CLAS
\cite{clas3} are in the kinematical region $Q^2=0.15-1.64 ~~\mathrm{GeV}^2$. 
The low $Q^2$ region is of particular interest because contributions due to
nonperturbative dynamics dominate here and thus the transition from soft to
hard physics can be studied. In fact the measurements in \cite{clas3} clearly
indicate a dominant contribution from the resonances and at higher $Q^2$
they are below the perturbative QCD  evolved scaling value of $g_1$.
 This in fact
illustrates the necessity of further investigation of $g_1 (x_B, Q^2)$ in the
transition region. In these fixed target experiments, low $Q^2$ is
associated with low values of $x_B$, thus the covered kinematical region is
smaller compared to the unpolarized data. Data on $g_1(x_B, Q^2)$ for small $x_B$ and
in the scaling region are missing due to the absence of polarized 
colliders so far (with the exception of RHIC, 
which has started operating in the polarized mode for $pp$ collisions 
only very recently). The small $x_B$ region is again interesting; 
it is the region
of high parton densities, and measurements in this region will provide
information about the effects of large $[ \alpha_s \ln^2{1/ x_B}]^k$  
resummation and DGLAP evolution, and also about the "soft" to "hard" scale transition
\cite{badel,reya,bass}.
A better understanding of $g_1 (x_B, Q^2)$ in this region is necessary  
in order to determine its first moment experimentally. The kinematics of 
QED Compton events is 
different from the one of inclusive deep inelastic scattering 
due to the radiated photon in the final state and thus it provides a novel 
way to  access $g_1(x_B, Q^2)$ in a kinematical region not well covered by 
inclusive measurements, as already  stated for $F_2(x_B, Q^2)$ in
the preceding chapters.

In Sections \ref{sec:polelastic} and \ref{sec:polinelas} we derive 
the analytic expressions of the cross section for the polarized QED 
Compton process in the elastic and inelastic channels, respectively. In
Section \ref{sec:vcs} we discuss the background coming from virtual Compton
scattering  and also the interference between the two processes. The
numerical results are presented in Section \ref{sec:num_pol}. 
A short summary is
given in Section \ref{sec:polsummary}. The kinematics of the QED
Compton  process
is described in Appendix D. 
The analytic expressions of the matrix elements can be
found in Appendix E.        

\section{Elastic QED Compton Scattering}
\label{sec:polelastic}
We consider QED Compton scattering in the elastic process:
\be
{\vec e}(l)+{\vec p}(P) \rightarrow e(l')+\gamma(k')+ p(P'),
\ee
where the incident electron and proton  are longitudinally polarized and 
the four-momenta of the particles are given in brackets. 
Instead of the electron, one can also 
consider a  muon beam (COMPASS); the analytic expressions will be the same.  
We make use of the invariants \eqref{comptonmandel} and   \eqref{pap1_invar},
$k=P-P'$ is the four-momentum of the virtual photon. As for the unpolarized
reaction, discussed in Chapter 5, we neglect the
electron mass everywhere except when it is necessary to avoid divergences in
the formulae and take the proton to be massive, $ P^2=P'^2=m^2 $. The
relevant Feynman diagrams for this process are shown in Figure 
\ref{fig:one}, with $X$ being a proton and $P_X=P'$. 
The matrix element squared can be written as
\be
{\mid \Delta M^{{\rm QEDCS}}_{\mathrm{el}} \mid }^2={1\over t^2} 
H^{\alpha\beta {\rm A}}_{\mathrm{el}}(P,P') T^{\rm A}_{\alpha \beta}
(l;l',k'),
\ee
$H^{\alpha\beta {\rm A}}_{\mathrm{el}}(P,P')$  
and $T^{\rm A}_{\alpha \beta}(l;l',k')$ 
being the antisymmetric 
parts of the hadronic and leptonic tensors respectively, which
contribute to the polarized cross section. As before  we use the
notation (\ref{eq:lorentz})
for the Lorentz invariant $N$-particle phase-space element. 
The spin dependent counterpart of \eqref{eq:sigmaintel} reads
\be
\Delta\sigma_{\mathrm{el}}(s)={1\over {2 (s-m^2)}} \int {\der\hat s\over 2 \pi}
\,\der PS_2
(l+P;l'+k',P') {1\over t^2} H^{ \alpha \beta {\rm A}}_{\mathrm{el}} 
(P,P') X^{\rm A}_{\alpha\beta}(l,k)~.
\label{deltasigma}
\ee
The tensor $X^{\rm A}_{\alpha\beta}$  is antisymmetric in the
indices $\alpha$, $\beta$  and is defined as
\be
X^{\rm A}_{\alpha \beta}(l,k)=\int \der PS_2(l+k;l',k') 
T^{\rm A}_{\alpha \beta}
(l,k;l',k'),
\ee
where $T^A_{\alpha\beta}(l,k;l',k')$ is the antisymmetric part of 
the leptonic tensor \eqref{eq:leptonict},
\be
T^{\rm A}_{\alpha\beta}(l,k;l',k')= -{4  i e^4\over \hat s \hat u}\, 
\varepsilon_{\alpha
\beta \rho \sigma}  \Big [ (\hat s-t) l^\rho +(\hat u-t) l'^\rho
\Big ]\, k{^\sigma}.
\label{lep_pol}
\ee
$X^{\rm A}_{\alpha \beta}$ contains all the informations about the 
leptonic part of the process and can be expressed in terms of the  
Lorentz scalar $\Delta \tilde{X}$:
\be
X^{\rm A}_{\alpha \beta}=-{ i \over (\hat s -t)} 
\,\varepsilon_{\alpha \beta \rho \sigma}
k^\rho l^\sigma \,\Delta\tilde{X}(\hat s, t),
\label{xmunu}
\ee
with
\be
\Delta  \tilde{X}(\hat s, t)=- 2 X^{\rm A}_{\alpha \beta} P^{\alpha \beta {\rm A}},
\ee
$P^{\rm A}_{\alpha \beta}$ being the antisymmetric part of the
 photon polarization density matrix given in 
\eqref{eq:density}-\eqref{eq:dens2}.
We define the function $\Delta X (\hat s, t ,\hat t) $ as
\be
\Delta \tilde{X}(\hat s,t)= \int_{\hat t_{\mathrm{min}}}^
{\hat t_{\mathrm{max}}} \der \hat t \int_0^{2\pi}\der \varphi^*
\,  \Delta {X} (\hat s,t,\hat t);
\label{deltaX}
\ee
the integration limits of  $\hat t$ are the same as in   \eqref{thatlim} and
$\varphi^*$ is the azimuthal angle of the final electron-photon
system in the electron-photon center-of-mass frame.
Hence
$\Delta X(\hat s, t, \hat t)$ can be obtained from the relation
$\Delta X = -2\,T_{\alpha\beta}^{\rm A}P^{\alpha\beta {\rm A}}$, 
explicitly
\be
\Delta X(\hat s, t, \hat t)= {4 \alpha^2\over \hat s \hat u (\hat s-t)} \bigg [
{(\hat s -t)}^2+ {2 t \hat t (\hat u-t)\over \hat s-t} -  (\hat s +\hat t)^2
\bigg ].
\label{x2a}
\ee
The hadronic tensor for polarized scattering \eqref{poleltens}, i.e. 
\be
H_{\rm el}^{\alpha \beta {\rm A}}=-i e^2m \varepsilon^{\alpha \beta \rho \si}
 k_\rho \bigg [ 2 G_E G_M S_\si
-{G_M (G_M-G_E)\over  1+\tau} {k\cdot S\over m^2} P_\si \bigg ],
\label{polelten2}
\ee
is expressed  in terms of the  electric and magnetic form factors of the proton
$G_E$ and $G_M$; moreover $ \tau= {-t/ 4m^2}$ and $S$ is the spin
vector of the proton, which fulfils $S\cdot P = 0$ and $S^2 = -1$. 
If we express $S$ as
\begin{equation}
S_{\alpha} = N_S \bigg (P_{\alpha} - \frac{m^2}{l\cdot P}\,l_{\alpha} \bigg ),
\end{equation}
with $N_S$ given in \eqref{eq:normals} and \eqref{eq:normals2}, then
from \eqref{deltasigma}, \eqref{xmunu} and \eqref{deltaX}-\eqref{polelten2} 
we have
\be
\Delta \si_{\mathrm{el}}(s) &=& {\alpha\over 8\pi  (s-m^2)^2} \int_{m_e^2}^{(
\sqrt{S}-m)^2} \der\hat s \int_{t_{\mathrm{min}}}^{t_\mathrm{max}} {\der t\over
t} \int_{\hat t_{\mathrm{min}}}^{\hat t_{\mathrm{max}}} \der\hat t \,
\int_0^{2\pi}\der\varphi^* \,\Delta X(\hat s, t, \hat t) \nonumber\\&&~~~~~~~
\times\bigg [ \bigg ( 2 \,
{ s-m^2 \over \hat s-t }-1+{2 m^2\over t} {\hat s - t \over
 s-m^2} \bigg ) G_M^2(t)\nonumber\\&&~~~~~~~~~~
-2\bigg ({ s-m^2 \over \hat s-t }-1+
{m^2\over t} {\hat s-t\over s-m^2} \bigg ) {G_M (G_M-G_E)\over 1+\tau}
\bigg ],
\label{elsigg}
\ee   
the integration bounds being the same as in  \eqref{thatlim}  and 
\eqref{eq:tlimits}.  
These bounds are modified due to the experimental cuts which we
impose numerically. In the EPA limit, we neglect $\mid t\mid $ vs.
$\hat s$ and $m^2$ vs. $s$ and get
\be
\Delta X(\hat s,t,\hat t)\approx \Delta X (\hat s,0,\hat t)~=~
{4 \alpha^2\over \hat
s} \bigg ({\hat s\over \hat u}-{\hat u\over \hat s} \bigg )~=~ -{2 \hat s\over
\pi} {\der \Delta \hat \si \over \der \hat t} ,
\ee 
where ${\der \Delta \hat \sigma / \der\hat t} $ is the  polarized differential 
cross section for the real 
photoproduction process $e{\gamma}\to e\gamma$ \eqref{eq:polcompt}
and $ \Delta \tilde{X}(\hat s,0)=-4 \hat s \Delta \hat \si$. 
The elastic cross section then becomes
\be
\Delta\si_{\mathrm{el}} \approx \Delta\si_{\mathrm{el}}^{\mathrm{EPA}} = \int_{x_{\rm min}}^{(1-{m\over \sqrt S})^2} \der x
\int_{m_e^2-\hat s}^0 \,\der \hat t\, \Delta \gamma^p_{\mathrm{el}}(x) 
{\der\Delta \hat \si(xs,
\hat t) \over
\der \hat t}
\ee
where $m_e$ is the mass of the electron and 
$\Delta \gamma^p_{\mathrm{el}}(x)$ is the elastic contribution to the 
polarized equivalent photon distribution of the proton \eqref{eq17},
\be
\Delta \gamma^p_{\mathrm{el}}(x) = -{\alpha\over 2 \pi } \int_{t_{\mathrm{min}}}^{t_{\mathrm{max}}} {\der t\over t} \bigg [ \bigg (2-x
+{2 m^2 x^2\over t} \bigg ) G_M^2-2 \bigg ( 1-x+ {m^2 x^2\over t}\bigg ) {G_M
(G_M-G_E)\over 1+\tau} \bigg ]\nonumber \\
\ee
with $x={\hat s/ s}$, $t_{\rm min} = -\infty$ and 
$t_{\rm max} = -m^2x^2/(1-x)$.
\section{Inelastic QED Compton Scattering}
\label{sec:polinelas}
We next consider the corresponding inelastic process
\be
\vec {e}(l)+\vec {p}(P) \rightarrow e(l')+\gamma(k')+X(P_{X}).
\label{eq2}
\ee
We take  the invariant mass of the produced hadronic system
to be $W$; moreover  $Q^2=-k^2=-t$ and  the Bjorken variable $x_B$ 
is related to them via \eqref{eq:bjorkenx}. 
The cross section for inelastic scattering reads
\be
\Delta \sigma_{\mathrm{inel}}(s)= {1\over 32 \pi^3 (s-m^2)^2} 
\int_{W^2_{\mathrm{min}}}^{W^2_{\mathrm{max}}} \der W^2 
\int_{m_e^2}^{(\sqrt{S}-W)^2} \der 
\hat s \int_{Q^2_{\mathrm{min}}}^{Q^2_{\mathrm{max}}}
 {\der Q^2\over Q^4} \,W^{\alpha \beta {\rm A}} X_{\alpha \beta}^{\rm A},
\ee
where $X^A_{\alpha \beta}$ is given by   (\ref{xmunu}) 
and $W^{\rm A}_{\alpha \beta}$ is the hadronic tensor, see 
\eqref{eq:polhadrtens} together with  \eqref{eq:g1} and \eqref{eq:g2},
\be
W^A_{\alpha \beta}= 2\pi i e^2 { m\over P \cdot k} \,
\varepsilon_{\alpha \beta
\rho \sigma} k^\rho \bigg [ g_1(x_B, Q^2) S_\sigma +g_2 (x_B, Q^2) 
\bigg (S_\sigma-{k \cdot S\over k \cdot P} P_{\sigma}  \bigg ) \bigg ].
\ee
Hence the cross section takes the form
\be
\Delta \sigma_{\mathrm{inel}}(s) &=& {\alpha\over 4 \pi (s-m^2)^2} \int_{W^2_{\mathrm{min}}}^{W^2_{\mathrm{max}}} \der W^2 \int_{m_e^2}^{(\sqrt{S}-m)^2} \der \hat s \int_{Q^2_{\mathrm{min}}}^{Q^2_{\mathrm{max}}}
{\der Q^2\over Q^2} {1\over (W^2+Q^2-m^2)} \nonumber \\ &&\,\times\,\,\bigg \{ \bigg [-2  {s-m^2\over
\hat s+Q^2}  +{W^2+Q^2-m^2\over Q^2}+{2 m^2 \over Q^2} \bigg ({\hat
s+Q^2\over s-m^2} \bigg ) \bigg ] g_1 (x_B, Q^2) \nonumber\\&&~~~~~+\,\,{4 m^2\over W^2+Q^2-m^2} \,g_2 (x_B, Q^2) \bigg \}
\Delta \tilde{X}(\hat s, Q^2),     
\ee
and $\Delta \tilde{X} (\hat s, Q^2)$ is given by \eqref{deltaX}
 and (\ref{x2a}).
The limits of the $Q^2, W^2$ and $\hat t$   integrations  are given in
 \eqref{eq:q2limit}, 
\eqref{eq:w2limit} and \eqref{thatlim} respectively. 
In the region of validity of the EPA,  $s \gg m^2 $ and 
$\hat s \gg Q^2 $, the cross section  becomes
\be
\Delta \sigma_{\mathrm{inel}} \simeq \Delta \sigma_{\mathrm
{inel}}^{\mathrm{EPA}} =
\int_{x_{\mathrm{min}}}^{(1-m/\sqrt S)^2} \der x \, \int_{m_e^2 -\hat s}^0   
\der \hat t~ \Delta \gamma^p_{\mathrm{inel}}(x, x s) \,
\frac{ \Delta \der\hat\sigma(x s, \hat t)}{\der\hat t}~,
\label{epain}
\ee
where again $x={\hat s/s}$  and $\Delta \gamma^p_{\mathrm{inel}}(x, x s)$ 
is the inelastic contribution to the polarized e\-qui\-va\-lent photon 
distribution of the proton:
\be
\Delta \gamma^p_{\mathrm{inel}}(x, x s) = {\alpha\over 2 \pi} \int_x^1 {\der y
\over y} \int_{Q^2_{\mathrm{min}}}^{Q^2_{\mathrm{max}}} {\der Q^2\over Q^2} 
\bigg ( 2-y-{2m^2 x^2\over Q^2} \bigg ) 2 g_1 \bigg ({x \over y}, Q^2 \bigg ),
\label{gamin}
\ee
where we  take the scale $Q^2_{\rm max}$ to be $\hat s$
and $Q^2_{\rm min} = x^2m^2/(1-x)$. Here we
have neglected the contribution from $g_2(x_B, Q^2)$.
 Expressing $g_1(x_B, Q^2)$ in terms of the
polarized quark and antiquark  distributions, one can confirm that the above
expression reduces to that given in \eqref{eq12}. However, in this case,
one chooses the minimal (but not compelling) boundary condition $\Delta
\gamma^p_{\rm inel} (x, Q_0^2) =0$ at a scale $Q_0^2=0.26$ $\mathrm {GeV}^2$.
The expression   (\ref{gamin}) is free from
this particular boundary condition. 
\section{Background from Virtual Compton Scattering}
\label{sec:vcs}

The  processes  $ep\to e\gamma p$ and 
$ep\to e\gamma X$ receive contributions from the
virtual Compton Scattering (VCS), when the photon is emitted from the proton
side  as well as from the interference between the QED Compton
scattering (QEDCS) and VCS, see Figures \ref{fig:qedcs} and \ref{fig:vcs}. 
The polarized cross section for the elastic process, analogously to
\eqref{elsig_4}, reads
\be
\Delta \sigma_{\mathrm{el}}(s)={1\over {2(4\pi)^4  (s-m^2)^2}} 
\int_{m_e^2}^{(\sqrt{S}-m)^2} \der\hat s \int_{{t}_{\mathrm{min}}}^{t_{\mathrm{max}}}
\der t \int_{\hat t_{\mathrm{min}}}^{\hat t_{\mathrm{max}}} \der\hat t 
 \int_0^{2\pi} \der\varphi^* {1\over {(\hat s-t)}}
{{\mid {\Delta M_{\mathrm{el}}}\mid }^2},\nonumber \\
\label{elsig}
\ee
where 
\be
{{\mid {\Delta M_{\mathrm{el}}}\mid }^2}= {{\mid
{\Delta M^{\rm {QEDCS}}_{\mathrm{el}}}\mid
}^2}+{{\mid {\Delta M^{{\rm VCS}}_{\mathrm{el}}}\mid }^2} - 2\, 
{\Re{\bf{\it e}}}\,
 {\Delta M^{{\rm QEDCS}}_{\mathrm{el}}
\Delta M^{{\rm VCS} *}_{\mathrm{el}}}
\label{eq:polelastic_vcs}
\ee
is the matrix element squared of the process. 
The integration bounds are the same as in   (\ref{elsigg}). The 
interference term will have opposite sign if we consider a positron 
instead of an electron. 
The cross section of the VCS process is expressed in terms of generalized
parton distributions and one needs a realistic model for a quantitative
estimate of this background \cite{gpd}. Here, as in the preceding chapter,
in order to find the
cuts to suppress the VCS,
we make a simplified assumption: we take the proton to be a massive pointlike
 particle with an
effective $ \gamma^* p$ vertex, $ -i \gamma^\alpha F_1(t) $. The explicit
expressions for the matrix elements are given in Appendix E.  

Particularly interesting for our purpose of extracting the polarized photon 
distribution of the proton is the inelastic channel. 
Here we use a unified parton model  
to estimate the VCS and QEDCS rates, similar to  \eqref{eq:parton}.
The cross section within the parton model is given by
\be
\frac{\der \Delta \sigma_{\mathrm{inel}}}{\der x_B \,\der Q^2 \,\der \hat s\, 
\der \hat t \,\der\varphi^*}=\sum_{q} \, \Delta q(x_B,Q^2)\, \frac{\der \Delta \hat
\sigma^q}{\der \hat s\, \der Q^2 \,\der \hat t\, \der \varphi^*}~,
\ee
where $\Delta q(x_B,Q^2)$  are the  polarized quark and antiquark 
distributions of the initial
proton, $q = u,\, d, \,s, \,\bar{u}, \,\bar{d}, \,\bar{s}$ and 
$\der \Delta \hat \sigma^q$ is
the differential cross section of the subprocess
\be
{\vec e}(l)+{\vec q(p)} \rightarrow e(l')+\gamma(k')+q(p'),
\label{eqsub}
\ee
${\vec q}$ being a longitudinally polarized quark in a longitudinally
polarized proton and $q$  a quark in the final state.
The integrated cross section reads
\newpage
\be
\Delta \sigma_{\mathrm{inel}}(s)&\!\!\!\!=&\!\!\! \!
{1\over {2(4 \pi)^4 (s-m^2)^2}} \sum_q
\int_{W^2_{\mathrm{min}}}^{W^2_{\mathrm{max}}}
\der W^2 \int_{m_e^2}^{(\sqrt{S}-W)^2} \!\!\!\der\hat s
\int_{Q^2_{\mathrm{min}}}^{Q^2_{\mathrm{max}}}
 {\der Q^2 \over Q^2}
\int_{{\hat t}_{\mathrm{min}}}^{{\hat t}_{\mathrm{max}}}\!\!\!
\der\hat t \int_0^{2\pi}\!\!\! \der\varphi^* 
\nonumber\\&&~~~~~~~~~~~~\times ~\frac{1} {(\hat s+Q^2)}\, {{\mid {\Delta \hat M^q_{\mathrm{inel}}}\mid }^2} \Delta q(x_B,Q^2),
\label{insig}
\ee
where
\be 
{{\mid {\Delta \hat M^q_{\mathrm{inel}}}\mid }^2}= {{\mid
{\Delta \hat M^{q\,\rm {QEDCS}}_{\mathrm{inel}}}\mid
}^2}
+{{\mid {\Delta \hat M^{q\,{\rm VCS}}_{\mathrm{inel}}}\mid }^2} -2\, 
{\Re{\it e}} 
{\Delta \hat M^{q\,{\rm QEDCS}}_{\mathrm{inel}}
\Delta \hat M^{q\,\rm{VCS} *}_{\mathrm{inel}}}
\ee 
and the limits of integrations are given in (\ref{thatlim}), 
(\ref{eq:q2limit}) and (\ref{eq:w2limit}). 
The explicit expressions of the matrix elements are given in  Appendix E.
As for the study of the unpolarized process, it is useful to introduce 
the invariants $\hat S=(p'+k')^2$
and $\hat U= (p'-k)^2$, which can be written in terms of measurable 
quantities and satisfy the relation (\ref{eq:auxiliary}).


\begin{table}[t]
\begin{center}
\begin{tabular}{|c|c|c|}
\hline 
 HERMES  & COMPASS  & eRHIC   \\
\hline\hline 
& & $~~E_{p} = 250\, \mathrm{GeV}~~$ \\
$~~E_e = 27.5 \, \mathrm{GeV}~~ $ & $~~E_{\mu} =  160 \,\mathrm{GeV}~~ $ & $E_{e} = 10\, \mathrm{GeV}~~$ \\  $~~~ 0.04 < \theta_e, \, \theta_{\gamma} <  0.2~~~$ &$~~~ 0.04 < \theta_{\mu}, \, \theta_{\gamma} <  0.18~~~$    &
$~ 0.06 < \theta_e, \, \theta_{\gamma} < \pi - 0.06~$ \\
$~~E_e', \, E_{\gamma}' > 4 \, \mathrm{GeV}~~ $ & $~~E_{\mu}', \, E_{\gamma}' > 4 \,\mathrm{GeV}~~ $ &$~~E_e', \, E_{\gamma}' > 4 \, \mathrm{GeV}~~ $  \\
 $ \hat s > 1\,\mathrm{GeV}^2  $  & $ \hat s > 1\,\mathrm{GeV}^2  $ & $ \hat s > 1\,\mathrm{GeV}^2 $  \\
 $ \hat s > Q^2 $  & $ \hat s > Q^2 $ & $\hat s > Q^2 $  \\

\hline

\end{tabular}
\end{center}
\caption{Energies, angular acceptance and kinematical cuts for 
the HERMES, COMPASS and eRHIC experiments.}
\label{tableone}
\end{table}
\section{Numerical Results}
\label{sec:num_pol}

In this section we present our numerical results. The cuts used for
the  HERMES, COMPASS and eRHIC kinematics are given in Table \ref{tableone}. 
As for the spin independent process, the constraints on the energies and
polar angles of the detected particles reduce the background contributions
coming from the  radiative emissions (when the final state photon is emitted
along the incident or the final lepton line), because they prevent the
lepton propagators to become too small. The QED Compton 
events are singled out at HERA
by imposing a maximum limit on the acoplanarity angle $\phi$ 
defined in \eqref{eq:acopla}. We have observed in the preceding 
chapter that, 
instead of this limit on $\phi$, the
constraint $\hat s > Q^2$, which is applicable experimentally,  is more 
efficient in extracting the equivalent
photon distribution from the exact result. Here we use this constraint.

The unpolarized cross section has been calculated using the formulae in
Chapter 4; for the numerical estimates we have used the 
ALLM97 parametrization \cite{allm97} of the structure function $F_2(x_B, Q^2)$.
We have taken $F_L(x_B, Q^2)$ to be zero, assuming  the Callan-Gross relation.
In the polarized cross section, we have neglected the
contribution from $g_2(x_B, Q^2)$ and used the parametrization 
 \cite{bad} for
$g_1(x_B, Q^2)$. In this parametrization,  $g_1(x_B, Q^2)$ is described 
in the low-$Q^2$ region  by the {\it generalized vector meson dominance} (GVMD)
model together with the Drell-Hearn-Gerasimov-Hosoda-Yamamoto sum rule and,
for large $Q^2$, $g_1(x_B, Q^2)$ is expressed in terms of the NLO GRSV01 
\cite{grsv} parton
distributions (standard  scenario) in terms of a suitably defined scaling
variable 
\be
{\bar x}=\frac{Q^2+Q_0^2}{ Q^2+Q_0^2+W^2-m^2}
\ee
with $Q_0^2=1.2$ $\mathrm{GeV}^2$. The scale $Q^2$ is changed to 
$Q^2+Q_0^2$, so as to extrapolate to the 
low-$Q^2$ region. It is to be noted that for QED Compton scattering, the
effects of $F_L(x_B, Q^2)$ and $g_2 (x_B, Q^2)$ have to be taken into account 
in a more accurate
study as their effect may become non-negligible in the low-$Q^2$ region.
However, this is beyond the scope of the present work.
\begin{figure}[ht]
\begin{center}
\epsfig{figure=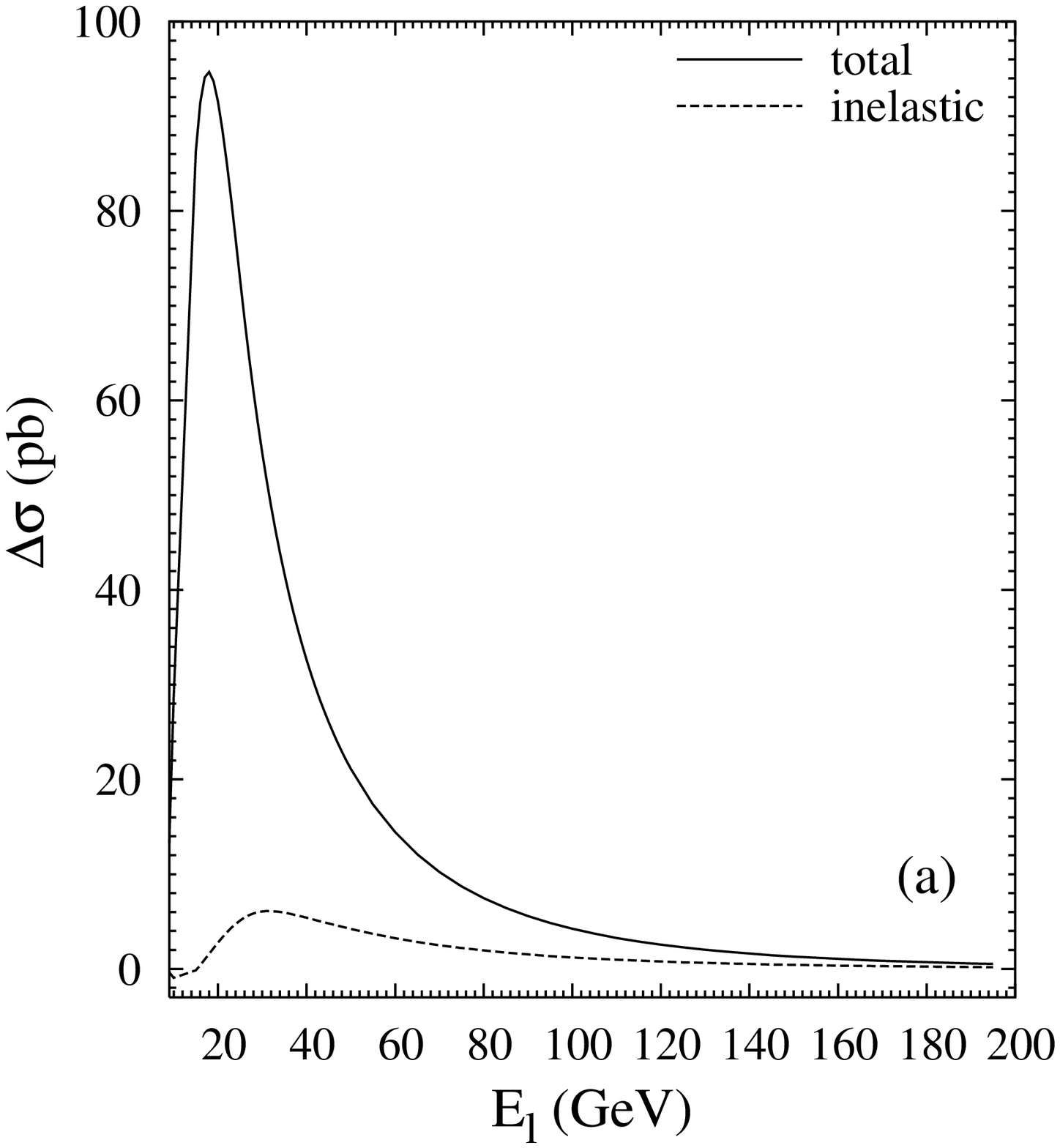,width=7.9 cm}
\epsfig{figure=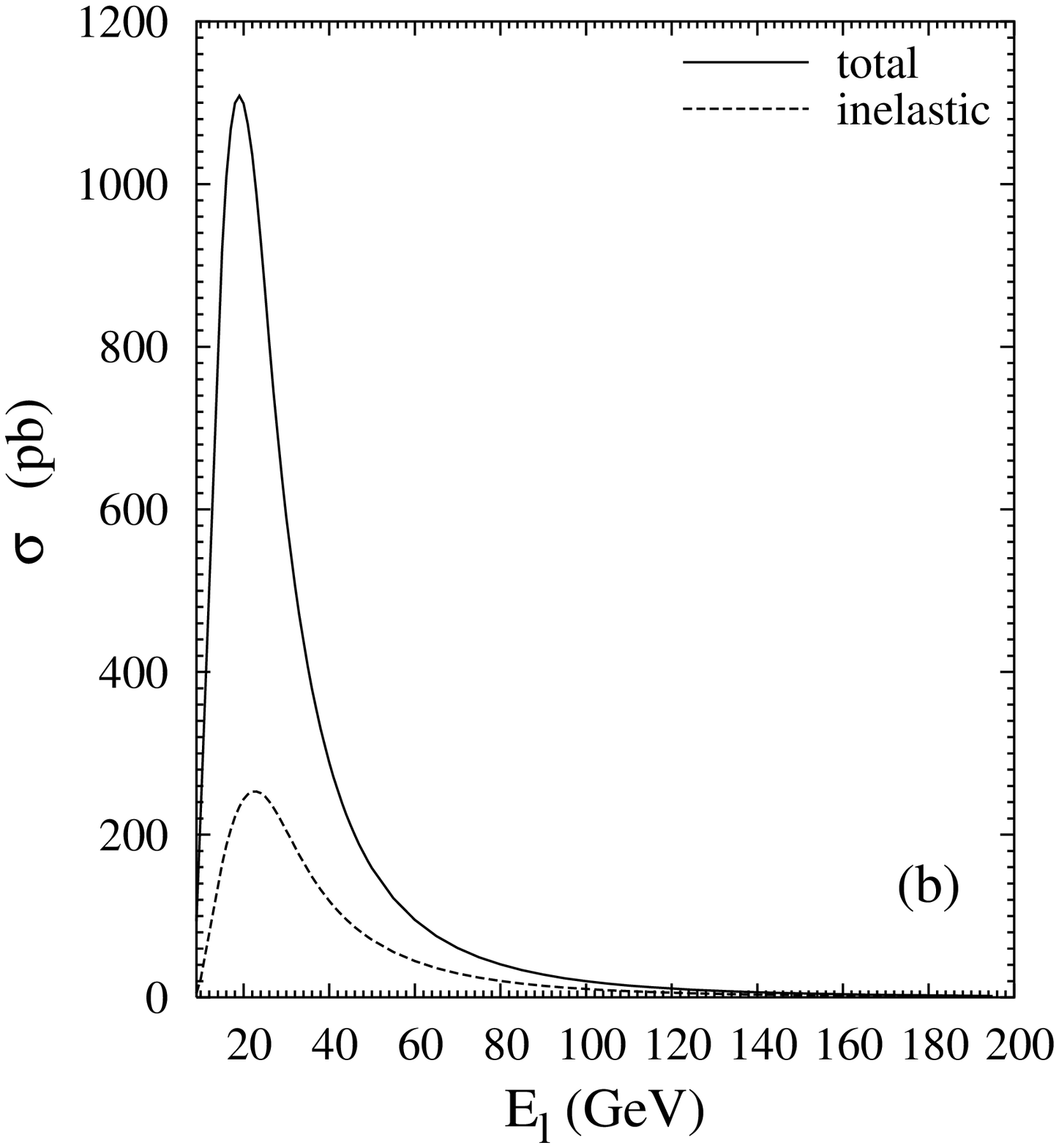,width=7.9cm}
\end{center}
\caption{QEDCS cross section vs. energy of the incident lepton; (a)
polarized, (b) unpolarized. The continuous line is the total cross section and 
the dashed line is the cross section in the inelastic channel. The cuts imposed
are given in the central column  of Table \ref{tableone}. 
We have used the ALLM97 parametrization of
$F_2$  and the Badelek {\it et al.} parametrization of $g_1$.}
\label{fig:3_pol}
\end{figure}
\begin{figure}
\begin{center}
\epsfig{figure=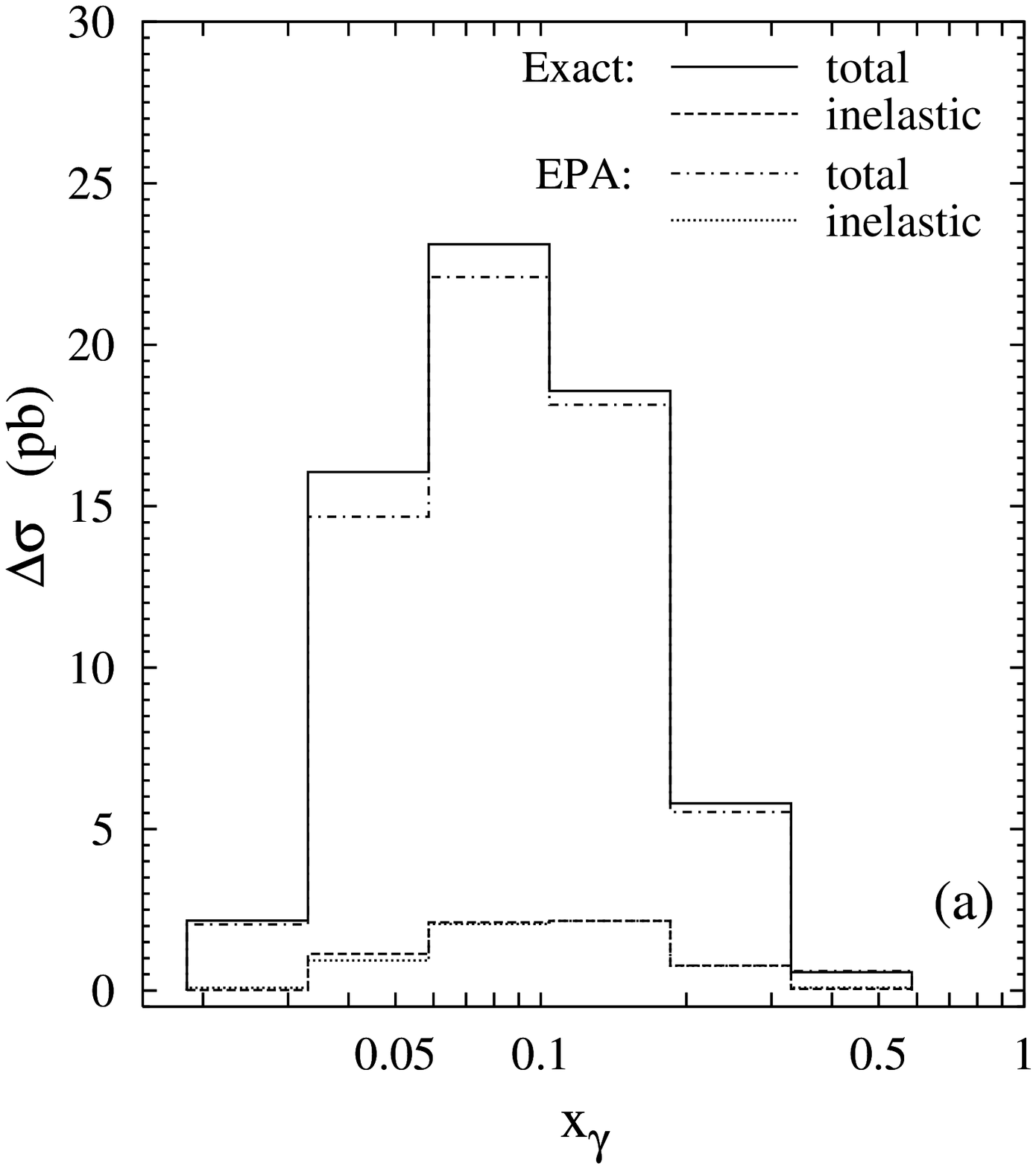,width=7 cm}
\epsfig{figure=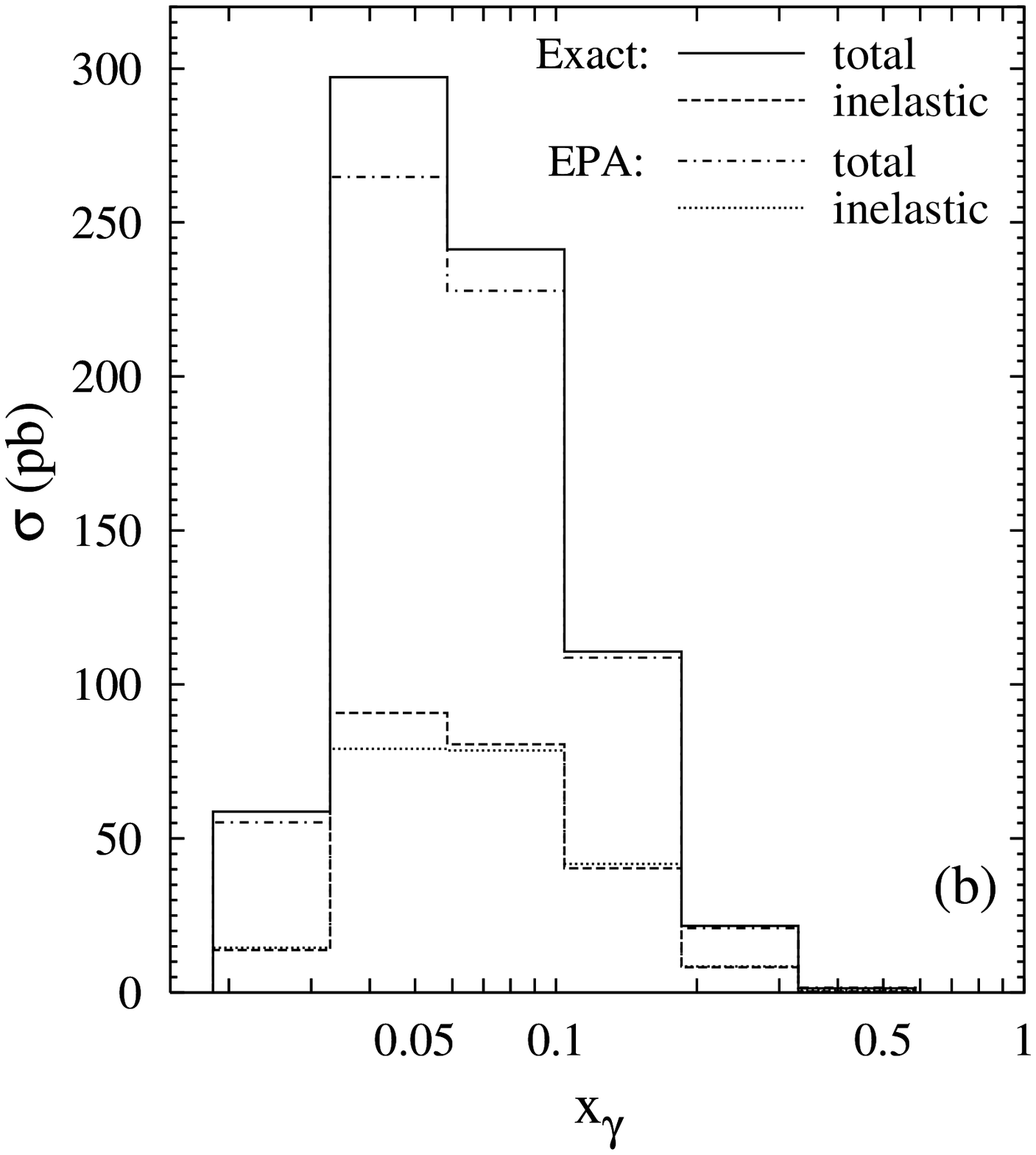,width=7 cm}

\vspace*{1cm}
\epsfig{figure=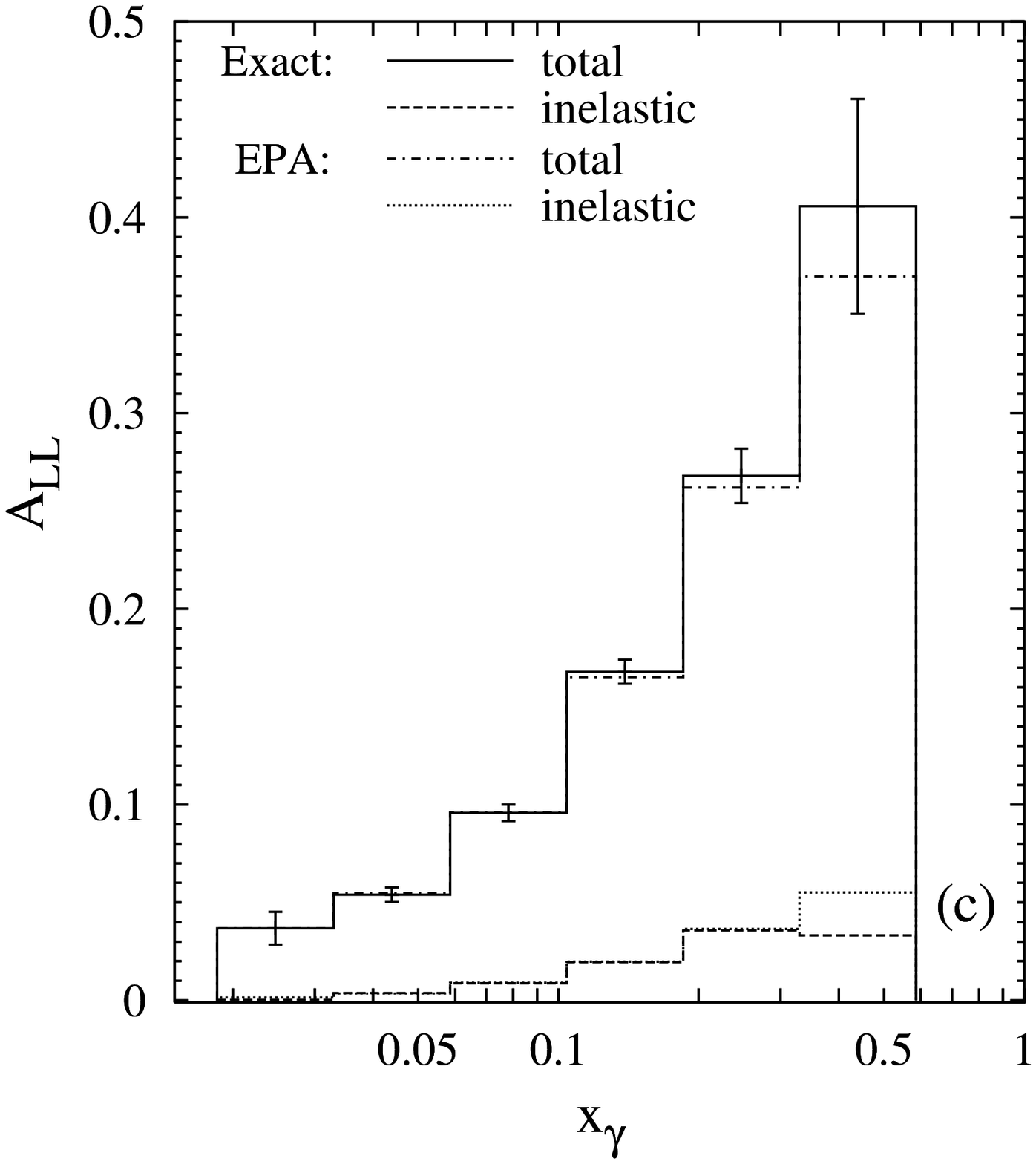,width=7 cm}
\epsfig{figure=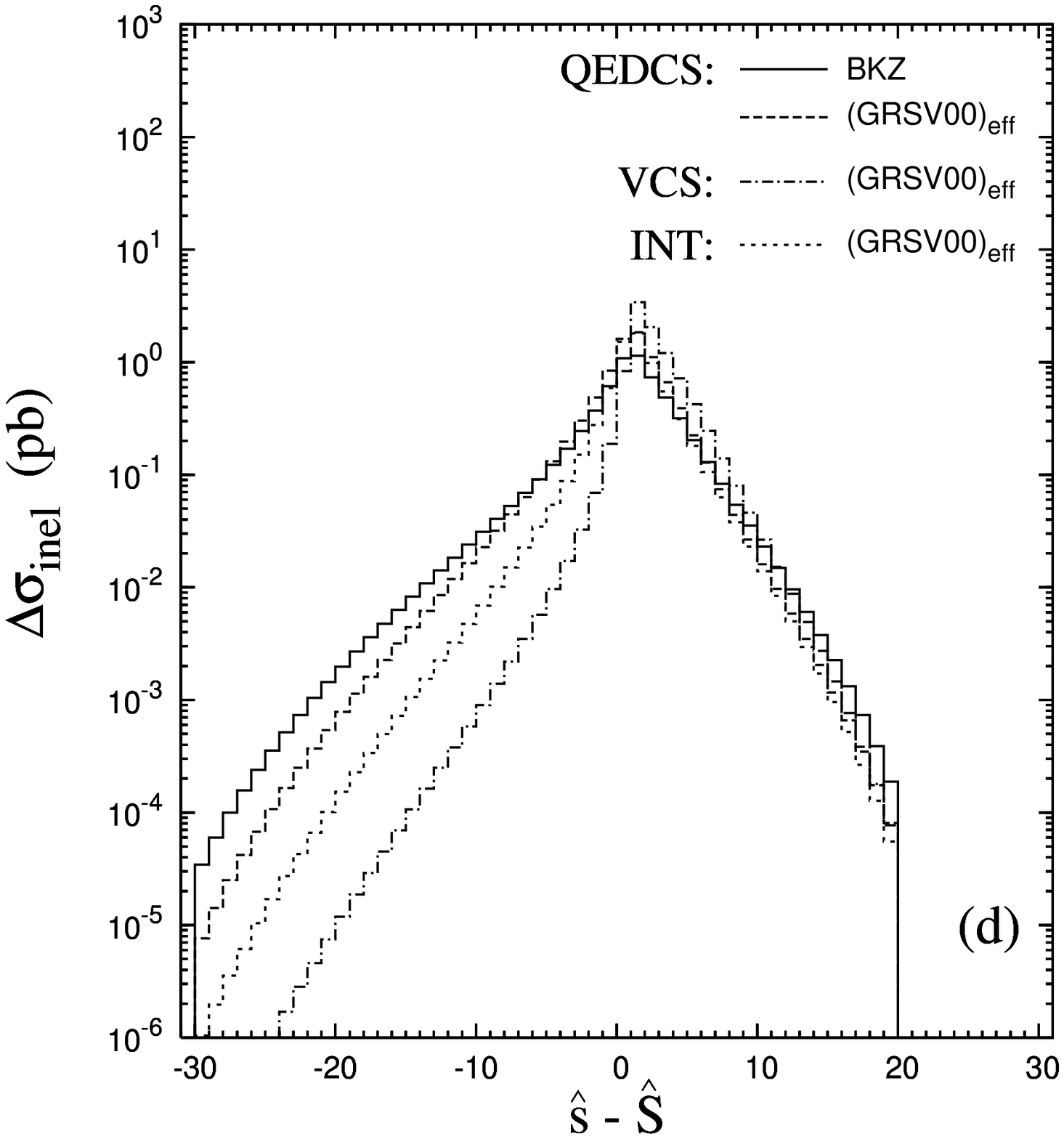,width=7 cm}
\end{center}   
\caption{Cross section for QED Compton scattering (QEDCS) at
HERMES in bins of $x_\gamma$ (a) polarized, (b) unpolarized, 
(c) the asymmetry; for the polarized cross section the Badelek {\it et al.} 
 parametrization of $g_1$ (BKZ) and for the unpolarized cross section the 
ALLM97 parametrization of $F_2$ have been used; (d) polarized inelastic cross 
section 
for QEDCS (long dashed), VCS
(dashed-dotted) and the interference (dashed)  at 
HERMES in the effective parton model.
The bins are in $\hat s- \hat S$, expressed in GeV$^2$. The
continuous  line is the QEDCS cross section using 
the BKZ parametrization of $g_1(x_B, Q^2)$. The constraints
imposed are given in Table \ref{tableone}.}
\label{fig:4_pol}
\end{figure}

Before discussing the results for specific experiments, it is interesting
to investigate some general properties of the total cross section. 
Figures \ref{fig:3_pol}
(a) and (b) show the total QEDCS cross sections, polarized and unpolarized
respectively, as functions of the incident lepton energy $E_l$. We have
imposed the constraints in the second column of Table \ref{tableone} on 
the energies and
angles of the outgoing particles, as well as those on $\hat s$. Both
polarized and unpolarized cross sections increase sharply with $E_l$, reach
a peak at around $E_l=20~\mathrm{GeV}$ and then start to decrease. The cross
sections in the inelastic channels are also shown, which have  similar trends 
except that the peak in the polarized case is broader.  

\subsection{HERMES}
Figures  \ref{fig:4_pol} (a) and (b) show the total (elastic+inelastic) 
polarized and unpolarized QED Compton scattering
cross sections respectively,  in bins of the variable $x_\gamma$,
defined in \eqref{eq:icsgamma}, 
for HERMES kinematics, subject to the
cuts of Table \ref{tableone}. We have taken the incident electron energy
to be $E_e=27.5$
$\mathrm{GeV}$. We also show the cross section calculated in the EPA. 
Furthermore the contributions due to the inelastic channel of the reaction  
are plotted.
The cross section, integrated over $x_\gamma$, agrees with the EPA within $7.1\%$ (unpolarized) and
$4.8 \%$ (polarized). From the figures
it is also clear that the agreement in the inelastic channel
($2.5 \%$ in the polarized case) is much better than for HERA kinematics
discussed in Chapter 5. 
This is because at HERMES $Q^2$
can never become too large (maximum $13.7$  $\mathrm{GeV}^2$), subject to our
kinematical cuts, which is expected in a fixed target experiment. 
The agreement is not so good without the constraint $\hat s >1$
 $\mathrm{GeV^2}$.  Figure  \ref{fig:4_pol} (c) shows the asymmetry, 
which is defined as
\be
A_{LL}={\sigma_{++}-\sigma_{+-}\over {\sigma_{++}+\sigma_{+-}}}
\ee
where $+$
and $-$ denote the helicities of the incoming electron and proton.  
They are calculated with the
same set of constraints. The asymmetry is quite sizable at HERMES and
increases in higher $x_\gamma$ bins. The asymmetry in the EPA is also shown.
It is interesting to note that the discrepancy between the exact cross
section and the one   evaluated in the EPA,
evident in Figures \ref{fig:4_pol} (a) and (b), actually gets canceled in the 
asymmetry; as a consequence Figure \ref{fig:4_pol} (c) shows an excellent 
agreement, 
except in the last bin, between exact and  approximated results.
We have also calculated the expected statistical errors for each bin, using 
the following formula, valid when the asymmetry is
 not too large \cite{wernerv},
\be
\delta A_{LL} \simeq {1\over \mathcal {P}_e \mathcal{P}_p  \sqrt {\mathcal {L}
\sigma_{\mathrm{bin}}}}~,
\label{err}
\ee     
where $\mathcal {P}_e$ and $\mathcal {P}_p$ are the polarizations of the
incident lepton and proton, respectively, $\mathcal {L}$ is the integrated
luminosity and $\sigma_{\mathrm{bin}}$ is the unpolarized cross
section in the corresponding $x_\gamma$ bin. We have taken $\mathcal {P}_e=
\mathcal {P}_p=0.7$ and $\mathcal {L}={1 ~{\rm fb}^{-1}}$ for HERMES. 
The expected
statistical errors increase in higher $x_\gamma$ bins, because the number of
events become smaller. However the asymmetries seem to be measurable 
at HERMES.                

The background
from virtual Compton scattering is reduced at HERA by the experimental
condition of no observable hadronic activity at the detectors. Basically the
electron and photon are detected in the backward detectors and the hadronic
system in the forward detectors. In the previous chapter, we have observed that
for unpolarized scattering at HERA, such a constraint is insufficient to
remove the VCS contribution for higher $x_\gamma$. We have proposed a new
constraint $\hat S \ge \hat s  $, where $\hat S$ and $\hat s$ can be measured experimentally,  
to be imposed on the
cross section. Here, we investigate the effect of this constraint on the polarized cross section.  To estimate the inelastic contribution
coming from VCS, we use   (\ref{insig}), together with an effective model
for the parton distribution of the proton. The effective parton distributions
are of the form 
\be
\Delta \tilde q(x_B, Q^2)= \Delta q({\bar x}, Q^2+Q_0^2),
\ee
$\Delta q (x_B, Q^2)$ being the NLO GRSV01 (standard scenario)  distribution 
functions \cite{grsv}. In the relevant  kinematical 
region, $Q^2$ can be very small and may become close to zero, where the
parton picture is not applicable. The parameter $Q_0^2=2.3\,\,\mathrm{GeV}^2$ 
prevents the scale of the parton
distribution to become too small, while ${\bar x}$ is a suitably 
defined scaling variable, 
\be
{\bar x}= {x_B (Q^2+Q_0^2)\over Q^2+x_B Q_0^2}~. 
\ee     
To estimate the unpolarized background
effect, we use the same expressions as in Chapter 6 with the
effective parton distributions given in  \eqref{fq}. 
Figure  \ref{fig:4_pol} (d) shows the polarized cross section 
in the inelastic
channel at HERMES, subject to the constraints of Table \ref{tableone}, in 
bins of $\hat
s-\hat S$ calculated in the effective parton model. The VCS and the
interference contributions are also shown. The QEDCS cross section using the
Badelek {\it et. al} parametrization of $g_1(x_B, Q^2)$ is also plotted. 
In fact, the
cross section in the effective parton model lies close to this. Within the 
parton model, the
VCS is suppressed when $\hat s < \hat S$, similar to the unpolarized case at
HERA. Unlike HERA, the interference between QEDCS and VCS is not
negligible at HERMES, although smaller than the QEDCS in the relevant region.
Since the interference term changes  sign when a  positron beam is used instead
of an electron beam, a combination of electron and positron scattering data can
eliminate this contribution. In order to estimate the VCS in the elastic
channel, one needs a suitable model for the polarized generalized parton
distributions. However, in the simplified approximation of a pointlike
proton  with an
effective vertex as described in Section \ref{sec:vcs}, the elastic VCS 
as well as the interference contribution is
much suppressed at HERMES. Similar observations hold for unpolarized
scattering.

\begin{figure}[t] 
\begin{center}
\epsfig{figure= 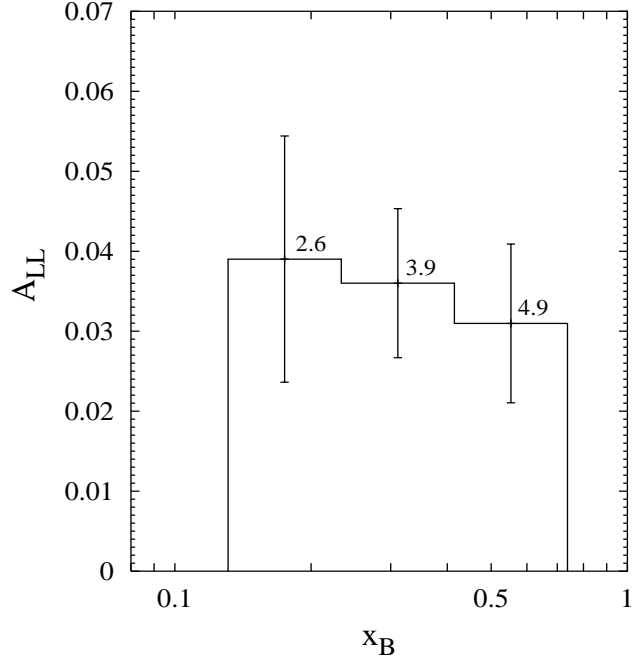,width=9cm, height = 8.5cm}
\end{center}
\caption{Asymmetry in the inelastic channel in bins of $x_B$ at HERMES. We
have used the Badelek {\it et al.} parametrization of $g_1$. The
constraints imposed are as in Table \ref{tableone} (except $\hat s > Q^2$), 
together with 
$\hat S - \hat s > 2$ {GeV}$^2$. The average $Q^2$ 
(in {GeV}$^2$) of  each bin is also shown.}  
\label{fig:5_pol}
\end{figure}

Figure  \ref{fig:5_pol} shows the asymmetries in the inelastic 
channel in bins of $x_B$.
In addition to the cuts mentioned above and shown in table \ref{tableone}, 
we have also chosen $\hat S -\hat s
> 2~~\mathrm{GeV}^2$ to suppress the background. The asymmetry is small but sizable and could
be a tool to access $g_1(x_B, Q^2)$ at HERMES. In fact, QED 
Compton events can be
observed at HERMES in the kinematical region $x_B=0.02-0.7$ and
$Q^2=0.007-7~ \mathrm{GeV}^2$ (small $Q^2$, medium $x_B$).  
However, from the figure 
it is seen that the asymmetry
is very small for $x_B$ below $0.1$. We have also shown the expected
statistical error in each bin. The average $Q^2$ value in $\mathrm{GeV}^2$ for the
polarized cross section for each
bin is shown, which has been calculated using the formula
\be
\langle Q^2\rangle={\int_{\mathrm{bin}} Q^2\, \der \Delta \sigma \over \int_{\mathrm{bin}} \der \Delta \sigma}~.
\ee       

\begin{figure}     
\begin{center}
\epsfig{figure=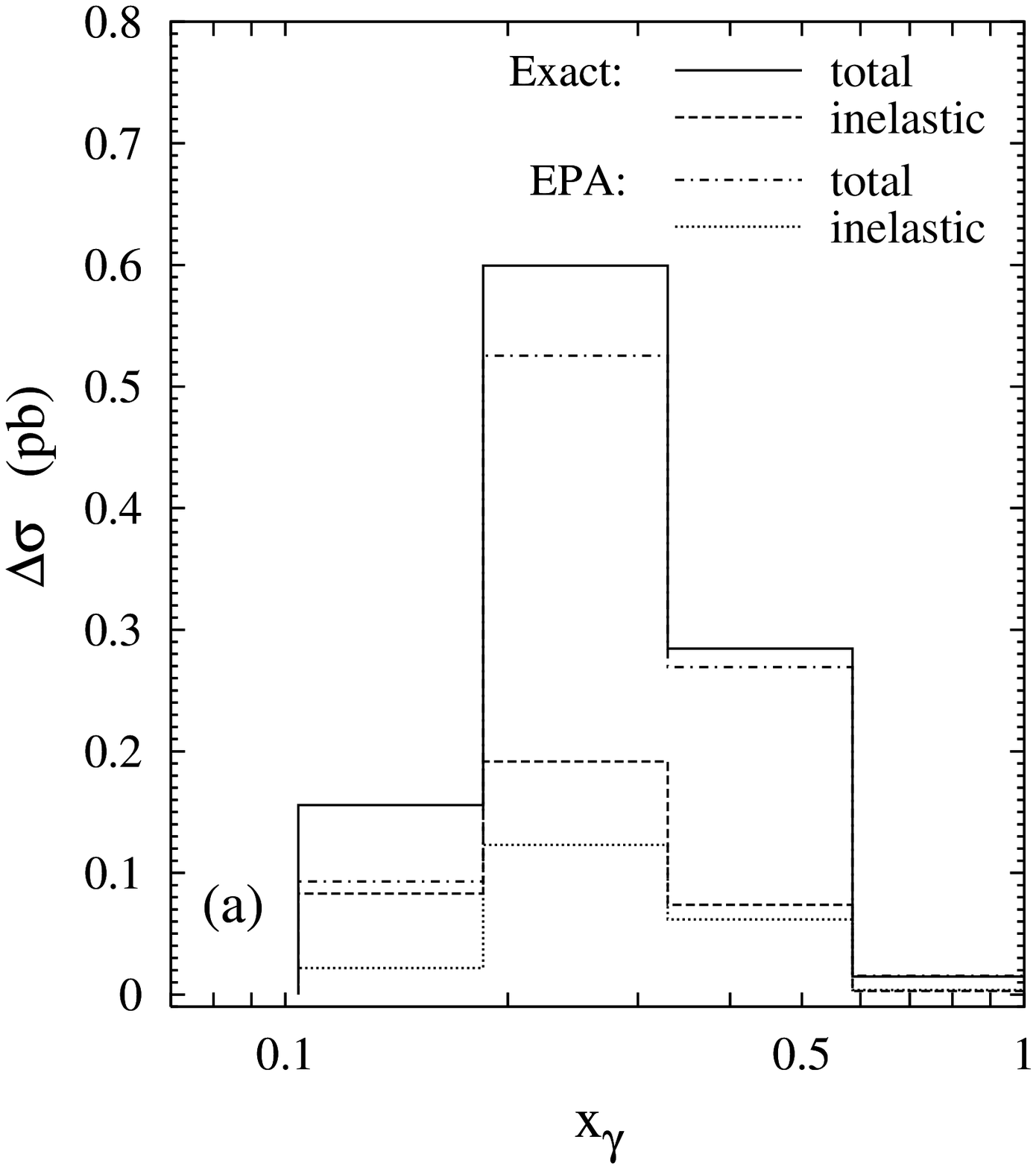,width=7 cm}
\epsfig{figure=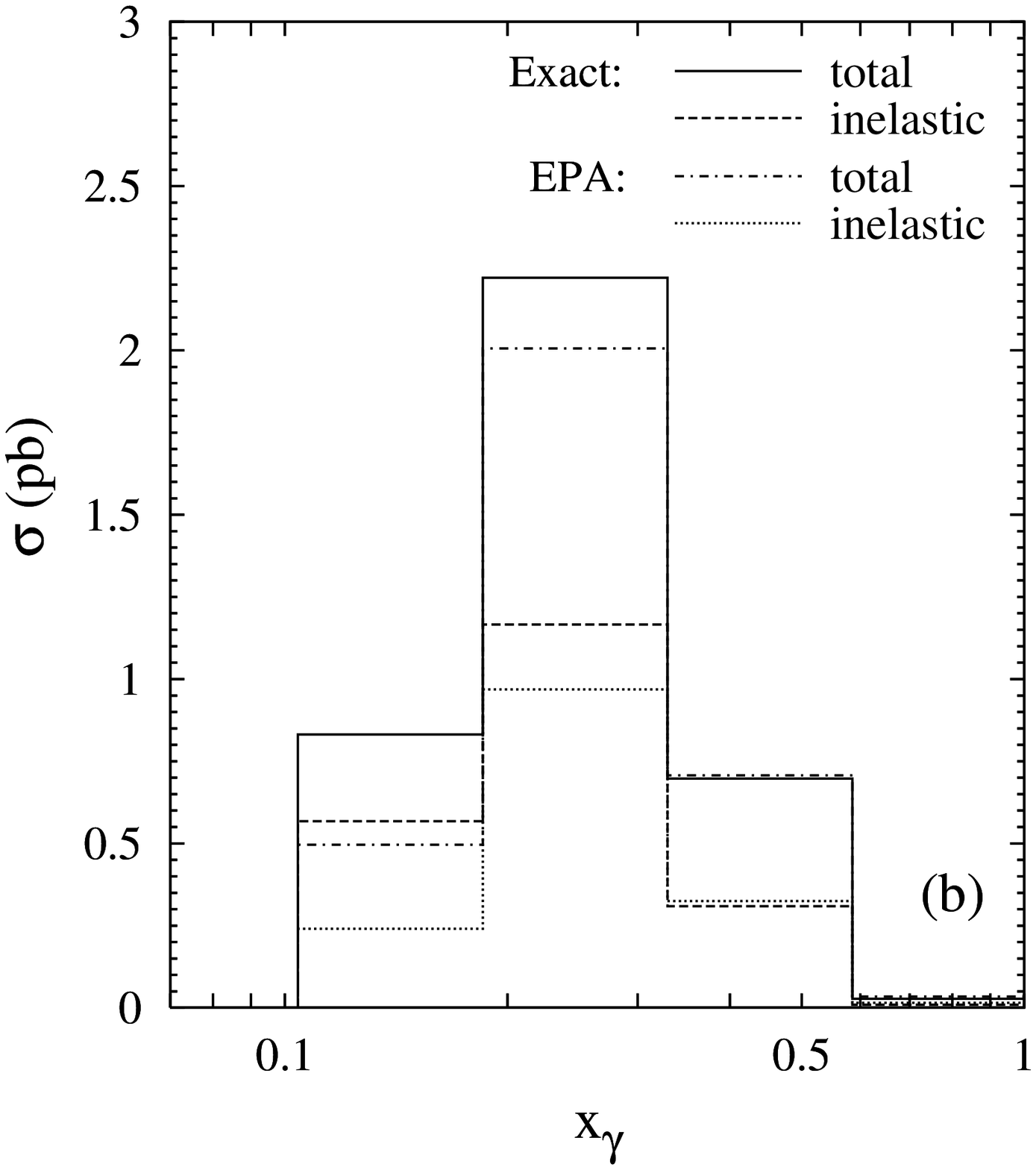,width=7 cm}

\vspace{1cm}
\epsfig{figure=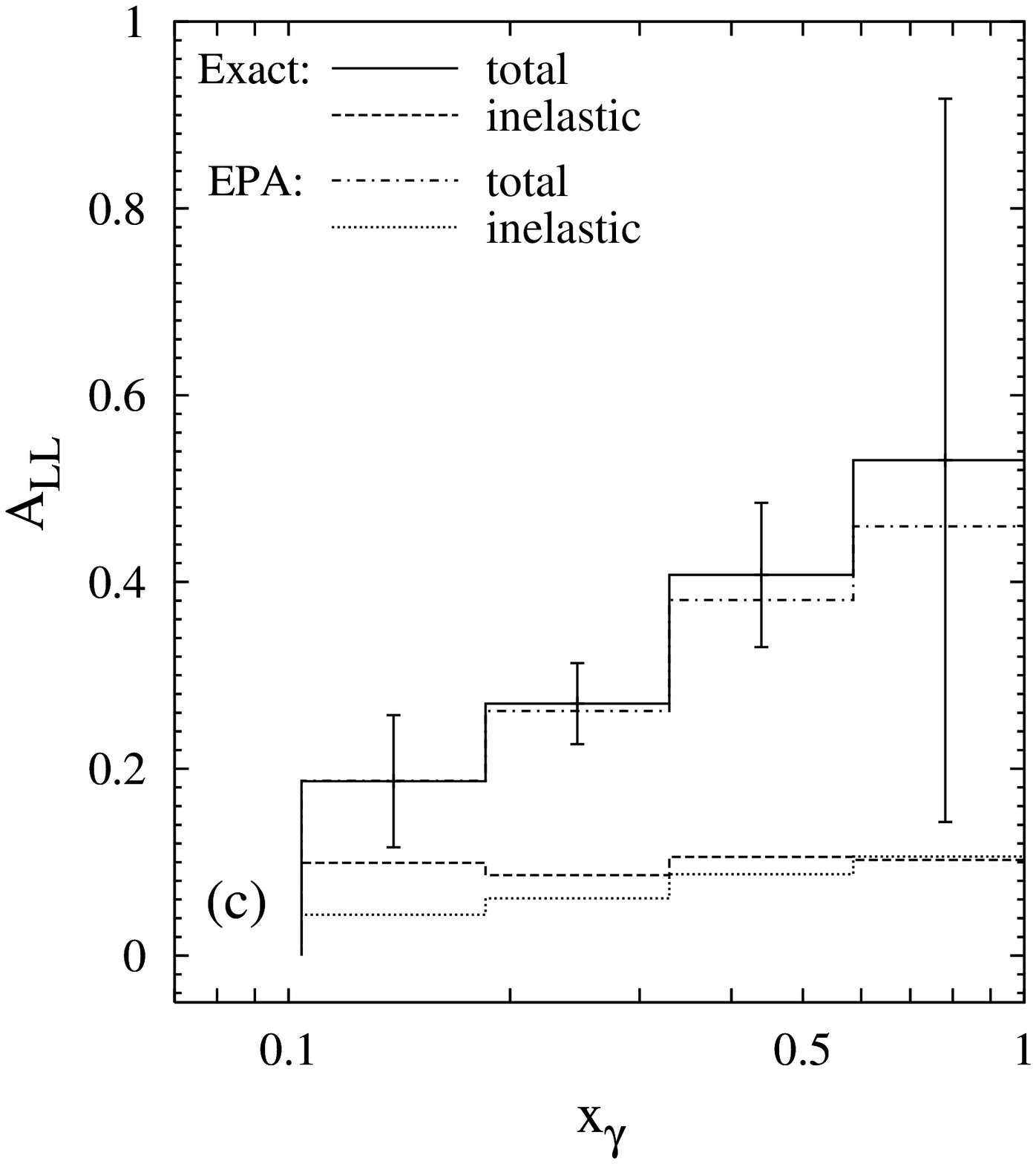,width=7 cm}
\epsfig{figure=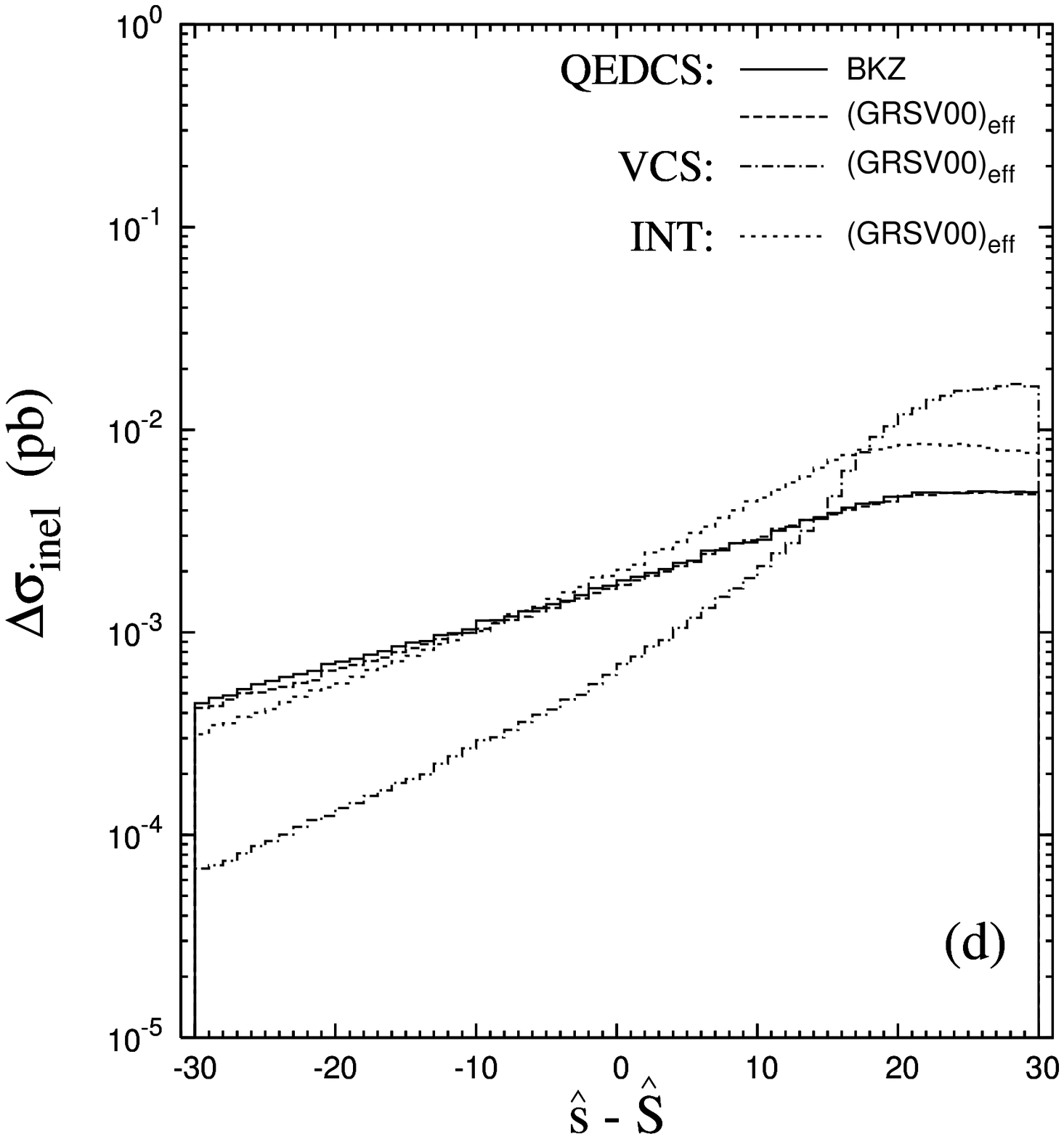,width=7 cm}
\end{center}   
\caption{(a), (b), (c) and (d) are the same as in Figure  \ref{fig:4_pol} 
but for COMPASS. The
constraints imposed are given in  Table \ref{tableone}.}
\label{fig:6_pol} 
\end{figure}
\begin{figure}[ht]
\begin{center}
\epsfig{figure= 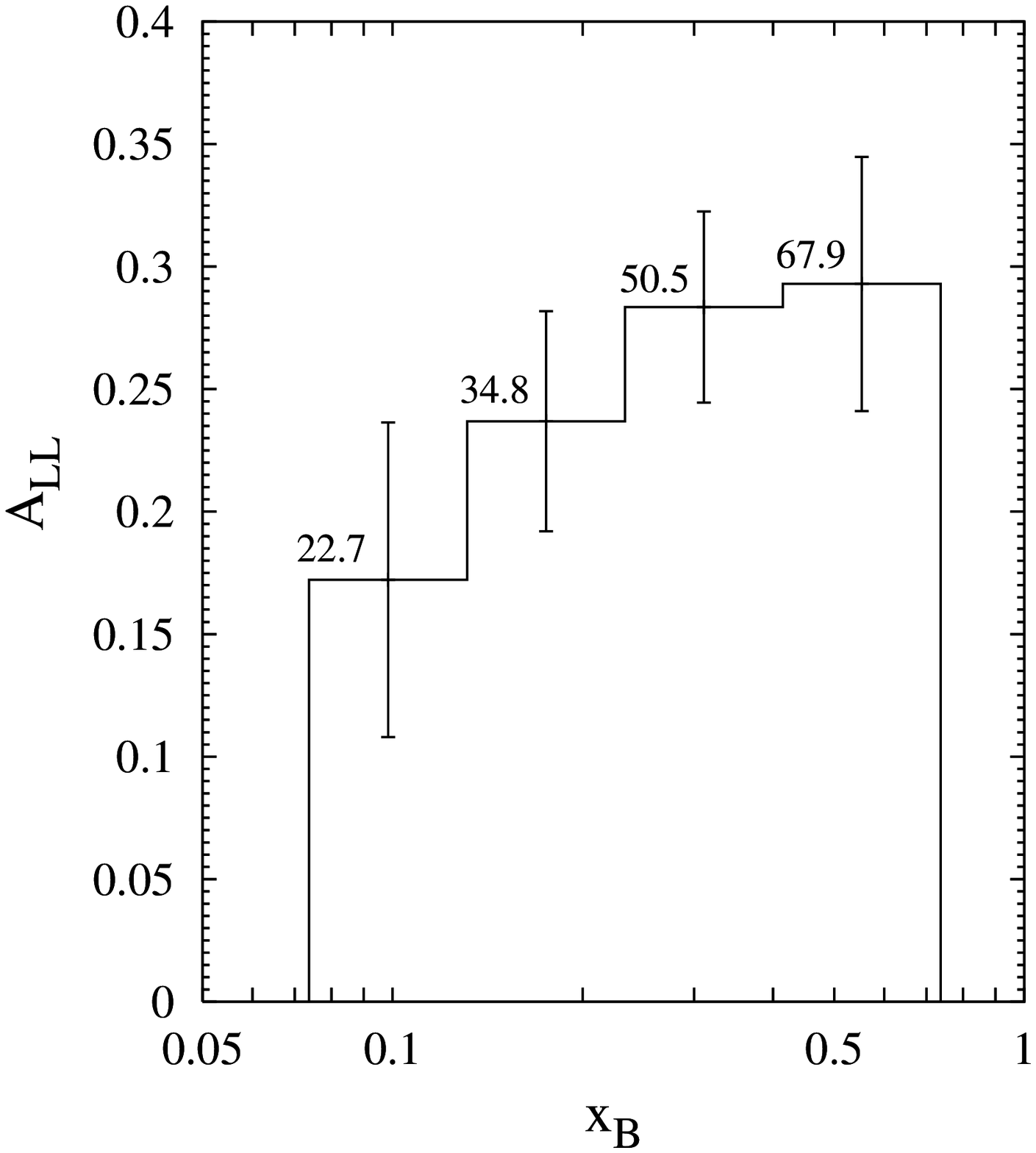,width=9cm,height=8.5cm}
\end{center}
\caption{Asymmetry in the inelastic channel in bins of $x_B$ at COMPASS. We
have used the Badelek {\it et al.} parametrization of $g_1$. 
The constraints imposed are as in  Table \ref{tableone} 
(except $\hat s > Q^2$), together with $\hat S - \hat s > 2$ GeV$^2$. 
The average $Q^2$ (in GeV$^2$) of  each bin is also shown.}
\label{fig:7_pol}  
\end{figure}
\begin{figure}      
\begin{center}
\epsfig{figure=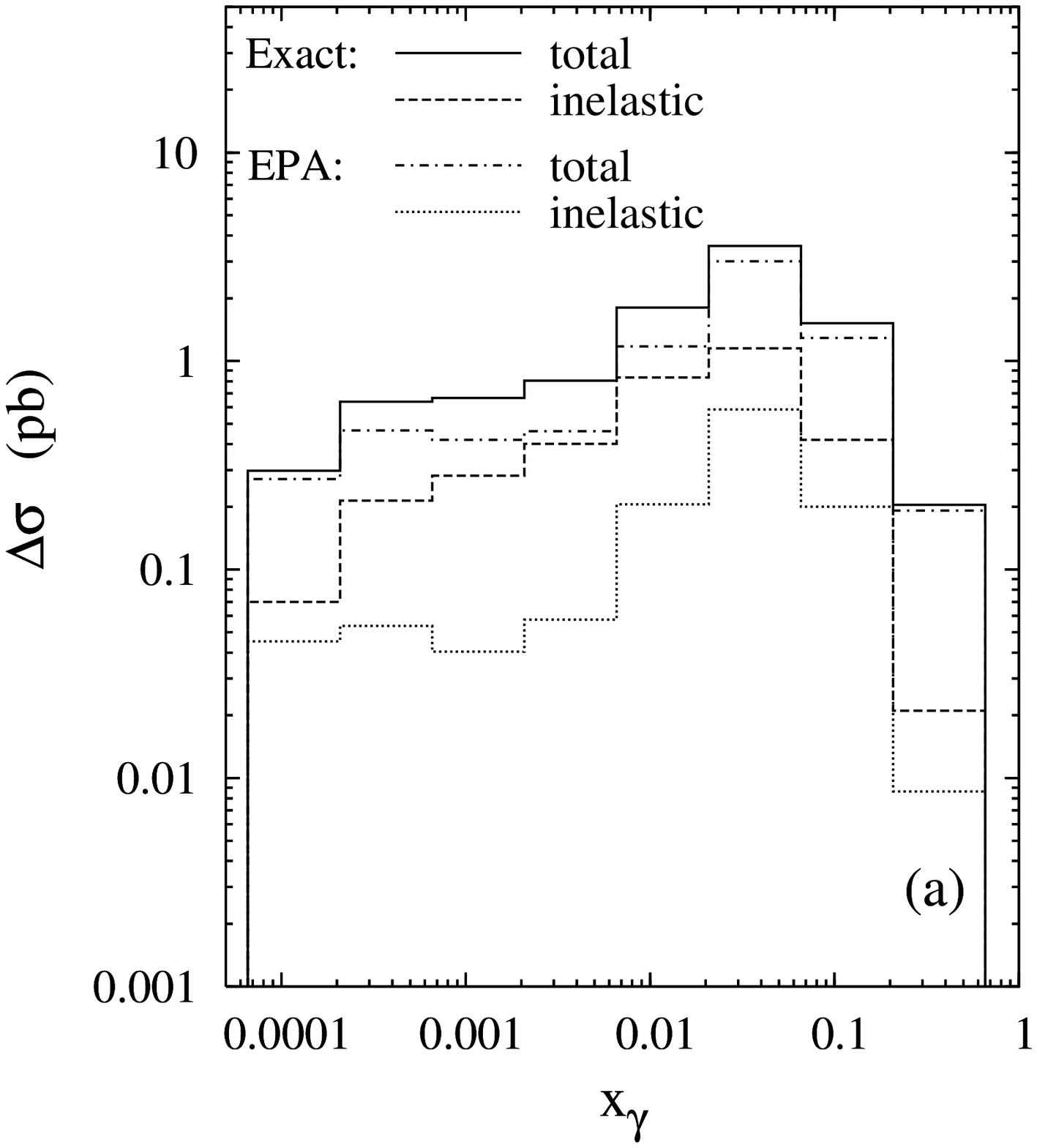,width=7 cm}
\epsfig{figure=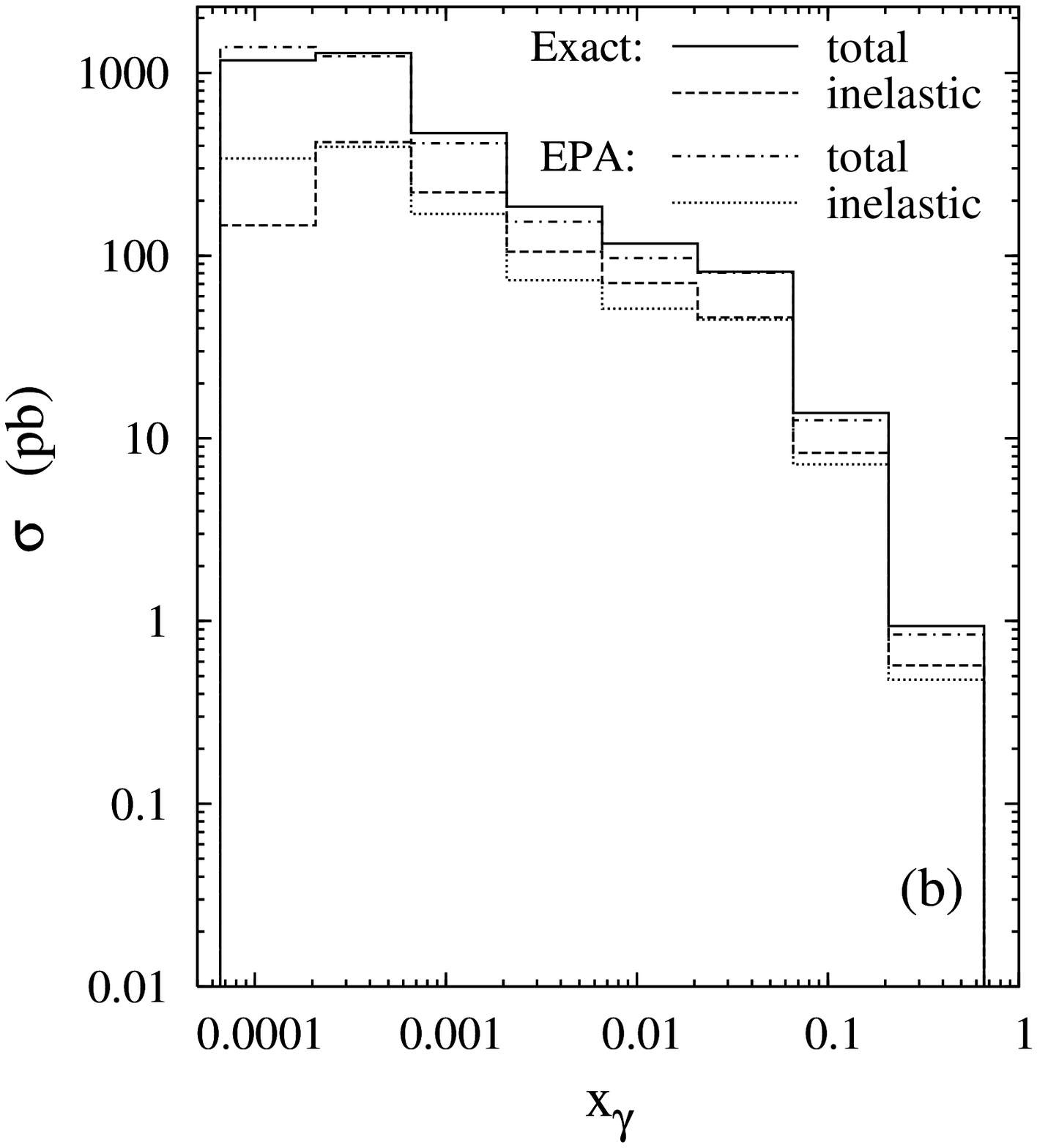,width=7 cm}

\vspace{1cm}
\epsfig{figure=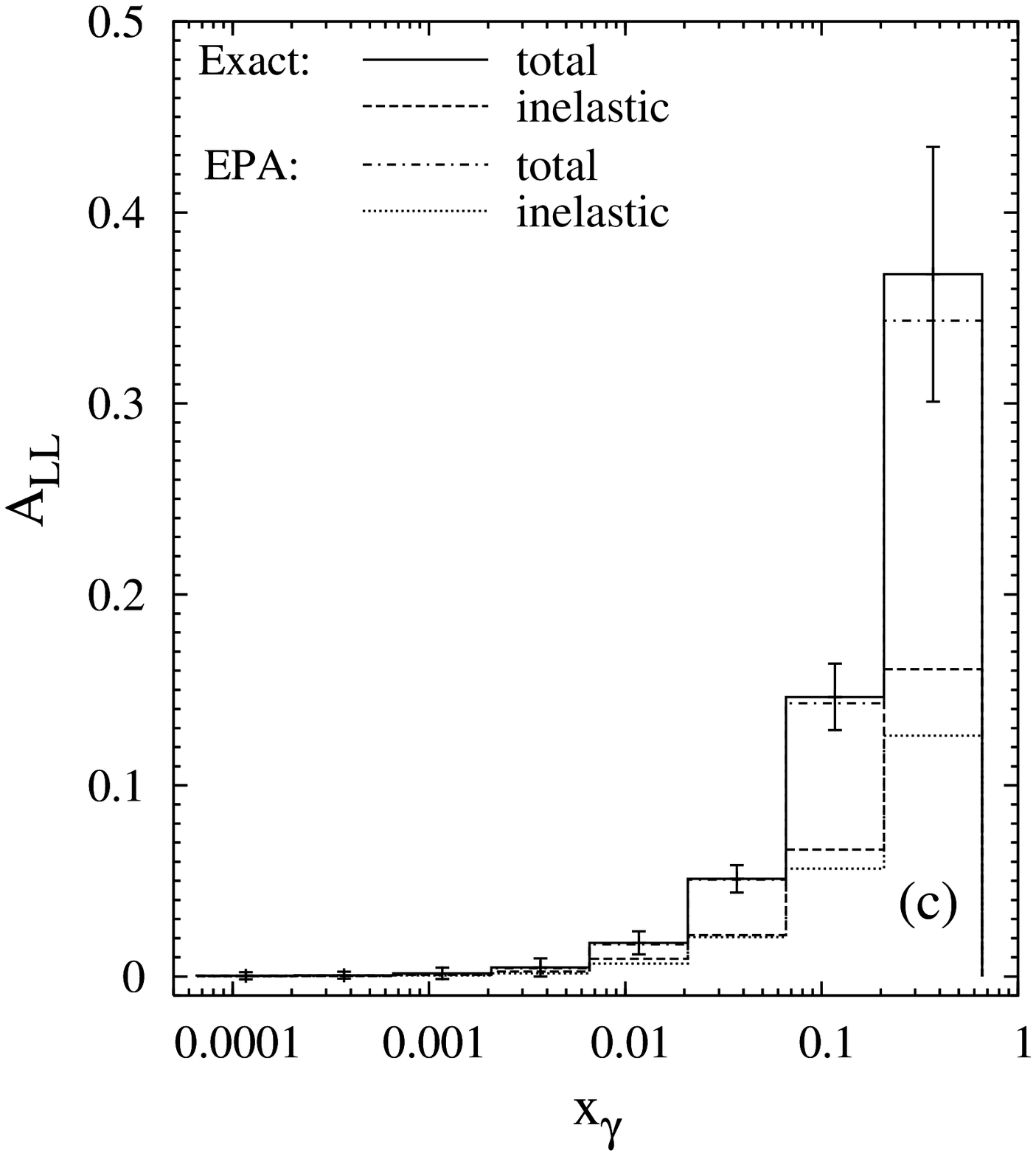,width=7 cm}
\epsfig{figure=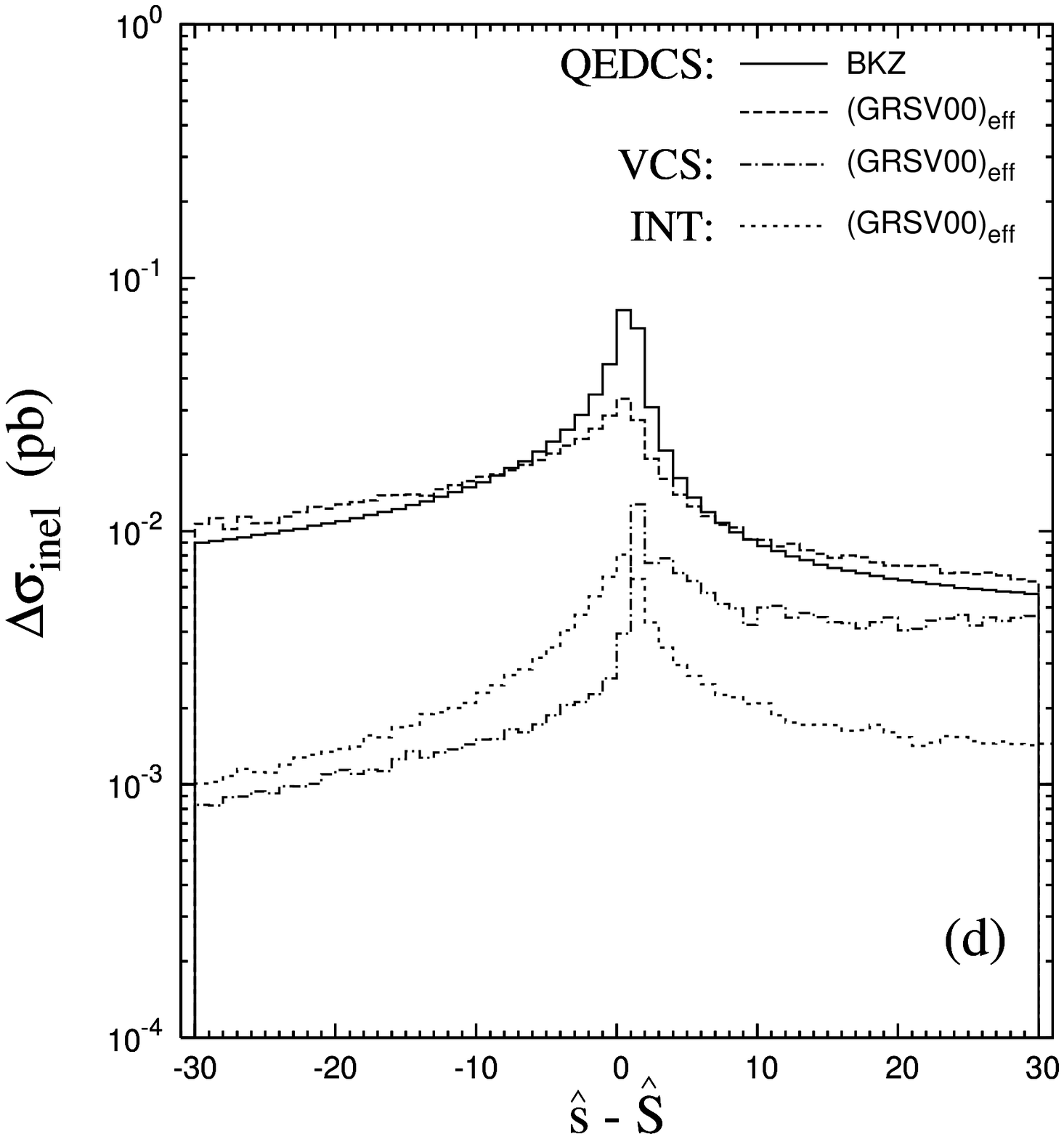,width=7 cm}
\end{center}   
\caption{(a), (b), (c) and (d) are the same as in Figure  \ref{fig:4_pol} 
but for eRHIC. The
constraints imposed are given in Table \ref{tableone}.}
\label{fig:8_pol} 
\end{figure}
\begin{figure}[ht]
\begin{center}
\epsfig{figure= 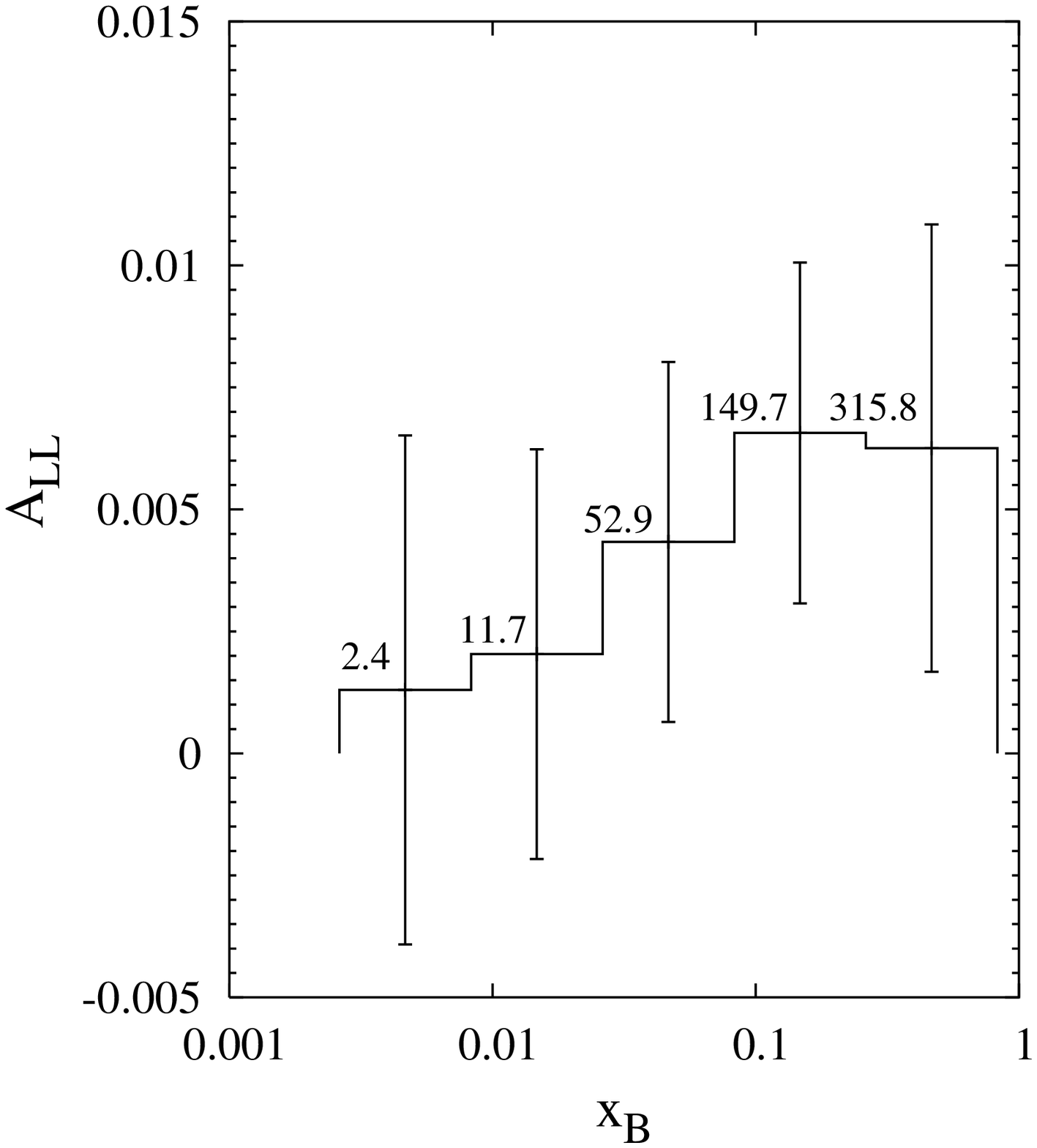,width=9.0cm,height=8.5cm}
\end{center}
\caption{Asymmetry in the inelastic channel in bins of $x_B$ at eRHIC. We
have used the Badelek {\it et al.} parametrization of $g_1$. The
constraints imposed are as in Table \ref{tableone} (except $\hat s > Q^2$), 
together with 
$\hat S > \hat s$. The average $Q^2$ (in GeV$^2$) of  each
 bin is also shown.}
\label{fig:9_pol}
\end{figure}
\begin{figure}[ht]
\begin{center}
\epsfig{figure= 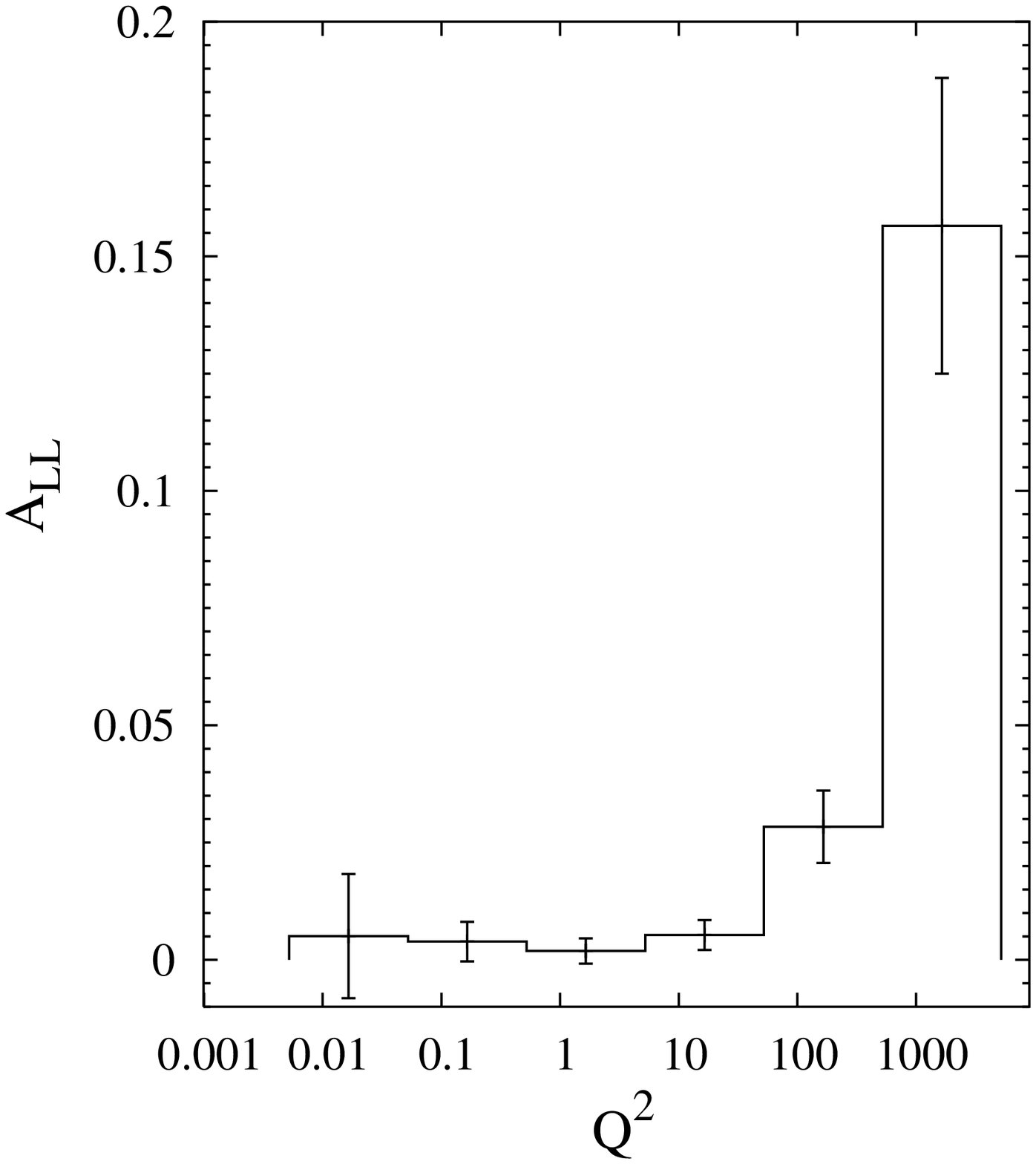,width=9.0cm,height=8.5cm}
\end{center}
\caption{Asymmetry  in bins of $Q^2$ (GeV$^2$) at eRHIC. We
have used the Badelek {\it et al.} parametrization of $g_1$. The
constraints imposed are the same as in Figure  \ref{fig:9_pol}.}  
\label{fig:10_pol}   
\end{figure}

\subsection{COMPASS}

Figures  \ref{fig:6_pol} (a) and (b) show the cross sections of the 
polarized  and unpolarized QEDCS process in bins of $x_\gamma$ 
for the kinematics of COMPASS. We take the energy of the
incident muon beam to be  $160$ $\mathrm{GeV}$, the target is a proton. 
The final muon and the photon are detected
in the polar angle region $ 0.04 < \theta_\mu,
\theta_\gamma < 0.18$. The cross sections in bins, subject to the kinematical 
constraints shown in Table \ref{tableone}, are much smaller than at HERMES, 
because 
they start to decrease with the increase of the incident lepton energy 
$E_l$ as $E_l$ becomes greater than about $20$ $\mathrm{GeV}$, as depicted
in Figure  \ref{fig:3_pol}. 
As before, the cuts remove the initial and final state radiative events. The
$x_\gamma$ integrated cross section agrees with the EPA within $14.2 \%$
(unpolarized) and $15.5 \%$ (polarized). The agreement thus is not as good
as at HERMES. From the figures it is seen that the cross section in the EPA 
actually lies below
the exact one, both for polarized and unpolarized cases. This discrepancy
is due to the fact that the EPA is expected to be a good approximation when
the virtuality of the exchanged photon is small. At COMPASS, with our
kinematical cuts, $Q^2$ can not reach a value below $0.07$ 
$\mathrm{GeV}^2$ and can
be as large as 144 $\mathrm{GeV}^2$, whereas for HERMES smaller values of 
$Q^2$ are
accessible (see the previous subsection). 
Figure  \ref{fig:6_pol} (c) shows the
asymmetry in bins of $x_\gamma$, also in its inelastic channel. 
The asymmetry is of the same order of magnitude as in HERMES and
is in good agreement with the EPA. We have also shown the expected
statistical error in each bin, calculated using    (\ref{err}). We have
taken  $\mathcal {P}_e=
\mathcal {P}_p=0.7$ and $\mathcal {L}={1~\rm{fb}^{-1}}$ for COMPASS. 
The statistical
error is large in higher $x_\gamma$ bins. Figure  
\ref{fig:6_pol} (d) shows the polarized
QEDCS, VCS and interference contributions (inelastic) calculated in the 
effective
parton model, in bins of $\hat s-\hat S$. As in HERMES, VCS is suppressed for
$\hat s < \hat S$. The interference term is not suppressed but using $\mu^+$
and $\mu^-$ beams this can be eliminated. We have also plotted the QEDCS
cross section using Badelek {\it et} {\it al.} parametrization of 
$g_1(x_B,Q^2)$.
The VCS and the interference contributions (elastic) are much suppressed in the
pointlike approximation of the proton with the effective vertex.

Figure \ref{fig:7_pol} shows the asymmetry at COMPASS in the 
inelastic channel plotted in
bins of $x_B$ with the same set of constraints and the additional cut $\hat S - \hat s > 2 \,\,\mathrm{GeV}^2$. The asymmetry is sizable and
can give access to $g_1(x_B, Q^2)$, the kinematically allowed range is 
$x_B > 0.07 $ . 
We have also shown the expected statistical errors in the
bins and the average $Q^2$ in each bin. Confronting Figure  \ref{fig:5_pol}
and \ref{fig:7_pol} one can see that there is no overlap in the 
kinematical region covered 
at HERMES and COMPASS. Higher values of $Q^2$ are probed at COMPASS in the 
same $x_B$ range as compared to HERMES.             
   
\subsection{eRHIC}

The cross sections for eRHIC kinematics, both polarized and unpolarized, are 
shown in Figure \ref{fig:8_pol} (a) and (b) respectively,
in bins of $x_\gamma$. We have taken the incident electron energy $E_e=10 $
GeV and the  incident proton energy $E_p= 250$ GeV. The cross section in the
EPA is also shown. The kinematic constraints are given in Table \ref{tableone}.
The polar angle acceptance of the detectors at eRHIC is not known. We have
taken the range of $\theta_e, \theta_\gamma $ to be the same as at HERA. We
have checked that the constraints on the energies and the polar angles of
the outgoing electron and photon are sufficient to prevent the electron
propagators to become too small and thus reduce the radiative contributions.
The unpolarized total (elastic+inelastic) cross section, 
integrated over $x_\gamma$
agrees with the EPA within $1.6 \%$. The agreement in the inelastic channel
is about $6.3 \%$. The polarized total cross section agrees with the 
EPA within $9.8 \%$. The
EPA in this case lies below the exact result in all the bins. 
The agreement in
the inelastic channel is about $19.6\%$. More restrictive constraints
instead of $\hat s > Q^2$, like $\hat s > 10$ $ Q^2$, makes the agreement
better, about $1.2 \%$ in the polarized case and $1.9 \%$ in the unpolarized
case. Figure  \ref{fig:8_pol} (c) shows the asymmetry for eRHIC, in bins 
of $x_\gamma$. The
discrepancy in the cross section cancels in the asymmetry, as a result good
agreement with the EPA is observed in all  bins except the last one at
higher $x_\gamma$. The asymmetry in the inelastic channel is also shown. We
have plotted the expected statistical error in the bins using  
(\ref{err}). For eRHIC, we have taken  $\mathcal {P}_e=
\mathcal {P}_p=0.7$ and $\mathcal {L}={1}$ ${\rm{fb}^{-1}}$. 
The expected statistical error  
increases in higher $x_\gamma$ bins. The asymmetry is very small for small
$x_\gamma$ but becomes sizable as $x_\gamma$ increases. 
Figure  \ref{fig:8_pol} (d) shows the
polarized cross section in the inelastic channel, in bins of $\hat s-\hat S$,
in the effective parton model for eRHIC. The VCS is suppressed in all
bins, especially for $\hat s <\hat S$. The interference contribution is
negligible, similar to HERA. The effective parton model QEDCS cross
section is also compared with the  one calculated using the 
Badelek {\it et. al.} parametrization for $g_1(x_B, Q^2)$. 
Similar effects are observed in the unpolarized
case. In the pointlike approximation of the proton with the effective vertex, 
as before, the elastic VCS as well as
the interference contributions are very much suppressed.

Figure  \ref{fig:9_pol} shows the 
asymmetry in bins of $x_B$ in the inelastic channel, which may be relevant for
the determination of $g_1(x_B, Q^2)$ using QEDCS at eRHIC. The asymmetry is small but
sizable, however the error bars are large and therefore good statistics is
needed. $x_B$ can be as low as $0.002$. A wide range of $Q^2$ can be
accessed at eRHIC starting from $0.008$ to $2000$ $\mathrm{GeV}^2$; 
the average $Q^2$ value in the bins ranges from $2.4$ to $315$ 
$\mathrm{GeV}^2$. 

Figure \ref{fig:10_pol} shows the total  asymmetry in $Q^2$ bins for eRHIC. 
The asymmetry in this case
is bigger in each bin and the error bars are smaller  than for the $x_B$ bins, 
except in the last $Q^2$-bin where the number of events is smaller.     
\section{Summary}
\label{sec:polsummary}

In this chapter we have analyzed the QED Compton process in
polarized $lp$ scattering, both in the elastic and inelastic channel. 
As for the unpolarized process, we have  shown that the cross section can
be expressed in terms of the equivalent photon distribution of the 
proton,
convoluted  with the real photoproduction cross section. 
Furthermore we have provided the necessary kinematical
constraints for the extraction of the polarized photon content of 
the proton by measuring the 
QED Compton process at HERMES, COMPASS and
eRHIC. We have shown that the
cross section and, in particular, the asymmetries are  accurately 
described
by the EPA. We have also discussed the possibility of suppressing  the major
background process, namely the virtual Compton scattering. 
We point out that such an
experiment can give access to the spin structure function $g_1(x_B, Q^2)$ in
the region of low $Q^2$ and medium $x_B$ in fixed target experiments and
over a broad range of $x_B,~ Q^2$ at the future polarized $ep$ collider,
eRHIC. Because of the different kinematics compared to the fully inclusive
processes, the QED Compton process can provide information on $g_1(x_B, Q^2)$ 
in a range
not well-covered by inclusive measurements and thus is a valuable tool to
have a complete understanding of the spin structure of the proton.

\clearemptydoublepage


\chapter{{\mbox{\boldmath$\nu W$}} {\bf Production in} {\mbox{\boldmath $e p \rightarrow \nu W X$}}}



In \cite{kniehl} the unpolarized elastic photon 
distribution was tested in the case of $\nu W$ production in the process 
$ep\rightarrow \nu W  p$. The relative error of the cross section as 
calculated in the EPA with respect to the exact result
was shown as a function of $\sqrt s$, in the range  
$100 \le \sqrt s \le 1800$ $\mathrm{GeV}$. The agreement turned out to be very 
good, the approximation reproducing the exact cross section within less 
than one percent. Motivated by this results,
following the lines of \cite{cris}, our aim here is to check if the
same holds in the inelastic channel. 

The process $ep \rightarrow \nu W X $ has been widely studied by several 
authors \cite{alt,gab,baur,neufeld,bohm}. Its relevance is related to the 
possibility of measuring
the three-vector-boson coupling $W W \gamma$, which is a manifestation 
of the non-abelian gauge symmetry upon which the Standard Model is based.
The observation of the vector boson self interaction would be a crucial test
of the theory. Furthermore, such a reaction is also an important background 
to a number of processes indicating the presence of new physics. 
The lightest Supersymmetric Standard Model particle has no charge and 
interacts very weakly with matter; it means that, exactly as the neutrino 
from the Standard Model, it escapes the detector unobserved and can be 
recognized only by missing momentum. This implies that a detailed study of the 
processes with neutrinos in the final states is necessary to distinguish 
between the new physics of the Supersymmetric Standard Model and the physics 
of the Standard Model. 
At the HERA collider energies ($\sqrt s = 318 $ 
GeV) the $e p \rightarrow \nu W  X $ cross section is 
much smaller than the one for $ep \rightarrow e W X $ \cite{gab,baur},
 also sensitive to the $WW\gamma$ coupling, due to the presence in the latter
of an additional 
Feynman graph  where an almost real photon  and a massless quark are 
exchanged in a $u$-channel configuration ($u$-channel pole). The 
dominance of the  process $ep\rightarrow e W X $   justifies the 
higher theoretical and experimental \cite{exp} attention that it has 
received so far, as compared to $ep\rightarrow \nu W X$. One way of improving
the problem of the low number of deep inelastic $\nu W$ 
events at HERA would be to consider also the elastic and quasi-elastic 
 channels of the reaction, as will be discussed in 
Section \ref{sec:cris_results}.  

It is worth mentioning that not all the calculations of  the $ep \rightarrow \nu W X$ event rates available in the literature, in which only the photon
exchange is considered (see Figure \ref{figure1}), are in agreement, as already pointed out in \cite{bohm}. In particular, the numerical
estimate of the cross section for HERA energies presented   in 
\cite{alt,gab}, obtained in the EPA approach, is one half of the one 
published in \cite{bohm}, obtained within the framework of the helicity 
amplitude
formalism without any approximation.  The value given in \cite{neufeld}
is even bigger than the one in \cite{bohm}: all these discrepancies cannot be
due to the slightly different kinematical cuts employed in the papers cited
above and stimulate a further analysis. Our results agree with \cite{alt,gab}. 

The plan of this chapter is as follows. In  Section \ref{sec:theory} 
 we calculate the exact 
cross section for the inelastic channel in a manifestly covariant way and we 
show in which kinematical region it is supposed to be well described by 
the EPA. The formulae for the corresponding elastic cross sections, both
the exact and the one evaluated in the EPA, are also given. 
The numerical results are discussed in Section \ref{sec:cris_results}. 
The summary is given in Section \ref{sec:cris_summary}.

\section{Theoretical Framework}
\label{sec:theory}

The $\nu W $ production from inelastic $e p $ scattering, 
\be
 e(l)+ p(P) \rightarrow \nu(l')+ W(k')+ X(P_X),
\ee
is described, considering only one photon exchange, by the Feynman 
diagrams depicted in Figure \ref{figure1}.
%
\begin{figure}[ht]
\begin{center}
\epsfig{figure=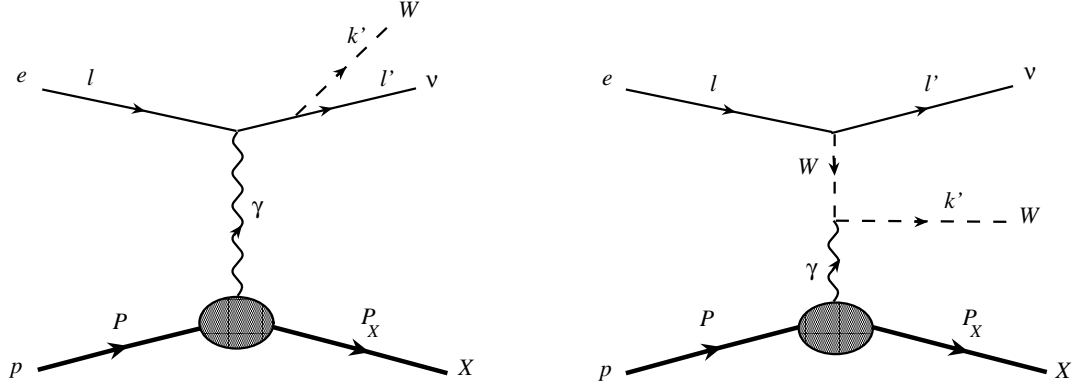, width=15cm}
\end{center}
\vspace*{-1.6cm}
\caption{Feynman diagrams for the process $ep \rightarrow \nu W X$.}
\label{figure1}
\end{figure}
The four-momenta of the particles are given
in  the brackets; $P_{X}=\sum_{X_i} P_{X_i}$ is the sum over all momenta 
of the produced hadronic system. We introduce the invariants 
\be
s=(P+l)^2, ~~~~~~\hat s=(l+k)^2, ~~~~~~ \hat t = (l-l')^2, ~~~~~~ Q^2 =-k^2,
\label{w_invar}
\ee
where $k = P-P_X$ is the four-momentum of the virtual photon.
If we denote with $\epsilon$ and $\epsilon_W^*$ the polarization vectors 
of the incoming photon and outgoing $W$ and with $u$ and $\bar{u}$ the 
Dirac spinors of the initial electron and  final neutrino,
the amplitude of the subprocess $e\gamma^*\to \nu W$ reads
\be
\hat M & = &-\frac{e^2}{2\sqrt{2} \sin\theta_W}\,
\epsilon^{\alpha}(k)\epsilon_W^{* \mu}(k')\bar{u}(l')
\left [ \,\frac{1}{\hat s} \,\gamma_{\mu}
(l'\!\!\!/ + k'\!\!\!\!/\,)\gamma_{\alpha} + \frac{1}{\hat t -M_W^2}\,
\gamma^\rho\Gamma_{\alpha\rho\mu}\, \right ](1-\gamma^5)u(l)\nonumber \\
&&\equiv \epsilon_{\alpha}(k)\hat M^{\alpha},
\label{amplitude_eq}
\ee
where $M_W$ is the mass of the $W$ boson, $\theta_W$ the weak-mixing 
angle and
\be
\Gamma_{\alpha\rho\mu} = g_{\alpha\rho}(2k-k')_\mu + g_{\rho\mu}
(2k'-k)_{\alpha} - g_{\mu\alpha}(k'+k)_{\rho}
\ee
describes the $WW\gamma$ vertex.

The integrated cross section of the full reaction $ep\rightarrow \nu W X$ 
has been calculated in \cite{cris} and can be written as (\ref{siin}), that is
\be
\sigma_{\mathrm{inel}}(s)\!\!\!&=&\!\!\!{\alpha\over 4 \pi (s-m^2)^2} \int_{W^2_{{\mathrm
{min}}}}^{W^2_{\mathrm{max}}}
\!\!\! \der W^2 \int_{\hat s_{\mathrm{min}}}^{(\sqrt
s-W)^2}\!\!\! \der\hat s \int_{Q^2_{\mathrm{min}}}^{Q^2_{\mathrm{max}}} {\der 
Q^2 \over Q^4} \int_{\hat t_{\mathrm{min}}}^{\hat
t_{\rm max}} \!\!\!\der \hat t \int_0^{2 \pi} \!\!\!\der \varphi^* 
\bigg \{ \bigg [\bigg ( 2 \,{{s-m^2}\over {\hat s+Q^2}}\nonumber\\&&
\times \bigg (1-{{s-m^2}\over {\hat s+Q^2}}\bigg )+(W^2-m^2) \bigg ( {2\,
(s-m^2)\over {Q^2 (\hat s + Q^2)}}-{1\over Q^2} +{m^2-W^2\over 2 \,Q^4}\bigg ) 
\bigg )
\nonumber\\&&~~\times [3
X_1(\hat s,Q^2,\hat t)+X_2(\hat s,Q^2,\hat t)]+\bigg ({1\over
Q^2}(W^2-m^2)+{(W^2-m^2)^2\over 2\, Q^4}+{2 m^2\over Q^2} \bigg  )
\nonumber\\&&~~~~~~~\times [X_1(\hat s,Q^2,\hat t)+X_2(\hat s,Q^2,\hat t)]-X_1(\hat
s,Q^2,\hat t)\bigg ] F_2(x_B,Q^2) {x_B\over
2}\nonumber\\&&~~~~~~~~~~~~~~~~~~~~~~~~~~~~~~~~~~-X_2(\hat s,Q^2,\hat
t)F_1(x_B,Q^2)\bigg
\},
\label{w_siin}
\ee   
where $W^2$ indicates the  invariant mass squared of the 
produced hadronic system $X$, $\varphi^*$ denotes the azimuthal angle of the
outgoing $\nu-W$ system in the $\nu-W$ center-of-mass frame, and
\be
x_B = \frac{Q^2}{W^2 + Q^2 - m^2} 
\ee
is the Bjorken variable. Furthermore,
$F_{1,2}(x_B, Q^2)$ are the structure functions of the proton and 
the two invariants $X_{1,2}(\hat s, Q^2, \hat t)$, which  contain all the 
information about the subprocess $e\gamma^* \rightarrow \nu W$, are given
by \eqref{x1}, \eqref{x2}:
\be
X_1(\hat s,Q^2,\hat t) & = & -\frac{Q^2\, l^\alpha l^\beta 
{\cal{W}}_{\alpha \beta}}{4 \pi^2 (\hat s+Q^2)^3}~, \nonumber \\
 X_2(\hat s,Q^2, \hat t) & = &\frac{g^{\alpha \beta} {\cal{W}}_{\alpha \beta}} 
{16 \pi^2 (\hat s+Q^2)}~.
\ee
The tensor ${\cal{W}^{\alpha\beta}}$ is obtained from 
\eqref{amplitude_eq}:
\be
{\cal{W}^{\alpha\beta}} = \frac{1}{2}\,\sum_{\rm spins}\,\hat M^{*\alpha}
\hat M^{\beta};
\ee
the sums over the electron and vector boson spins are performed by 
making use of the completeness relations \eqref{eq:compl_u} and 
\be
 \sum_{\lambda} \epsilon^{\lambda}_{\alpha}(k') 
\epsilon^{\lambda *}_{\beta}(k') = -g_{\alpha\beta} + 
\frac{k'_\alpha k'_{\beta}}{M_W^2}~.
\ee
The final result can be expressed as 
\be   
X_1 (\hat s, Q^2, \hat t)& = & \frac{\alpha G_F}{2 \sqrt 2 \pi}
\frac{Q^2 M_W^2}{ (Q^2 + \hat s)^3 \,(M_W^2 - \hat t)^2}   
[ (Q^2 + \hat s)^3 -\hat s (Q^2 + \hat s)^2 (Q^2 + \hat s + \hat t)  
\nonumber \\
&& ~~~~~~ +~ 2 (Q^2 + \hat s)^2 \hat t + 8
          (Q^2 + \hat s) \hat t^2 + 8 \hat t ^3]
\label{w_xoneex}
\ee
and
\be
 X_2(\hat s , Q^2, \hat t) & = & \frac{\alpha G_F}{2\sqrt 2 \pi}\frac{1}{\hat s ^2 (Q^2 + \hat s )(M_W^2 - \hat t)^2}  \{ 4 M_W^8 (Q^2 + \hat s) - 4 M_W^6 [3 \hat s (Q^2 + \hat s) \nonumber \\ && +  ~(2 Q^2 + \hat s) \hat t] 
+ 4 M_W^4 [\hat s (2 \hat s + \hat t)^2 + Q^2 (4 \hat s^2 +2 \hat s \hat t + \hat t^2)] - M_W^2 \hat s \nonumber \\&& ~~~~~~~\times~ [Q^4 \hat s + Q^2 (9 \hat s^2    + 2 \hat s \hat t  - 4 \hat t^2) 
+  4 (\hat s + \hat t ) (2 \hat s^2 + 2 \hat s \hat t + \hat t^2)] \nonumber \\&&~~~~~~~~~~~~~~~~+ Q^2 \hat s ^2 [\hat s (\hat s  + \hat t)
 +~ Q^2 (\hat s + 2 \hat t)]  \},
\label{w_xtwoex}
\ee
with 
\be
G_F = \frac{\sqrt{2} e^2}{8 M_W^2\sin^2\theta_W}~.
\ee
In   (\ref{w_siin}) the minimum value of $\hat s$ is given by the squared 
mass of the $W$ boson:
\be
\hat s_{\mathrm{min}} = M_W^2,
\label{w_smin}
\ee 
while the limits of the integration over $W^2$ are:
\be
W^2_{\mathrm{min}}=(m+m_{\pi})^2,~~~~~~~~W^2_{\mathrm{max}}=
(\sqrt s-\sqrt {\hat s_{\mathrm{min}}}\, )^2,
\ee
where $m_{\pi}$ is the mass of the pion.
The limits $Q^2_{\mathrm{min},\mathrm{max}}$ are given by:
\be
Q^2_{{{\mathrm{min}},{\mathrm{max}}}}& = & -m^2-W^2+{1\over 2 s} \Big 
[(s+m^2) (s-\hat s+W^2) \nonumber \\ 
&& ~~~~~~\mp (s-m^2){\sqrt{(s-\hat s+W^2)^2-4 s W^2}} \Big ],
\label{w_q2lim}
\ee  
and the extrema of $\hat t$ are \be
\hat t_{\mathrm{max}}=0, ~~~~\hat t_{\mathrm{min}} = -\frac{(\hat s + Q^2) (\hat s -M_W^2)}{\hat s}~.
\label{w_thatlim}
\ee
Integrating $X_{1,2} (\hat s, Q^2, \hat t)$ over $\varphi^*$ and 
$\hat t$, with the limits 
in   (\ref{w_thatlim}), one recovers   (4.1) and (4.2) of \cite{kniehl} 
respectively, times a factor of two due to a different normalization.   
The EPA consists of considering the exchanged photon as real; it is 
possible to get the approximated cross section $\sigma_{\mathrm{inel}}^{\mathrm{EPA}}$ from the exact one,   (\ref{w_siin}),   in a 
straightforward way, following again Chapter \ref{ch:qedcs}.  
We neglect $m^2$ 
compared to $s$ and $Q^2$ compared to $\hat s$
then, from   (\ref{w_xoneex})-(\ref{w_xtwoex}),
\be
{X}_1(\hat s, Q^2,\hat t) \approx {X}_1(\hat s,0, \hat t)=0,
\label{w_xone}
\ee
and \be
{X}_2(\hat s, Q^2, \hat t)\simeq {X}_2(\hat s,0, \hat t) = -{{2 \hat s}\over{\pi}}\, {{\der\hat{\sigma}(\hat s, \hat t)}\over{\der\hat t}}~,
\label{w_xtwo}
\ee
where we have introduced the differential cross section for the real 
photoproduction process $e \gamma \rightarrow \nu W$:
\be
{\der \hat \sigma (\hat s, \hat t)\over \der\hat t}=-\frac{ \alpha G_F M_W^2}{ \sqrt 2 \hat s^2} \bigg (1 - \frac{1}{1 + \hat u /\hat s} \bigg )^2 ~ \frac{\hat s^2 + \hat u^2 + 2 \hat t M_W^2}{\hat s \hat u}
\label{w_realsi} 
\ee
with $\hat u = (l-k')^2 =  M_W^2 -\hat s -\hat t$. Equation
   (\ref{w_realsi}) agrees
with the analytical result already presented in \cite{alt,gab}, obtained 
using  the helicity amplitude technique.  
Using   (\ref{w_xone}) and ({\ref{w_xtwo}), we can write
\be
\sigma_{\mathrm{inel}}(s) \simeq \sigma_{\mathrm{inel}}^{\mathrm{EPA}} =
\int_{x_{\mathrm{min}}}^{(1-m/\sqrt s)^2} \der x \, \int_{M_W^2-\hat s}^0
\der\hat t~
\gamma^p_{\mathrm{inel}}(x, x s) \,{{\der \hat\sigma(x s, \hat t)}
\over{\der\hat t}}~,
\label{w_epain}
\ee
where  $x={\hat s/s}$  and $\gamma^p_{\mathrm{inel}}(x, x s)$ is the 
inelastic component of the equivalent photon distribution of the proton
given in \eqref{gammain}.
As already pointed out (see also  \cite{alt}) there is some ambiguity in the 
choice of the scale of $\gamma^p_{\rm inel}$, here taken to be $\hat s = x s$,
which is typical of all leading logarithmic 
approximations, and any other quantity of the same order of magnitude of 
$\hat s$, like $-\hat t$ or $-\hat u$, would be equally acceptable 
within the limits of the EPA. The numerical effects
related to the scale dependence of the inelastic photon distribution are 
discussed in the next section.

The cross section relative to the elastic
channel, $ ep \rightarrow \nu W p$, has been calculated in \cite{kniehl}; 
it can be written as \eqref{sigel}, namely 
\be
\sigma_{\mathrm{el}}(s)\!\!\!&=&\!\!\!{\alpha\over 8 \pi (s-m^2)^2} \int_{\hat s_{\mathrm
{min}}}^{(\sqrt
s-m)^2} \der\hat s \int_{t_{\mathrm{min}}}^{t_{\mathrm{max}}}{\der t \over t} 
\int_{\hat t_{\mathrm{min}}}^{\hat t_{{\mathrm{max}}}} \der \hat t \int_0^{2 \pi} \der \varphi
^* \bigg\{ \bigg [ 2\, {s-m^2\over \hat s-t}
\bigg ( {s-m^2\over \hat s-t}-1 \bigg ) \nonumber\\&& \times ~[3 X_1(\hat s,t,\hat
t)+X_2 (\hat s,t,\hat t)] +{2 m^2\over t} [X_1(\hat s,t,\hat t)+X_2(\hat
s,t,\hat t)]+X_1(\hat s,t,\hat t)\bigg ]
H_1(t)\nonumber\\&&~~~~~~~~~~~~~~~~~~~~+~X_2(\hat s,t,\hat t) G_M^2(t) \bigg \},
\label{w_sigel}
\ee
with $t = -Q^2$, integrated over the range already defined by   
(\ref{w_q2lim}),  and $\hat s_{\mathrm{min}}$ given by   (\ref{w_smin}). 
The limits of integration of $\hat t $ are the same
as  in   (\ref{w_thatlim}) and the invariant $H_{1}(t)$ is given in
\eqref{eq:h12}
in terms of $G_E(t)$ and  $G_M(t)$, the  electric and magnetic form 
factors of the proton, respectively.

Again, in the limit $s \gg m^2$ and 
$\hat s \gg -t $, the cross section factorizes and is given by
\be
\sigma_{\mathrm{el}}(s) \simeq \sigma_{\mathrm{el}}^{\mathrm{EPA}} =
\int_{x_{\mathrm{min}}}^{(1-{m/ 
\sqrt s})^2}\, \der x \,\int_{M_W^2 -\hat s}^0
\der\hat t  \,\gamma^p_{\mathrm{el}} (x) \,{\der \hat \sigma (x s, \hat t)\over \der\hat t} ,
\label{w_epael}
\ee
where $x={\hat s/ s}$ and $\gamma^p_{\rm el}$ 
is the universal, scale independent, elastic component of the photon 
distribution of the proton introduced in \eqref{eqtwo}.

\section{Numerical Results}
\label{sec:cris_results}
In this section, we present a numerical estimate of the cross sections 
for the reactions $e p \rightarrow \nu W X$ and $ep \rightarrow \nu W p$, 
calculated both exactly and in the EPA, in the range 
$ 100 \le  \sqrt s \le 2000 $ GeV.   We take $M_W = 80.42$ GeV for the 
mass of the $W$ boson and $G_F = 1.1664 \times 10^{-5} $ $\mathrm{GeV}^{-2}$ 
for the Fermi coupling constant \cite{data}. All the integrations 
are performed numerically. In the evaluation of   (\ref{w_siin}) and  
(\ref{gammain}) we assume the  LO Callan-Gross relation 
\eqref{eq:callangross}
and we use the ALLM97 parametrization of the proton structure function 
$F_2(x, Q^2)$ \cite{allm97}, which is expected to hold over
the range  of $x_B$ and $Q^2$ studied so far, namely 
$3 \times 10^{-6} < x_B < 0.85 $ 
and $ 0 \le Q^2 < 5000$ $\mathrm{GeV}^2$. As for the 
QED Compton scattering process discussed in the preceding chapters,  
we do not consider the resonance contribution 
separately but
we extend the ALLM97 
parametrization from the continuous ($W^2 > 3$ $\mathrm{GeV}^2$) down to
the resonance  domain ($(m_{\pi}+ m)^2<W^2< 3$ $\mathrm{GeV}^2$): 
in this way it is possible to agree with the experimental data averaged 
over each resonance. In our analysis, the average value of $x_B$ always lies
within the kinematical region mentioned above,  where the experimental 
data are available. 
On the contrary, the avarage value of $Q^2$  becomes larger
than $5000$ $\mathrm{GeV}^2$ when $\sqrt{s} \gtrsim 1200$ GeV, so we need to
extrapolate the  ALLM97 parametrization  beyond the region where the
data have been fitted.   
Our conclusions do not change if we utilize
a parametrization of $F_2(x_B, Q^2)$ whose behaviour at 
large $Q^2$ is constrained by the Altarelli-Parisi evolution
equations, like GRV98  \cite{grv98}.
The electric and magnetic form factors, necessary 
for the determination of the elastic cross sections in   (\ref{w_sigel})
and (\ref{w_epael}),
 are empirically parametrized as dipoles, according to \eqref{eq:emp_dipole}. 

At the HERA collider, where the electron and the proton beams 
have energy $E_e = 27.5 $ GeV and $E_p = 920$ GeV respectively, 
the cross section is dominated by the inelastic channel: 
$\sigma_{\mathrm{el}} = 2.47 \times 10^{-2} $ pb, while 
$\sigma_{\mathrm{inel}} = 3.22 \times 10^{-2} $ pb; therefore the expected 
integrated  luminosity of $200$ $\mathrm{pb}^{-1}$ would yield  a total 
of about 11 events/year.
\begin{figure}
\begin{center}
\epsfig{figure=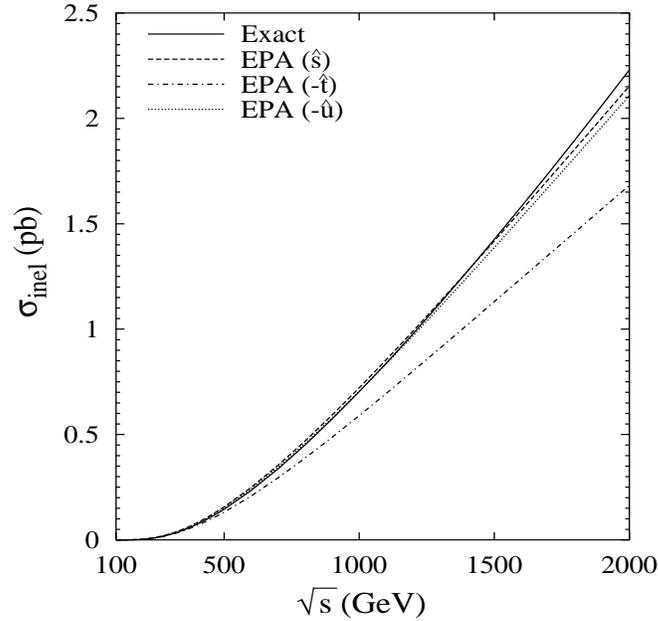, width = 9 cm,height = 8cm}  
\caption{Exact and approximated (EPA) inelastic cross sections of the process 
$ep\rightarrow \nu W X$ as functions of $\sqrt s$. The different scales  
utilized in the calculation of the approximated cross section are written
in the brackets.}
\end{center}
\label{figure2}
\end{figure}

Figure \ref{figure2}  shows a comparison of the inelastic cross 
section calculated
in the EPA, $\sigma^{\mathrm{EPA}}_{\mathrm{inel}}$, with the exact one, 
$\sigma_{\mathrm{inel}}$, as a function of $\sqrt s$, where several scales for 
$\sigma^{\mathrm{EPA}}_{\mathrm{inel}}$ are proposed, namely
$Q^2_{\mathrm{max}} = \hat s $,\, $-\hat u $, $-\hat t $ in  
(\ref{gammain}). It turns out 
that the choice of $-\hat t$ does not provide an adequate description
of $\sigma_{\mathrm{inel}}$, while $\hat s$ and $-\hat u$ are 
approximatively equivalent in reproducing $\sigma_{\mathrm{inel}}$.  
In particular, the choice of
 $- \hat u $ is slightly better in the  range
$ 300 \lesssim  \sqrt s \lesssim 1000$ GeV, while $\hat s $ guarantees 
a more accurate description of the exact cross section for 
$\sqrt s \gtrsim  1000$ GeV.  At HERA energies, 
$\sigma^{\mathrm{EPA}}_{\mathrm{inel}} = 3.64 \times 10^{-2}$ pb,
$3.51 \times 10^{-2}$ pb and
$3.07 \times 10^{-2}$ pb for $Q^2_{\mathrm{max}} = \hat s$, $-\hat u$ and
$-\hat t $, respectively.
In the following we will fix the scale to be 
$\hat s$, in analogy to our previous studies about the QED Compton 
scattering process in $ep \rightarrow e \gamma X$ \cite{pap1,pp2,pol}.
  
In \cite{pp2,pol} it was suggested that the experimental selection of only
those events for which $\hat s > Q^2$ restricts the kinematics of the
process to  the region of validity of the EPA and 
improves the  extraction the equivalent photon distribution from the 
exact cross section. 
\begin{figure}
\begin{center}
\epsfig{figure=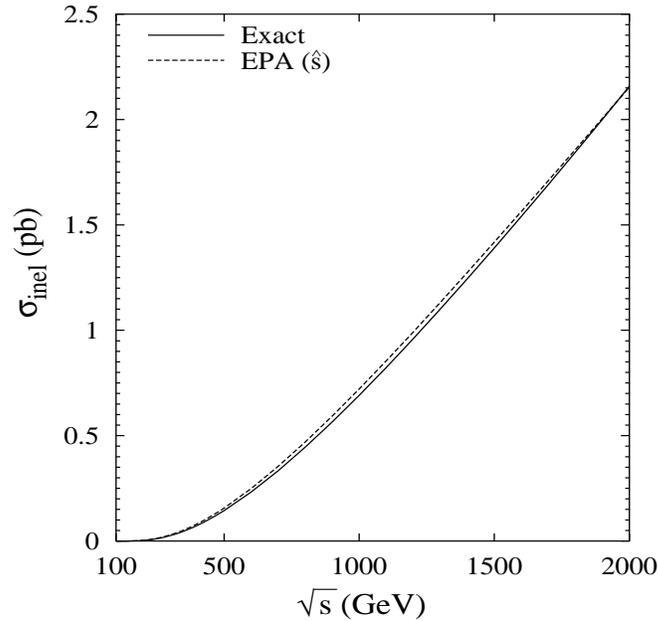, width = 9cm,height = 8cm}
\end{center}   
\caption{Exact and approximated (EPA) inelastic cross sections of the process 
$ep\rightarrow \nu W X$ as functions of $\sqrt s$. The scale $\hat s$  is 
utilized in the calculation of the approximated cross section and the 
kinematical cut $\hat  s > Q^2 $ is imposed in the exact one.}
\label{figure3}
\end{figure}
%
%
%
The effect of such a cut on the reaction 
$ep \rightarrow \nu W X $ is shown in Figure \ref{figure3}
 and the reduction of the 
discrepancy is evident at large $\sqrt s$, but   not
at  HERA energies, where  $\sigma_{\mathrm{el}} $ and 
$\sigma_{\mathrm{inel}} $ are unchanged. 
\begin{figure}
\begin{center}
\epsfig{figure=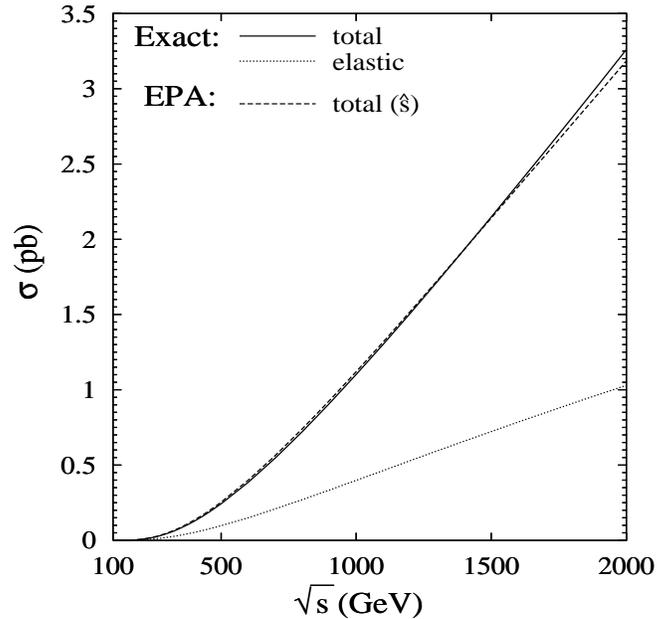, width = 9 cm,height = 8cm}  
\end{center}
\caption{
Exact and approximated (EPA) total ( = elastic +  inelastic) cross 
sections of the process 
$ep\rightarrow \nu W X$ as functions of $\sqrt s$.
The exact elastic component, which is indistinguishable from the approximated one,
is shown separately.}
\label{figure4}
\end{figure}

In Figure \ref{figure4} the total (elastic + inelastic) exact cross section 
is depicted
as a function of $\sqrt s$, together with the approximated one. 
Here the kinematical constraint $\hat s > Q^2$ is 
{\it not} imposed on the exact cross section.  The average discrepancy is 
reduced to be about $2 \%$, due to the inclusion
of the elastic channel, better described by the EPA (average 
discrepancy $ 0.05  \% $). 
The elastic component is also shown separately, and it agrees with 
the curve presented in Figure 3 of \cite{kniehl}. For $\sqrt s = 318$ GeV,
$\sigma^{\mathrm{EPA}}_{\mathrm{el}} = 2.47 \times 10^{-2}$ pb, in
perfect agreement with the exact value $\sigma_{\mathrm{el}}$. 

We compare now our results with the ones already published. In \cite{bohm},
taking into account the photon exchange only (Figure \ref{figure1}) and with no further 
approximation, fixing $M_W = 83.0$ GeV, 
$\sin ^2\theta_W = 0.217$, $E_e = 30$ GeV, 
$E_p = 820$ GeV and using the parton distributions \cite{owens} (Set 1),
together with the cuts $Q^2 > 4$  $\mathrm{GeV}^2$ and  $ W^2> 10$ 
$\mathrm{GeV}^2$, the value $\sigma_{\mathrm{inel}} = 3.0 \times 10^{-2}$ pb  
was obtained. This is  in contrast to 
$\sigma_{\mathrm{inel}} = 1.5 \times 10^{-2}$ pb, calculated using   
(\ref{w_siin}) with the same sets of cuts, values of the energies, $M_W$ and 
parton distributions utilized in \cite{bohm}. The authors of  \cite{bohm} 
also report the value $\sigma_{\mathrm{inel}} = 4.0 \times 10^{-2}$ pb, 
obtained in \cite{neufeld} with a similar analysis at the same energies,
using $M_W = 78$ GeV and $\sin ^2\theta_W = 0.217$. The lower limit on $Q^2$  
was taken to be ${\cal{O}}(1)$ $\mathrm{GeV}^2$, but not explicitly mentioned. 
Even with the ALLM97 parametrization, which allows us to use no cutoff
on $Q^2$,  we get  $\sigma_{\mathrm{inel}} = 3.1 \times 10^{-2}$ pb, 
far below  $4.0 \times 10^{-2}$ pb.  No analytical 
expression of the cross section is provided in \cite{neufeld,bohm}, which 
makes it difficult to understand the source of the discrepancies. 

Finally, an estimate of the the 
$ep \rightarrow \nu W X $ cross section is also given  in \cite{alt,gab}, 
utilizing an inelastic equivalent photon distribution   
slightly different
from  the one in   (\ref{gammain}), which can be written in the form
\be
\tilde{\gamma}^p_{\mathrm{inel}} (x, Q^2_{\mathrm{max}}) = \frac{\alpha}{2\pi} \int_x^1 
\der y\, F_2\bigg(\frac{x}{y}, \langle Q^2 \rangle \bigg ) 
~\frac{1 + (1-y)^2}{x\,y} ~\ln \frac{Q^2_{\mathrm{max}}}{Q^2_{\mathrm{cut}}}~,
\label{gabepa}
\ee
where
\be
\langle Q^2 \rangle = \frac{Q^2_{\mathrm{max}} - Q^2_{\mathrm{cut}}}{
\ln {\frac{Q^2_{\mathrm{max}}}{Q^2_{\mathrm{cut}} }}}~,
\ee
$ Q^2_{\mathrm{max}} = x_B s -M_W^2$ and $Q^2_{\mathrm{cut}} = 1$ 
$\mathrm{GeV}^2$. Equation  (\ref{gabepa})
can be obtained from   (\ref{gammain}) neglecting the mass term and 
approximating the integration over $Q^2$. In the 
calculation performed in \cite{gab,alt},  $\tilde{\gamma}^p_{\mathrm{inel}} 
(x, Q^2_{\mathrm{max}})$ is convoluted with the differential cross section
for the real photoproduction process in   (\ref{w_realsi}).  
At $\sqrt s = 300$ GeV, fixing $M_W = 84$ GeV,  $\sin ^2\theta_W = 0.217$
and using the parton distribution parametrization 
\cite{owens} (Set 1), we get $\sigma_{\mathrm{inel}} = 1.6 \times 10^{-2}$ pb, 
very close to the value $ 1.5 \times 10^{-2}$ pb published in \cite{alt,gab}.

\section{Summary}
\label{sec:cris_summary}
We have calculated the cross section for the inelastic  
process $ep\rightarrow \nu W X$,  both exactly and using the 
equivalent photon approximation of the proton, in
order to test its accuracy in the inelastic channnel and complete the 
study  initiated in \cite{kniehl}, limited to the  elastic process 
$ep\rightarrow \nu W p$. The relative error of the approximated result
with respect to the exact one is scale dependent; fixing 
the scale to be $\hat s$, it decreases from about $10 \%$ at HERA energies
down to $0.5 \%$  for $\sqrt s = 1500$ GeV, then it slightly increases up 
to $3\%$ for $\sqrt s = 2000$ GeV.  In conclusion, even if not so 
remarkable as for the elastic channel, in which the deviation is always 
below one percent \cite{kniehl}, the approximation can be considered 
quite satisfactory.
We have  compared our calculations with previous ones in the literature
and found that they are in agreement with \cite{alt,gab}, but
disagree with \cite{neufeld,bohm}. Furthermore,
we have estimated the total number of $\nu W$ events expected at the HERA 
collider, including the elastic  and quasi-elastic channels of the 
reaction. The production rate turns out to be quite small, 
about 11 events/year, 
assuming a luminosity of 200 $\mathrm{pb}^{-1}$, but the process 
could still be detected.

\clearemptydoublepage

\chapter{\bf{Summary and Conclusions}}

In a photoproduction reaction initiated by  a nucleon $N$ ($= p,$ $n$), 
the nucleon can be considered to be equivalent to a beam of photons, 
whose distribution can be computed theoretically. 
It was shown, see \eqref{eq:unpol}  and  \eqref{factor:pol}, 
that the cross section of the 
reaction  factorizes, being given by the convolution of  
the universal (process independent)  photon distribution of 
the nucleon, with the corresponding  photon-nucleon cross section.\\
The advantage of this model is that the nucleon's photon content   
can be utilized  to simplify the
calculations  of  photon-induced subprocesses in elastic and deep inelastic 
$eN$ and $NN$ collisions, commonly described in terms of the 
electromagnetic form factors,
structure functions
and  parton distributions $(\Delta) f(x, \mu^2)$
(with $f = q$, $\bar q$, $g$) introduced in Chapter 2.

The polarized and unpolarized photon distributions $(\Delta)\gamma(x, \mu^2)$ 
were evaluated in the equivalent photon approximation (EPA) in Chapter 3. 
Both of them consist of two components: 
the elastic ones, $(\Delta)\gamma_{\rm el}$, due to $N\rightarrow \gamma N $, 
and the inelastic ones, $(\Delta)\gamma_{\rm inel}$,
due to $N\rightarrow \gamma X$, with $X \neq N$. 
The elastic photon components turned out to be scale independent
and uniquely determined by the  electromagnetic form factors. 
The inelastic photon components  were radiatively generated in a model  
where they vanish at a low resolution scale $\mu_0^2$.
At large scales $\mu^2$ the resulting distributions
are rather insensitive to details at the input $\mu_0^2$, therefore
such not compelling vanishing boundary conditions  are supposed to
yield reasonable results for $(\Delta )\gamma_{\rm inel}(x, \mu^2)$.
However at the low scales characteristic   of fixed target experiments,
like HERMES at DESY, 
$(\Delta)\gamma_{\rm {inel}}(x,\mu^2)$ 
depend obviously
on the assumed details at  $\mu_0^2$.
We did not investigate any effect due to 
$\gamma_{\mathrm{inel}}(x,\mu_0^2) \ne 0$;
this should rather be examined experimentally if our expectations 
based on the  vanishing boundaries  turn out to be
in disagreement with observations.

Chapter 4 was devoted to demonstrate the feasibility of the
measurement of $(\Delta)\gamma (x, \mu^2)$.
We  studied muon pair production in electron-nucleon 
collisions $e N \to e \mu^+ \mu^- X$  via the subprocess 
$\gamma^e \gamma^N \to \mu^+ \mu^-$ and the QED Compton process
$e N \to e \gamma X$ via the subprocess 
$e \gamma^N \to e \gamma$ for the HERA collider 
and the polarized and unpolarized fixed target HERMES experiments. 
The muon pair  production process
was evaluated in the leading order equivalent 
photon approximation, hence we  considered just the simple $2\to 2$ 
subprocess \mbox{$\gamma^e\gamma^N\to \mu^+\mu^-$}, instead
of  the full $2\to 3$ or (even more involved) $2\to 4$ subprocesses 
$\gamma^e q\rightarrow \mu^+\mu^-q$ and $e q\to \mu^+\mu^-e q$. 
Similarly, the analysis of 
the deep inelastic QED Compton
process was
reduced to the study of the  $e\gamma^N\to e\gamma$ scattering
instead of  $eq\to e\gamma q$.
We concluded that the  production rates of lepton-photon and dimuon pairs
would be sufficient to 
facilitate the extraction of the polarized and unpolarized
photon distributions  in the available kinematical regions.

The   full $2 \rightarrow 3$ QED
Compton process  in $e p\rightarrow e\gamma p$
and  $ep \rightarrow e\gamma X$, with unpolarized incoming electron 
and proton, 
was calculated in a manifestly covariant way in Chapter 5 by employing 
an appropriate 
parametrization of the structure function $F_{2}$ \cite{allm97} and
assuming the Callan-Gross relation  \eqref{eq:callan} for $F_L$.
These ``exact'' results were compared with the aforementioned
 ones based on the EPA,
as well as with the experimental data  and theoretical estimates for the 
HERA collider given in \cite{thesis}.
Although the cross section in the elastic channel is
accurately described by the EPA (within $1\%$), this is 
not the case in the inelastic channel.
It turned out that the agreement with the EPA is slightly better in 
$x_{\gamma}$  bins, where $x_{\gamma}$, defined
in \eqref{eq:icsgamma}, is the fraction of the 
longitudinal momentum of the proton carried by the virtual photon, compared 
to the bins in the leptonic variable $x_l$ \eqref{xl},  
used in \cite{thesis}.   
In addition the  results obtained by an 
iterative approximation procedure \cite{kessler,h1}, commonly 
used \cite{thesis}, were found to deviate  from our analysis in certain 
kinematical regions.

The virtual Compton scattering in unpolarized  
  $e p \rightarrow e\gamma p $ and 
$e p \rightarrow e \gamma X$, where the photon is emitted from the 
hadronic vertex, was  studied Chapter 6. 
It represents the major background to the 
QED Compton scattering.
New kinematical cuts were suggested in order to suppress the virtual Compton  
background and improve the extraction of the equivalent photon content of 
the proton at the HERA collider. The total (elastic + inelastic) discrepancy 
of the exact cross section with the approximate one was reduced to be 
about $2\%$, which should be compared with the value $14\%$ relative to 
the cuts \cite{thesis} discussed in Chapter 5.

The $2\to 3$ QED Compton   process in longitudinally 
polarized lepton-proton scattering was  analysed
in Chapter 7, both in its elastic and inelastic channels.
For our numerical estimates we utilized the BKZ parametrization \cite{bad}  
of
the spin dependent structure function $g_1$ 
and we neglected
$g_2$. It has to be noted that the effects of $g_2$, as well as of $F_L$
in the unpolarized cross section, may become relevant at low scales
and it would be interesting to take them into account in future studies. 
The kinematic cuts  necessary to extract the polarized 
photon content of the proton  and to
suppress the major background  coming from 
virtual Compton scattering were provided for  HERMES, COMPASS and the 
future eRHIC experiments.
For the  HERMES and eRHIC kinematics, we found 
that the total cross section can be described by the EPA with
an error estimated to be less than $10\%$. 

The reliability of the EPA in reproducing the cross section of the
process $ep\to \nu W X$ was investigated in Chapter 8. 
In order to examine this issue, both the subprocesses
$e\gamma^p\to\nu W$ and  $e q\rightarrow \nu W q$  were studied. 
The relative error of the approximate result with respect to the 
exact one was found to be less than $10 \%$ in the inelastic channel 
and less than 
$1\%$ in the elastic one, this last value being 
in agreement with  \cite{kniehl}.

To conclude, we suggested how to access the photon 
distributions $(\Delta)\gamma(x, \mu^2)$ in $eN$ collisions, 
studying the  accuracy of the equivalent photon approximation of the nucleon
in describing different processes and trying to  identify the  kinematical 
regions of its validity. 
Our findings should now be confronted with experiments.
As already mentioned,  these measurements would not only be interesting 
on their own, but would provide 
the opportunity of getting additional and independent  informations 
concerning the  structure functions $F_{1,2}$ and $g_{1,2}$, underlying
the (inelastic) photon distributions. 
In particular, being  
the kinematics of QED Compton events  
different from the one of inclusive deep inelastic scattering, 
due to the radiated photon in the final state,  
it provides a novel 
way to  access the structure functions in a kinematical region not 
well covered by 
inclusive measurements \cite{blu,gpr2,pol,thesis,lend,f2h1}.
Hence it represents a valuable 
complementary tool to
have a complete understanding of the  structure of nucleons.

\clearemptydoublepage

\begin{appendix}

\chapter{\bf{Notations and Conventions}}
\label{app:notations}

{\bf Units} 

Natural units $\hbar=c=1$, with $\hbar=h/{2\pi}$, are used throughout this
thesis, where $h$ and $c$ denote the Planck constant
and the speed of light, respectively.\\
  
\noindent
{\bf Relativistic conventions} 

The metric tensor $g_{\alpha\beta}=g^{\alpha\beta}$, with $\alpha, \beta = 0, 1, 2, 3$, 
is given by
\begin{equation}
g^{00} = +1,~~~g^{11}=g^{22}=g^{33}=-1,~~~{\mathrm{otherwise}} = 0.
\end{equation}
The contravariant vectors of the space-time coordinate and energy-momentum
of a particle of mass $m$ and energy $E$ are
given by
\be
x^{\alpha} = (t, {\mbox{\boldmath $x$}}), ~~~~~~~~~~
p^{\alpha} = (E, {\mbox{\boldmath $p$}}),   
\ee
where 
\be
E = \sqrt{{\mbox{\boldmath $p$}}^2 + m^2}~.
\ee
The covariant vectors are
\be
x_{\alpha} & = & g_{\alpha\beta}x^{\beta} = (t, -{\mbox{\boldmath $x$}}),\\
p_{\alpha} & = & g_{\alpha\beta}p^{\beta} = (E, -{\mbox{\boldmath $p$}}),
\ee
and hence their scalar product is defined by
\be
x\cdot p = x_{\alpha}p^{\alpha} = g_{\alpha\beta}x^{\alpha}p^{\beta} = tE -{\mbox{\boldmath $x p$}},
\ee
where it is understood that repeated indices are summed.
Furthermore the totally antisymmetric tensor $\varepsilon^{\alpha\beta\rho\sigma}=
 -\varepsilon_{\alpha\beta\rho\sigma}$ is defined so that 
\be
\varepsilon_{0123} = +1.
\ee

\noindent
{\mbox{\boldmath $\gamma$}} {\bf matrices}

The Dirac $\gamma$ matrices $\gamma^{\alpha}= (\gamma^0, \gamma^i)$, 
with $i = 1, 2, 3$, satisfy the anticommutation relation
\be
\gamma^{\alpha}\gamma^{\beta} + \gamma^{\beta}\gamma^{\alpha} = 2 g^{\alpha\beta}.
\ee 
The matrix $\gamma^5$ is defined as
\be
\gamma^5 = \gamma_5 = {i}\, \gamma^0\gamma^1\gamma^2\gamma^3.
\label{eq:gamma5}
\ee
The hermitian conjugate of $\gamma^{\alpha}$ is taken to be
\be
\gamma^{\alpha \dag} = \gamma^0\gamma^{\alpha}\gamma^0,
\ee
so that according to the definition  \eqref{eq:gamma5} one has
\be
\gamma^{5\dag} = \gamma^5.
\ee
The scalar product of the $\gamma$ matrices and any four-vector $A$ is
defined as
\be
A\!\!\!/ = \gamma^{\alpha}A_{\alpha} = \gamma^0A^0 -\gamma^1A^1-\gamma^2A^2 
-\gamma^3A^3.
\ee

\noindent
{\bf Spinors and normalizations}

The normalization of the one particle state $| p, \alpha \rangle$ with momentum
$p$ and other quantum numbers $\alpha$ is taken to be
\be
\langle p, \alpha \,| \,p' \alpha' \rangle = (2\pi)^3 2 E\delta^3  ({\mbox{\boldmath $p$}} -{\mbox{\boldmath $p$}}')  \delta_{\alpha\alpha'}. 
\label{eq:norma}
\ee
The Dirac spinors for fermion, $u(p, s)$, and antifermion, 
$v(p, s)$, with momentum $p$, spin  $s$ and
mass $m$ satisfy
\be
0 & = & (p\!\!\!/-m)u(p, s) = \bar{u}(p, s)(p\!\!\!/-m)\label{eq:u} \\
  & = & (p\!\!\!/+m)v(p,s) = \bar{v}(p, s)(p\!\!\!/+m),\label{eq:v}
\ee
where  $\bar{u}(p, s) = u^{\dag}(p, s)\gamma^0$. 
The Dirac spinors   are normalized in such a way that 
\begin{equation}
\bar{u}(p, s)u(p, s') = 2m\delta_{ss'},~~~~~
\bar{v}(p, s){v}(p, s') =-2m\delta_{ss'},
\label{eq:norm_spinor}
\end{equation}
which hold also for  massless particles and antiparticles. 
Accordingly the completeness relations read
\be 
\sum_{s}{u}(p,s)\bar{u}(p, s) = (p\!\!\!/+m),\label{eq:compl_u}\\
\sum_{s}{v}(p, s)\bar{v}(p, s) = (p\!\!\!/-m).\label{eq:compl_v}
\ee

\noindent
{\bf Spin projection operators}

The polarization vector  $s$ of a relativistic spin $\frac{1}{2}$ particle or
antiparticle with momentum $p$ is a pseudovector which fulfils 
\be
s^2 = -1,~~~~~~s\cdot p = 0.
\ee  
The projection operators onto a state with polarization $\pm s$ are known to be
\be
{\cal{P}}(\pm s)& =& \frac{1}{2}\,(1 \pm \gamma^5s\!\!\!/)~~~~~~   
{\rm for~ particles},\label{eq:spin_prj}\\
{\cal{P}}(\pm s)& = &\frac{1}{2}\,(1\mp\gamma^5s\!\!\!/)~~~~~~   
{\rm for~antiparticles}.
\ee
For massless particles and antipartcles only longitudinal polarization 
is possible and the operators introduced above become {\em helicity} 
projectors: 
\be
{\cal{P}}_{\pm} &= &\frac{1}{2}\,(1 \pm \gamma^5)~~~~~~   
{\rm for~ particles},\label{eq:hel1}\\
{\cal{P}}_{\pm}& = &\frac{1}{2}\,(1\mp\gamma^5)~~~~~~   
{\rm for ~antiparticles},\label{eq:hel2}
\ee
where the subscript $+$ indicates positive helicity, that is spin parallel to
the momentum, and the subscript $-$ indicates negative helicity, that is spin
antiparallel to the momentum. 

\newpage

\clearemptydoublepage
\chapter{\bf{Photon-induced Cross Sections}}

The well-known  unpolarized and polarized cross sections 
for  dimuon production in $\gamma\gamma\to\mu^+\mu^-$ 
and  for the Compton scattering reaction $e\gamma\to e\gamma$ 
are recalculated in a manifestly covariant way in  Appendices  \ref{app:dimuon}
and \ref{app:compton} respectively.  

\section{Dimuon Production}
\label{app:dimuon}

The process of photon-photon annihilation into  a muon-antimuon pair
\begin{equation}
\gamma(k_1)+ \gamma(k_2) \to \mu^+(l_1)+ \mu^-(l_2)
\end{equation}
is described to lowest order in QED by the two Feynman diagrams in 
Figure \ref{muon}.
\begin{figure}[h]
\vspace{0.5cm} 
\begin{center} 
\epsfig{figure=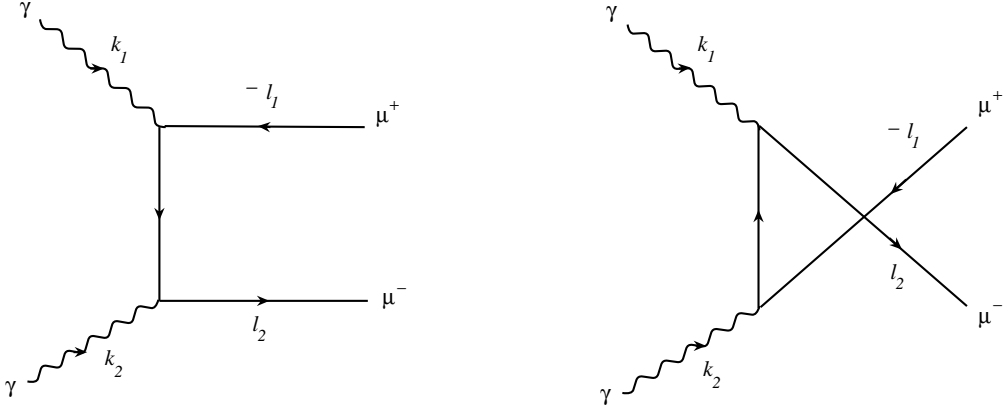, width= 14cm} 
\vspace{-0.5cm}
\caption{Feynman diagrams for $\gamma\gamma\to\mu^+\mu^-$.}
\label{muon}
\end{center}
\end{figure} 

\noindent
We define the  Mandelstam variables 
\begin{equation}
\hat{s} = (k_1+k_2)^2,~~~\hat t = (k_1-l_1)^2,~~~\hat u = (k_2-l_1)^2,
\label{eq:mandel}
\end{equation}
which satisfy the constraint
\begin{equation}
\hat s + \hat t + \hat u = 0,
\label{eq:massless}
\end{equation} 
as the masses of the particles are neglected. 
The scattering amplitude reads 
\begin{eqnarray}
\hat M = - e^2\epsilon^{\lambda_1}_\alpha (k_1)\epsilon^{\lambda_2}_\beta (k_2)\bar u(l_2) \bigg [ \,  \frac{1}{(k_1-l_1)^2}\,\gamma^\beta 
(k_1\!\!\!\!\!/-l_1\!\!\!\!/)\gamma^\alpha + \frac{1}{(k_2-l_1)^2}\,
\gamma^\alpha (k_2\!\!\!\!\!/ -l_1\!\!\!\!/)\gamma^\beta      \,\bigg ]v(l_1),
\nonumber\\ 
\label{eq:amplitude_mu}
\end{eqnarray}  
where $e$ is the proton charge and  $\epsilon^{\lambda_i}_{\alpha}(k_i)$, 
with $i=1,$ $2$, denotes the 
polarization vector of an incoming photon with momentum $k_i$ and helicity 
$\lambda_i$. The Dirac spinors, whose spin dependence is not shown
explicitely here, were introduced in \eqref{eq:u} and \eqref{eq:v}. \\

\noindent
{ \bf Unpolarized cross section}

Making use of the completeness relations \eqref{eq:compl_u}, \eqref{eq:compl_v}
and
\be
\sum_{\lambda}\epsilon^{\lambda}_{\alpha}(k)\epsilon^{\lambda}_{\beta}(k)^*
= -g_{\alpha\beta},
\label{eq:compl_g}
\ee
one can perform the spin sums and get 
\begin{eqnarray}
\overline{|\hat M|^2} &=& \frac{1}{4}\sum_{\rm{spins}} |\hat M|^2
  = \frac{1}{2} \, \bigg (|\hat{M}_{++}|^2 +|\hat{M}_{+-}|^2 \bigg )
  \nonumber \\
    &=& e^4\bigg \{\bigg [ \frac{1}{(k_1-l_1)^4}+\frac{1}{(k_2-l_1)^4}\bigg ]\,
    {\rm{Tr}} [\,{l_2\!\!\!\!/} \,{k_1\!\!\!\!\!/} \, {l_1\!\!\!\!/} \, 
      {k_1\!\!\!\!\!/} \,] -4 \, \frac{k_1\cdot k_2 -l_1\cdot k_2 -
        l_1\cdot k_1 }{(k_1-l_1)^2
         (k_2-l_1)^2}\,\mathrm{Tr}[\,l_2\!\!\!\!/\, 
          l_1\!\!\!\!/\,]\,\bigg \},\nonumber\\
\label{eq:tracmuon}
\end{eqnarray}
with the subscripts $\pm$ referring to the two possible 
values $\pm 1$ of the helicity $\lambda_i$ of each photon.  
The traces can be easily calculated and expressed in terms of the variables 
\eqref{eq:mandel}, so that \eqref{eq:tracmuon} reduces to the simple form
\begin{eqnarray}
\overline{|\hat{M}|^2}  =  2e^4 \bigg (\frac{\hat t}{\hat u} + \frac{\hat u}
{\hat t} \bigg ),
\label{eq:amplit}
\end{eqnarray}
the last (interference) term in \eqref{eq:tracmuon} being zero due to 
\eqref{eq:massless}. From the definition \eqref{eq:dsdt2}, that is
\begin{equation}
\frac{\der\hat\sigma}{\der\hat t} = \frac{1}{16\pi\hat s^2}
\,\overline{ |\hat{M}|^2},
\label{eq:dsdt}
\end{equation}
one gets the final result for the unpolarized cross section:
\begin{equation}
\frac{\der\hat\sigma}{\der\hat t}=  \frac{2\pi\alpha^2}{\hat s^2} 
\bigg (\frac{\hat t}{\hat u} + \frac{\hat u}{\hat t} \bigg ). 
\end{equation}

\noindent
{\bf Polarized cross section}

It can be shown (see discussion below \eqref{eq:pol_final}) 
that the two photons annihilate into a muon-antimuon pair  only if 
they have opposite helicities, that is
\begin{equation}
|\hat {M}_{++}|^2 = 0 ,
\label{eq:ampl_zero}
\end{equation} 
which, substituted in the first line of \eqref{eq:tracmuon}, together with
\eqref{eq:amplit}, gives the following result
\begin{eqnarray}
|\Delta \hat{M}|^2  = \frac{1}{2}\,\bigg (|\hat{M}_{++}|^2 -|\hat{M}_{+-}|^2 
\bigg )=  
-2e^4 \bigg (\frac{\hat t}{\hat u} + \frac{\hat u}{\hat t} \bigg ),
\label{eq:pol_ampl}
\end{eqnarray}
needed for the calculation of the polarized cross section,
\begin{equation}
\frac{\der\Delta\hat\sigma}{\der\hat t} =\frac{1}{16\pi\hat s^2}\,
{ |\Delta\hat{M}|^2}.
\label{eq:pol_sigma}
\end{equation}
Substituting  \eqref{eq:pol_ampl} into \eqref{eq:pol_sigma}, one obtains
\begin{equation}
\frac{\der\Delta\hat\sigma}{\der\hat t} = -\frac{\der\hat\sigma}{\der\hat t}
= -\frac{2\pi\alpha^2}{\hat s^2} 
\bigg (\frac{\hat t}{\hat u} + \frac{\hat u}{\hat t} \bigg ).
\label{eq:pol_final}
\end{equation}

\small
To demonstrate the validity of \eqref{eq:ampl_zero}, one has to inspect
the helicity structure of the amplitude \eqref{eq:amplitude_mu}. Making
use of the helicity projection operators \eqref{eq:hel1}, \eqref{eq:hel2}, 
one can perform the following substitutions in \eqref{eq:amplitude_mu}
\be
\bar{u}(l_2)&\longrightarrow& \bar{u}_{\pm}(l_2) =\frac{1}{2}\,\bar{u}(l_2)(1\mp\gamma^5),
\nonumber\\
v(l_1)&\longrightarrow & v_{\pm}(l_1) = \frac{1}{2}\,(1\mp\gamma^5)\,v(l_1), 
\ee
where the subscripts on the spinors denote the helicities of the
corresponding fermions. In this way one notices that, when the helicities 
of the fermions are the same, the amplitude vanishes because of a 
mismatch in the projection operators $(1\pm \gamma^5)/2$. Hence
\be
\hat{M}_{++} = \hat{M}_{++;+-} + \hat{M}_{++;-+}~,
\label{eq:helicities}
\ee
and, in particular, 
\begin{eqnarray}
\hat{M}_{++;+-} & = & - \frac{1}{2}\,e^2\bar u(l_2)(1-\gamma^5) \bigg 
[ \,\frac{1}{(k_1-l_1)^2}\,{\epsilon \!\! /}^+(k_2) 
(k_1\!\!\!\!\!/-l_1\!\!\!\!/){\epsilon \!\! /}^+(k_1) \nonumber \\ 
& & ~~~~~+ \frac{1}{(k_2-l_1)^2}\,
{\epsilon \!\! /}^+(k_1) (k_2\!\!\!\!\!/ -l_1\!\!\!\!/)
{\epsilon \!\! /}^+(k_2)        \,\bigg ]v(l_1).
\label{eq:hell}
\end{eqnarray}  
It is possible to introduce a  representation for the polarization states
of the photons without choosing any specific frame. This means that the 
calculation of the helicity amplitudes in \eqref{eq:helicities} can be   
performed in a covariant way. 
The circularly polarized states of a photon with four-momentum $k$ 
can be expressed as
\be
\epsilon_{\mu}^{\pm} =\mp \frac{1}{\sqrt{2}}\,(\epsilon^{||}_{\mu}  
\pm i\epsilon^{\perp}_{\mu}),
\ee
where the vectors $\epsilon^{||}$ and $\epsilon^{\perp}$ satisfy the 
conditions
\be
(\epsilon^{||})^2 & = & (\epsilon^{\perp})^2 = -1,\label{eq:norm} \\
k \cdot \epsilon^{||} & = & k\cdot \epsilon^{\perp} = 
\epsilon^{||}\cdot \epsilon^{\perp} = 0.
\ee
If we introduce two arbitrary vectors, $p$ and $q$, for convenience  taken 
to be light-like ($p^2= q^2= 0$), then we can write  
\be
\epsilon^{||}_{\mu} & = &2\sqrt{2}N[\,(q\cdot k)p_{\mu} -(p\cdot k)q_{\mu}],\\
\epsilon^{\perp}_{\mu} & = & 
2\sqrt{2}N\varepsilon_{\mu\alpha\beta\gamma}q^{\alpha}
p^{\beta}k^{\gamma},
\ee
where the normalization factor, which is the same for both polarization 
vectors, is fixed by \eqref{eq:norm}  to be
\be
N = \frac{1}{4}\,[(p\cdot q)(p\cdot k)(q\cdot k)]^{-\frac{1}{2}}.
\ee
In  the amplitude \eqref{eq:amplitude_mu} the photon polarization vectors only 
appear in the combination $\epsilon \!\!/$; using the identity  
\be
i\gamma^{\mu}\epsilon_{\mu\alpha\beta\gamma} = (\gamma_{\alpha}\gamma_{\beta}
\gamma_{\gamma}-\gamma_{\alpha}g_{\beta\gamma}+\gamma_{\beta}g_{\alpha\gamma}
-\gamma_{\gamma}g_{\alpha\beta}){\gamma^5},
\ee
one can write \cite{Gastmans:1990xh}: 
\be
\epsilon^{\pm}\!\!\!\!\!\!/\,\, = \pm N [\, k\!\!\!/ \,p\!\!\!/ \,q\!\!\!/\,(1\pm \gamma^5)
-p\!\!\!/\,q\!\!\!/\,k\!\!\!/\,(1\mp \gamma^5) \mp 2 (p\cdot q)k\!\!\!/\,
\gamma^5\,].
\ee
The choice of the four-vectors $p$ and $q$ is  arbitrary.
When the photon line is next to an external fermion or antifermion  line,
one usually  exploits this freedom of choice and takes either $p$ or $q$ equal
to the fermion momentum. Use of the  equations \eqref{eq:u}, 
\eqref{eq:v}  for massless fermions  leads to great simplifications. 
For the process under study we fix $p = l_2$ and $q = l_1$, so that
\be
{\epsilon \!\!/}^{+}(k_i) = N\,
 [\, k_i\!\!\!\!/ \,\,l_2\!\!\!\!\!/ \,\,\,l_1\!\!\!\!\!/\,\,(1
+ \gamma^5)
-l_2\!\!\!\!\!/\,\,\,l_1\!\!\!\!\!/\,\,k_i\!\!\!\!\!/\,\,(1- \gamma^5) 
-2(l_1\cdot l_2) k_i\!\!\!\!/ \,\,\gamma^5 ],~~~
\label{eq:polfinal}
\ee
with $i = 1,$ $2$  and $ N  = (2\hat s \,\hat t \,\hat u)^{-\frac{1}{2}}$. 
The factor $(1-\gamma^5)$  in \eqref{eq:hell} only selects the first
and the third term in the RHS of  \eqref{eq:polfinal}, but the first one
does not contribute because $l_1\!\!\!\!\!/\,\,v(l_1)= 0$. Therefore  
\be
\hat{M}_{++;+-} & = & -\frac{1}{2}\,e^2N^2\hat s^2\bar{u}(l_2)(1-\gamma^5)
\bigg 
[ \,\frac{k_2\!\!\!\!\!/\,\, (k_1\!\!\!\!\!/-l_1\!\!\!\!/)k_1\!\!\!\!\!/\,\,
}{(k_1-l_1)^2} + \frac{k_1\!\!\!\!\!/\,\, (k_2\!\!\!\!\!/ -l_1\!\!\!\!/)
 k_2\!\!\!\!\!/}{(k_2-l_1)^2} \,\bigg ]v(l_1) 
\nonumber \\
& = &-\frac{1}{2}\,e^2N^2\hat s^2 \bar{u}(l_2)(1-\gamma^5)
({l_2\!\!\!\!\!/}\,\, +{ l_1\!\!\!\!\!/}\,\,\,)\,v(l_1)\nonumber \\
 & = &  0,
\label{eq:ampl_0}
\ee
where the last equality follows from $\bar u(l_2)l_2\!\!\!\!\!/\,\,= 0$ 
and
$l_1\!\!\!\!\!/\,\,v(l_1)= 0$.
Analogously one can show that $\hat M_{++;-+} =0$, which, together
with \eqref{eq:ampl_0},  gives \eqref{eq:ampl_zero}. 

\normalsize

\section{Compton Scattering}
\label{app:compton}

The Compton scattering process
\begin{equation}
e(l) + \gamma (k) \to e(l') + \gamma(k')
\end{equation}
can be described in terms of the Mandelstam variables 
\begin{equation}
\hat{s} = (l+k)^2,~~~\hat t = (l -l')^2,~~~\hat u = (k-l')^2,
\label{eq:mandel2}
\end{equation}
which satisfy \eqref{eq:massless}, since the photons are taken to
be on shell and the electron mass is neglected.
The corresponding amplitude to lowest order in QED reads, see Figure 
\ref{compton},
\begin{eqnarray}
\hat{M}&= & 
e^2\epsilon_{\alpha}\epsilon_{\beta}^* \,\bar{u}(l')
  \bigg [\gamma^{\beta}\frac{l\!\!/+k\!\!\!/}{(l+k)^2}\,\gamma^{\alpha} 
    +\gamma^{\alpha}
        \frac{l\!\!/-k'\!\!\!\!/}{(l-k')^2}\,\gamma^{\beta} \bigg ] 
         \frac{1}{2}\, (1+\gamma^5 ) \,u(l)\nonumber \\
           & \equiv &\epsilon_{\alpha}\hat {M}^{\alpha},
\end{eqnarray}
where $\epsilon$, $\epsilon^*$ are the polarization vectors of the incoming
and outgoing photons; $u$, $\bar{u}$ are the Dirac spinors of the 
initial and final electrons respectively. We have assumed that the incoming 
electron has right-handed helicity, i.e. its spin is  parallel to 
the direction of motion. 

\begin{figure}[h]
\vspace{0.5cm}
\begin{center} 
\epsfig{figure=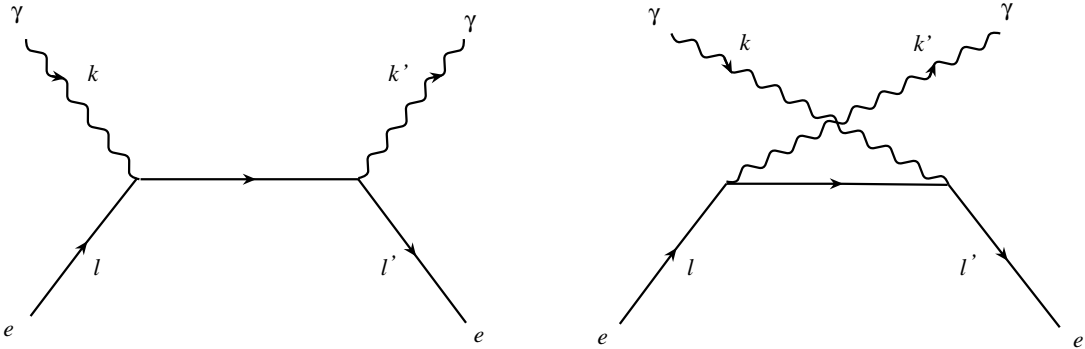, width= 15cm} 
\vspace{-1.5cm}
\caption{Feynman diagrams for $e\gamma\to e\gamma$.}
\label{compton}
\end{center}
\end{figure} 

\noindent
The leptonic tensor is given by
\begin{eqnarray}
T_{\alpha\beta} = \sum_{\mathrm{spins}} \hat {{M}}_{\alpha}^*\hat{M}_{\beta}
     = T_{\alpha\beta}^{\rm S} + T_{\alpha\beta}^{\rm A},
\end{eqnarray}
$T_{\alpha\beta}^S$  and $T_{\alpha\beta}^A$ being its symmetric and antisymmetric parts.
Using the completeness relations \eqref{eq:compl_u}, \eqref{eq:compl_g}   
one finds
\begin{eqnarray}
T_{\alpha\beta}(l;l',k') &=& e^4\,\mathrm{Tr}\bigg \{ \frac{1}{(l'+k')^2}\,[(1+\gamma^5)l\!\!/ \,\gamma_{\alpha}\,k' \!\!\!\!\!/\,\, \gamma_{\beta}\,] 
    - \frac{1}{(l-k')^2} \,[(1+\gamma^5)\,k'\!\!\!\!\!/\,\,\gamma_{\alpha}\,l'\!\!\!\!/\,\gamma_\beta \,] \nonumber \\
     & & ~~~~~~+ ~\frac{1}{(l'+k')^2(l-k')^2}[(1+\gamma^5)\,
      l'\!\!\!\!/\,l\!\!/ \gamma_{\beta}(l'\!\!\!\!/\,l\!\!/-l'\!\!\!\!/
       \,k'\!\!\!\!/ +k'\!\!\!\!/\,        l\!\!/)\gamma_{\alpha} 
\nonumber \\
       && ~~~~~~~~~~~+~ (1+\gamma^5)\,l'\!\!\!\!/\,\gamma_{\beta}\,(l\!\!/l'
\!\!\!\!/-k'\!\!\!\!/\,l'\!\!\!\!/+ l\!\!/\,k'\!\!\!\!/\,)
\,\gamma_\alpha\,l\!\!/\,]
\bigg \},
\label{eq:leptonict}
\end{eqnarray}
which is valid also for an incoming virtual photon.\\

\noindent
{\bf Unpolarized cross section}

In terms of the variables \eqref{eq:mandel2},
the symmetric part of the leptonic tensor reads, for real photons,
\be
T_{\alpha \beta}^{\rm S}(l;l',k')& =&\,\frac{4 e^4} {\hat{s}\hat{u}}\,\bigg
\{ \,\frac{1}{2}\,
g_{\alpha\beta}\,(\hat s^2 + \hat u^2) + 2\hat s\, l_{\alpha}l_{\beta}
+ 2 \hat u\,l'_{\alpha}l'_{\beta}~~~~~~~~~~~~~~~~~~\nonumber \\
&&~~~~~~~~~~~~~~~~+~\hat t\,(l_{\alpha}l'_{\beta}+l_{\beta} l'_{\alpha}) -{\hat s}\,
(l_{\alpha}k'_{\beta}+l_{\beta}k'_{\alpha})  \nonumber \\ 
&&~~~~~~~~~~~~~~~~~~~~~~~~~~~~~~~~~~~~~~~~~~+~
\hat u  \, (l'_{\alpha}k'_{\beta} +l'_{\beta}k'_{\alpha})\bigg\}.
\label{leptoo}    
\ee
From \eqref{leptoo}, one can determine the amplitude squared and averaged 
over the spins of the incoming photon
\begin{eqnarray}
\overline{|\hat{M}|^2}  &=& \frac{1}{2}\sum_{\rm{spins}} 
\epsilon_{\alpha}\epsilon_{\beta}^* T^{\alpha\beta} = -\frac{1}{2} 
   \,g^{\alpha\beta}T_{\alpha\beta}^{\rm S}\nonumber \\
&= & -2e^4\bigg (\frac{\hat s}{\hat u} + \frac{\hat u}{\hat s} \bigg), 
\end{eqnarray}
hence  the cross section \eqref{eq:dsdt} for the process
$e \gamma \to e \gamma$ is given by
\be
\frac{\der \hat \sigma}{\der \that} 
= - \frac{2 \pi \alpha^2}{\shat^2}
\left(\frac{\shat}{\uhat}+ \frac{\uhat}{\shat}\right).
\label{eq:sub-compton}
\ee

\noindent
{\bf Polarized cross section}

From the antisymmetric part of the leptonic tensor,
\be
T^{\rm A}_{\alpha \beta}(l;l',k')&=&-\frac{4  i e^4}{\hat s \hat u}\,\varepsilon_{\alpha
\beta \rho \sigma} \, ( \, \hat s \,l^\rho+ \hat u\, l'^\rho \,)
   \,k^\sigma ~,
\label{lepoo}
\ee
with $k = l'+k'-l$, one gets
\begin{eqnarray}
{|\Delta \hat{M}|^2} & =& \frac{1}{2}\,(\epsilon_{\alpha}\epsilon^*_{\beta} 
-\epsilon^*_{\alpha}
\epsilon_{\beta})\,T^{\alpha\beta {\rm A}} = P^{\rm A}_{\alpha\beta}T^{\alpha\beta {\rm A}}\nonumber \\
& = & -2e^4 \bigg (\frac{\hat s}{\hat u} - \frac{\hat u}{\hat s} \bigg ),
\end{eqnarray}
where $P^{\rm A}_{\alpha\beta}$  is the antisymmetric part of the photon polarization
density matrix \eqref{eq:density}, and the cross section \eqref{eq:pol_sigma} 
for the 
process $e \gamma\rightarrow e\gamma$, where the incoming electron is 
longitudinally
polarized and the incoming  photon circularly polarized, reads
\begin{equation}
\frac{\der\Delta\hat\sigma}{\der\hat t}  
= -\frac{2\pi\alpha^2}{\hat s^2}\,
\bigg ( \frac{\hat s}{\hat u} - \frac{\hat u}{\hat s} \bigg ).
\label{eq:polcompt}
\end{equation}

\newpage

\clearemptydoublepage
\chapter{\bf{Kinematics of the QED Compton Scattering Process}}
\label{app:kinel1}

The kinematics of the QED Compton scattering
process in $ep\to e\gamma p$ and $ep \to e\gamma X$ 
is described in the center-of-mass frame of the outgoing
$e-\gamma$ system and in the laboratory frame, both for a collider and 
a fixed-target experiment. 

\section{Elastic Channel}
\label{apps:kinel1}

In the elastic channel the process reads
\be
e(l)+p(P) \rightarrow e(l')+\gamma(k')+ p(P'),
\ee
with $P^2 = P'^2 = m^2$, $k'^2 = 0$, and $l^2 = l'^2 \simeq 0$. \\

\noindent
{\bf {Electron-photon center-of-mass frame}}

In the $e-\gamma$ center-of-mass frame, we choose the $z$ axis to 
be along the direction of the virtual photon exchanged in the
reaction, see Figure \ref{fig:one}. The four-momenta of the 
initial electron and proton are given by:
\be
 l = (E_e^*,\,0,\,0,\,-E_e^*),~~~~~~~~~
 P =  (E_p^*,\,P_p^* {\sin{\theta_p^*}},\,0,\,P_p^* {\cos{\theta_p^*}}),
\ee
where $P_p^*=\sqrt{{E_p^*}^2-m^2}$. For the outgoing electron and photon
we have
\be
l' & = & E'^* (1,\,\sin \theta^* \cos \varphi^*,\,
\sin \theta^* \sin \varphi^*,\, \cos \theta^* ), \\
 k' & = & E'^* (1,\,-\sin \theta^* \cos 
\varphi^*,\,-\sin \theta^* \sin \varphi^*,\, -\cos \theta^*).
\ee
The four-momentum of the virtual photon is  
\be
 k= l'+k'-l = (E_k^*,\,0,\,0,\,E_e^*), 
\ee
with $k^2=t$ and 
\be
E_k^*=\sqrt{{E_e^*}^2+t}~.
\label{ek}
\ee
The overall momentum conservation allows us to write the four-momentum of the
final  proton as 
\be
P'=l+P-l'-k'.
\ee
We introduce the following Lorentz invariants 
\be
\hat s= (l'+k')^2= 4 {{E'}^*}^2,
\ee
\be
\hat t= (l-l')^2=-2 {E_e^*} {E'}^* (1+\cos \theta^*),
\ee
\be
\hat u= (l-k')^2=-2 {E_e^*} {E'}^* (1-\cos \theta^*),
\ee
\be
s= (l+P)^2= m^2 + 2 E_p^* E_e^*+2 E_e^* P_p^* \cos \theta_p^*,
\ee
\be
T= (P-l')^2=m^2-2 {E'}^*(E_p^*-P_p^* \sin \theta^* \sin \theta_p^*\cos 
\varphi^*-P_p^* 
\cos\theta^* \cos \theta_p^*).
\label{cmT}
\ee   
\be
{U}= (P-k')^2=m^2-2 {E'}^*(E_p^*+P_p^* \sin \theta^* \sin \theta_p^* \cos 
\varphi^*+P_p^*\cos\theta^* \cos \theta_p^*),
\label{cmU}
\ee
In addition they satisfy:
\be
\hat s+\hat t+\hat u=t, ~~~~~~s+T+U=-t+3 m^2.
\ee     
Using the relations above, it is possible to write the energies of the
particles in the laboratory frame in terms of the integration variables $\hat s$,
 $\hat t$ , $t$ and the constant $s$:
\be
E_e^*={\hat s-t\over 2 \sqrt {\hat s}}, ~~~~~E_k^*={\hat s+t\over 2 \sqrt 
{\hat s}},
\ee
and
\be
E_p^*={s-m^2+t\over 2 \sqrt {\hat s}}, ~~~~~P_p^*= {\sqrt{(s-m^2+t)^2-4 
\hat s m^2}\over 2 \sqrt {\hat s}},
 ~~~~~E'^* = {{\sqrt{\hat{s}}}\over{2}}.
\label{cmenergy}
\ee
Similarly, for the angles we have
\be
\cos \theta^* = { t- \hat s -2\hat t\over \hat s-t}~,
\label{cmangle1}
\ee
\be
\cos \theta_p^*={{2\hat s \,(s-m^2) -(\hat s -t)(s-m^2 + t)}
\over (s-t) [(s+t -m^2)^2 -4\hat s m^2]^{1\over 2}}~.
\label{cmangle2}
\ee
In particular  (\ref{cmT}) and (\ref{cmU}), through
 (\ref{cmenergy})-(\ref{cmangle2}), express $T$ 
and $U$ in terms of our integration variables  $\hat s$, $t$,
$\hat t$, $\varphi^*$ and we have used them to relate the laboratory frame 
variables to the integration ones, as shown in the following. \\

\noindent
{\bf{Laboratory  frame}}

We  choose the laboratory frame such that the initial
proton moves  along the $z$ axis and the outgoing electron has zero
azimuthal angle. The four-momenta of the particles are given by:
\be
l = (E_e,\,0,\,0,\,-E_e),~~~~~~~~ P = (E_p,\,0,\,0,\,P_p),
\label{hera_kin}
\ee
where $P_p =\sqrt{{E_p}^2-m^2}$, and
\be
l' & =  & E_e'(1,\,\sin \theta_e,\,0,\,\cos\theta_e),\nonumber \\
k' & =  & E_{\gamma}'(1,\,\sin \theta_{\gamma} 
\cos\phi_{\gamma},\,\sin \theta_{\gamma} \sin \phi_{\gamma},\,
\cos\theta_{\gamma}).
\label{eq:outgoing}
\ee
The Lorentz invariants are:
\be
\hat s=(l'+k')^2=2 E_e' E_{\gamma}' (1-\sin \theta_e \sin \theta_
{\gamma} \cos \phi_{\gamma}-\cos\theta_e \cos \theta_{\gamma}),
\label{shatl}
\ee
\be
\hat t= (l-l')^2=-2 E_e E_e' (1+ \cos \theta_e),
\ee
\be
\hat u=(l-k')^2=-2 E_e E_{\gamma}' (1+\cos \theta_{\gamma}),
\ee
\be
s=(l+P)^2=m^2+2 E_e (E_p+P_p),
\ee
\be
T=(P-l')^2=m^2-2 E_e' (E_p-P_p \cos \theta_e).
\ee
\be
U =(P-k')^2=m^2-2 E_{\gamma}' (E_p-P_p \cos \theta_{\gamma}),
\label{ul}
\ee
The polar angles in the laboratory frame can be written in terms of 
the invariants and the incident energies:
\be
\cos \theta_e={E_p ~\hat t -E_e ~(T-m^2)\over  P_p ~\hat t 
+E_e ~(T -m^2)}~,
\label{thetae}
\ee
\be
\cos \theta_{\gamma}={E_p~(t-\hat s-\hat t)-E_e~(U  -m^2)\over P_p~
(t-\hat s -\hat t)+E_e~(U -m^2)}~.
\ee
In the same way, for the energies of the final electron and photon we
have: 
\be
E_e'=-{~\hat t~ P_p +E_e ~(T -m^2)\over s-m^2}~,
\ee
\be
E_{\gamma}'={P_p~(\hat s-t+\hat t)-E_e~(U -m^2)\over s-m^2}~.
\ee
The azimuthal angle of the outgoing photon is: 
\be
\cos\phi_{\gamma}={2 {E_e'} {E_{\gamma}'} (1-\cos\theta_e\cos\theta_
{\gamma})-\hat s\over 2 
{E_e'} {E_{\gamma}'}
\sin \theta_e \sin \theta_{\gamma} }~,
\label{phig}
\ee
which is related to the acoplanarity angle by  $\phi=|\,\pi-
\phi_{\gamma}\,|$. Using  (\ref{cmT}) and (\ref{cmU}) together with 
(\ref{cmenergy})-(\ref{cmangle2}), the formulae above for $\cos \theta_e$, 
$\cos \theta_{\gamma}$, ${E_e'}$, ${E_{\gamma}'}$ and $\cos\phi_{\gamma}$ 
can be  expressed in terms of the 
invariants and the incident energies. 
Equations (\ref{thetae})-(\ref{phig}) are
needed to implement numerically the kinematical region under study, since 
they relate the laboratory variables (energies and angles) 
to the ones used for the integration. 

For a fixed-target experiment, \eqref{hera_kin} should be replaced by 
\be
l = (E_e,\,0,\,0,\,E_e),~~~~~~~~ P = (m,\,0,\,0,\,0),
\ee
assuming that now the $z$ axis is along the incoming electron direction.
Keeping \eqref{eq:outgoing} unchanged, \eqref{shatl}-\eqref{phig} are still 
valid, provided one performes the following replacements
\be 
E_p\to m,~~~~~~~~~\theta_{e,\gamma}\to  \pi -\theta_{e,\gamma}~.
\ee

\section{Inelastic Channel}
\label{app:kininel1}

For the inelastic channel of the process,
\be
e(l)+p(P) \rightarrow e(l')+\gamma(k')+ X(P_X),
\ee
most of the expressions remain the same as given in Appendix 
\ref{apps:kinel1}.\\

\noindent
{\bf Electron-photon center-of-mass frame}

The relations among the invariants are now:
\be
\hat s+\hat t+\hat u=-Q^2,~~~~~~~~~~~~ s+T + U=
3 m^2+{Q^2\over x_B}~,
\label{inv}
\ee
where $Q^2=-t$ and $x_B$ is given in \eqref{eq:bjorkenx} with 
$W^2 = (P+l-l'-k')^2$.
The only formulae which are different
from the elastic channel are the ones involving $E_p^*$ and
$\cos\theta_p^*$. So   (\ref{cmenergy}) will be replaced by   
\be
E_p^*={s-Q^2-W^2\over 2 \sqrt {\hat s}}~, ~~~~~P_p^*={\sqrt{(s-Q^2-W^2)^2 -
4 \hat s m^2}\over 2 \sqrt {\hat s}}~,
 ~~~~~E'^* = {{\sqrt{\hat{s}}}\over{2}}~,
\label{cmenergyin}
\ee
while for the angles:
\be
\cos \theta^* = -{ Q^2 + \hat s +2\hat t\over \hat s + Q^2}~,
\label{cmangle1in}
\ee
and
\be
{\cos\theta_p^*}= {2 \hat s\, (s-m^2)-(s-Q^2-W^2) (\hat s+Q^2)
\over (\hat s+Q^2)
{[(s-Q^2-W^2)^2-4 \hat s m^2]}^{1\over 2}}~.
\label{cmangle2in}
\ee
Equations (\ref{cmenergyin})-(\ref{cmangle2in}) reduce to 
(\ref{cmenergy})-(\ref{cmangle2}) of the elastic channel for $W = m$ 
and $Q^2 = -t$.\\

\noindent
{\bf{Laboratory frame}}

The definition of the  invariants for  the inelastic channel is the same as 
for the elastic one: (\ref{shatl})-(\ref{ul});
  (\ref{inv}) describes the relation among them.
The expressions of $\cos \theta_e$, $\cos \theta_{\gamma}$,  
$E_e'$, $E_{\gamma}'$ and $\cos \phi_{\gamma}$, in terms of the integration 
variables $W^2$, $\hat s$, $Q^2$, $\hat t$, $\varphi^*$  are given by  
(\ref{thetae})-(\ref{phig}) together with  (\ref{cmU}), (\ref{cmT}) 
as before, but now  (\ref{cmenergyin})-(\ref{cmangle2in}) will replace 
 (\ref{cmenergy})-(\ref{cmangle2}).

\clearemptydoublepage
\chapter{\bf{Matrix Elements for the Unpolarized QEDCS and VCS Processes}}

In this Appendix we give the explicit expressions of the matrix elements 
relative to QED Compton scattering (QEDCS) and virtual Compton scattering 
(VCS) in $ep\to e\gamma p$ and $ep \to e\gamma X$, with
incoming unpolarized electron and proton.

\section{Elastic Channel}
\label{sec:matrix_el}
The amplitude squared of the process
\be 
e(l) + p(P)\to e(l') +\gamma(k') + p(P')
\ee
 can be written as in \eqref{amplielasunp}, i.e.
\be
\overline {{\mid {M_{\mathrm{el}}}\mid }^2}= 
\overline {{\mid {M^{\rm{QEDCS}}_{\mathrm{el}}}\mid
}^2}
+\overline {{\mid {M^{\rm{VCS}}_{\mathrm{el}}}\mid }^2}- 2\, {\Re{\bf{\it e}}}\,\overline {M^{\rm{QEDCS}}_{\mathrm{el}}
M^{\rm{VCS} *}_{\mathrm{el}}}.
\ee
We will make use of the Lorentz invariants \eqref{comptonmandel}, 
\eqref{pap1_invar}, 
\be
s &= &(l+P)^2, ~~~~~t ~=~-Q^2 ~=~ k^2,~~~~~ \nonumber \\
\hat s & = & (l+k)^2,~~~~~~\hat t ~=~ (l-l')^2,~~~~~~~\hat u~ =~ (l-k')^2,
\label{mandel:app} 
\ee 
with $k = l'+k'-l$, and
\be
S = (P'+k')^2,~~~~~~~~~~  U = (P-k')^2,
\label{SU}
\ee
which satisfy the relation  $S  =  -(\hat s + \hat u + U - 2m^2)$. 
According to the approximation explained in Section \ref{sec:vcsel}, we have
\be 
\overline {{\mid {M^{\rm{QEDCS}}_{\mathrm{el}}}\mid }^2 } = 
\frac{1}{t^2}\,{\cal{L}}_{\alpha\beta}(P;P')T^{\alpha\beta}(l;l',k'),
\label{qedunpol_ampl}
\ee
$T_{\alpha\beta}$ being given by \eqref{lept}, and
\begin{eqnarray}
{\cal{L}}^{\alpha\beta}(P;P')  
 =  2 \,e^2F_1^2(t) \{ P^{\alpha}P'^{\beta} + P'^{\alpha}P^{\beta} -g^{\alpha\beta} (P\cdot P'-m^2)\},
\end{eqnarray}
with $F_1$ denoting the Dirac form factor of the proton. Furthermore,
\be 
\overline {{\mid {M^{\rm{VCS}}_{\mathrm{el}}}\mid }^2 } = 
\frac{1}{\hat{t}^2}\,L_{\alpha\beta}(l;l'){\cal{V}}^{\alpha\beta}(P;P',k'),
\label{vcsunpol_ampl}  
\ee
where 
$L_{\alpha\beta}$ 
is the leptonic tensor of the non-radiative deep  
inelastic scattering $ep\rightarrow eX$ introduced in \eqref{eq:leptonic} and  
\begin{eqnarray}
{\cal{V}}_{\alpha\beta}(P;P',k')
& =& e^4F_1^2(\hat {t}^2)\,
\bigg \{ \frac{1}{S'^2} \,{\rm{Tr}} \left [\,(P\!\!\!\!/+m)  
\gamma_{\alpha}(P'\!\!\!\!\!/+k'\!\!\!\!/+m)(P'\!\!\!\!\!/-2m) \,
(P'\!\!\!\!\!/+k'\!\!\!\!/+m) \,\gamma_{\beta} \right ]\nonumber \\
&& -~\frac{1}{2 U'^2}\,{\rm Tr} \left [ (P\!\!\!\!/+m)  \gamma_\mu (P\!\!\!\!/-k'\!\!\!\!/+m) \gamma_{\alpha}(P'\!\!\!\!\!/+m)\gamma_\beta (P\!\!\!\!/-k'\!\!\!\!/+m)\gamma^\mu \right ]\nonumber \\
&&~-~\frac{1}{2 S'U'}\,{\rm Tr}[ (P\!\!\!\!/+m)  \gamma_\mu (P\!\!\!\!/-k'\!\!\!\!/+m) \gamma_\alpha (P'\!\!\!\!\!/+m)
\gamma^\mu(P'\!\!\!\!\!/+k'\!\!\!\!/+m)\gamma_\beta]\nonumber \\
&&~~-~\frac{1}{2S'U'}\,{\rm Tr}[ (P\!\!\!\!/+m)  \gamma^\alpha
(P'\!\!\!\!\!/+k'\!\!\!\!/+m) \gamma_\mu (P'\!\!\!\!\!/+m)\gamma_{\beta}
(P\!\!\!\!/-k'\!\!\!\!/+m)
\gamma^\mu ]  \bigg \}.\nonumber \\
\label{eq:tensorel} 
\end{eqnarray}
Explicit calculation gives
\be 
\overline {{\mid {M^{\rm{QEDCS}}_{\mathrm{el}}}\mid }^2} = \frac {4e^6}{t ~\hat s~\hat u} ~\bigg [{A +  \frac{2 m^2}{t} B}\bigg ]
~F_1^2(t),
\label{qel_unp}
\ee
\be 
\overline {{\mid {M^{\rm{VCS}}_{\mathrm{el}}}\mid }^2 } =  
\frac{4e^6}{\hat t\, U'\, S'}
 \, \bigg [ A - \frac{2 m^2}{\hat t~ U'~ S'}\, C \bigg ] ~ F_1^2(\hat t),
\label{vel}
\ee
where 
\be
A&=& 2~t^2 - 2~t~(\hat s - 2~s' - U')  
+ \hat s^2 - 2~\hat s~s' \nonumber \\ && +\,\, 4~s'^2 + 2~s'~\hat u + \hat u^2
+ 4~s'~U' + 2~\hat u~U' + 2~U'^2,
\ee
\be
B & = & 2~ t^2 - 2 ~t ~(\hat s + \hat u) + \hat s^2 + \hat u^2,
\ee
\be
C&=& (\hat s + \hat u)^2~[t^2 + \hat s^2 - 2~t~(\hat s - s') - 2~\hat s~s'
\nonumber\\&&+ 
      2~s'^2 + 2~s'~\hat u + \hat u^2 - 2~m^2~( \hat s + \hat u -t)] 
   \nonumber\\&&+ 2~(\hat s + \hat u)~[t^2 -t~\hat s + \hat u~(-\hat s + 2~s' + \hat u)]~U' \nonumber\\&& + 2~[t^2 + \hat u^2 -t~(\hat s + \hat u)]~U'^2.
\label{eq:c}
\ee
In \eqref{eq:tensorel}-\eqref{eq:c} we have used the 
notations $s'= s-m^2$, $S' =  S-m^2$,  
$U' = U-m^2$ for compactness and the electron mass has been neglected.
 
The interference term between QEDCS and VCS  is given by
\be
2\, {\Re\it{e}} \,\overline {M^{\rm{QEDCS}}_{\mathrm{el}}M^{\rm{VCS} *}_{\mathrm{el}}}= -\frac{1}{2t\hat t} \,{\cal{I}}^{\rho}_{\alpha\beta}(l;l',k')
{\cal{M}}^{\alpha\beta}_{\rho}(P;P',k'),
\ee
with
\be
{\cal{I}}^{\rho}_{\alpha\beta}(l;l',k') = e^3\bigg \{ \frac{1}{\hat s}\,
{\rm{Tr}}[\,l'\!\!\!/\gamma^{\rho}
(l'\!\!\!/+k'\!\!\!\!/)\gamma_{\alpha}l\!\!/\gamma_{\beta}\,]
+\frac{1}{\hat u}\,{\rm{Tr}}[\,l'\!\!\!/\gamma_{\alpha}(l\!\!\!/-k'\!\!\!\!/)
\gamma^{\rho}l\!\!/\gamma_{\beta}\,] \bigg \}
\label{eq:elint}
\ee
and 
\be
{\cal{M}}^\rho_{\alpha\beta}(P;P',k') &=& e^3F_1(t)F_1(\hat t)
\bigg \{ \frac{1}{S'}\,{\rm Tr}[\,(P'\!\!\!\!\!/+m) \gamma_\alpha\,(P\!\!\!\!/+m) \gamma^\rho(P'\!\!\!\!\!/+k'\!\!\!\!/+m)
\gamma_{\beta}\,] \nonumber \\
& &~~~~+ \frac{1}{U'}\,{\rm Tr}[\,(P'\!\!\!\!\!/+m) \gamma_\alpha\,(P\!\!\!\!/+m) \gamma^\rho(P\!\!\!\!/-k'\!\!\!\!/+m)\gamma_{\beta}
 \,] \bigg \}.
\ee
The final result can be written as
\be
2\, {\Re\it{e}} \,\overline {M^{\rm{QEDCS}}_{\mathrm{el}}M^{\rm{VCS} *}_{\mathrm{el}}}=4e^6\,\frac{D +2m^2E}
{ t~\hat s~\hat u~\hat t~U'~ S'}~F_1(\hat t) F_1(t),
\ee
with
\be
D&=&\{ (\hat s + \hat u)~[t~\hat u + s'~(\hat s + \hat u)] + 
   [\hat s~(\hat s + \hat u) -t~(\hat s - \hat u)]~U'\}~ \nonumber \\&&~~~
[2~t^2 
+ \hat s^2 -2~\hat s~s' +  
   4~s'^2 + 2~s'~\hat u + \hat u^2 - 2~t~(\hat s - 2~s' - U')\nonumber\\ && 
~~~~~~~~~~~~+ 4~s'~U' + 2~\hat u~U' + 2~U'^2],
\ee
\be
E&=& -s'~\hat u^3 - \hat s^3~(s' - 2~\hat u + U') - 
   \hat s^2~\hat u~(7~s' + 2~U') -   \hat s~\hat u^2~(7~s' + 2~\hat u + 5~U')
  \nonumber \\ && ~~~+ 
   2~t^2~[\hat s~(\hat u - U') + \hat u~(\hat u + U')] - 
   t~ (\hat s + \hat u)~[ \hat s~(-2~s' + 3~\hat u - 3~U') \nonumber \\ 
&&~~~~~~~+ \hat u~(-2~s' + \hat u + U')].
\ee

\section{Inelastic Channel}
\label{sec:matrix_inel}

The amplitude squared of the electron-quark scattering 
process \eqref{eqsub_4},
\be
e(l)+q(p) \rightarrow e(l')+\gamma(k')+q(p'),
\label{quarks}
\ee
which is needed, according to \eqref{insig_4},  for the calculation of the 
inelastic reaction 
\be
e(l)+ p(P)\to e(l') + \gamma(k') +  X(P_X)
\ee
is given by
\be 
\overline {{\mid {\hat{M}^q\mid} }^2}= \overline {{\mid {\hat{M}^{q 
\,{\rm QEDCS}}_{\mathrm{ }}}\mid}^2}
+\overline {{\mid {\hat{M}^{q \,{\rm VCS}}_{\mathrm{ }}}\mid }^2}  
-2\, {\Re{\it e}} \overline {\hat{M}^{q \,{\rm QEDCS}}_{\mathrm{ }}
\hat{M}^{q \,{\rm VCS} *}_{\mathrm{ }}}.
\ee
Within the framework of the parton model, we assume 
that the initial quark is collinear with the parent proton, $p = x_B P$.
If we define 
\be
\hat S = (p'+k')^2,~~~~~~\hat U = (p-k')^2,
\label{SUh}
\ee
then $\hat U  = x_B U'$,  $\hat S = -(\hat s + \hat u + x_B\, U')$, 
with $U' = U-m^2$; $U$ and the other invariants useful to describe
the process being given in \eqref{mandel:app}  and \eqref{SU}.
Neglecting the masses of all the interacting particles in \eqref{quarks}, 
the QED Compton scattering amplitude squared can be written as
\be
\overline{{\mid {\hat{M}^{q\,\rm{QEDCS}}}\mid }^2} = 
\frac{1}{Q^4}\, \,
w_{\alpha\beta}(p;p')T^{\alpha\beta}(l;l',k'), 
\label{eq:qedcs_sq}
\ee 
where $T_{\alpha\beta} $ is given in \eqref{lept} 
and  
$w_{\alpha\beta}$ is the quark tensor,
\be
w_{\alpha\beta}(p;p') = 2 \,e^2\,e^2_q\, ( p^{\alpha}p'^{\beta} + p'^{\alpha}p^{\beta} -g^{\alpha\beta}p\cdot p'),
\ee
with $e_q$ being the charge of the quark in units of the proton charge $e$.
Analogously,
for the VCS amplitude squared one has
\be
\overline{{\mid {\hat{M}^{q\,\rm{VCS}}}\mid }^2} = 
\frac{1}{\hat t^2}\, T^q_{\alpha\beta}(p;p',k') L^{\alpha\beta}(l;l'),
\label{eq:vcs_sq}
\ee 
where 
\be
T^q_{\alpha \beta}(p;p',k') =\,\frac{4 e^4e^4_q} {\hat{S}\hat{U}}\,\bigg
\{ \,\frac{1}{2}\,
g_{\alpha\beta}\,(\hat S^2 + \hat U^2 + 2 \hat t t) + 2\hat S\, p_{\alpha}p_{\beta}
+ 2 \hat U\,p_{\alpha}'p_{\beta}'~~~~~~~~~~~~~~~~~~\nonumber \\
~~~+\,(\hat t+t)(p_{\alpha}p_{\beta}'+p_{\beta} p_{\alpha}') 
-({\hat S}-\hat t)\,(p_{\alpha}k_{\beta}'+p_{\beta}k_{\alpha}')
~~~~~~~~\nonumber \\ 
+\,(\hat U -\hat t) \, (p_{\alpha}'k_{\beta}' +p_{\beta}'p_{\alpha}')\bigg\}~~
\ee
and  
$L_{\alpha\beta}$ is again the leptonic tensor  \eqref{eq:leptonic}.
The explicit form of \eqref{eq:qedcs_sq}  and \eqref{eq:vcs_sq}   is 
\be
\overline {{\mid {\hat M^{q\,{\rm QEDCS}}_{\mathrm{inel}}}\mid }^2}&=& 
-4\, e^6e_q^2 \,\frac{F}{ Q^2\, \hat s \,\hat u}~,
\label{qin}
\ee
\be
\overline {{\mid {\hat M^{q\,{\rm VCS}}_{\mathrm{inel}}}\mid }^2}&=&4 
\,e^6e_q^4\,\frac{F} {\hat t ~\hat U~\hat S}~ ,
\label{vin}
\ee 
with
\be
F &=&  \hat s^2 + \hat u^2 + 2 \,\{ Q^4+ Q^2~[\hat s - (2~s' +
U')~x_B] + x_B\, (s'~\hat u + \hat u~U'\nonumber\\&&- \hat s~s') 
+ x_B^2\,(2~s'^2 + 2~s'~U' + U'^2)\}.
\ee
 
The interference term reads
\be
2\, {\Re{\it e}} \overline {\hat{M}^{q \,{\rm QEDCS}}
\hat{M}^{q \,{\rm VCS} *}} = \frac{1}{2Q^2\hat t}\, 
{\cal{I}}^{\rho}_{\alpha\beta}(l; l', k')
{\cal{T}}^{\alpha\beta}_{\rho}(p; p', k'),
\label{eq:int_sq} 
\ee
${\cal{I}}^{\rho}_{\alpha\beta}$ being the same as in  \eqref{eq:elint}
and
\be
{\cal{T}}^{\alpha\beta}_{\rho} = e^3e^2_q\bigg \{ \frac{1}{\hat S}\,
{\rm{Tr}}[\,p'\!\!\!\!/\gamma^{\alpha}p\!\!\!/ \gamma^{\beta}
(p'\!\!\!\!/+k'\!\!\!\!/)\gamma_{\rho}\,]
+\frac{1}{\hat U}\,{\rm{Tr}}[\,p'\!\!\!\!/\gamma^{\alpha}p\!\!\!
/\gamma_{\rho}(p\!\!\!/-k'\!\!\!\!/)\gamma^{\beta})] \bigg \}.
\label{eq:tensorint}
\ee
Therefore we have, from  \eqref{eq:elint},
\eqref{eq:int_sq} and \eqref{eq:tensorint},
\be
2 \,{\Re\it{e}}\,\overline {\hat M^{q\,{\rm QEDCS}}_{\mathrm{inel}}
\hat M^{q\, {\rm VCS} *}_{\mathrm{inel}}} &=&- 4\,e^6 e_q^3\,\frac{G}
{Q^2~\hat s~\hat u~\hat t~\hat U~\hat S}
\ee
with 
\be
G & =& \{-Q^2~\hat u~(\hat s + \hat u) 
+ Q^2~(\hat s - \hat u)~U'~x_B + 
   (\hat s + \hat u)~[s'~\hat u \nonumber\\&&+ \hat s~(s' + U')]~x_B\}~\{2~Q^4 + \hat s^2 + \hat u^2 - 
   2~\hat s~s'~x_B + 2~Q^2~[\hat s - (2~s' \nonumber\\&&+ U')~x_B]+ 
   2~x_B~[\hat u~(s' + U') + (2~s'^2 + 2~s'~U' + U'^2)~x_B]\}.
\ee
The analytic form of the interference term agrees with \cite{brodsky} 
but differs from \cite{metz}
in the massless case slightly, in particular in   (15) of \cite{metz}, $8$
in the first line should be replaced by $4$ and $(-8)$ in the sixth line
should be replaced by $(-16)$. However we have checked that this does not 
affect our numerical results for HERA kinematics.

\clearemptydoublepage
\chapter{\bf{Matrix Elements for the Polarized QEDCS and VCS Processes}}

In this Appendix we give the explicit expressions of the matrix elements 
relative to QED Compton scattering (QEDCS) and virtual Compton scattering 
(VCS) in $ep\to e\gamma p$ and $ep \to e\gamma X$, with initial
longitudinally polarized electron and proton.

\section{Elastic Channel}

According to \eqref{eq:polelastic_vcs},
\be
{{\mid {\Delta M_{\mathrm{el}}}\mid }^2}= {{\mid
{\Delta M^{\rm {QEDCS}}_{\mathrm{el}}}\mid
}^2}+{{\mid {\Delta M^{{\rm VCS}}_{\mathrm{el}}}\mid }^2} - 2\, 
{\Re{\bf{\it e}}}\,
 {\Delta M^{{\rm QEDCS}}_{\mathrm{el}}
\Delta M^{{\rm VCS} *}_{\mathrm{el}}}
\ee
is the matrix element squared of the process
\be
\vec{e}(l) + \vec{p}(P)\to e(l') +\gamma(k') + p(P'),
\ee 
where the incoming electron and proton are longitudinally polarized. 
In terms of the Lorentz invariants \eqref{mandel:app} and \eqref{SU},   
the spin dependent counterparts of \eqref{qedunpol_ampl} and
\eqref{vcsunpol_ampl}  read
\be 
\overline {{\mid {M^{\rm{QEDCS}}_{\mathrm{el}}}\mid }^2 } = 
\frac{1}{t^2}\,{\cal{L}}^{\rm A}_{\alpha\beta}(P;P')
T^{\alpha\beta {\rm A}}(l;l',k'),
\label{elpolqedcs}
\ee
\be 
\overline {{\mid {M^{\rm{VCS}}_{\mathrm{el}}}\mid }^2 } = 
\frac{1}{\hat{t}^2}\,L^{\rm A}_{\alpha\beta}(l;l')
{\cal{V}}^{\alpha\beta {\rm A}}(P;P',k'),
\ee
$L^{\rm A}_{\alpha\beta}$ and $T^{\rm A}_{\alpha\beta}$  being given by 
\eqref{eq:antilept} and \eqref{lep_pol} respectively, while
\be
{\cal{L}}^A_{\alpha\beta}(P, S;P') = 
2ie^2F_1^2(t)m\varepsilon_{\alpha\beta\rho\sigma}S^{\rho}
(P-P')^{\sigma}~,
\ee
where $S$ is the spin four-vector of the proton, which satisfies
the conditions $S^2 = -1$ and $S\cdot P = 0$.
The tensor
${\cal{V}}^{\rm A}_{\alpha\beta}$ can be calculated from \eqref{eq:tensorel},
inserting the polarization operator of the proton $\gamma^5 S\!\!\!\!/$ 
next to the term $(P\!\!\!\!/+m)$ inside each trace. In \eqref{elpolqedcs}
$F_1$ is the Dirac form factor of the proton. 
Finally one obtains
\be
{{\mid {\Delta M^{{ \rm QEDCS}}_{\mathrm{el}}}\mid }^2} &=&   \frac{4}{t \,\hat{s}\, \hat u}\, \bigg [ -\Delta A + \frac{2\,m^2}{t\, s'} \,\Delta B \bigg ]
\, F_1^2(t), 
\ee
\be
{{\mid {\Delta M^{ {\rm VCS}}_{\mathrm{el}}}\mid }^2} &=& -\frac{4}{\hat t \,U'\, {S}'}\, \bigg [\Delta A + \frac{2\,m^2}{{S}'\, s' U' } \,\Delta C \bigg ]\, F_1^2(\hat t), 
\ee
with
\be
\Delta A  =  2 t^2 + (\hat s - 2 s' - \hat u) (\hat s + \hat u) - 
2 t (\hat s - 
2 s' - U') - 2 \hat u U',
\ee
\be
\Delta B =  - 2 t^3  + \hat{s}^3 - \hat s \hat {u}^2 + 2 t^2 
(2 \hat s + \hat u) - t (3 \hat {s}^2 + \hat{u}^2),
\ee
\be
\Delta C  & = & (\hat s + \hat u)^2 \,[-2 s'^2 + \hat s \hat u - 2 s' \hat u - \hat{u}^2 
+ 2 m^2 ( \hat {s} + \hat {u}- t) - 
     t (s' + \hat{u})]  \nonumber \\ & & \,\,- (\hat{s} + \hat{u}) \,[2 t^2 - 3 t \hat{s} + \hat{s}^2 + \hat{s} (s' - 2 \hat u) + 
     3 \hat u (s' + \hat u)]\, U'\nonumber \\ &&\,\,\,\,- \,[2 t^2 + s^2 - \hat s \hat u + 2 \hat{u}^2 - t (3 \hat s + \hat u)] \,U'^2.
\ee
We have  used the notations $s'=s-m^2, ~U'=U-m^2,~ S'=S-m^2$  
for compactness.

The interference term between QEDCS and VCS is given by
\be
2\, {\Re\it{e}} \,{\Delta M^{\rm{QEDCS}}_{\mathrm{el}}
\Delta M^{\rm{VCS} *}_{\mathrm{el}}}= -\frac{1}{2t\hat t} 
\,{\cal{I}}^{\rho {\rm A}}_{\alpha\beta}(l;l',k')
{\cal{M}}^{\alpha\beta {\rm  A}}_{\rho}(P;P',k'),
\ee
with
\be
{\cal{I}}^{\rho {\rm A}}_{\alpha\beta}(l;l',k') 
= e^3\bigg \{ \frac{1}{\hat s}\,
{\rm{Tr}}[\,\gamma^5l'\!\!\!/\gamma^{\rho}
(l'\!\!\!/+k'\!\!\!\!/)\gamma_{\alpha}l\!\!/\gamma_{\beta}\,]
+\frac{1}{\hat u}\,{\rm{Tr}}[\,\gamma^5
l'\!\!\!/\gamma_{\alpha}(l\!\!\!/-k'\!\!\!\!/)
\gamma^{\rho}l\!\!/\gamma_{\beta}\,] \bigg \}
\label{eq:ia}
\ee
and 
\be
{\cal{M}}^{\rho {\rm A}}_{\alpha\beta}(P;P',k') &=& e^3F_1(t)F_1(\hat t)
\bigg \{ \frac{1}{S'}\,{\rm Tr}[\,
\gamma^5 S\!\!\!\!/
\,(P\!\!\!\!/+m) \gamma^\rho(P'\!\!\!\!\!/+k'\!\!\!\!/+m)
\gamma_{\beta}(P'\!\!\!\!\!/+m) \gamma_\alpha\,] \nonumber \\
& &~~~~+ \frac{1}{U'}\,{\rm Tr}[\,
\gamma^5 S\!\!\!\!/\,(P\!\!\!\!/+m) \gamma^\rho(P\!\!\!\!/-k'\!\!\!\!/+m)\gamma_{\beta}(P'\!\!\!\!\!/+m) \gamma_\alpha
 \,] \bigg \}.
\ee
Explicit calculation gives
\be
 2\, {\Re{\bf{\it e}}}\,
 {\Delta M^{{\rm QEDCS}}_{\mathrm{el}}\Delta M^{{\rm VCS} *}_{\mathrm{el}}} &=& -\frac{4e^6}{ t\, \hat s \, \hat u\, \hat t\, U'\, {S}'}   \,\bigg [ {\Delta D + \frac{2 m^2 \Delta E}{s'}} \bigg ]\,F_1(\hat t) \,F_1(t),  
\label{qel}
\ee
where $\Delta D$ and $\Delta E$ read:
\be
\Delta D & = & [ 2 t^2 + (\hat s - 2 s' - \hat u) (\hat s + \hat u) - 2 t (\hat s - 2 s' - U') -2 \hat u U' ]
 \nonumber \\
 & & \,\,\times  \{ (\hat s + \hat u) [t \hat u + s' (\hat s + \hat u)] + 
 [t  (\hat s - \hat u) + \hat s (\hat s + \hat u)] U'\},
\ee
\be
\Delta E & = & [\hat s ( \hat s - t)^2 (\hat s - 2 t) + ( 2 t ^3  - t^2  \hat s - \hat{s}^3) \hat u + (-2 t^2  - 3 t \hat{s} + \hat{s}^2) \hat{u}^2 + ( t +  3 \hat s) 
\hat{u}^3]\, U'\nonumber \\ 
 & & - (\hat s + \hat u) \{-2 t^3 \hat u - (\hat s + \hat u) 
[\hat{s}^2 (s' - 2 \hat{u}) + s' \hat{u}^2 + 2 \hat{s} \hat{u} (s' + \hat u)] \nonumber \\  &&
 \,\,- t [-7 \hat{s} s' \hat{u} + \hat{u}^2 (-2 s' + \hat u)  + 
        \hat{s}^2 (-3 s' + 5 \hat u)] + t^2 [2 \hat{u} ( \hat u -s') + \hat s 
( 5 \hat u - 2 s')] \}.\nonumber \\
\ee

\section{Inelastic Channel}

For the corresponding inelastic channel the  matrix elements
relative to the electron-quark scattering process
\be
\vec{e}(l)+\vec{q}(p) \rightarrow e(l')+\gamma(k')+q(p')
\ee
read:
\be
{{\mid \Delta{\hat{M}^{q\,\rm{QEDCS}}}\mid }^2} = 
\frac{1}{Q^4}\, \,
w^{\rm A}_{\alpha\beta}(p;p')T^{\alpha\beta {\rm A}}(l; l', k')
\label{eq:qedcs_sq_pol}
\ee 
and
\be
{{\mid {\hat{M}^{q\,\rm{VCS}}}\mid }^2} = 
\frac{1}{\hat t^2}\, T^{q \rm A}_{\alpha\beta}(p; p', k') 
L^{\alpha\beta \rm A}(l;l'),
\label{eq:vcs_sq_pol}
\ee 
where $T^{\rm A}_{\alpha\beta} $ and  $L^{\rm A}_{\alpha\beta}$ 
are given in  \eqref{lep_pol} and in \eqref{eq:antilept}
respectively.  
Furthermore, $w^{{\rm A}}_{\alpha\beta}$ is the antisymmetric part of 
the quark tensor,
\be
w^{\rm A}_{\alpha\beta}(p;p') = -2 i e^2 e^2_q\, 
\varepsilon_{\alpha\beta\rho\sigma}p^{\rho}p'^{\sigma},
\ee
$e_q$ being the charge of the quark in units of the proton charge $e$, and
\be
T^{q \rm A}_{\alpha\beta}(p;p',k')= -{4  i e^4 e^4_q\over \hat S \hat U}\, 
\varepsilon_{\alpha
\beta \rho \sigma}  \Big [ (\hat S-\hat t) p^\rho +(\hat U- \hat t) p'^\rho
\Big ]\, (p-p'){^\sigma}.
\ee
The final results can be written as
\be
{{\mid
{\Delta \hat M^{q\,{\rm QEDCS}}_{\mathrm{inel}}}\mid}^2}  &=&  4 \,e^6e_q^2 
\,\frac{\Delta F}
{Q^2\,\hat s \,\hat u}~,
\ee
and
\be
{{\mid {\Delta \hat M^{q\,{\rm VCS}}_{\mathrm{inel}}}\mid }^2} & = & 
- 4 \,e^6e_q^4 \,\frac{\Delta F}{\hat t\,\hat S \,\hat U}~,
\ee
where, in terms of the invariants   defined in \eqref{mandel:app}-\eqref{SUh}, 
\be 
\Delta F & = & \hat{s}^2 - \hat{u}^2 + 2 Q^4 + 2 Q^2 \hat{s} - 
  2 x_B [\hat{s} s' + \hat{u} (s' + U') + Q^2 (2 s' + U')].
\ee

The spin dependent counterpart of \eqref{eq:tensorint} reads 
\be
2\, {\Re{\it e}}  {\hat{M}^{q \,{\rm QEDCS}}
\hat{M}^{q \,{\rm VCS} *}} = \frac{1}{2Q^2\hat t}\, 
{\cal{I}}^{\rho \rm A}_{\alpha\beta}{\cal{T}}^{\alpha\beta \rm A}_{\rho},
\label{eq:int_sq_pol} 
\ee
with  ${\cal{I}}^{\rho \rm A}_{\alpha\beta}$ being the same as
 in \eqref{eq:ia} and 
\be
{\cal{T}}^{\alpha\beta \rm A}_{\rho} = e^3e^2_q\bigg \{ \frac{1}{\hat S}\,
{\rm{Tr}}[\,\gamma^5\,p'\!\!\!\!/\gamma^{\alpha}p\!\!\!/ \gamma^{\beta}
(p'\!\!\!\!/+k'\!\!\!\!/)\gamma_{\rho}\,]
+\frac{1}{\hat U}\,{\rm{Tr}}[\,\gamma^5\,p'\!\!\!\!/\gamma^{\alpha}p\!\!\!
/\gamma_{\rho}(p\!\!\!/-k'\!\!\!\!/)\gamma^{\beta})] \bigg \}.
\label{eq:tensorint_pol}
\ee
The final result is given by
\be
2\, {\Re{\it e}} 
{\Delta \hat M^{q\,{\rm QEDCS}}_{\mathrm{inel}}
\Delta \hat M^{q\,{\rm VCS} *}_{\mathrm{inel}}} &=&  
\,-4\,e^6 e_q^3\, \frac{\Delta G\, \Delta H}{Q^2\,\hat s\, \hat u 
\,\hat t \,\hat S\, \hat U},
\ee
with
\be
\Delta G & = & 2 Q^4 + \hat{s}^2 - 2 \hat s s' x_B - \hat u [\hat u + 2 (s' + U') 
x_B] +    2 Q^2 [\hat s - (2 s' + U') x_B],
\ee
\be
\Delta H = Q^2 [\hat s (\hat u - \hat U) + \hat u (\hat u + \hat U)]   
- x_B\,(\hat s + \hat u)[s' \hat u + \hat s (s' + U')].
\ee 

\end{appendix}
\clearemptydoublepage




\clearemptydoublepage

\chapter*{{\bf Acknowledgements}}

It is a great pleasure to thank Prof. Dr. E. Reya and Prof. Dr. M. Gl\"uck
for suggesting this  interesting research topic to me and for their
continuous and very instructive scientific guidance.

I am  particularly grateful to Dr. A. Mukherjee and Dr. I. Schienbein
for the fruitful collaboration on this project.
I also warmly acknowledge Dr. V. Lendermann and Dr. W. Vogelsang
for helpful discussions and comments.

I would like to express my gratitude to Prof. Dr. M. Anselmino for 
his friendly and illuminating advices.

Finally, my thanks go to all the members of TIV and  TIII and to  our
secretary Mrs. S. Laurent for a pleasant working atmosphere.

\clearemptydoublepage

\end{document}